\documentclass[aps,rmp,reprint,floatfix]{revtex4-2}

\usepackage[english]{babel}
\usepackage[utf8x]{inputenc}
\usepackage[T1]{fontenc}
\usepackage{hepnames}
\usepackage{heppennames}

\usepackage[a4paper,top=3cm,bottom=2cm,left=3cm,right=3cm,marginparwidth=1.75cm]{geometry}

\usepackage{amsmath}
\usepackage{graphicx}
\usepackage[colorinlistoftodos]{todonotes}
\usepackage[colorlinks=true, allcolors=blue]{hyperref}
\usepackage{subfiles}
\usepackage{xspace}
\usepackage{relsize}

\DeclareRobustCommand{\plusplus}{\raisebox{0.2ex}{\smaller ++}}
\newcommand{\MCatNLO}{M\protect\scalebox{0.8}{C}@N\protect\scalebox{0.8}{LO}\xspace}

\newcommand{\POWHEG}{P\protect\scalebox{0.8}{OWHEG}\xspace}
\newcommand{\Powheg}{\POWHEG}

\newcommand{\LOPS}{L\scalebox{0.8}{O}P\scalebox{0.8}{S}\xspace}
\newcommand{\NLOPS}{N\scalebox{0.8}{LO}P\scalebox{0.8}{S}\xspace}
\newcommand{\NNLOPS}{NN\scalebox{0.8}{LO}P\scalebox{0.8}{S}\xspace}
\newcommand{\UNLOPS}{UN\scalebox{0.8}{LO}P\scalebox{0.8}{S}\xspace}
\newcommand{\UNNLOPS}{UN\ensuremath{{}^2}\scalebox{0.8}{LO}P\scalebox{0.8}{S}\xspace}
\newcommand{\MENLOPS}{ME\protect\scalebox{0.8}{NLO}PS\xspace}
\newcommand{\MEPSatNLO}{M\protect\scalebox{0.8}{E}P\protect\scalebox{0.8}{S}@N\protect\scalebox{0.8}{LO}\xspace}

\newcommand{\MINLO}{MIN\protect\scalebox{0.8}{LO}\xspace}
\newcommand{\FxFx}{F\protect\scalebox{0.8}{X}F\protect\scalebox{0.8}{X}\xspace}
\newcommand{\Alpgen}{A\protect\scalebox{0.8}{LPGEN}\xspace}
\newcommand{\Madgraph}{M\protect\scalebox{0.8}{AD}G\protect\scalebox{0.8}{RAPH}\xspace}

\newcommand{\MGaMC}{M\protect\scalebox{0.8}{AD}G\protect\scalebox{0.8}{RAPH5}\_aM\protect\scalebox{0.8}{C}@N\protect\scalebox{0.8}{LO}\xspace}

\newcommand{\Herwig}{H\protect\scalebox{0.8}{ERWIG}\xspace}

\newcommand{\Herwigpp}{\Herwig{}\plusplus\xspace}

\newcommand{\Pythia}{P\protect\scalebox{0.8}{YTHIA}\xspace}
\newcommand{\PythiaSix}{\Pythia{}6\xspace}
\newcommand{\PythiaEight}{\Pythia{}8\xspace}

\newcommand{\CKKW}{C\protect\scalebox{0.8}{KKW}\xspace}
\newcommand{\MLM}{M\protect\scalebox{0.8}{LM}\xspace}
\newcommand{\Geneva}{G\protect\scalebox{0.8}{ENEVA}\xspace}
\newcommand{\aMCatNLO}{aM\protect\scalebox{0.8}{C}@N\protect\scalebox{0.8}{LO}\xspace}
\newcommand{\aMCatLO}{aM\protect\scalebox{0.8}{C}@L\protect\scalebox{0.8}{O}\xspace}

\newcommand{\BlackHat}{B\protect\scalebox{0.8}{LACK}H\protect\scalebox{0.8}{AT}\xspace}

\newcommand{\OpenLoops}{O\protect\scalebox{0.8}{PEN}L\protect\scalebox{0.8}{OOPS}\xspace}

\newcommand{\LOOPSIM}{L\protect\scalebox{0.8}{OOP}S\protect\scalebox{0.8}{IM}\xspace}

\newcommand{\MCFM}{M\protect\scalebox{0.8}{C}F\protect\scalebox{0.8}{M}\xspace}

\newcommand{\Jetphox}{J\protect\scalebox{0.8}{ET}P\protect\scalebox{0.8}{HOX}\xspace}

\newcommand{\Vbfnlo}{V\protect\scalebox{0.8}{BF}N\protect\scalebox{0.8}{LO}\xspace}
\newcommand{\Sherpa}{S\protect\scalebox{0.8}{HERPA}\xspace}

\newcommand{\HEJ}{H\protect\scalebox{0.8}{EJ}\xspace}

\newcommand{\Rivet}{R\protect\scalebox{0.8}{IVET}\xspace}

\newcommand{\HepData}{H\protect\scalebox{0.8}{EP}D\protect\scalebox{0.8}{ATA}\xspace}

\newcommand{\Photos}{P\protect\scalebox{0.8}{HOTOS}\xspace}

\newcommand{\FEWZ}{F\protect\scalebox{0.8}{EWZ}\xspace}

\newcommand\tevatron{T\protect\scalebox{0.8}{EVATRON}\xspace}
\newcommand\Tevatron{\protect\tevatron}

\newcommand\cdf{CDF\xspace}
\newcommand\CDF{\cdf}
\newcommand\dzero{D\O\xspace}
\newcommand{\DO}{\dzero}

\newcommand\LHC{L\protect\scalebox{0.8}{HC}\xspace}
\newcommand\ATLAS{\atlas}
\newcommand\atlas{A\protect\scalebox{0.8}{TLAS}\xspace}
\newcommand\CMS{\cms}
\newcommand\cms{C\protect\scalebox{0.8}{MS}\xspace}

\newcommand\LHCb{\lhcb}
\newcommand\lhcb{L\scalebox{0.8}{HC}b\xspace}

\newcommand{\done}{\ensuremath{\mathrm{d}}}

\newcommand{\order}{\ensuremath{\mathcal{O}}}

\newcommand{\nnb}{\nonumber}

\newcommand{\ppbar}{\ensuremath{p\bar{p}}\xspace}
\newcommand{\pp}{\ensuremath{pp}\xspace}

\newcommand{\W}{\ensuremath{W}\xspace}
\newcommand{\Z}{\ensuremath{Z}\xspace}

\newcommand{\Zjets}{\ensuremath{Z+\text{jets}}\xspace}
\newcommand{\Zjet}{\ensuremath{Z+\text{jet}}\xspace}
\newcommand{\Vjets}{\ensuremath{V+\text{jets}}\xspace}
\newcommand{\Vjet}{\ensuremath{V+\text{jet}}\xspace}
\newcommand{\Wjets}{\ensuremath{W+\text{jets}}\xspace}
\newcommand{\Wjet}{\ensuremath{W+\text{jet}}\xspace}
\newcommand{\WZjets}{\ensuremath{W/Z+\text{jets}}\xspace}
\newcommand{\WZjet}{\ensuremath{W/Z+\text{jet}}\xspace}
\newcommand{\gammajet}{\ensuremath{\gamma+\text{jet}}\xspace}
\newcommand{\gammajets}{\ensuremath{\gamma+\text{jets}}\xspace}

\newcommand{\kT}{\ensuremath{k_\mathrm{T}}\xspace}

\newcommand{\pT}{\ensuremath{p_\mathrm{T}}\xspace}

\newcommand{\qT}{\ensuremath{q_\mathrm{T}}\xspace}

\newcommand{\HT}{\ensuremath{H_\mathrm{T}}\xspace}

\newcommand{\ET}{\ensuremath{E_\mathrm{T}}\xspace}

\newcommand{\alphaS}{\ensuremath{\alpha_\text{s}}\xspace}

\newcommand{\nf}{\ensuremath{n_f}}

\newcommand{\Qcut}{\ensuremath{Q_\text{cut}}}

\newcommand{\PAp}{\ensuremath{\bar{\Pp}}}
\newcommand{\PQb}{\ensuremath{\text{b}}}

\begin{document} 

\title{Vector Bosons and Jets in Proton Collisions}
\author{Paolo Azzurri}
\affiliation{INFN Pisa, Largo B. Pontecorvo  3, 56127 Pisa, Italy}
\author{Marek Sch{\"o}nherr}
\affiliation{Institute for Particle Physics Phenomenology, Durham University, Durham, DH1 3LE, United Kingdom} 
\author{Alessandro Tricoli}
\affiliation{Brookhaven National Laboratory, Upton, New York 11973, U.S.A.} 
\date{\today} 

\begin{abstract}
Events with vector bosons produced in association with jets have been extensively studied at hadron colliders and provide high-accuracy tests of the Standard Model. A good understanding of these processes is of paramount importance for precision Higgs physics, as well as for searches for new physics. In particular, associated production of $\gamma$, $W$ or $Z$ bosons with light-flavor and heavy-flavor jets is a powerful tool for testing perturbative QCD calculations, Monte Carlo event generators, and can also constrain the parametrizations used to describe the parton content of the proton. Furthermore, events with a $W$ or $Z$ boson produced with two well-separated jets can be used to distinguish between electroweak and strong production mechanisms, and to set limits on contributions of physics beyond the Standard Model. This review summarises the historical theoretical developments and the state-of-the-art in the modeling of vector-boson-plus-jet physics, while focusing on experimental results by \LHC collaborations in Run-1 and Run-2, and including comparisons with recent measurements at the \Tevatron. 
\end{abstract}

\maketitle 
\tableofcontents

\section{Introduction}
\label{sec:intro}

\begin{figure}[htb!]
  \centering
  \includegraphics[width=0.49\textwidth]{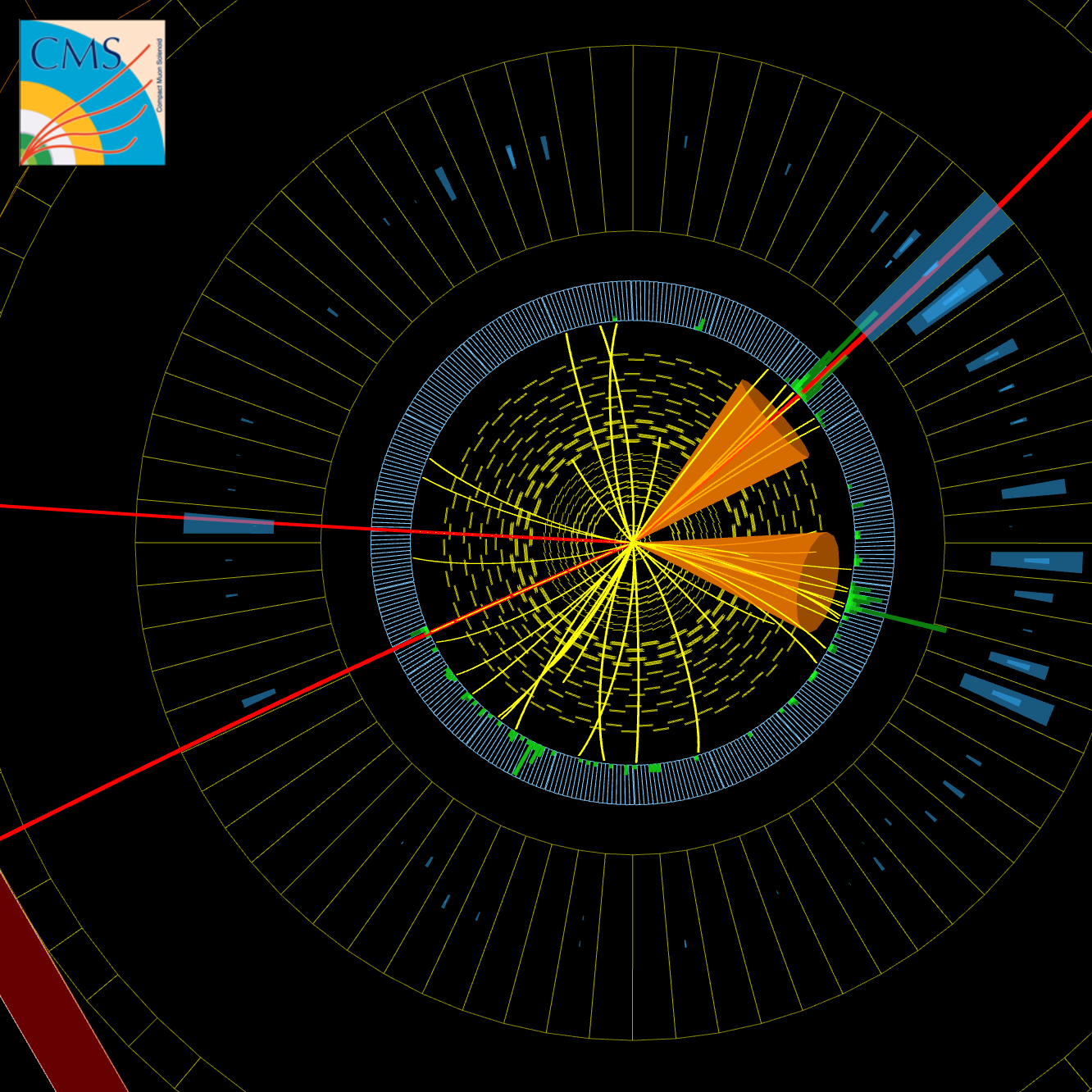}
  \caption{A vector boson plus jets event recorded by the CMS experiment in proton collisions at 13~TeV center-of-mass energy. The two solid lines on the left correspond to two reconstructed muons from the decay of a Z boson. The two cones on the right contain two collimated sprays of particles reconstructed as hadronic jets. 
  The internal composition of the jets indicate that both are likely to have originated from charm quarks, and one of them contains a muon from a displaced hadronic decay. Figure taken from~\cite{CMSevent}.
    \label{fig:CMS_Vjets_event}
  }
\end{figure}

Vector boson production in association with hadronic jets is  
one of the most important classes of processes that can be measured 
at hadron colliders. 
While the vector bosons, i.e. photons, Z, and W bosons, are the carriers of electroweak 
interactions, associated hadronic jets stem from the presence of strong interactions,
as a result of the process of fragmentation and hadronization of energetic partons (quarks and gluons). 
Figure~\ref{fig:CMS_Vjets_event} shows a proton collision event recorded by the CMS experiment with a Z boson produced in association with two jets. 
The two jets were identified to be likely originating from charm quarks. 
Each jet is a spray of hadronic particles collimated in the general direction 
of the initial parton, carrying the bulk of its total energy and transverse momentum. 

Therefore the study of \Vjets ($V=\gamma,W,Z$) events
constitutes an ideal probe for testing quantum 
chromodynamics (QCD) and electroweak (EW) interactions 
as well as a major source of backgrounds 
to searches for new physics. Measurements of \Vjets also validate the adequacy of the approximations 
used in theoretical calculations and models used for background estimates in 
precision measurements. For example, the modeling of \Vjets has a significant impact on studies of the Higgs boson and top-quark sectors of the Standard 
Model (SM), or in searches for physics beyond the Standard Model.
Thanks to their large cross sections, the colorless nature of the 
$\gamma$, $W$ and $Z$ bosons as carriers of electromagnetic and weak forces respectively,
and their high sample purities, accurate studies of \Vjets are 
of paramount importance for the success of a hadron collider physics program.

More specifically, a precise understanding of \gammajet production and its modeling
plays an important role in new physics searches, as \gammajets constitute a background, for example, to the production of high-mass 
resonances in the search for excited quarks in quark-compositeness 
models 
 or quantum black holes in models of extra spatial 
dimensions. 
 Similarly, reliable \WZjet calculations are important to correctly model the 
SM backgrounds in many searches of new particles in processes producing a single or two charged leptons\footnote{In the following sections 'charged lepton' refers to an electron or a muon, unless explicitly stated, since \WZjet measurements have primarily focused on electronic and muonic decay channels of the $W$ and $Z$ bosons.} with associated multi-jets, as they occur for example in supersymmetric theories.

Vector boson production processes in association with jets containing 
heavy-flavor hadrons play important roles in 
several measurements at hadron colliders. 
From them, the dynamics of the underlying heavy-flavor quark processes 
can be inferred.
In particular, they can give access to the heavy-flavor content 
of the proton, which is a limiting factor in several analyses at the 
Large Hadron Collider (\LHC). 
In addition, a detailed understanding of heavy-flavor quark dynamics 
in these processes was of paramount importance to the recent observation of the Higgs boson decay into a $b$-quark pair~\cite{Aaboud:2018zhk,Sirunyan:2018kst}, and
is vital for many new physics searches.

\begin{figure*}[t]
  \centering
  \includegraphics[width=\textwidth]{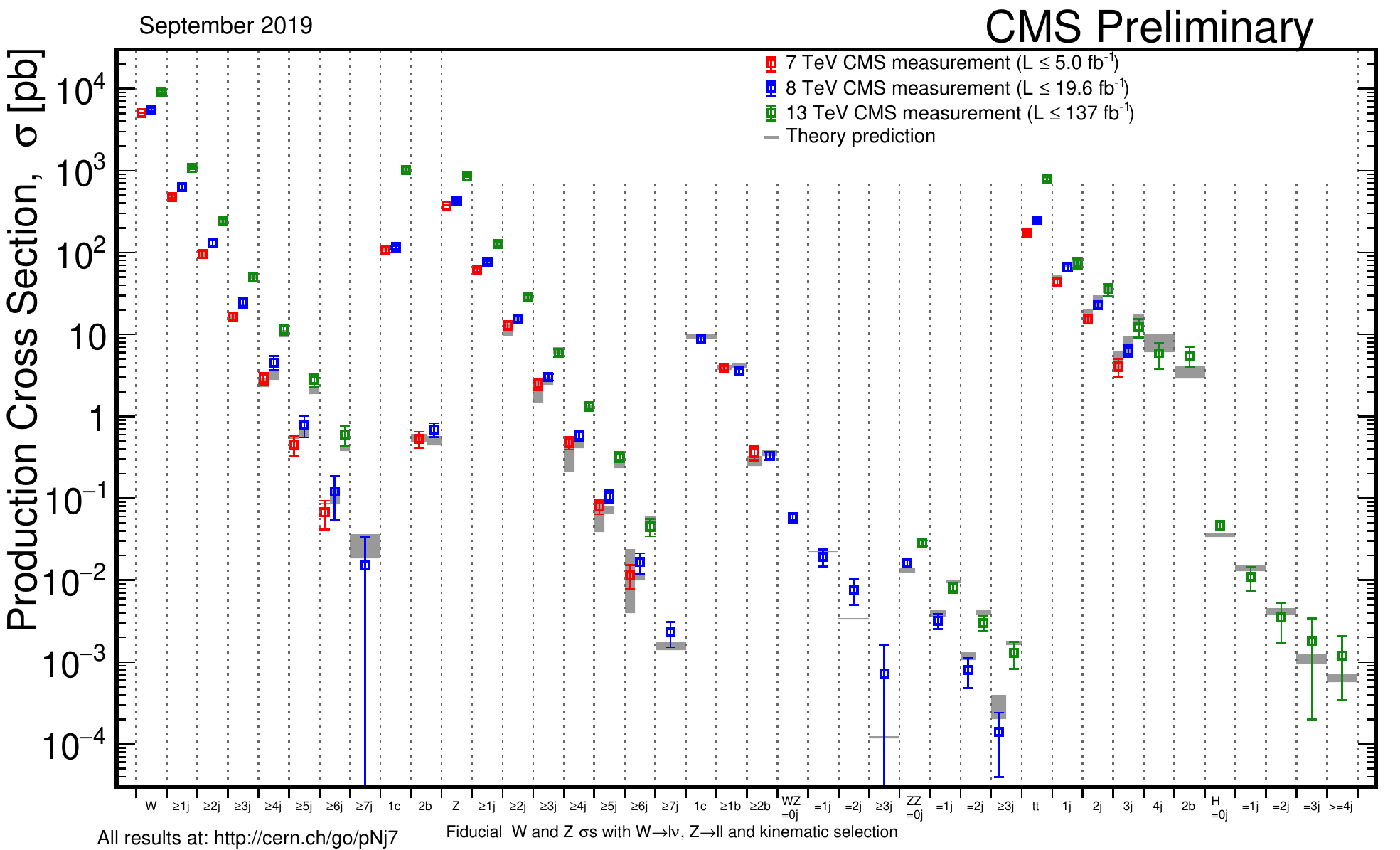}
  \caption{\label{fig:CMS_Summary_Vjets}
    Summary of production cross sections for processes with a $W$ or a $Z$ 
    boson produced in association with 
    light or heavy-flavor jets studied at different center of mass energies at 
    the \LHC, and their cross sections compared to other SM processes. 
    Figure taken from \cite{CMSxstab}.
  }
\end{figure*}

In the last 30 years many measurements of \Vjet event properties have been carried 
out starting from UA1 and UA2 experiments~\cite{UA1aa,UA1a,UA2a,UA1b,UA2b,UA1c,UA1d,UA2c,UA2d} 
at the Sp\ensuremath{\bar{\text{p}}}S.
Extensive measurements of such processes by the \CDF and the \DO collaborations at 
the \Tevatron have prompted significant development in the understanding of the 
underlying QCD dynamics, including new techniques for calculating high-precision theoretical predictions as well as non-perturbative modeling in Monte Carlo (MC) event generators. 
\Tevatron data provided an important stepping-stone for \Vjet analyses at the \LHC, 
despite the large difference in center-of-mass energies between the two colliders. 
The measurements carried out by the \LHC collaborations, \ATLAS, \CMS and 
\LHCb, in the \LHC Run-1 and Run-2 have motivated further developments in the theoretical 
description of such processes both in the QCD and the EW sectors.
The wide range of center-of-mass energies, from 1.96\,TeV in \ppbar collisions 
at the \Tevatron to 7, 8 and 13\,TeV in \pp collisions at the \LHC, allows to explore 
QCD dynamics in different energy regimes over a broad range of energy scales.
The ranges of \Vjet cross sections at the \LHC, compared to inclusive vector boson 
production and other SM processes, are shown in Figure~\ref{fig:CMS_Summary_Vjets}.

The underlying physics for the production of \Vjet processes at the \LHC cannot be 
considered as a simple rescaling of scattering processes at the \Tevatron. 
The different beam types and center-of-mass energies between the two colliders 
lead to a different relative importance of the various underlying production mechanisms 
and their associated phenomenology. 
More specifically, the \LHC reaches far larger energy scales $Q$ than the \Tevatron 
thanks to the higher beam energies, while it can simultaneously probe a lower 
Bjorken-$x$ range.
The inclusive production of a $W$ boson at the \LHC is dominated by events 
with a Bjorken-$x$ in the range $10^{-4}$ -- $10^{-1}$ and $Q^{2}\approx M^{2}_{W}$, 
while the exclusive production of a $W$ boson and at least one jet is shifted to larger 
values of $x$, with the majority of the events in the $x$ range of 
$10^{-2}$ -- $3\cdot ~10^{-1}$ and larger $Q^{2}$ values. 

As a consequence the two colliders are sensitive to particulars of the parameterizations of the 
parton densities inside the proton and their collision events are subject to different production mechanisms. 
For instance, at the \Tevatron, \Vjet processes have a significant valence-quark 
contribution from $q \bar{q}$ interactions, where the quark ($q$) is originating from 
the colliding proton and the anti-quark ($\bar{q}$) from the colliding anti-proton. 
On the other hand, at the \LHC there are significant gluon ($g$) and sea-quark 
contributions, including  $\bar{q}g$ , $q g$ and $gg$ interactions.  
For example, the $Z$+ 2 jet production, i.e. $qg \rightarrow Z+qg$ process, contributes around $75\%$ at the 
\LHC, while only around $25\%$ at the \Tevatron. 
At the \LHC the relative fraction of sub-processes initiated by $qq$/$q\bar{q}$/$\bar{q}\bar{q}$, $gq$/$g\bar{q}$, and $gg$ interactions varies with the number of associated jet production in the final state. In \Wjet processes the fraction of $qq$/$q\bar{q}$/$\bar{q}\bar{q}$ ($gq$/$g\bar{q}$)  sub-processes increases (decreases) from 18 (82)$\%$ with 1 jet to 21 (73)$\%$, 23 (70)$\%$ and 25 (67)$\%$ with 2, 3 and 4 jets, respectively, while for $gg$ sub-processes it increases from 0$\%$ with 1 jet to 6$\%$, 7$\%$ and 8$\%$ with 2, 3 and 4 jets, cf.~\cite{Kom:2010mv}.

As a result of the \LHC sensitivity to large contributions of sea-quark and gluon densities,
 \Vjet cross sections are far 
larger at the \LHC than at the \Tevatron. For example the $W +$ 4-jet cross 
section at the \LHC is 500 times larger than that at the \Tevatron  with similar 
kinematic selection, while the inclusive $Z + b$ cross section at the \LHC is 50 
times larger than that at the \Tevatron ~\cite{Campbell:2003dd}. 

This article reviews the evolution of the theoretical developments in the description of \Vjets physics for the \Tevatron and the \LHC, focusing on the recent achievements for the \LHC Run-1 and Run-2. An overview of experimental analysis techniques to identify \Vjet events and reconstruct their kinematics will be given together with a selection of measurements at different center-of-mass energies. The review of experimental results of \WZjets will focus on the leptonic decays of the $W$ and $Z$ bosons. The comparisons between experimental measurements and cutting-edge theoretical predictions will be highlighted. This review starts with the discussion of the production of vector bosons in association with light-flavor jets in Section~\ref{sec:VLF}, which is followed by the presentation of the EW production of vector bosons in Sections~\ref{sec:VBF} and the associated production of a vector boson and heavy-flavor jets, i.e., $V + b ~{\rm or}~ c$ jets,  in Section~\ref{sec:VHF}. Each of these sections starts with the discussion of the theoretical predictions for the different \Vjet production modes, specifically highlighting the most recent developments on higher-order calculations and MC event generators, then proceeds with the discussion of the experimental results and their comparison with the theory. In this review the authors do not attempt to accomplish the arduous task of exhaustively presenting all available measurements, but they have striven to highlight selected examples of measurements that give specific insight into \Vjet production dynamics. The numerous measurements of \Vjets make it also difficult to select a representative set of results, therefore the authors have chosen to provide a balanced representation of the different types of measurements and by the different experiments at the \Tevatron and the \LHC.
Section~\ref{sec:QCD_interpretation} illustrates some examples of analyses and interpretations of experimental \Vjet results, for example for tuning of  MC generators, constraining the proton parton densities and for setting limits on anomalous contributions to SM interactions. The article concludes in Section~\ref{sec:conclusions} with a summary and an outlook for \Vjet analyses at future \LHC runs and future colliders.

%

\section{Associated production of a vector boson and light-flavor jets}
\label{sec:VLF}
\subsection{Theoretical predictions}
\label{sec:VLF:theory}

\begin{figure*}
  \centering
  \begin{minipage}{0.22\textwidth}
    \centering
    \includegraphics[width=\textwidth]{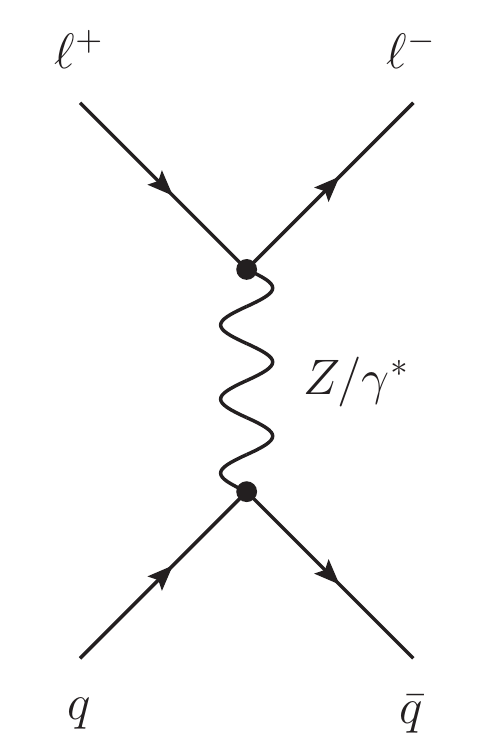}\\
    (a)
  \end{minipage}
  \hfill
  \begin{minipage}{0.44\textwidth}
    \centering
    \includegraphics[width=0.5\textwidth]{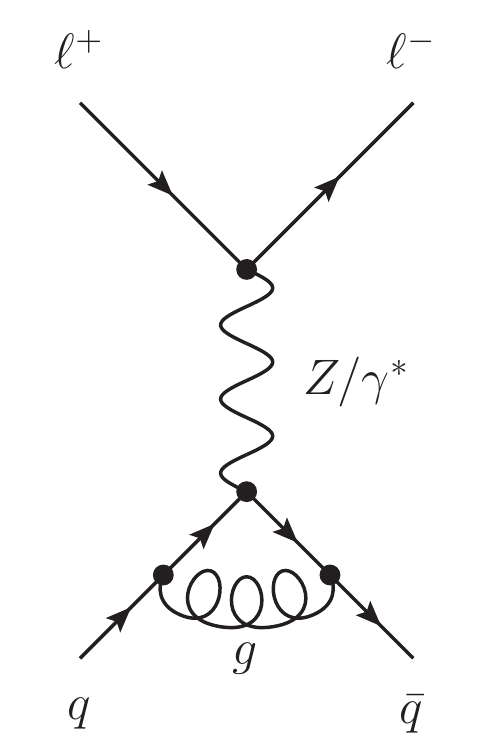}
    \hspace*{-0.05\textwidth}
    \includegraphics[width=0.5\textwidth]{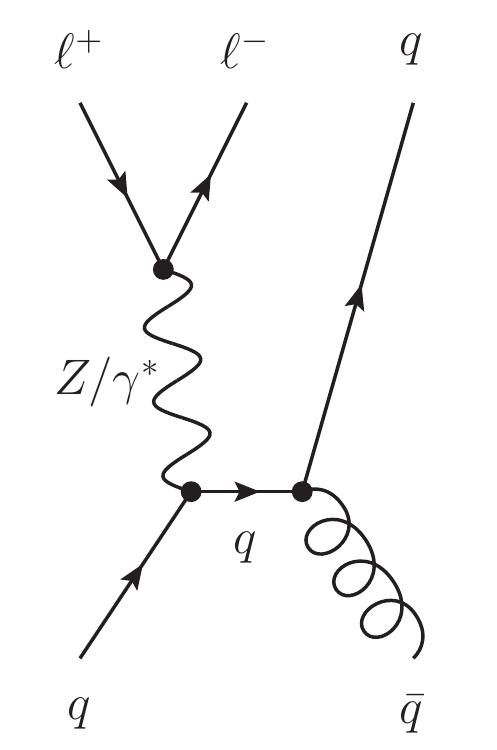}\\
    (b)
  \end{minipage}
  \hfill
  \begin{minipage}{0.22\textwidth}
    \centering
    \includegraphics[width=\textwidth]{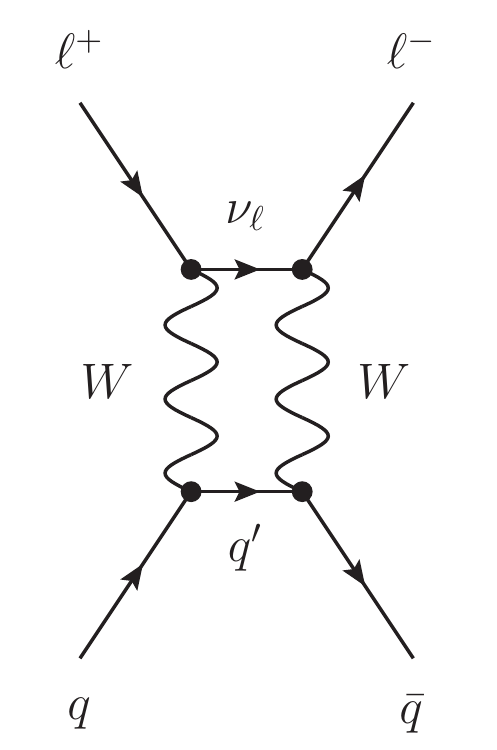}\\
    (c)
  \end{minipage}
  \caption{\label{fig:V_diagrams}
    Representative Feynman diagrams for the production of a pair of charged leptons at LO (a) and 
    representative contributions at NLO QCD (b) and NLO EW (c).
  }
\end{figure*}

\begin{figure*}
  \centering
  \begin{minipage}{0.22\textwidth}
    \centering
    \includegraphics[width=\textwidth]{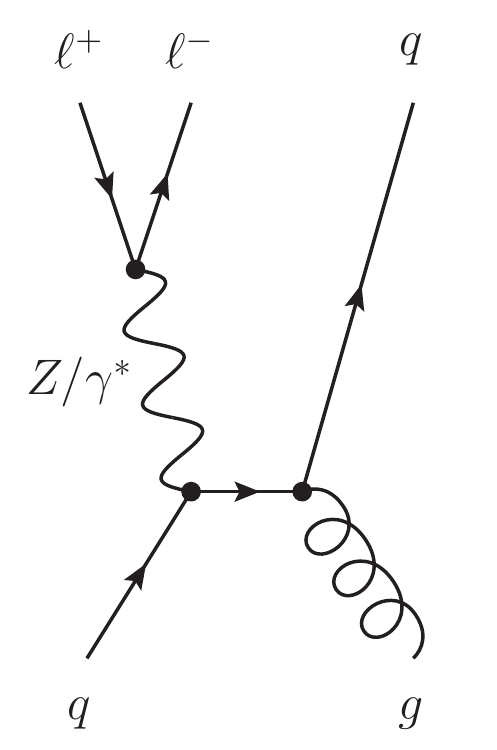}\\
    (a)
  \end{minipage}
  \hfill
  \begin{minipage}{0.44\textwidth}
    \centering
    \includegraphics[width=0.5\textwidth]{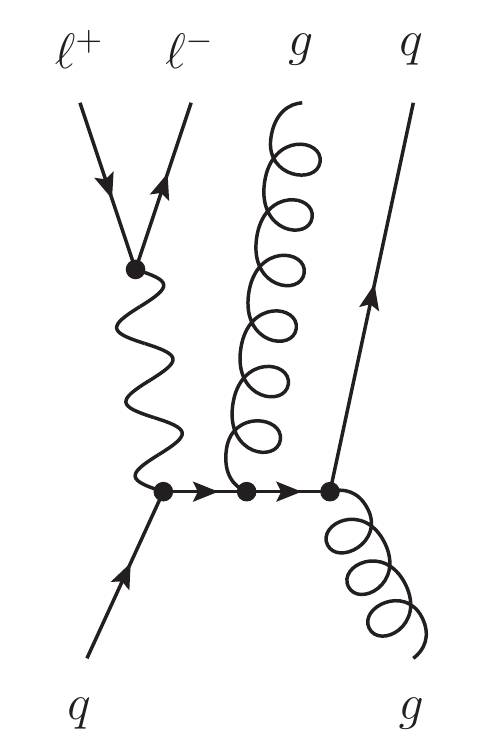}
    \hspace*{-0.05\textwidth}
    \includegraphics[width=0.5\textwidth]{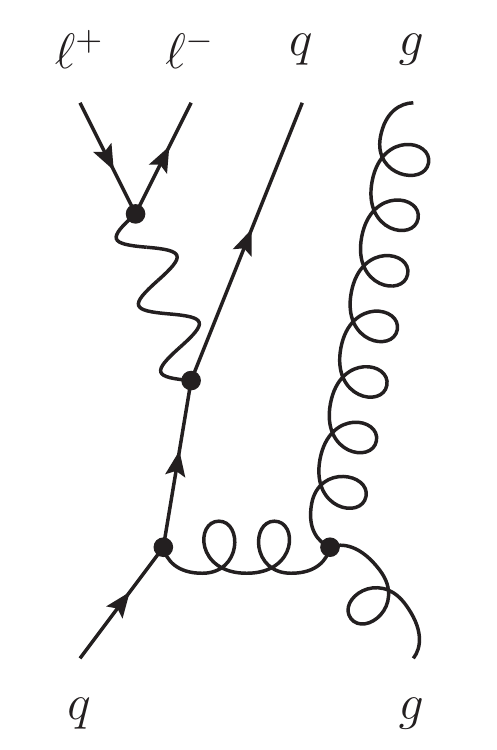}\\
    (b)
  \end{minipage}
  \hfill
  \begin{minipage}{0.22\textwidth}
    \centering
    \includegraphics[width=\textwidth]{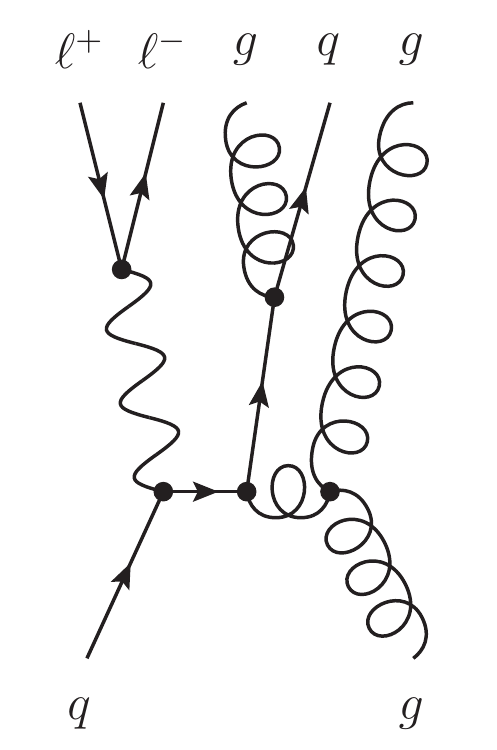}\\
    (c)
  \end{minipage}
  \caption{\label{fig:Vjets_diagrams}
    Representative Feynman diagrams for the production of a pair of charged leptons in association with one jet (a), 
    two jets (b), and three jets (c) at LO.
  }
\end{figure*}

\begin{figure*}
  \centering
  \begin{minipage}{0.22\textwidth}
    \centering
    \includegraphics[width=\textwidth]{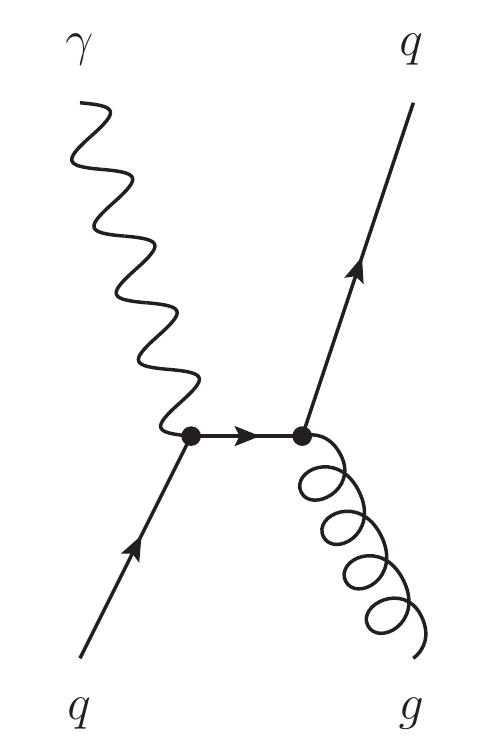}\\
    (a)
  \end{minipage}
  \hfill
  \begin{minipage}{0.44\textwidth}
    \centering
    \includegraphics[width=0.5\textwidth]{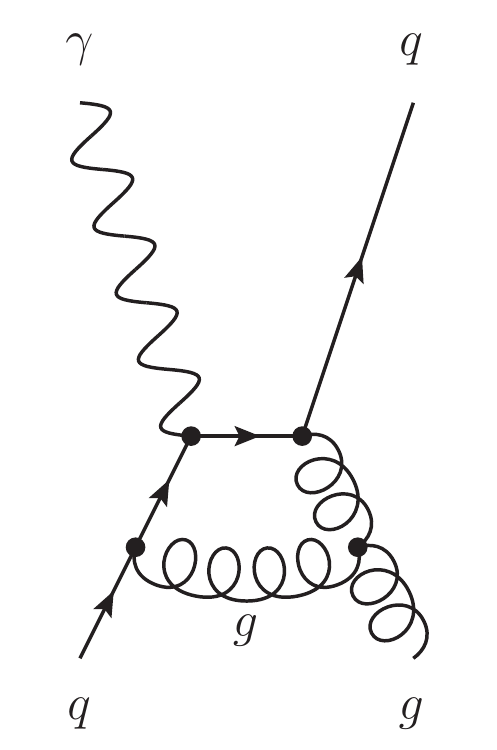}
    \hspace*{-0.05\textwidth}
    \includegraphics[width=0.5\textwidth]{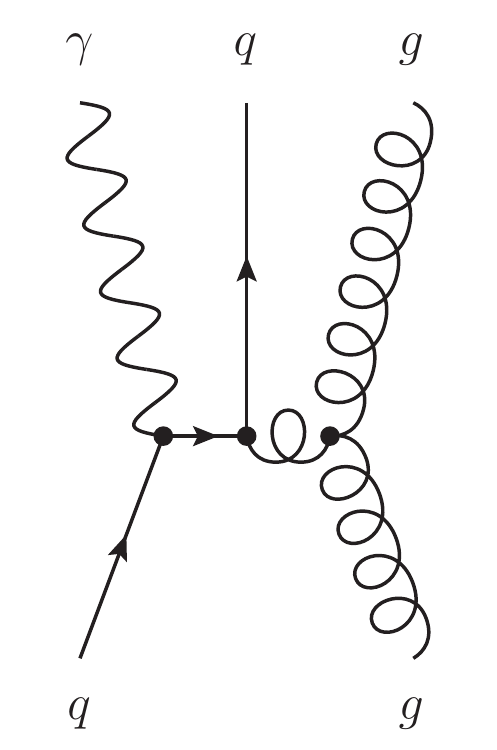}\\
    (b)
  \end{minipage}
  \hfill
  \begin{minipage}{0.22\textwidth}
    \centering
    \includegraphics[width=\textwidth]{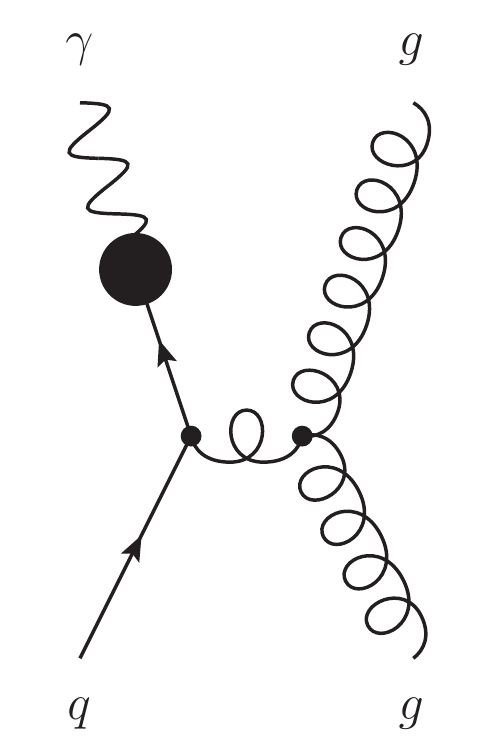}\\
    (c)
  \end{minipage}
  \caption{\label{fig:aj_diagrams}
    Representative Feynman diagrams for the production of a photon at LO (a), 
    representative contributions at NLO QCD (b) and photon production from 
    quark fragmentation (c) that enters at the same order.
  }
\end{figure*}

The production of a single vector boson, $\gamma$, $W$ or $Z$, 
is the class of processes with the largest 
cross sections among the electroweak processes at hadron colliders. 
While the massive vector bosons, $W$ and $Z$, can be produced 
by simple quark-anti-quark annihilation without additional final-state 
partons at leading order (LO), photons, as the massless gauge boson of QED, 
are only measurable when they have finite transverse momentum and 
therefore always need at least one parton to recoil against.

To set the stage, 
the Feynman diagrams of the dominant leading and 
next-to-leading order (NLO), in both the strong and electroweak couplings $\alphaS$ and $\alpha$, production processes of inclusive massive gauge bosons 
are detailed in Fig.\ \ref{fig:V_diagrams}. 
A substantial fraction of all events with single massive vector 
bosons are accompanied by additional hadronic jet activity. 
These processes are of specific interest due to their 
clean signature and the relative precision with which they 
can be calculated in the Standard Model. 
Representative leading order Feynman diagrams are shown in 
Fig.\ \ref{fig:Vjets_diagrams}.

Single photon production, on the other hand, is always accompanied 
by hadronic recoil, as discussed above. 
Figure \ref{fig:aj_diagrams} shows representative diagrams of photon 
production in association with jets.

\subsubsection{Higher order computations}
\label{sec:VLF:theory:ho}

\begin{figure*}[t!]
  \centering
  \includegraphics[width=0.47\textwidth,height=0.45\textwidth]{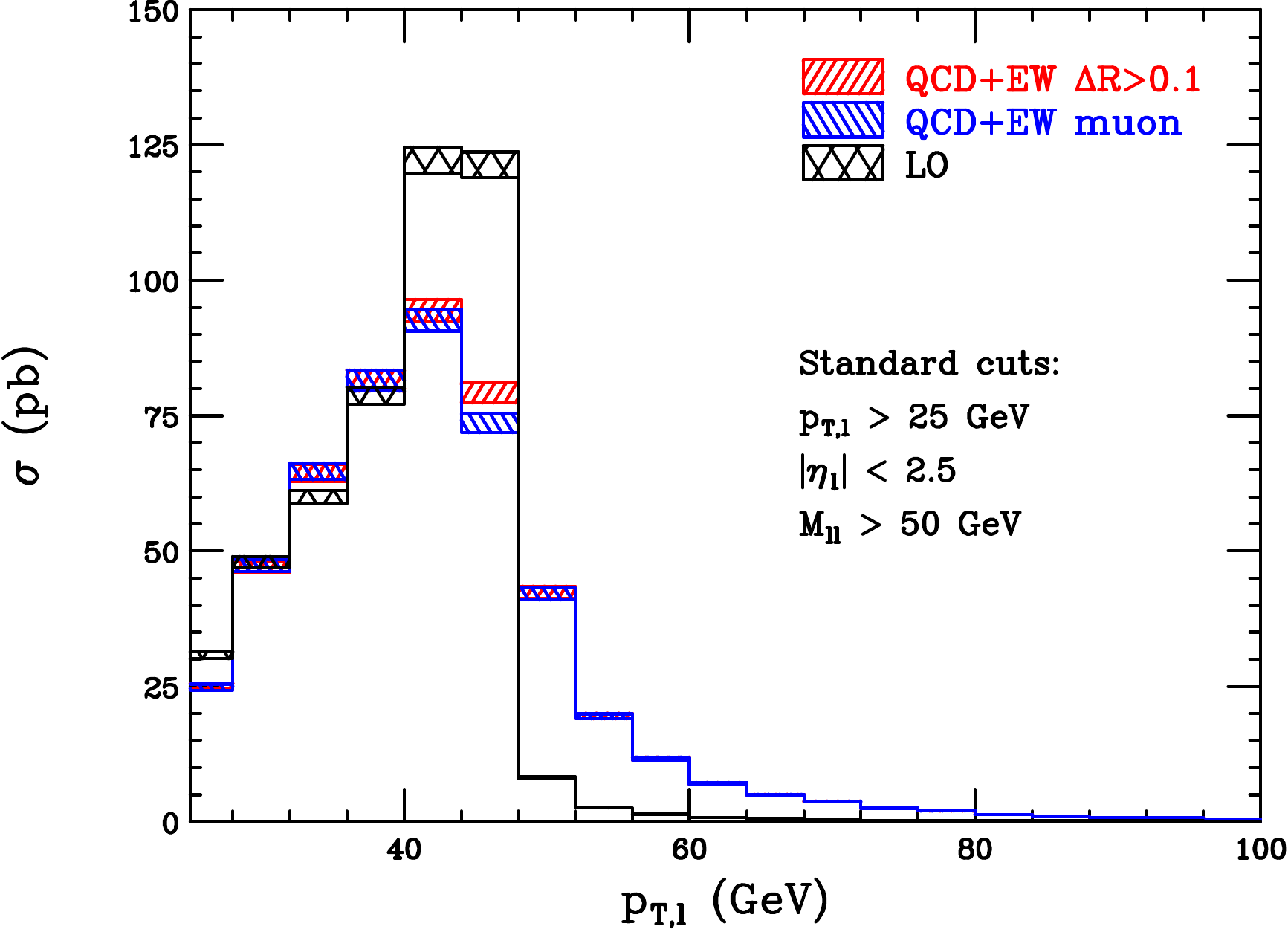}
  \hfill
  \includegraphics[width=0.47\textwidth]{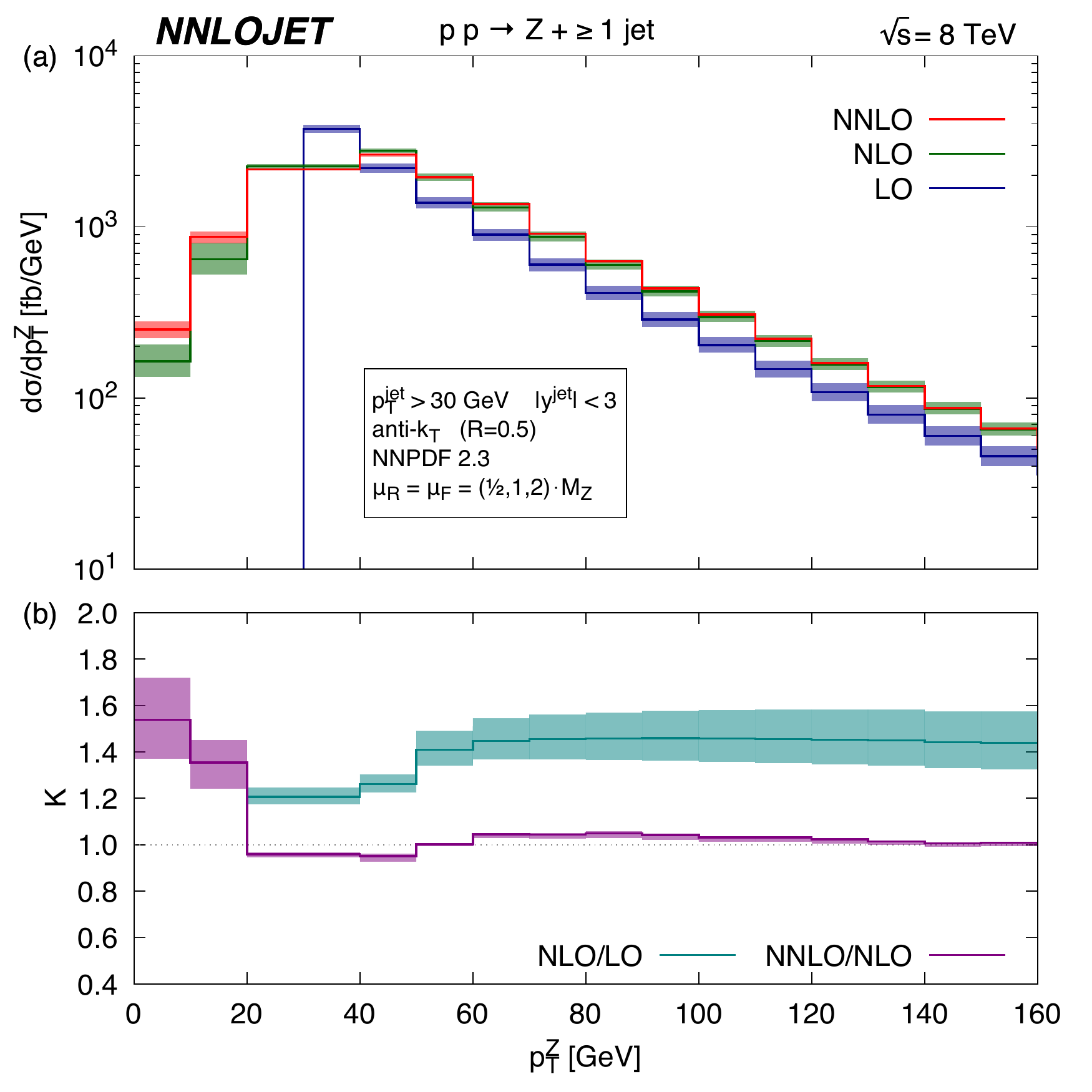}
  \caption{\label{fig:VLF:theory:nnlo}
    Charged lepton transverse momentum in the inclusive production of two charged leptons 
                   at NNLO QCD plus NLO EW (left), figure taken from \cite{Li:2012wna}.
    Transverse momentum of two charged leptons in the production of two charged leptons in
                    association with at least one jet at LO, NLO and NNLO QCD (right), figure taken 
                    from \cite{Ridder:2015dxa}.
  }
\end{figure*}

The effort to increase the available accuracy of the theory predictions\footnote{
  Theoretical uncertainties are typically estimated by varying all 
  unphysical scales of the calculation, e.g., the renormalization 
  and factorization scale, etc.
  The resulting quoted error estimate has, however, no statistical 
  interpretation.
} 
for inclusive \W and \Z boson production processes started early. 
While the first results beyond leading order accuracy date back forty years 
\cite{Altarelli:1979ub},
the current standard sees the inclusive cross sections determined at next-to-next-to-leading order (NNLO) in QCD with NLO EW corrections, i.e.,
NNLO QCD + NLO EW \cite{Hamberg:1990np,Anastasiou:2003ds,Melnikov:2006kv,Melnikov:2006di,Catani:2009sm,
Gavin:2010az,Li:2012wna}. 
The NNLO QCD-EW mixed contributions of $\order(\alphaS\alpha)$ are 
known in the pole approximation \cite{Dittmaier:2014qza}, which is valid 
in the region where the invariant mass of the charged lepton pair is close to
the respective gauge boson's mass.
Figure~\ \ref{fig:VLF:theory:nnlo} displays on its left hand side such a 
state-of-the-art calculation for the inclusive transverse momentum of 
the positively-charged lepton in the inclusive production of a charged-lepton pair, 
showing the relatively large corrections received through the 
higher-order QCD corrections with respect to the leading order calculation.

Another key inclusive experimental observable is the transverse momentum 
distribution of the vector boson itself. 
Its description, however, depends on an accurate description of its 
recoil. 
It thus vanishes identically at leading order and only starts at 
$\order(\alphaS)$ or $\order(\alpha)$, respectively. 
Its precise description is further complicated by large logarithms in 
the small-transverse-momentum region, as a result of an infrared divergence 
at $\pT=0$, which spoil the convergence of the perturbative expansion in the coupling parameters. 
As the same logarithms reappear order-by-order, however, they can be 
resummed, and the respective results will be detailed in 
Sec.\ \ref{sec:VLF:theory:res}. 
In the medium- to large-transverse-momentum region, a fixed-order 
expansion is sufficient to achieve percent-level accuracy. 
While the QCD two-loop amplitudes have been available for some time 
\cite{Garland:2001tf,Garland:2002ak,Moch:2002hm},
the inception of novel subtraction formalisms 
\cite{Gaunt:2015pea,Boughezal:2015aha,Kosower:1997zr,
      GehrmannDeRidder:2005cm,Currie:2013vh,
      Catani:2007vq,Catani:2019iny} 
along with the computational frameworks that are able to deal with 
the complexity of the infrared structure of such a calculation were 
only recently available. 
They paved the way for precise NNLO QCD calculations for this kinematic region 
\cite{Ridder:2015dxa,Ridder:2016nkl,Gehrmann-DeRidder:2016jns,
      Gauld:2017tww,
      Boughezal:2015ded,Boughezal:2016isb,Campbell:2017dqk,
      Gehrmann-DeRidder:2017mvr,Gehrmann-DeRidder:2019avi,
      Boughezal:2015dva,Boughezal:2016dtm},
as is detailed in the right hand side of Fig.\ \ref{fig:VLF:theory:nnlo}. 

\nocite{Rubin:2010xp}

At the same time, the electroweak corrections for both the 
vector boson transverse momentum and vector-boson-plus-jet 
production in general are also known 
\cite{Kuhn:2004em,Kuhn:2005az,Denner:2011vu,Actis:2012qn,Denner:2012ts,
      Denner:2014ina,Hollik:2015pja,Kallweit:2015fta,Kallweit:2015dum,
      Lindert:2017olm,Kuhn:2007qc,Kuhn:2007cv,Kallweit:2014xda}. 
In their combination with the higher-order QCD corrections 
typically two schemes are followed. 
They can be combined additively, corresponding to a strict 
next-to-leading order expansion, commonly denoted as NLO QCD+EW. 
Or they can be combined multiplicatively, referred to as NLO QCD$\times$EW, 
which assumes a factorization of both effects and is especially suited 
if the typical scales of both processes are well separated. 
The difference is formally of higher order and can be used 
to estimate the potential size of the mixed QCD$-$EW NNLO corrections.

Conversely, the inclusive photon production cross section, as it is always 
accompanied by hadronic activity, starts at $\order(\alpha_s\alpha)$ at the 
Born level and is thus of the same level of complexity as $W/Z+1\text{jet}$ 
production. 
Hence, all pieces of the calculation at NNLO accuracy in the strong coupling 
have only been computed recently 
\cite{Campbell:2016lzl,Campbell:2017dqk,Chen:2019zmr} while the NLO 
EW corrections have been known for a slightly longer period 
\cite{Kuhn:2005gv,Kallweit:2015fta,Lindert:2017olm}.

In addition, due to their nature as massless gauge bosons, photons 
can be produced both promptly by the hard interaction and emerge 
from a fragmentation process, cf.\ Fig.\ \ref{fig:aj_diagrams}.
Thus, unless the photon is identified using the smooth cone 
isolation procedure \cite{Frixione:1998jh}, which completely 
removes the fragmentation component by construction, 
such additional fragmentation processes  
have to be considered starting at the next-to-leading order 
\cite{Aurenche:2006vj,Gluck:1994iz}. 
It remains to be noted that the smooth cone isolation cannot be adopted 
by the experiments due to finite detector resolution, while the standard cone isolation, that is typically used experimentally, necessitates the use of fragmentation 
functions in theory calculations. 
Thus, for all calculations using the smooth cone isolation 
the correspondence of its parameters to approximately match the 
experimentally used ones needs to be confirmed. 
Mixed schemes, e.g., a smooth/standard cone hybrid \cite{Siegert:2016bre}
can be used to mediate such differences.

\begin{figure*}[t!]
  \centering
  \includegraphics[width=0.47\textwidth]{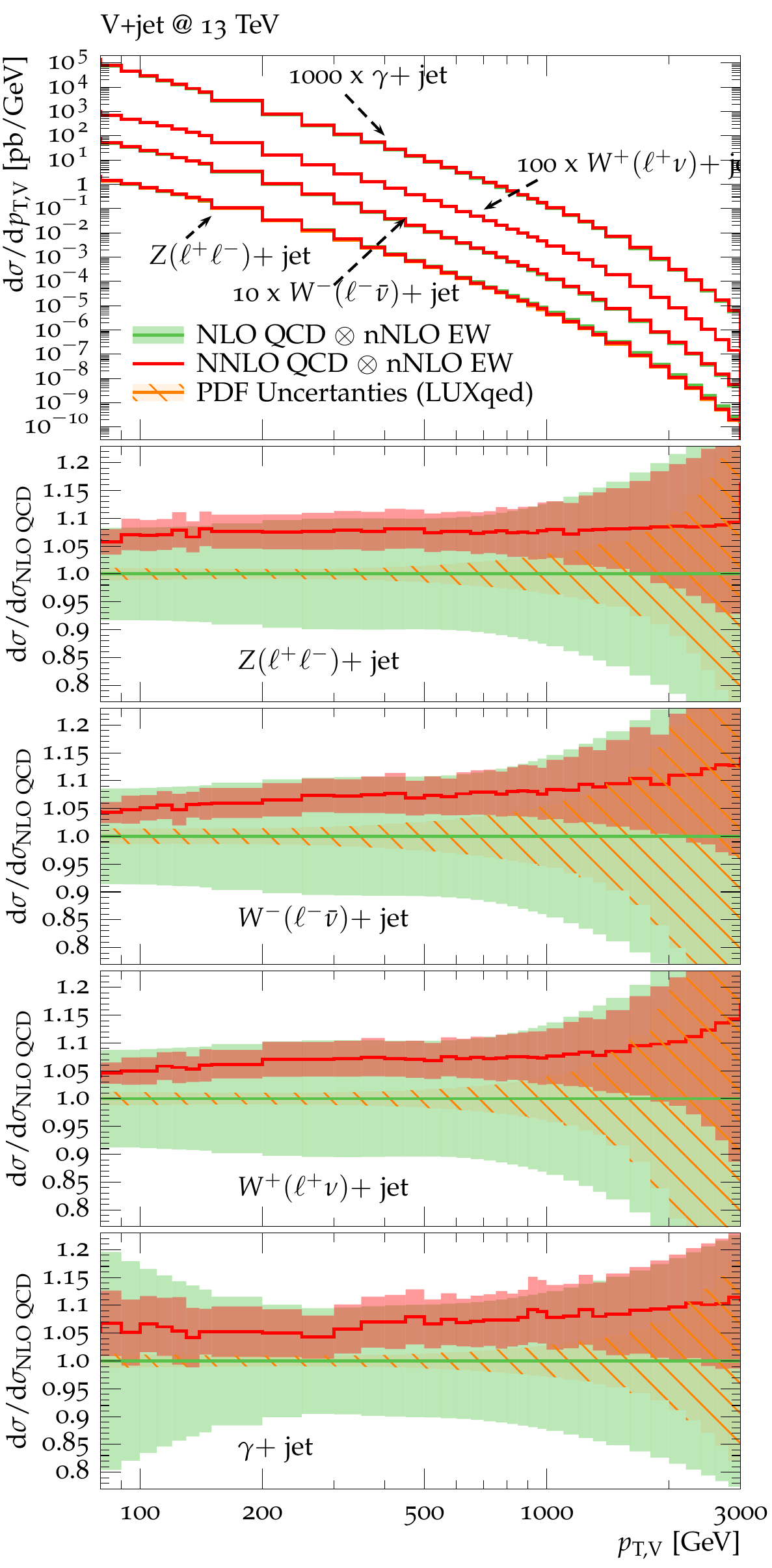}
  \hfill
  \includegraphics[width=0.47\textwidth]{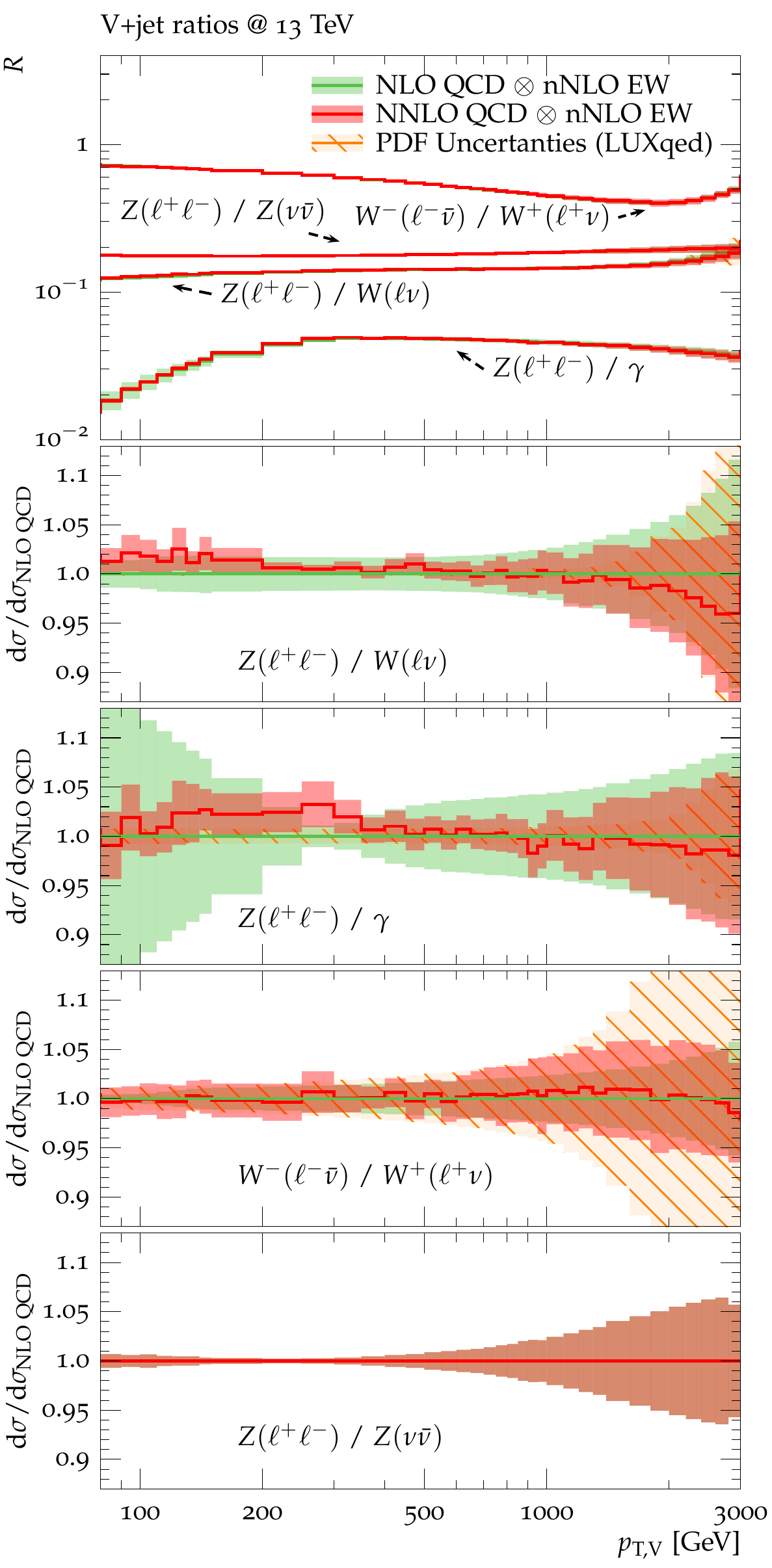}
  \caption{\label{fig:VLF:theory:pT-ratio}
    Transverse momentum of the reconstructed vector boson (left, from top to 
                   bottom $Z$, $W^-$, $W^+$ and $\gamma$) in inclusive vector boson 
                   production at NLO and NNLO QCD plus nNLO EW, 
                   figure taken from \cite{Lindert:2017olm}.  
    Pairwise ratios of the differential cross section in the transverse 
                    momentum of the reconstructed vector bosons in inclusive vector 
                    boson production at NLO and NNLO QCD plus nNLO EW (right), 
                    figure taken from \cite{Lindert:2017olm}.
  }
\end{figure*}

The transverse momentum distribution of a $Z$ boson decaying 
into neutrino pairs is of particular interest to new physics searches 
at the \LHC, e.g., as a background to searches for Dark Matter, as it has a similar topology and features particles that are invisible to the \LHC detectors. 
Due to this very fact, however, measuring this Standard Model background 
independently proves to be challenging. 
Typically it is indirectly inferred by measuring the transverse momentum 
of a leptonically decaying $Z$ or $W$ boson or a photon instead, 
relying on theory predictions 
to estimate to the sought after $Z\to\nu\bar{\nu}$ distribution. 
The needed ratios of
production cross sections of the different processes can now be predicted with high precision 
\cite{Bern:2011pa,Campbell:2017dqk,Lindert:2017olm,Gehrmann-DeRidder:2017mvr,Bizon:2019zgf}, thanks to the recently available NNLO QCD plus NLO EW accurate predictions, as discussed above, and an understanding of the correlations across processes. 
Figure~\ \ref{fig:VLF:theory:pT-ratio} details the predictions for 
these ratios, which are found to have percent-level accuracy for 
vector boson transverse mementa up to around 1\,TeV, growing to 
5-10\% in the regions beyond.

\begin{figure*}[t!]
  \centering
  \includegraphics[width=0.47\textwidth,height=0.62\textwidth]{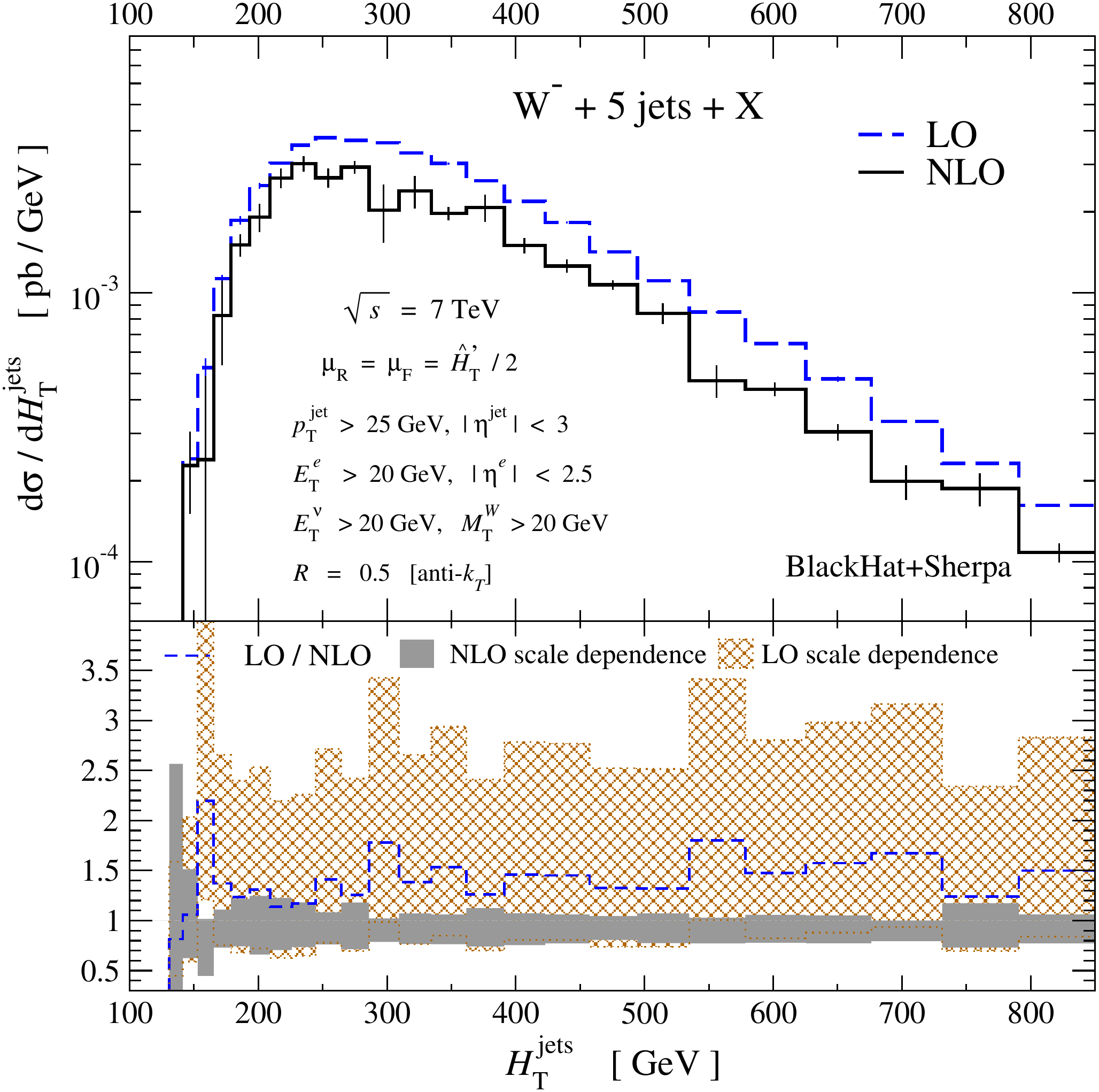}
  \hfill
  \includegraphics[width=0.47\textwidth]{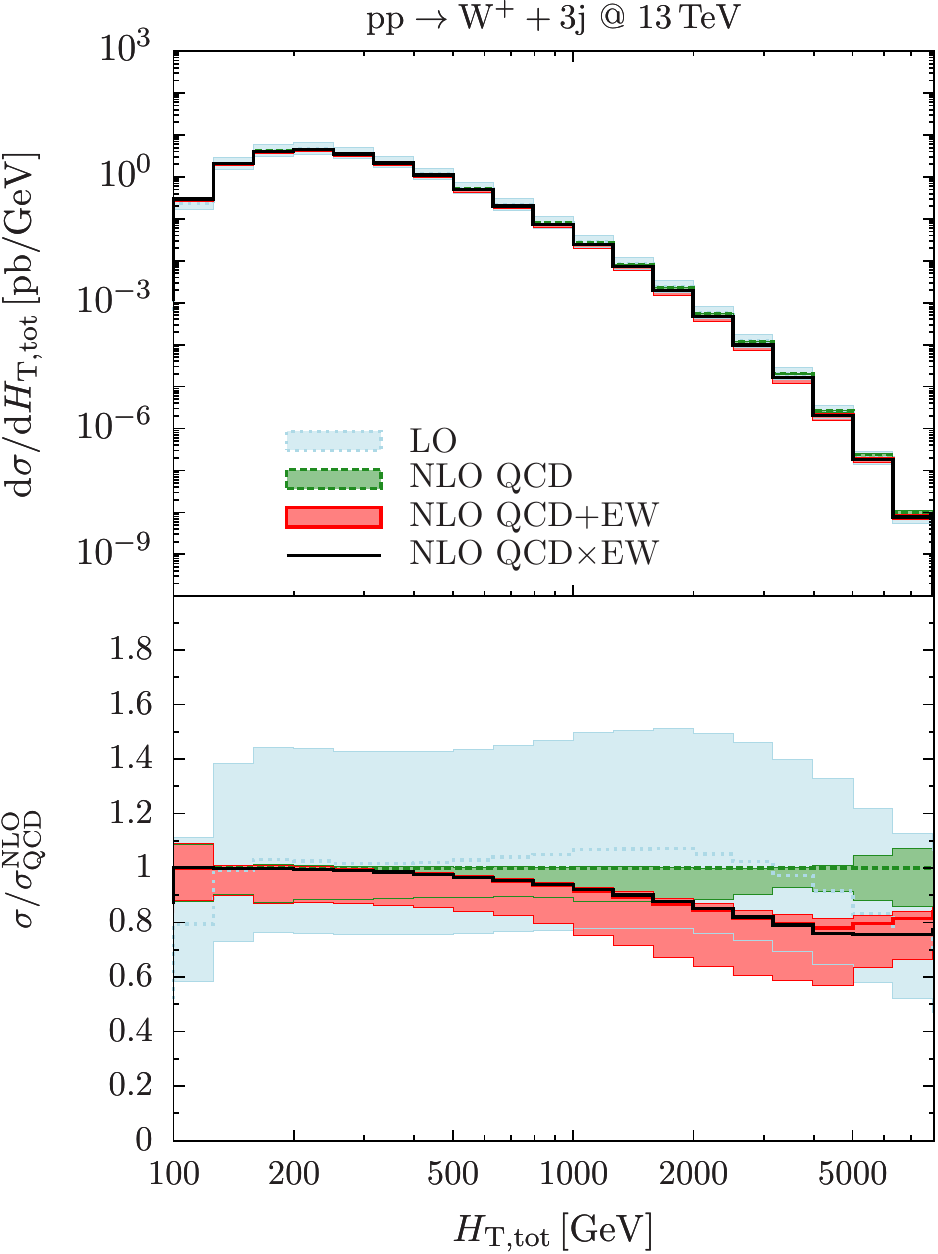}
  \caption{\label{fig:VLF:theory:nlo}
    Scalar sum of jet transverse momenta in single charged lepton production in 
                   association with missing transverse momentum and at least five jets 
                   at LO and NLO QCD (left), figure taken from \cite{Bern:2013gka}.  
    Scalar sum of jet transverse momenta in single charged lepton production in 
                    association with missing transverse momentum and at least three jets at LO and 
                    NLO QCD plus NLO EW (right), figure taken from \cite{Kallweit:2014xda}.
  }
\end{figure*}

Finally, also vector boson production in association with multiple 
jets is of interest. 
Not only are such final states frequently measured at the \LHC, 
they also often constitute backgrounds to searches for particular new physics 
models. 
With the adoption of the anti-$k_\text{t}$ jet definition \cite{Cacciari:2008gp}, 
a sequential recombination algorithm of the longitudinally-invariant-$k_\text{t}$ 
family \cite{Catani:1993hr,Ellis:1993tq,Dokshitzer:1997in}, by the \LHC experiments 
as their default jet definition, longstanding issues around the compromised 
infrared safety of the \Tevatron era jet algorithms 
\cite{Blazey:2000qt} were resolved. 
This allows for high-precision higher-order calculations to be made 
for the precise observable that is measured.

The NLO QCD predictions are available for $W$ plus up to five jets 
\cite{Campbell:2002tg,Berger:2009zg,KeithEllis:2009bu,Berger:2010zx,Bern:2013gka} and 
$Z$ plus up to four jets \cite{Campbell:2002tg,Berger:2010vm,Ita:2011wn}, 
and approximate NNLO corrections, dubbed ${\rm \bar{n}}$NLO, can be calculated through the 
\LOOPSIM method \cite{Rubin:2010xp,Maitre:2013wha}.
At the same time, NLO EW corrections are known 
for fully off-shell production only up to two jets \cite{Denner:2014ina,Kallweit:2015dum}, 
and in the on-shell approximation for up to three jets \cite{Kallweit:2014xda,Chiesa:2015mya}.
The effect of these corrections are detailed in Fig.\ \ref{fig:VLF:theory:nlo}.
Processes with up to nine jets can, however, be calculated at LO accuracy 
\cite{Hoeche:2019rti}.

\subsubsection{Resummation calculations}
\label{sec:VLF:theory:res}

\begin{figure*}[t!]
  \centering
  \includegraphics[width=0.47\textwidth]{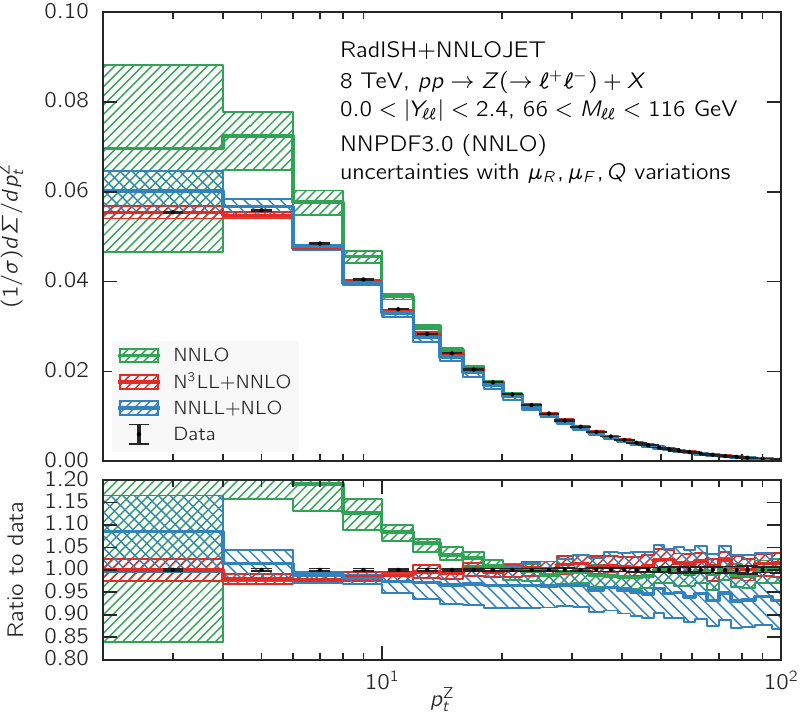}
  \hfill
  \includegraphics[width=0.47\textwidth]{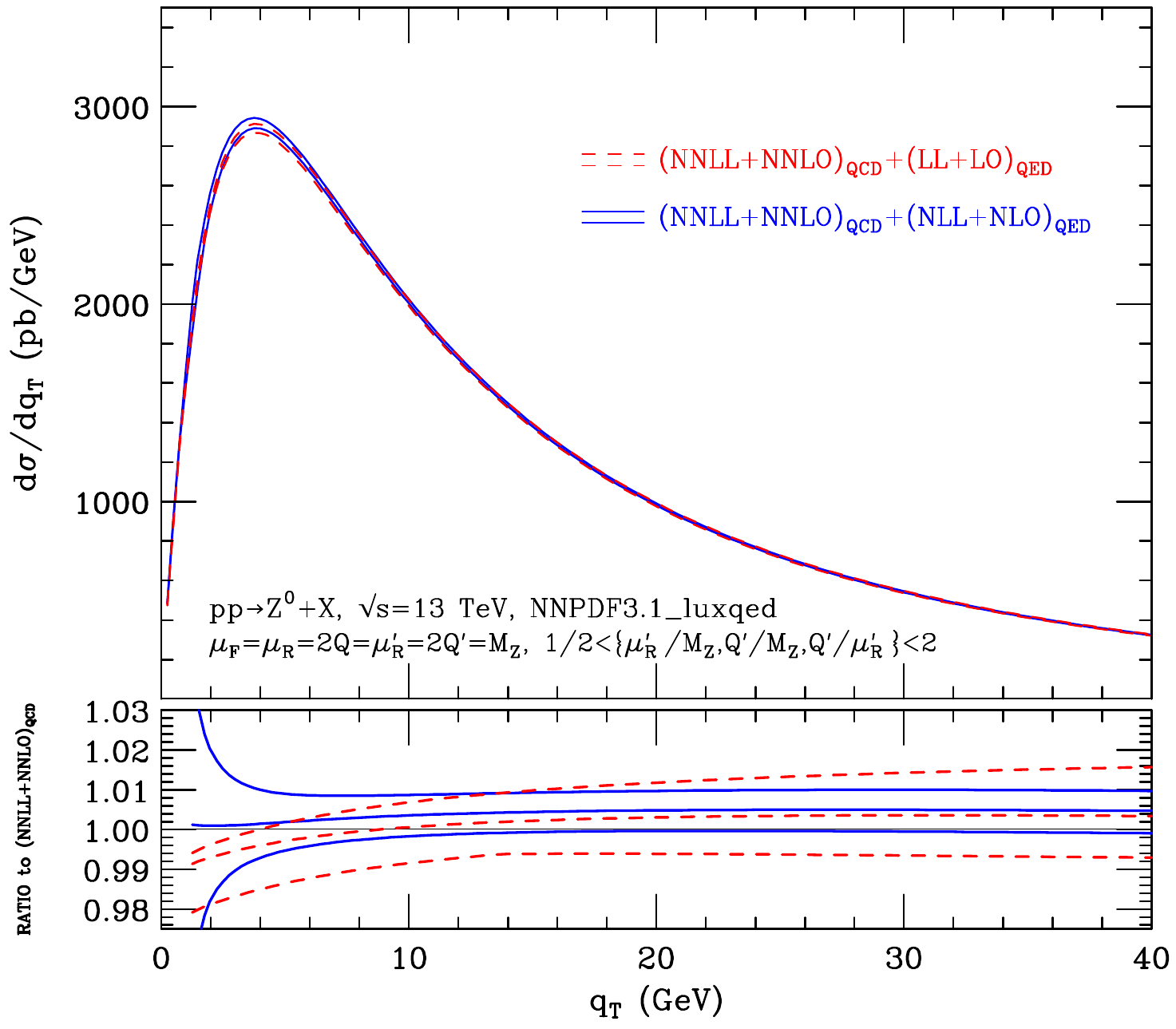}
  \caption{\label{fig:VLF:theory:res}
    Transverse momentum of the charged lepton pair in the inclusive production of two charged leptons 
                   at NNLO, NNLL+NLO and N$^3$LL+NNLO QCD (left), showing the QCD scale uncertainties, 
                   figure taken from \cite{Bizon:2018foh}.  
    Transverse momentum of the charged lepton pair in the inclusive production of two charged leptons 
                    at NNLL+NNLO QCD plus LL+LO and NLL+NLO QED (right), showing the QED-related uncertainties only, 
                    figure taken from \cite{Cieri:2018sfk}.
  }
\end{figure*}

Fixed-order calculations fail to yield reliable cross-section 
predictions in phase space regions where large, typically logarithmic, 
terms appear at every order of the perturbative expansion. 
Consequently, a truncation of the perturbative series fails to 
converge quickly enough after the first, second or third order. 
To render a truncation after any finite number of orders meaningful,
a resummation of the terms spoiling the convergence is mandated. 
As the functional form of these convergence-impairing terms is dependent 
of the phase space region probed by the respective observable, all 
resummation formulations are observable specific. 
In the literature, several approaches to identify and resum the 
relevant logarithms have been formulated, cf.\  
\cite{Collins:1984kg,Ladinsky:1993zn,Balazs:1997xd,
      Bozzi:2010xn,Catani:2013tia,Becher:2010tm,Mantry:2010bi,
      Ebert:2016gcn,Monni:2016ktx,Bizon:2018foh,
      Coradeschi:2017zzw,Martinez:2019mwt,Bacchetta:2019sam}. 
This is in particular relevant for electroweak precision measurements 
like the $W$ mass, or for probing so-called intrinsic transverse momentum 
of the partons inside the proton.

For the production of vector bosons in association with light-flavor jets, 
the transverse momentum of the vector boson is of particular interest. 
In this observable, resummation is required to accurately describe 
the small-\pT region, whereas the large-\pT region suffers no such 
effects and is most accurately described by fixed-order perturbation 
theory. 
Thus, to achieve the best predictions for the spectrum, the resummed 
calculation at small \pT has to be matched to the fixed-order calculation 
at large \pT. 
Several solutions, fitting the various resummation procedures, 
are available. Figure~\ \ref{fig:VLF:theory:res} displays two examples. 
On the left hand side, the state-of-the-art next-to-next-to-next-to leading logarithmic (N$^3$LL) resummation 
matched to NNLO fixed-order prediction in QCD is shown \cite{Bizon:2018foh}. 
The effect of resumming large logarithms in the expansion 
of the cross-section in the strong coupling, $\alphaS$, as well as 
including the large logarithms appearing in the simultaneous 
expansion in the electroweak coupling, $\alpha$, is shown on the 
right hand side of Fig.~\ \ref{fig:VLF:theory:res}~\cite{deFlorian:2018wcj,Cieri:2018sfk}. 
Although the effects are somewhat small in this case, they are 
needed for a meaningful interpretation of the high-precision 
data to be taken in the future \LHC runs.

Other resummed calculations are available for related observables, 
e.g., the $\phi^*$-distribution \cite{Banfi:2011dx} that can be measured with superior 
experimental precision \cite{Vesterinen:2008hx,Banfi:2010cf},
or the jet veto efficiency \cite{Stewart:2010pd,Tackmann:2012bt,Banfi:2012jm}.

\subsubsection{Monte Carlo event generators}
\label{sec:VLF:theory:mc}

\begin{figure*}[t!]
  \centering
  \includegraphics[width=0.47\textwidth]{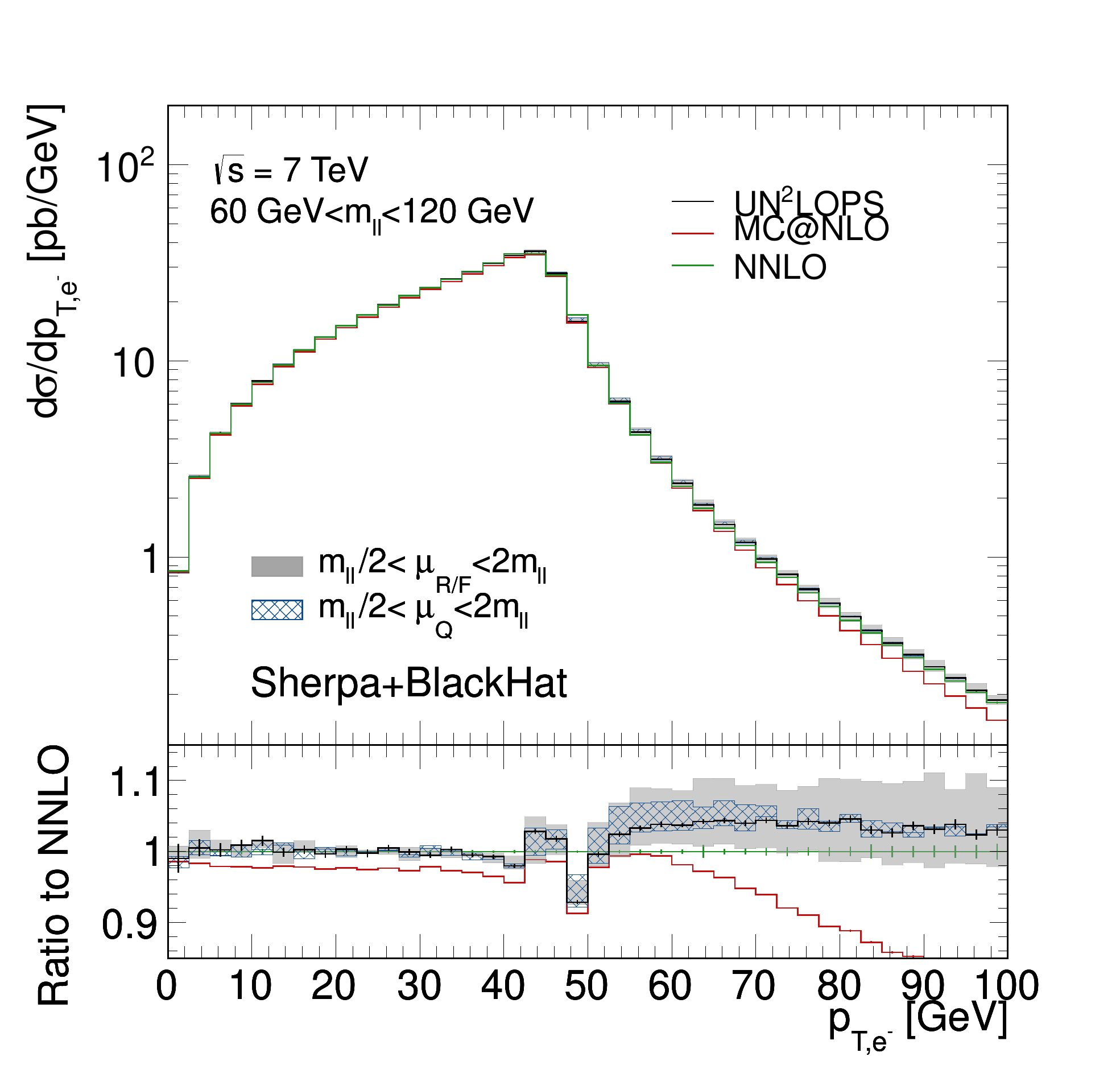}
  \hfill
  \includegraphics[width=0.47\textwidth,height=0.42\textwidth]{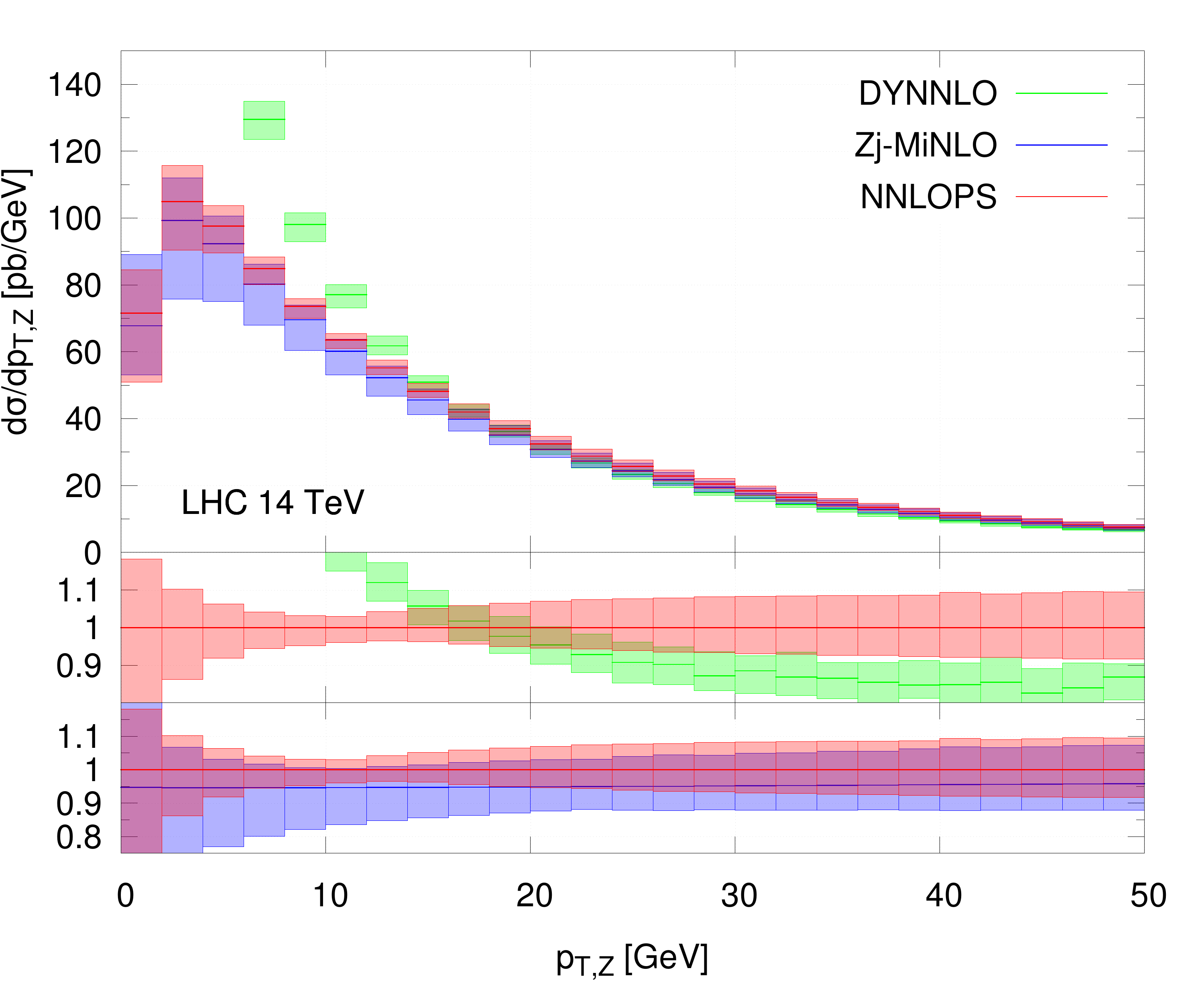}
  \caption{\label{fig:VLF:theory:nnlops}
    Lepton transverse momentum in the inclusive production of a charged lepton pair 
                   calculated at NNLO QCD accuracy matched to the parton shower 
                   using the \protect\UNNLOPS method as implemented in \protect\Sherpa, compared to \MCatNLO and NNLO calculations (left), 
                   figure taken from \cite{Hoeche:2014aia}.  
    Reconstructed $Z$ boson transverse momentum in the production of a charged lepton pair 
                    calculated at NNLO QCD accuracy matched to the 
                    parton shower using the \protect\MINLO method as implemented 
                    in \protect\Powheg interfaced to \protect\PythiaEight, compared to DYNNLO and NNLOPS calculations (right),
                    figure taken from \cite{Karlberg:2014qua}.
  }
\end{figure*}

The previously-discussed high-precision calculations suffer from one important short-coming: they are parton-level 
calculation and do not fully account for parton evolution or 
non-perturbative effects. 
Thus, to arrive at either particle-level\footnote{
  Particles with a lifetime of $c\tau>10\,\text{mm}$ 
  are considered stable by the typical collider experiments.
  The stage of event evolution where all remaining particles are
  stable on this scale is referred to as particle-level.
} predictions that can be directly 
compared to detector-corrected experimental data, 
or at simulated detector read-outs 
to derive the detector corrections in the first place, these 
high-precision calculations need to be interfaced to parton-shower 
calculations, multiparton interaction and hadronisation 
models, as well as hadron decays. 
This is implemented in so-called Monte Carlo event generators, for example
\Herwig \cite{Bellm:2015jjp}, \Pythia \cite{Sjostrand:2014zea}, and \Sherpa \cite{Bothmann:2019yzt}.
They can produce fully 
differential calculations, i.e., results that explicitly provide 
the flavor and four momentum of every particle that is 
produced in a high-energy collision. 
This allows the predictions to be projected onto arbitrary 
observables a posteriori.

Within the Monte Carlo event generators, the parton showers (PS)
provide a fully-differential resummation of the 
parton splitting process in terms of their respective 
evolution variable, albeit at a lower theoretical accuracy, 
as compared to the inclusive observable-specific resummations 
discussed in the previous section.
They are matched to fixed-order expressions for the hard scattering at 
the leading, next-to-leading and next-to-next-to-leading orders to improve the 
precision of the calculation outside the strongly hierarchical 
regime. 
While the matching to LO matrix elements is trivial, there 
exist two general variants of matching strategies for the 
combination with next-to-leading order matrix elements, 
\Powheg \cite{Nason:2004rx,Frixione:2007vw,Hoche:2010pf} 
and \MCatNLO \cite{Frixione:2002ik,Hoeche:2011fd,Alwall:2014hca}. 
Both are formulated in a generic way and, especially the \MCatNLO 
method, can be applied in an automated way to arbitrary processes. 
These methods have been applied to inclusive vector boson 
production \cite{Frixione:2004wy,Alioli:2008gx,Hamilton:2008pd,
  Hoche:2010pf} as well as vector boson production in association 
with up to three jets \cite{Alioli:2010qp,Frederix:2011ig,Hoeche:2011fd,
  Hoeche:2012ft,Re:2012zi,Campbell:2013vha,Jezo:2016ypn,Siegert:2016bre}. 
In addition, at least in the \Powheg approach for the inclusive Drell-Yan production, also next-to-leading 
order electroweak corrections have been matched to co-evolving 
QCD+QED parton showers \cite{Bernaciak:2012hj,Barze:2012tt,Muck:2016pko,Barze:2013fru}. 
They constitute the state-of-the-art tools to calculate 
the Standard Model predictions for electroweak precision 
measurements, such as the $W$ mass, the angular coefficients 
and charged-lepton asymmetry in lepton-pair production.

The current logarithmic accuracy of parton showers, however, 
only allows them to be matched to matrix elements at NNLO 
accuracy with the most trivial color structure. 
Thus, results are only available for inclusive $W$ and $Z$ 
boson production, but not for their production in association 
with a jet. 
Here, again, two different schemes exist: for \NNLOPS matched 
calculations, \MINLO \cite{Hamilton:2012np,Hamilton:2012rf} 
simulations reweighted to inclusive NNLO distributions 
\cite{Karlberg:2014qua,Monni:2019whf}, and \qT-slicing combined with 
\MCatNLO predictions in the \UNNLOPS scheme \cite{Hoeche:2014aia}. 
Figure~\ref{fig:VLF:theory:nnlops} details the results 
of both approaches, including their uncertainties, and compares 
them to the fixed-order results of the same accuracy. 
In both approaches, the advantages of combining the 
resummation properties of the parton shower with the 
fixed-order matrix element become apparent throughout the 
respective spectrum. The otherwise unphysical description of 
the low transverse momentum region of the weak boson is 
now described in a reliable way.
For prompt photon production, as it is always accompanied 
by a jet at leading order, no \NNLOPS description is available.

\begin{figure*}[t!]
  \centering
  \includegraphics[width=0.47\textwidth,height=0.47\textwidth]{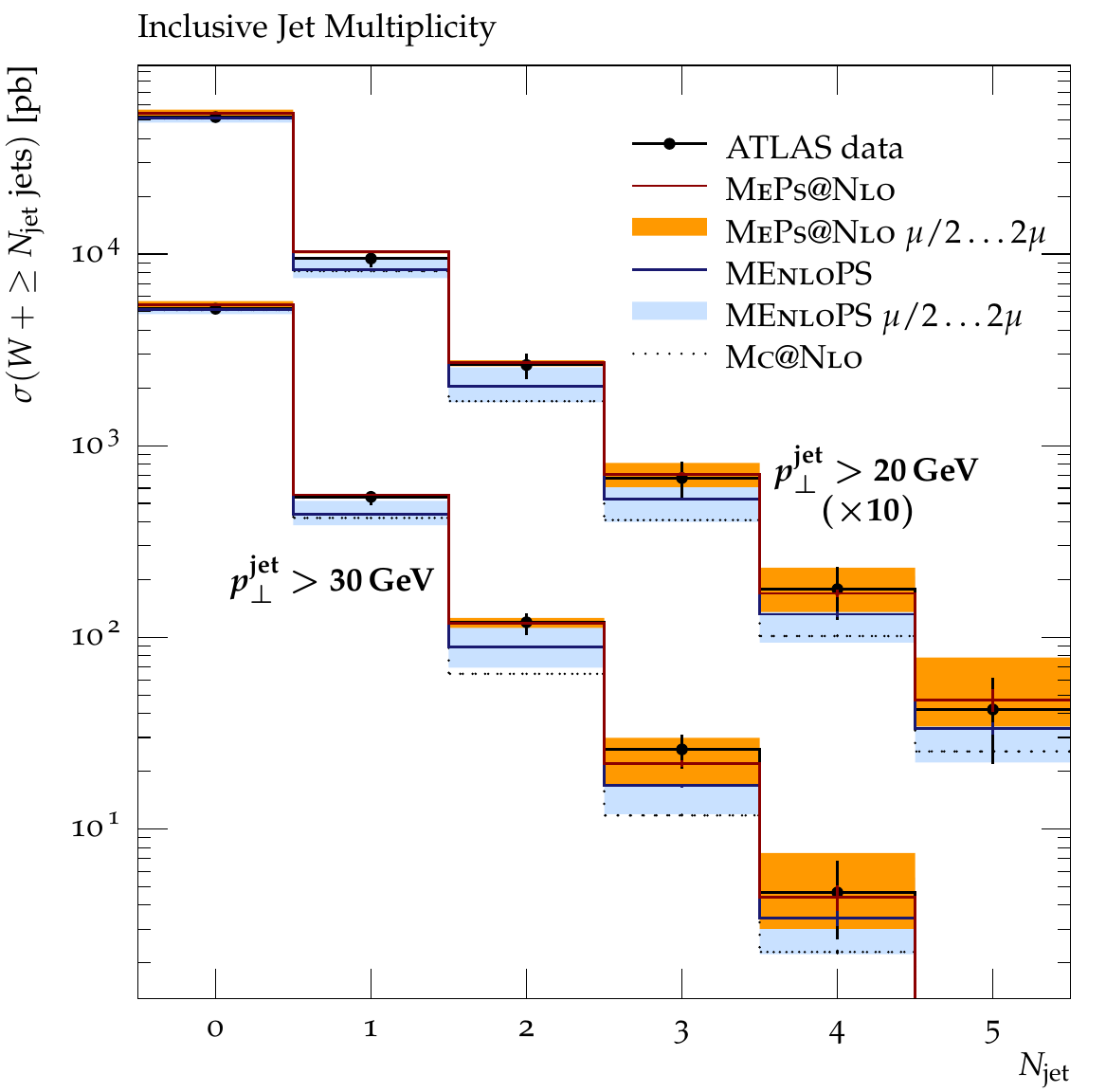}
  \hfill
  \includegraphics[width=0.47\textwidth,height=0.47\textwidth]{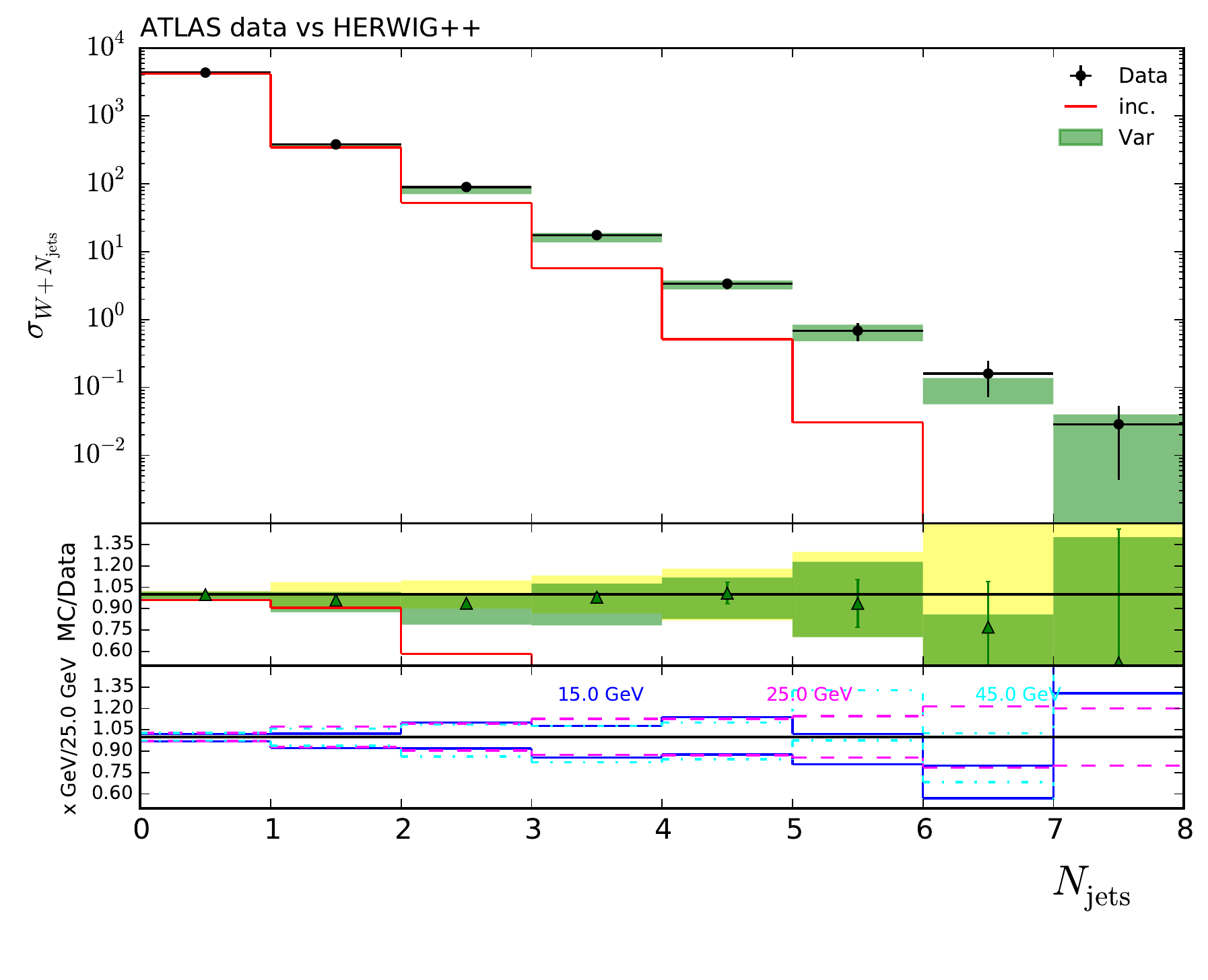}
  \caption{\label{fig:VLF:theory:merged}
    Inclusive jet multiplicity in the production of a charged lepton and a neutrino  
                   in association with jets 
                   using the \protect\MEPSatNLO method 
                   and merging up to two jets at NLO and 
                   four jets at LO accuracy as implemented in \protect\Sherpa, compared to \protect\MENLOPS  predictions and \ATLAS data (left),
                   figure taken from \cite{Hoeche:2012yf}.  
    Exclusive jet multiplicity in the production of a charged lepton and a neutrino 
                    in association with jets 
                    using the \protect\FxFx method
                    and merging up to two jets at NLO accuracy 
                    as implemented in \protect\MGaMC 
                    and \protect\Herwigpp (right), 
                    figure taken from \cite{Frederix:2015eii}.
  }
\end{figure*}

Beyond the description of a fixed multiplicity at the highest 
possible accuracy, the inclusive production of a vector boson with any 
number of jets is of prime interest to the experiments.
Thus, multijet-merged calculations aim to combine the advantages 
of both the high-precision descriptions of hard and wide-angle 
radiation through fixed-order matrix elements 
with the excellent description of the soft-collinear intrajet dynamics 
offered by the parton shower. 
Prescriptions to merge multiple LO-accurate parton shower (\LOPS) 
calculations of successive jet 
multiplicities into inclusive calculations were derived about 
twenty years ago. 
They can be grouped in the \CKKW-like methods \cite{Catani:2001cc,
  Lonnblad:2001iq,Lavesson:2007uu,Hoeche:2009rj,Hamilton:2009ne,
  Hamilton:2010wh,Hoche:2010kg,Lonnblad:2011xx,Lonnblad:2012ng} 
on the one hand side, and the \MLM-like approaches \cite{Mangano:2001xp,
  Alwall:2007fs} on the other. 

The \CKKW-like approaches split the emission phase-space of a 
lower-order process into a matrix element region and a 
parton-shower region using the merging scale \Qcut\ as a 
separator. 
While the soft and collinear phase space is populated by the 
parton shower acting on lower-multiplicity matrix-elements, 
radiation into the matrix element region is vetoed.
This veto, as it is determined by the parton-shower 
emission probability, now provides the correct Sudakov weight 
for the higher-multiplicity matrix element to correctly 
include the respective resummation properties in this region. 

Conversely, the \MLM-like prescriptions attach an unconstrained 
parton shower to the matrix element configuration of each 
individual jet multiplicity, letting it run its course without 
any awareness of the merging scale which was used to define 
the matrix element region initially. 
Only at this point of the generation, jets are reconstructed using a given jet algorithm 
and matched, both in direction and transverse momentum, to 
their counterparts in the originating matrix element. 
If non-matching jets are found in either configuration the 
event is discarded, in this way providing the needed 
Sudakov weights. 
Although both methods yield comparable results \cite{Alwall:2007fs}, 
it should be noted that a formal proof of the mathematical 
correctness only exists for the \CKKW-like approaches 
\cite{Hoeche:2009rj}.

These methods have then subsequently been promoted 
to merging \NLOPS-matched calculations, both in 
the \CKKW-like approach, see Refs. \cite{Lavesson:2008ah,Hoeche:2012yf,
  Gehrmann:2012yg,Hoeche:2014rya} for the \MEPSatNLO variant,  
Refs. \cite{Lonnblad:2012ix,Platzer:2012bs,Bellm:2017ktr} for the \UNLOPS variant, and Refs. \cite{Alioli:2015toa} for the \Geneva variant, as well as in the \MLM-like prescription \cite{Frederix:2012ps} (\FxFx). 
In the former, possibilities to complement the \NLOPS-accurate prescriptions of the lowest few multiplicities 
with \LOPS-accurate higher multiplicities have been 
formulated \cite{Hamilton:2010wh,Hoche:2010kg,Gehrmann:2012yg,Hoeche:2014rya}.
Figure\ \ref{fig:VLF:theory:merged} displays the 
results of these state-of-the art computations compared 
to data taken by the \ATLAS experiment at the \LHC at 
7\,TeV. 
When contrasted, the NLO-accurate predictions prove 
to be superior in both their central values and their 
uncertainties. 

\begin{figure*}[t!]
  \centering
  \includegraphics[width=0.47\textwidth]{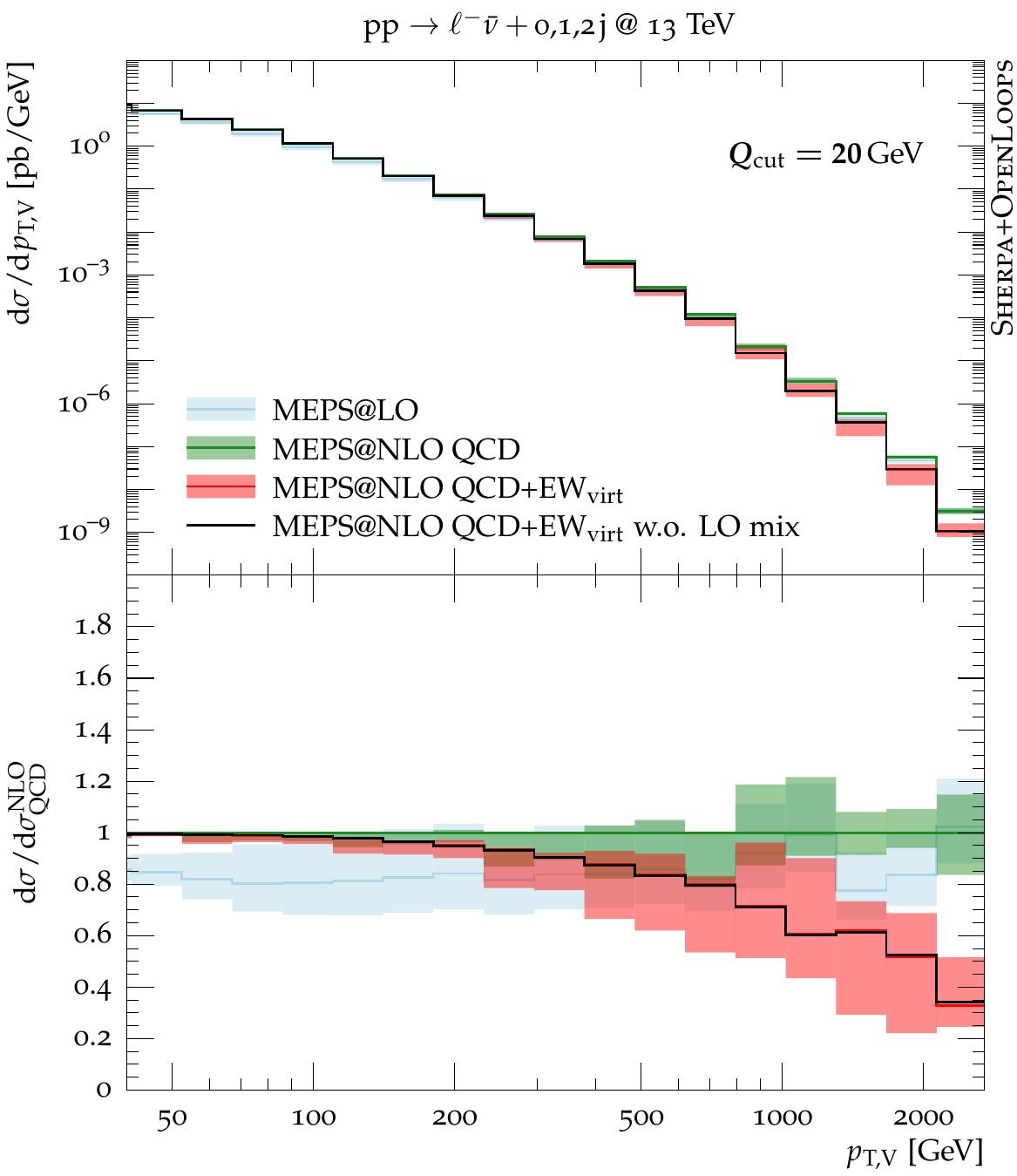}
  \hfill
  \includegraphics[width=0.47\textwidth]{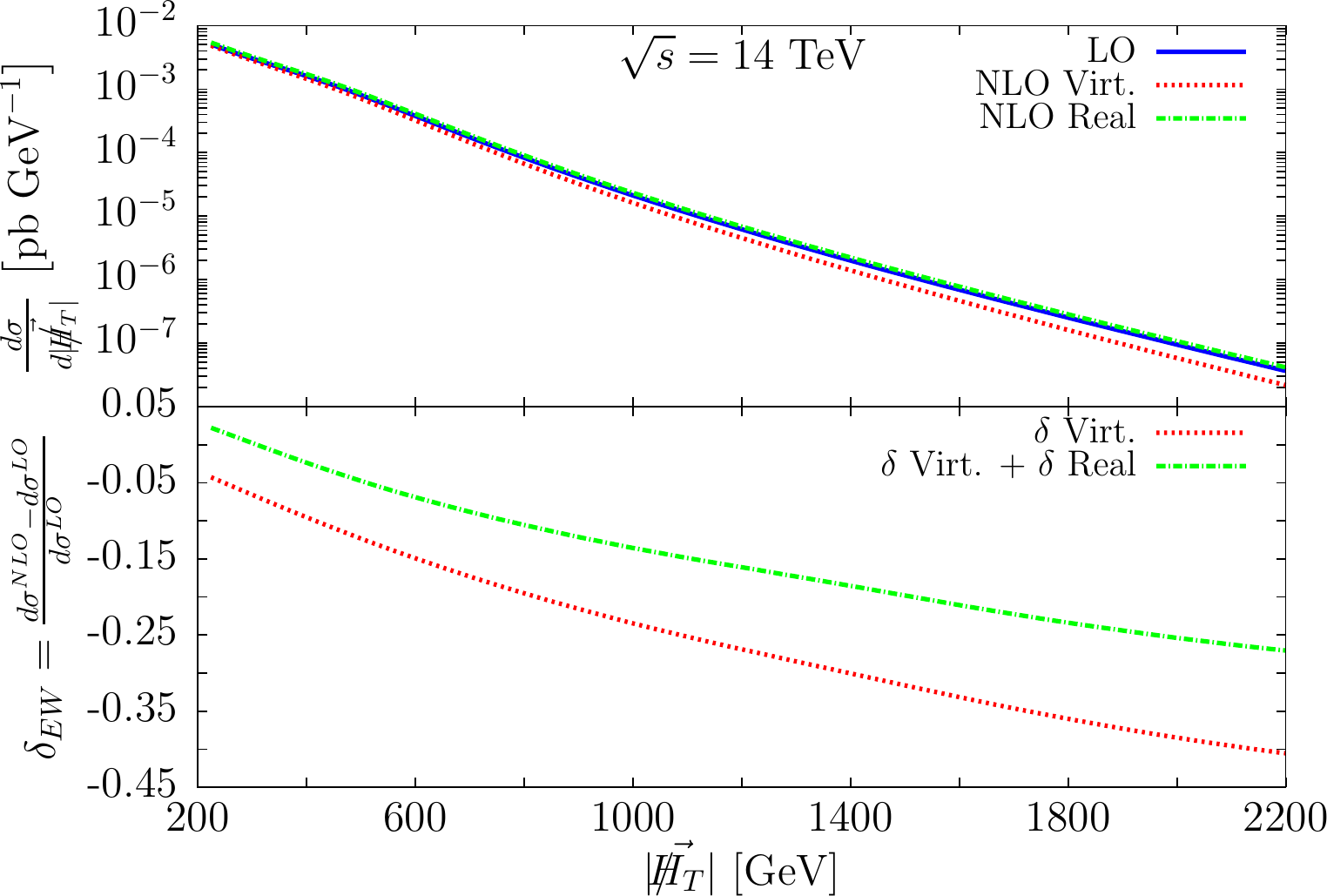}
  \caption{\label{fig:VLF:theory:merged_ew} 
    Reconstructed $W$ boson transverse momentum in the production of a charged lepton and a neutrino 
                   in association with jets calculated 
                   using the \protect\MEPSatNLO method including approximate electroweak 
                   corrections as implemented in \protect\Sherpa (left), 
                   figure taken from \cite{Kallweit:2015dum}.  
    Missing transverse momentum in neutrino-pair production in association 
                    with jets at leading order in QCD including EW corrections in the 
                    Sudakov approximation as implemented in \protect\Alpgen and \protect\Herwig (right), 
                    figure taken from \cite{Chiesa:2013yma}.
  }
\end{figure*}

As discussed in Section \ref{sec:VLF:theory:ho}, electroweak 
corrections are not only important for precision measurements 
but also in observables probing regions of large momentum transfers. 
So far, however, no solution to incorporate the exact NLO 
electroweak corrections in the TeV regime has been formulated. 
Nonetheless, there exist two methods to incorporate the 
dominant electroweak correction in this region in an approximate way. 
The first method \cite{Chiesa:2013yma} supplements the leading order 
matrix elements used in a LO-accurate multi-jet merged prediction with 
multiplicative EW Sudakov form factors \cite{Denner:2000jv,Denner:2001gw}. 
Conversely, the second method \cite{Kallweit:2015dum} completes 
the NLO QCD components of NLO-accurate multi-jet merged calculations 
with exact NLO EW virtual corrections and approximate NLO EW real 
emission corrections integrated over the single-emission phase space. 
The latter can, where needed, be supplemented with subleading LO 
corrections to account for further relevant contributions. 
Results for both methods are shown in Fig.\ \ref{fig:VLF:theory:merged_ew}, 
and the general feature of a logarithmic suppression of the production 
cross section, the so-called EW Sudakov correction, can be observed, 
e.g., reaching several tens of percent for the transverse momentum of 
the vector boson.

\begin{figure*}[t!]
  \centering
  \includegraphics[width=0.45\textwidth]{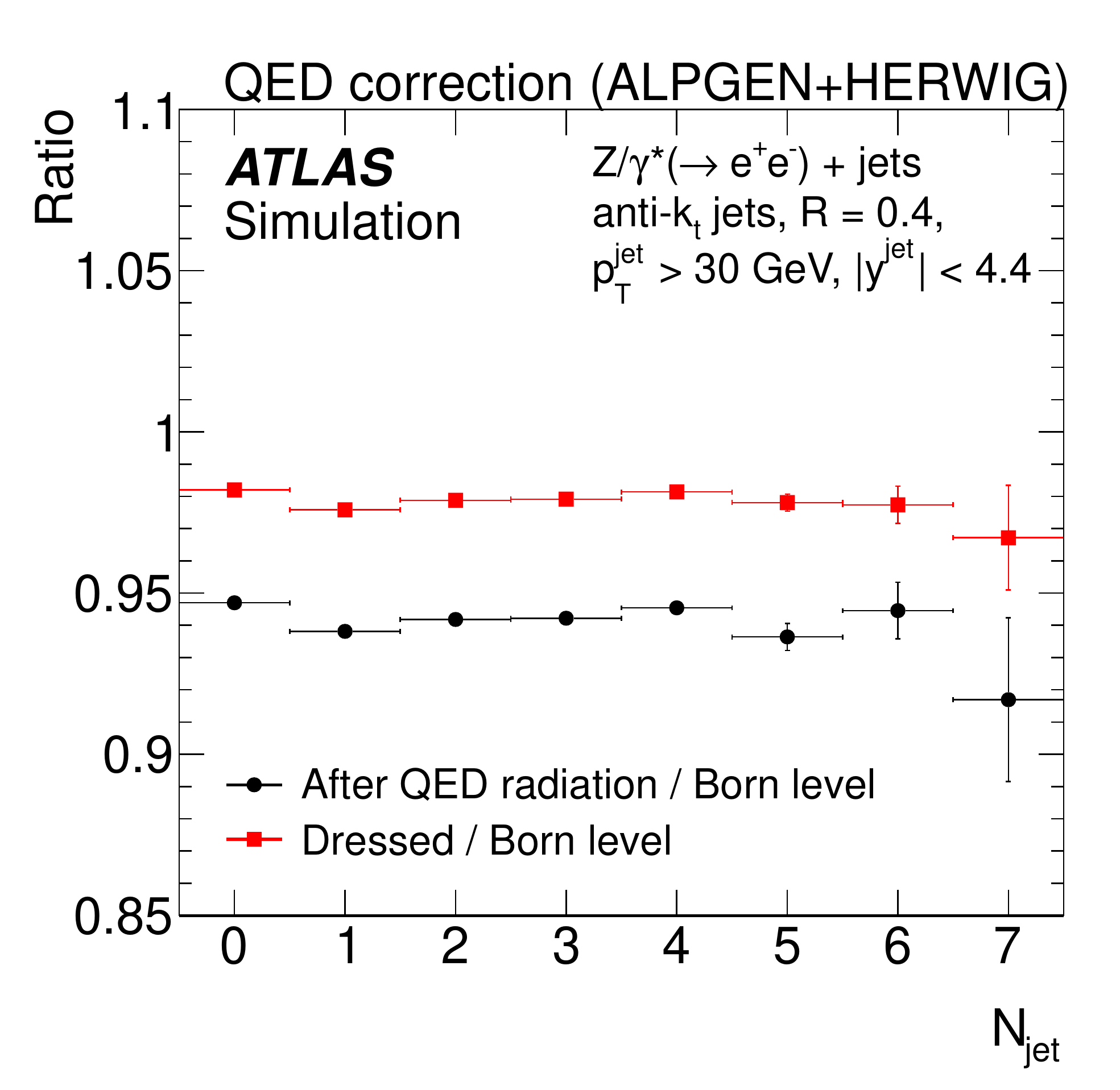}
  \includegraphics[width=0.45\textwidth]{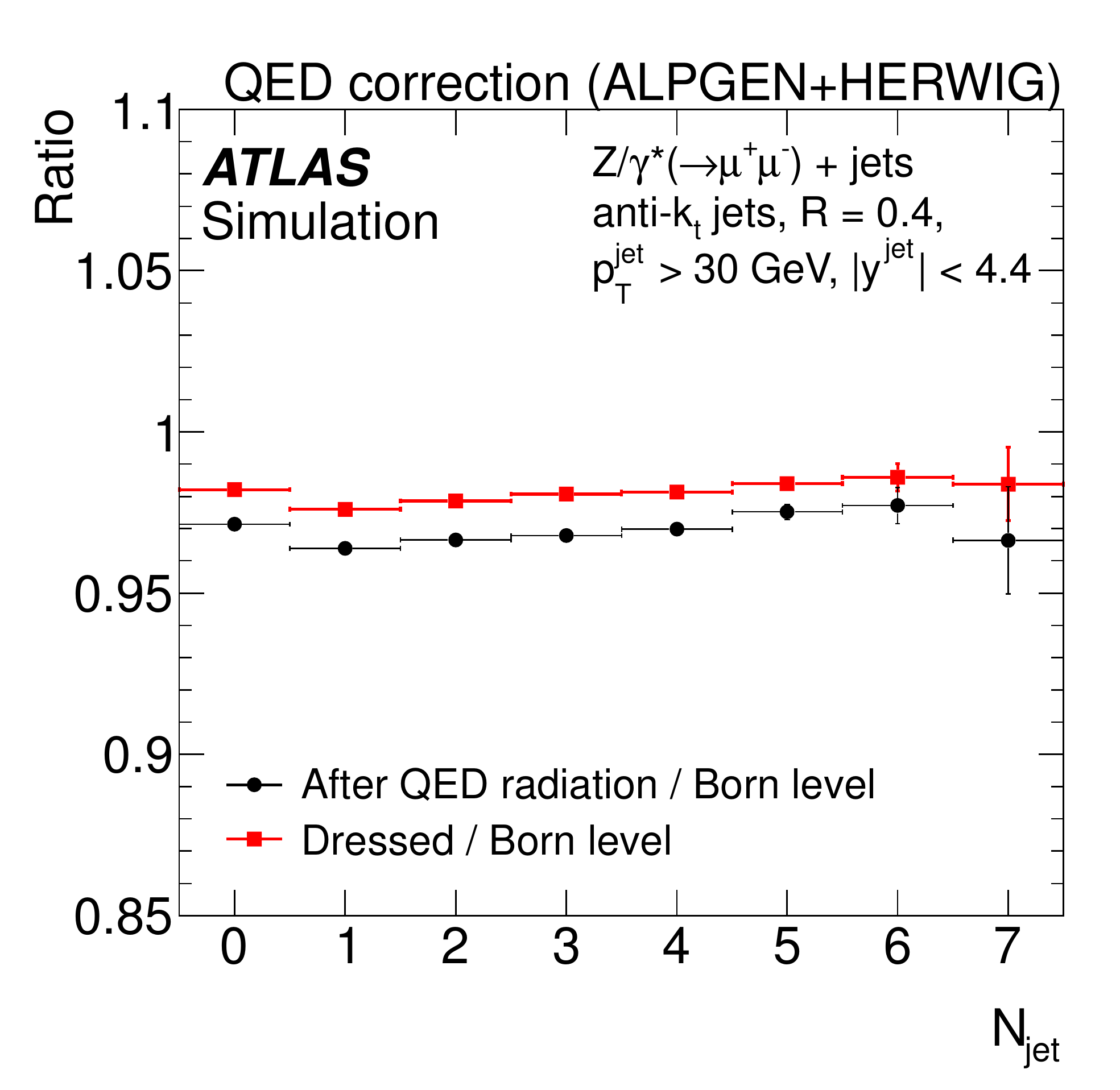}
  \caption{\label{fig:ATLAS_Zjets_7TeV_QEDcorr}
    QED final state correction factors from Born-level to the infrared-safe 
    bare and dressed charged lepton definitions as calculated using \Photos 
    \cite{Davidson:2010ew} 
    for electrons 
    (left) and 
    muons (right) as a function of the number of accompanying jets. Figure taken from 
    \cite{Aad:2013ysa}. 
  }
\end{figure*}

Electroweak effects also become relevant through the radiation of either 
massive weak bosons \cite{Christiansen:2014kba,Krauss:2014yaa} (again, mostly relevant for 
TeV scale objects), or photon bremsstrahlung 
\cite{Bloch:1937pw,Yennie:1961ad,Barberio:1990ms,Seymour:1991xa,Hamilton:2006xz,Schonherr:2008av}.
The latter mainly affects observables that depend on the charged lepton kinematics, ranging 
from a few percent on charged-lepton \pT-spectra to $\order(1)$ effects on 
invariant mass distributions below resonance peaks or threshold-induced shoulders.
In particular, since QED is an infrared-free theory, various different 
charged-lepton definitions can be and are used in measurements at various 
colliders: 
bare charged-leptons take the final-state charged-lepton at face value, 
and dressed charged-leptons recombine all photonic energy in a cone of size $\Delta R$ 
with the bare charged-lepton. 
While the bare charged-lepton definition demands a charged-lepton mass carried through at 
least some parts of the calculation \cite{Dittmaier:2008md}, the 
dressed charged-lepton definition is suitable also for calculations with 
massless leptons through its insensitivity to collinear radiation. 
The historic and occasionally still used Born charged-lepton definition relies on 
event record documentation entries\footnote{
  In event generators, before NLO EW parton-shower matched calculations were available and, in fact, 
  in most cases it is still the case now, the charged-lepton kinematics are generated first
  at LO accuracy, before dressing the interaction by photon radiation.
  The Born charged-lepton definition then relates the physical bare, or dressed 
  charged-lepton, to its Born-level counterpart by using the recorded technical 
  details of how the above calculation was carried out. 
} and is not infrared safe at any higher order. 
It thus should be abandoned, for precision measurements in particular.
Figure~\ref{fig:ATLAS_Zjets_7TeV_QEDcorr} details the corrections effected through photon bremsstrahlung 
where the dominant effect originates in the changes of the charged-lepton transverse momentum and invariant mass distributions, which influence the efficiency of the kinematic selection.

\begin{figure*}[t!]
  \centering
  \hspace*{-0.05\textwidth}
  \includegraphics[width=0.35\textwidth,height=0.38\textwidth]{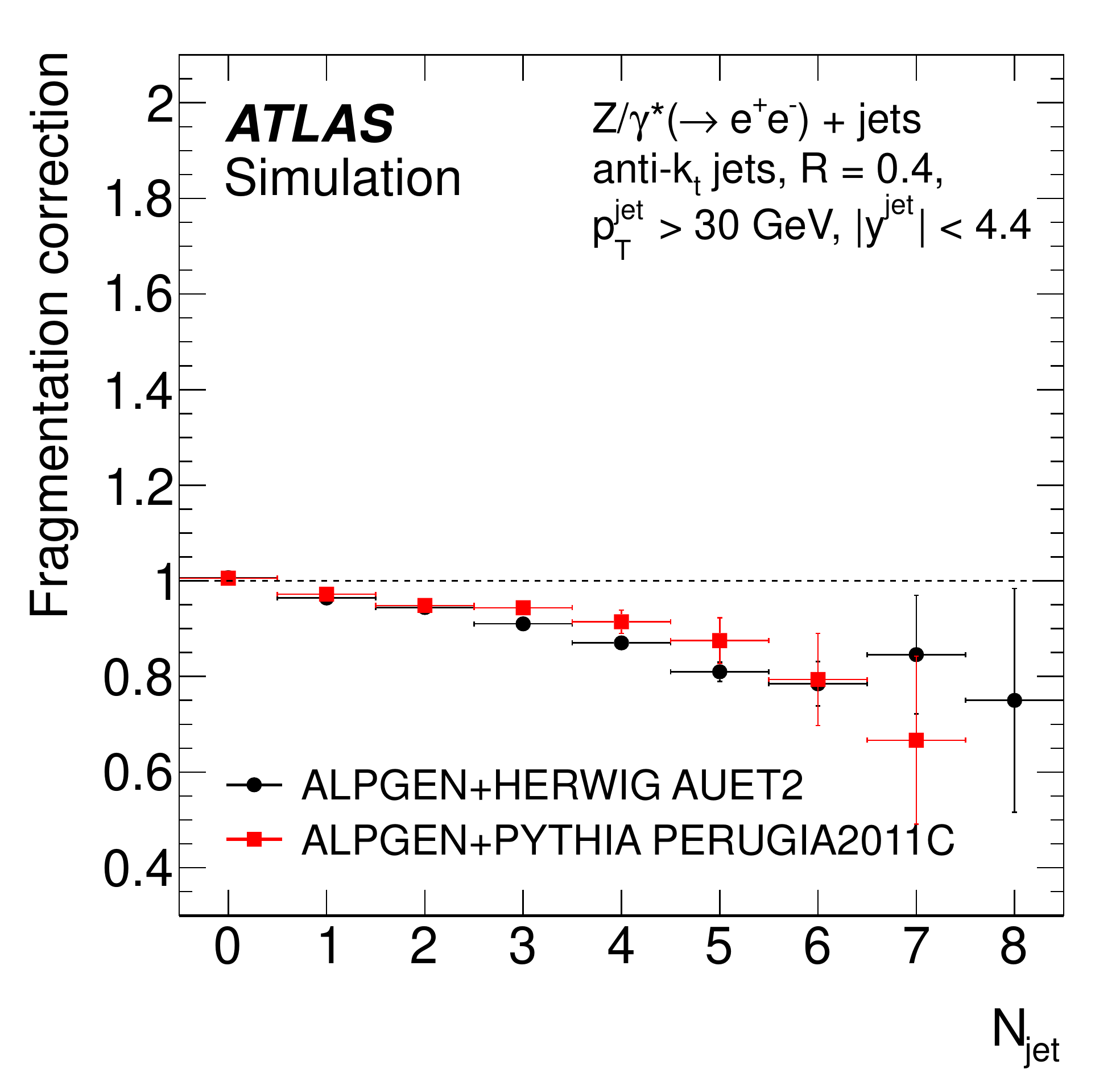}
  \hspace*{-0.02\textwidth}
  \includegraphics[width=0.35\textwidth,height=0.38\textwidth]{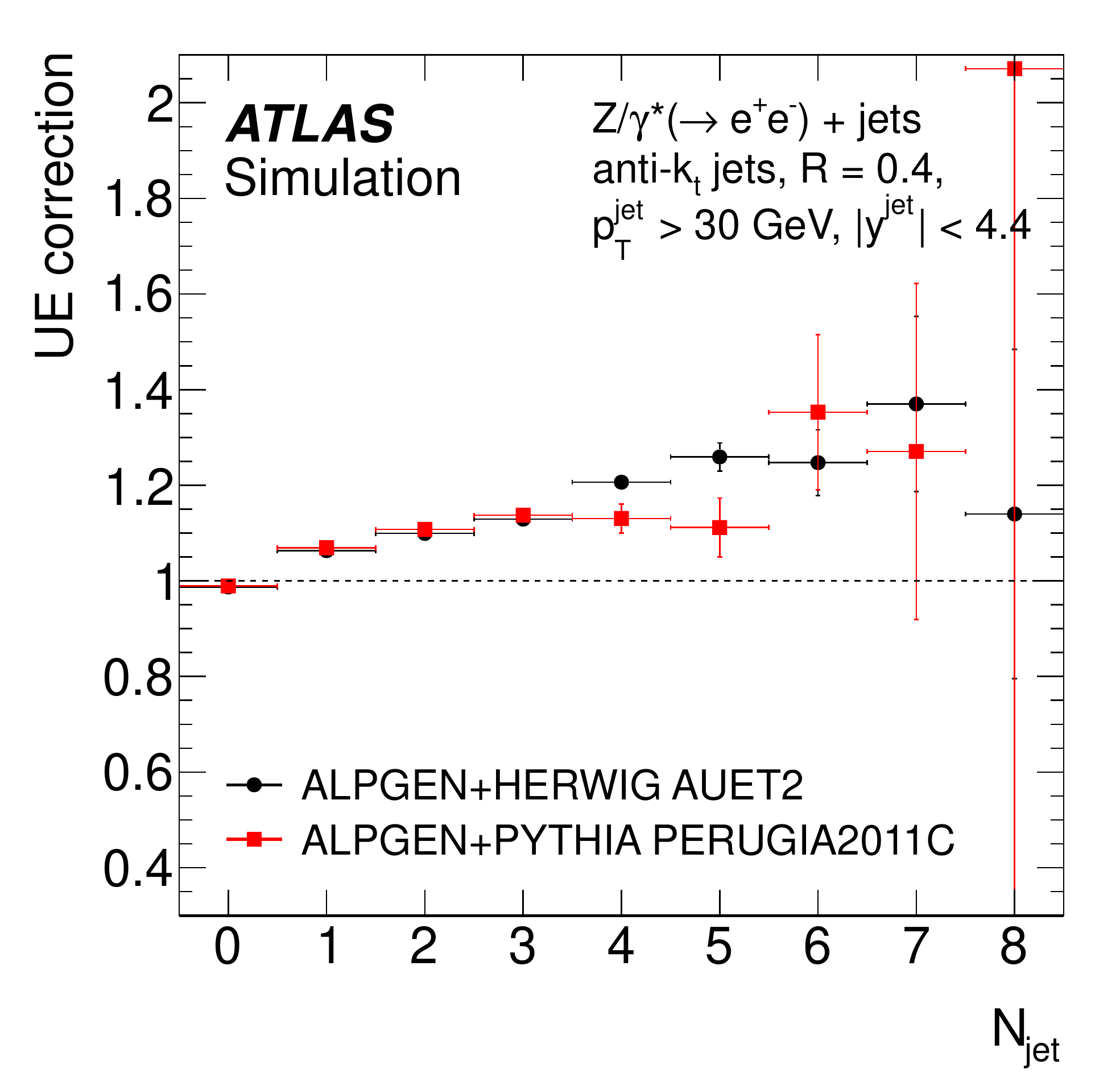}
  \hspace*{-0.02\textwidth}
  \includegraphics[width=0.35\textwidth,height=0.38\textwidth]{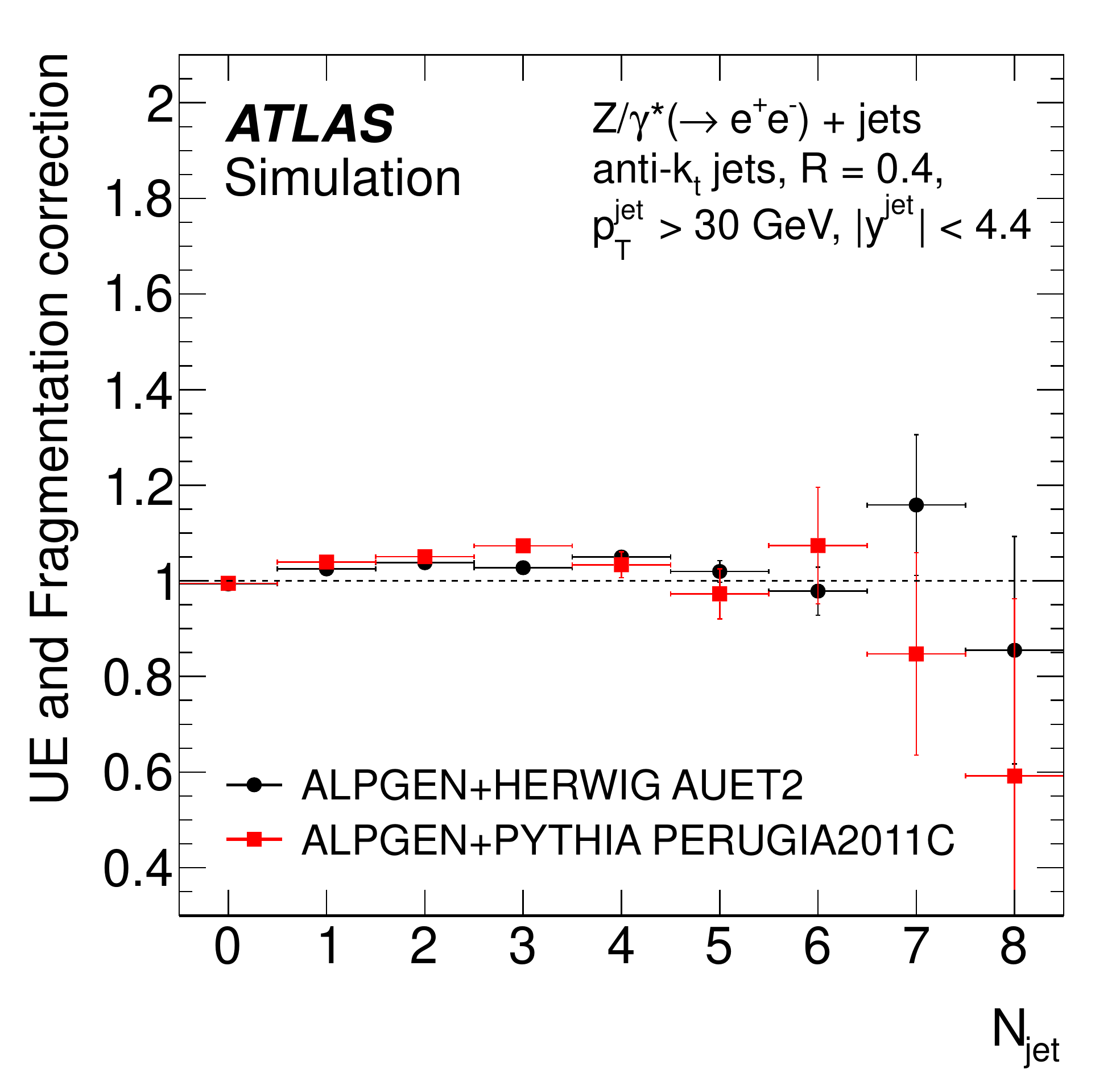}
  \caption{\label{fig:ATLAS_Zjets_7TeV_NPcorr}
    Non-perturbative correction factors as calculated using \Herwig and \Pythia 
    for the \ATLAS $Z+\text{jet}$ measurement in the electron-pair decay channel 
    as a function of the number of accompanying jets: fragmentation (left), 
    underlying event (middle) and the result of the two (right). Figure taken from 
    \cite{Aad:2013ysa}. 
  }
\end{figure*}

Besides providing a fully exclusive event description at parton level, 
event generators comprise tools to calculate non-perturbative effects 
like multi-parton interactions
\cite{Sjostrand:1987su,Sjostrand:2004pf,Corke:2009tk,
      Butterworth:1996zw,Bahr:2008dy,Gieseke:2016fpz}, 
parton-to-hadron transition \cite{Field:1976ve,Field:1982dg,
  Andersson:1983ia,Webber:1983if,Winter:2003tt} 
and hadron decays to arrive at 
a fully differential event description at particle level. 
Since methods to calculate these effects 
on timescales of $\order(1\,\text{CPU-s})$ per event are currently not available, phenomenological models with tunable, 
a priori unknown, parameters are used. 
These parameters, believed to be universal, have to be determined by a 
finite set of measurements 
in dedicated phase space regions, to be used in all other calculations.
The size of these non-perturbative corrections is typically estimated 
using \Herwig and \Pythia, and Fig.~\ref{fig:ATLAS_Zjets_7TeV_NPcorr} shows an example of these corrections for an \ATLAS $Z+\text{jet}$ study at the \LHC. 
These corrections are usually applied on partonic calculations, e.g., fixed-order NNLO QCD calculations described in Sec.\ \ref{sec:VLF:theory:ho}, to be compared to data.

\subsection{Experimental results}
\label{sec:VLF:exp}

\begin{figure*}
\centering
\includegraphics[width=0.73\textwidth]{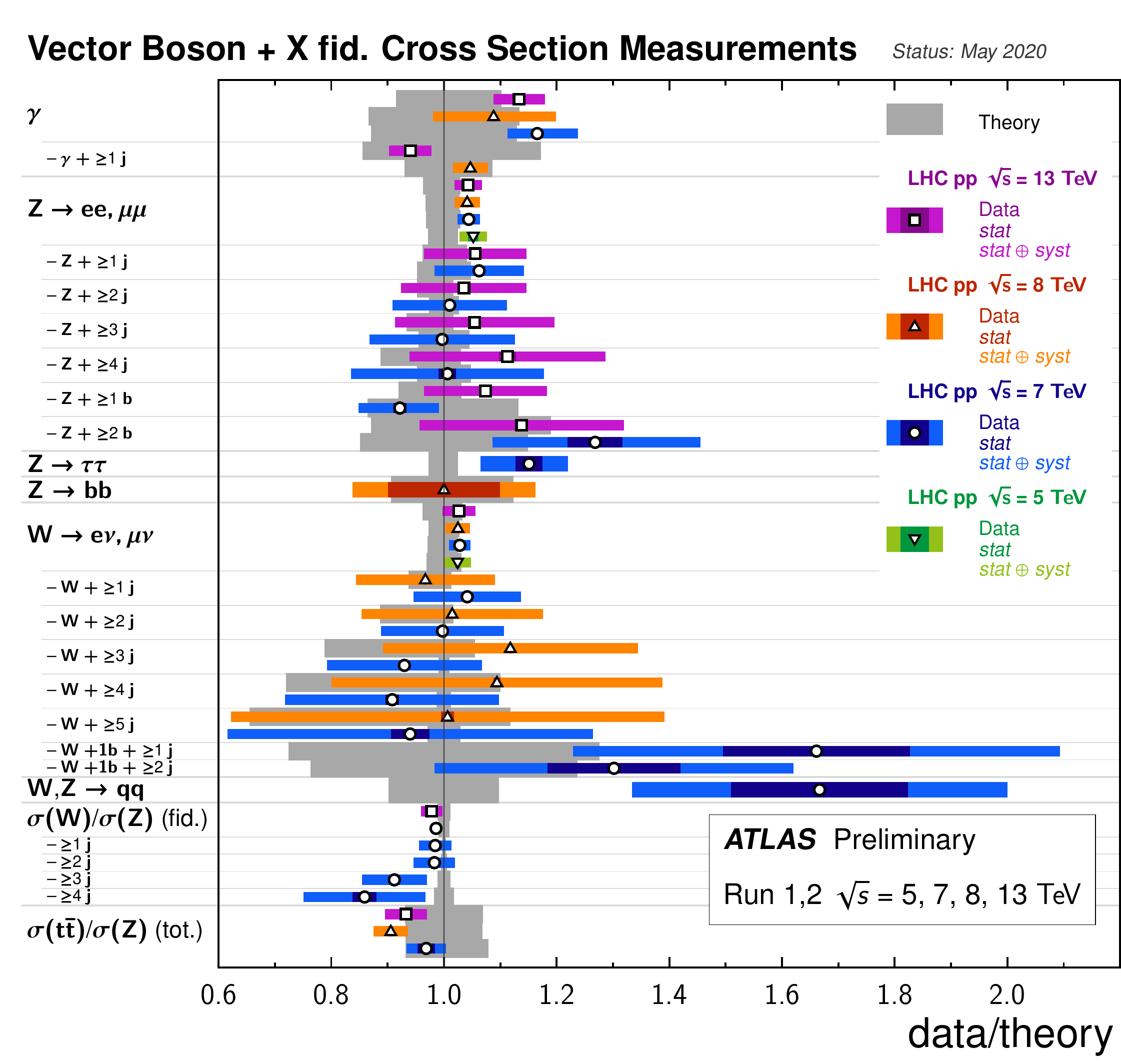}
\caption{\label{fig:ATLAS_Summary_Vjets} Summary of ratios of vector boson $+ \text{X}$ cross section measurements and predictions at 5, 7, 8 and 13 TeV center-of-mass energies in \pp collisions at the \LHC. Figure taken from~\cite{ATL-PHYS-PUB-2019-024}.}
\end{figure*}

 The extensive program for measurements of \Vjet processes at the \Tevatron has provided a critical incentive for the development of sophisticated higher-order calculations and MC generators. In the first years of the \LHC data-taking, \LHC experiments benefited from the availability of accurate calculations and MC generators tuned to \Tevatron data, however the production of \Vjet events at the \LHC is not a simple rescaling of \Tevatron scattering, therefore a new program for extensive measurements of \Vjet processes was setup early on in the \LHC physics program and \Vjet papers were among the first published by the \LHC collaborations~\cite{Aad:2010sp, Khachatryan:2010fm,Aad:2010ab, Chatrchyan:2011ne}. While the first \LHC measurements of \Vjet processes established SM measurements and assessed the validity of theoretical predictions at the \LHC energy scales, later measurements considerably improved the experimental precision, reaching the percent level, and were thus able to expose data-prediction discrepancies and shortcomings in calculations. Such precision measurements highlighted the need for the development of more precise higher-order calculations in QCD and electroweak physics at the \LHC, as detailed in Sec.~\ref{sec:VLF:theory}. 

With the high precision achieved by \LHC experiments the accurate definitions of the quantities that are experimentally measured is of great importance, as they must be theoretically sound, i.e., independent of the order of the theoretical approximation used, 
 they must be related to fundamental physical quantities rather than parameters in theoretical models 
and as close as possible to experimental definitions to minimize model-dependent extrapolations. These general guidelines allow for accurate comparison of experimental results with theoretical predictions and ensure that results can be compared with future predictions without prior knowledge of the experimental apparatus or possibly dated theoretical models.
In this spirit, \Vjets measurements at the \Tevatron and the \LHC are primarily reported in fiducial phase spaces.
Cross section measurements for \WZjet processes at the \LHC are reported with decay charged leptons defined at ''dressed-level'', and corrections to a Born-level definition are often provided (see Sec.\ref{sec:VLF:theory:mc}). The cross sections measured in different
decay channels of the $W$ or $Z$ bosons can be combined when the charged leptons are defined at Born level, however channel combination with charged leptons defined at dressed level is also done with a per-mille accuracy, i.e., below the experimental precision of the measurements. For a discussion of phase space and particle definitions at the \LHC see Ref~\cite{ATL-PHYS-PUB-2015-013}. For the measurements of the \gammajet production cross sections, isolation requirements are imposed on the photon to improve the identification at detector-level and to suppress the contribution of photons from the fragmentation of quarks and gluons at particle level (see Sec.~\ref{sec:VLF:theory}).

Experimental measurements include absolute or normalized differential cross sections in a fiducial phase space. The differential cross sections are measured as a function of several observables, i.e., event-based observables (jet multiplicity $N_{\rm jets}$, boson transverse momentum \pT, \HT that is the scalar sum of the \pT of clustered jets, event-shapes etc.), and  
jet-based observables ($n^{\rm{th}}$-jet \pT or rapidity $y$). Measurements also include angular correlations between final-state objects (jet-jet, lepton-jet, $Z$-jet or $\gamma$-jet etc.), for example the azimuthal difference $\Delta \phi$, the rapidity difference $\Delta y$, $\Delta R=\sqrt{\Delta\phi^2 + \Delta\eta^2}$, or the invariant mass of the two leading jets $m_{jj}$. 

Particle-level measurements are finally compared to theoretical predictions from MC simulations or to fixed-order calculations.  While MC simulations provide particle-level final states, fixed-order calculations (e.g., by \BlackHat~\cite{Bern:2013gka,Berger:2008sj,Berger:2010vm,Ita:2011wn,Berger:2009zg,Berger:2010zx}, \MCFM~\cite{Campbell:2010ff} or \Jetphox~\cite{Aurenche:2006vj,Catani:2002ny}) are at parton level and are often corrected for non-perturbative effects, such as underlying event and hadronisations ($3-4\%$ corrections), as discussed in Sec.~\ref{sec:VLF:theory:mc}.
Uncertainties in fixed-order NLO calculations in perturbative QCD (pQCD) due to missing higher-order terms are (conventionally) estimated by variations of the scales (renormalization and factorization), and are typically found in the $4-20\%$ range. These are followed by uncertainties on parton densities in the $1-4\%$ range, and on $\alpha_s$ in the $1-3\%$ range.

Figure~\ref{fig:ATLAS_Summary_Vjets} shows the breadth of \Vjet measurements at different center-of-mass energies at the \LHC and gives an overview of the level of agreement between measurements and state-of-the-art theoretical predictions. It is impressive to see such a level of agreement overall, however in the most precise experimental measurements, e.g., the ratios of \Wjets and \Zjets, discrepancies are visible. These discrepancies become more significant, up to two standard deviations or greater, in some regions of phase space in differential cross section measurements. 
These measurements and comparisons with theoretical predictions will be presented in greater detail in following sections.

\subsubsection{Experimental event reconstruction}
\label{sec:VLF:exp:physobj}

In experimental analyses, \Vjet events are reconstructed by identifying particles, such as photons, leptons, and clusters of particles such as jets, and applying selection requirements to purify the data samples. After background subtraction and corrections for the detector efficiency and resolution ({\it unfolding}), the production cross sections are measured inclusively or differentially in a fiducial phase space at particle level that is defined as close as possible to the detector-level kinematic selection. The unfolding of experimental results is an important part of the process of extraction of experimental measurements as it allows for the direct comparison with theoretical predictions
with no prior knowledge of the detector layout, efficiency or resolution.

\paragraph{Particle reconstruction: photons, electrons, muons, missing transverse momentum and jets}

\mbox{}\newline

The selection of photons is based on energy clusters reconstructed in the electromagnetic calorimeter, and depending on the number of matching tracks in the tracker they can be classified as unconverted or converted photons.
Photons are reconstructed within the tracker acceptance, typically of $|\eta|<1.0$ at the \Tevatron  and $|\eta|<2.37-2.5$ in the \ATLAS and \CMS detectors at the \LHC. See~\cite{Aad:2019tso,Khachatryan:2015iwa} and references therein for a representative selection of recent articles on photon reconstruction, calibration and identification strategies and performance at \ATLAS and \CMS. Since the reconstruction of the photon momentum relies on measurements of energy deposits in cells in the calorimeter system, the transverse momentum of the photon is often reported as  transverse energy (\ET).
Measurements of prompt-photon production require isolation of photons to avoid the large contribution from neutral-hadron decays into photons. While a smooth cone isolation criterion (see Sec.~\ref{sec:VLF:theory}) is used for photon isolation in theoretical calculations, experimentally a cone-base isolation technique is used as most suitable for finite-granularity detectors: the photon is required to be isolated based on the amount of transverse energy in a cone of typical size $\Delta R=0.4$ around the photon.
 
An electron is reconstructed as a charged-particle track geometrically associated with energy clusters in the electromagnetic calorimeter, while a muon is identified as a track segment in the muon system consistent with a track in the inner tracker, and can be associated with a minimum ionization signature in the calorimeters. Both electrons and muons are reconstructed within the inner tracker acceptance and are required to be isolated to further suppress background from misidentified objects, such as hadrons, or semi-leptonic heavy-flavor decays. 
The isolation requirements are tuned so that the electron or muon isolation efficiencies are high for signal, typically greater than $90\%$.
The reader is referred to the following articles and references therein for details on electron and muon reconstruction, calibration, and identification as well as their performance at the \LHC~\cite{Aad:2019tso,Khachatryan:2015hwa,Aad:2016jkr,Sirunyan:2019yvv,Aaij_2019,Aaij:2014jba,Aaij:2019uij}. Small correction factors, typically within $1\%$ are applied to correct differences in the photon, muon and electron efficiencies between data and simulation.

Jets are clustered from energy deposits in the calorimeters in the \ATLAS detector~\cite{Aad:2020flx,Aad:2015ina}
and from particle candidates reconstructed by a particle flow algorithm in the \CMS detector~\cite{Khachatryan:2016kdb,Sirunyan_2017,Sirunyan:2020foa}. Different  jet  algorithms  are  used  at  the  \Tevatron  and  the  \LHC.  At  the \Tevatron, iterative cone algorithms, e.g., MidPoint, with split-merge prescriptions to resolve cases of overlapping stable cones, are used with a typical cone radius of $R=0.4-0.7$.  The experiments at the \LHC migrated to the 
anti-$k_t$ infra-red and collinear-safe jet algorithm, that also produces cone-shaped clustered jets. 
Jets are calibrated based on the jet \pT response in MC simulations, and pileup (the particle production from multiple interactions per bunch crossing) contribution to the jet energy is subtracted on an event-by-event basis in the calibration process using data-driven techniques. In situ measurements of the momentum balance  in  dijet,  \gammajet,  \Zjet,  and  multi-jet  events  are  used  to  correct  for  any  residual difference in jet energy scale between data and simulation.

In \Wjet analyses with the $W$ decaying leptonically, the event selection purity can be improved by imposing a requirement on the presence of missing transverse momentum in the event since the neutrino escapes direct detection. The missing transverse momentum is calculated as the negative vectorial sum of the transverse momenta of final-state particles. In the \CMS experiment particles are reconstructed by a particle flow algorithm and are used as inputs to the computation of the missing transverse momentum, while in the \ATLAS experiment the selected final state particles, e.g., electrons, muons, photons and jets, are used together with soft particles that are not associated with any other selected object, i.e., low-energy deposits in the calorimeter or low-momentum tracks associated with the primary vertex~\cite{Aaboud:2018tkc,Sirunyan:2019kia,Aaij:2014jba}.

\paragraph{Event reconstruction: photon+jets}
\mbox{}\newline

Events with a photon and jets reconstructed in the final states are recorded using highly efficient (close to $100\%$ efficiency) single-photon triggers, see Refs.~\cite{Aad:2019wsl,Khachatryan:2016bia} and references therein for a representative selection of articles on photon trigger architecture and performance at the \LHC.
Despite the application of the tight identification and isolation requirements on the photon, a non-negligible background originating from hadrons misidentified as photons contaminates the selected sample. The signal purity is typically higher for photon reconstructed centrally in the detector, increases as the photon \ET or jet \pT increase, and can reach values in the 70–-$90\%$ range at high photon \ET.
The background is subtracted using data-driven methods based on signal-suppressed control regions. The photon reconstruction and selection efficiencies depend on the photon \ET and $\eta$ and are in the approximate range of 70–-$100\%$. They decrease with the increasing number of jets in the event, primarily due to the photon isolation requirement.

Photons are selected in a broad range of minimum \ET requirements, approximately 20 -- 200 GeV, similarly, the minimum jet \pT requirement varies and ranges between about 15 GeV and 100 GeV. 

The \gammajet cross section measurements are dominated by experimental systematic uncertainties, such as photon calibration and identification, and jet energy scale, at the level of few percent over a broad range of jet or photon transverse momenta. Only around the TeV energy scales, statistical uncertainty in data becomes the leading contribution at the \LHC.

\paragraph{Event reconstruction: \texorpdfstring{\WZjets}{W/Z+jets}}
\mbox{}\newline

Events with a $W$ or $Z$ boson provide clean experimental signatures in the leptonic decay channels that can be triggered by single high-\pT electron or muon, or low-\pT dilepton (electron and muon) triggers~\cite{Aad:2019wsl,Khachatryan:2016bia,Aaij:2018jht,Aaij:2019uij}. The two leptonic channels with one (in $W$ events) or two (in $Z$ events) electrons or muons provide useful cross-checks of the results and can provide additional information to constrain experimental uncertainties in the combination of the cross sections. 
In addition to requirements on the charged lepton and jet transverse momenta and (pseudo-)rapidity acceptance, in \Wjet events further requirements on the missing transverse momentum or on the transverse invariant mass are applied, while for \Zjet event reconstruction a requirement on the dilepton invariant mass is imposed in a window around the nominal $Z$ boson mass.
Typical selection requirements for \WZjet events include electrons or muons with a minimum \pT in the range 20--30 GeV within the tracker acceptance, jets with distance parameter in the range $R=0.4$--$0.7$ with \pT requirement in the range 20--50 GeV and in rapidity ranges that vary from about 2 to 5 units, with a minimal separation between the lepton and jets of $\Delta R$(lepton,jet) > 0.4--0.7. 
At the \lhcb experiment the weak boson decay charged leptons  are reconstructed in the forward pseudo-rapidity region, in the range $2.0 <  \eta  < 4.5$, while jets are reconstructed in the pseudo-rapidity range $2.2<\eta_{\rm{jet}}<4.2$.
%
\begin{figure}
\centering
\includegraphics[width=0.45\textwidth] {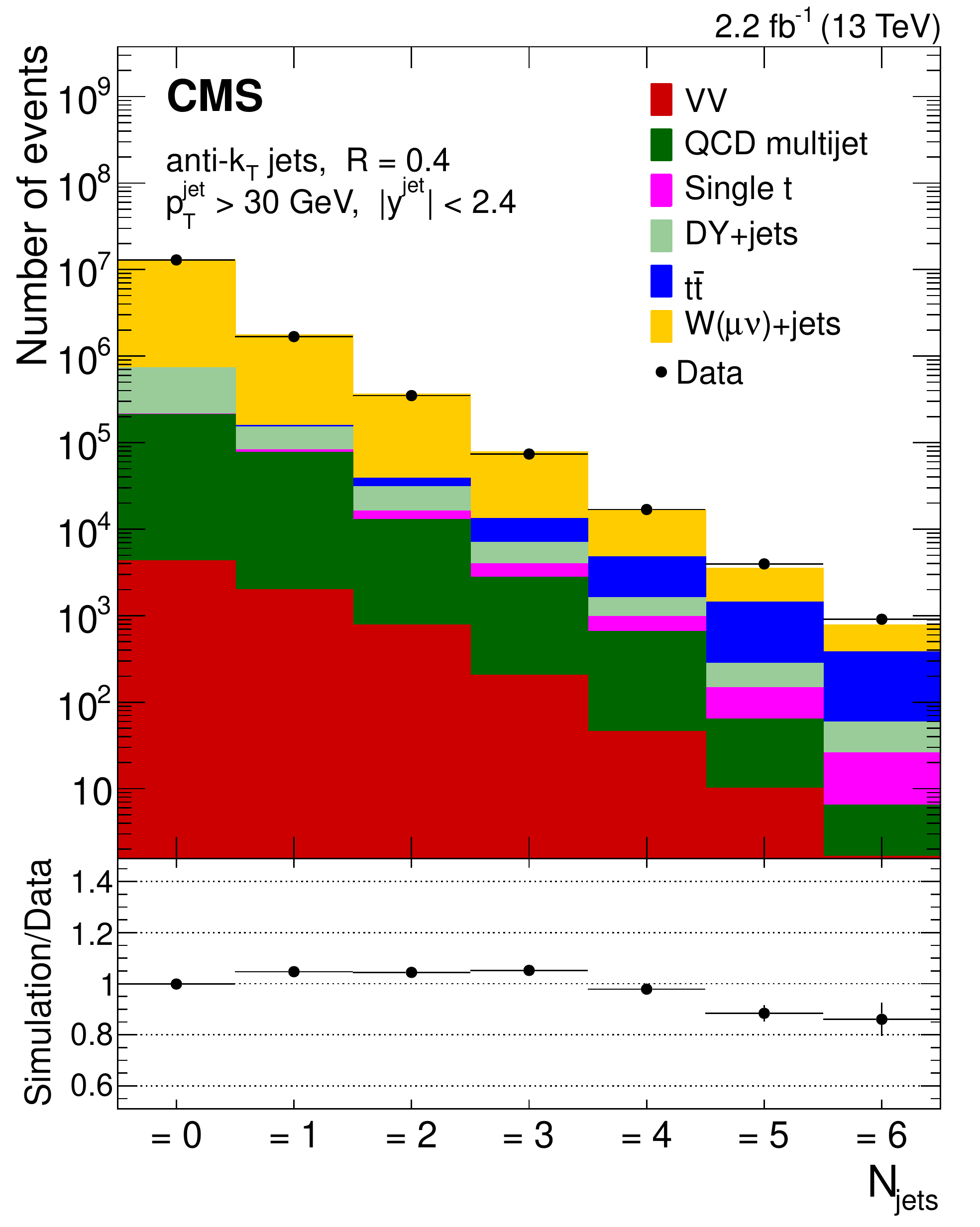}
\newline
\includegraphics[width=0.45\textwidth]{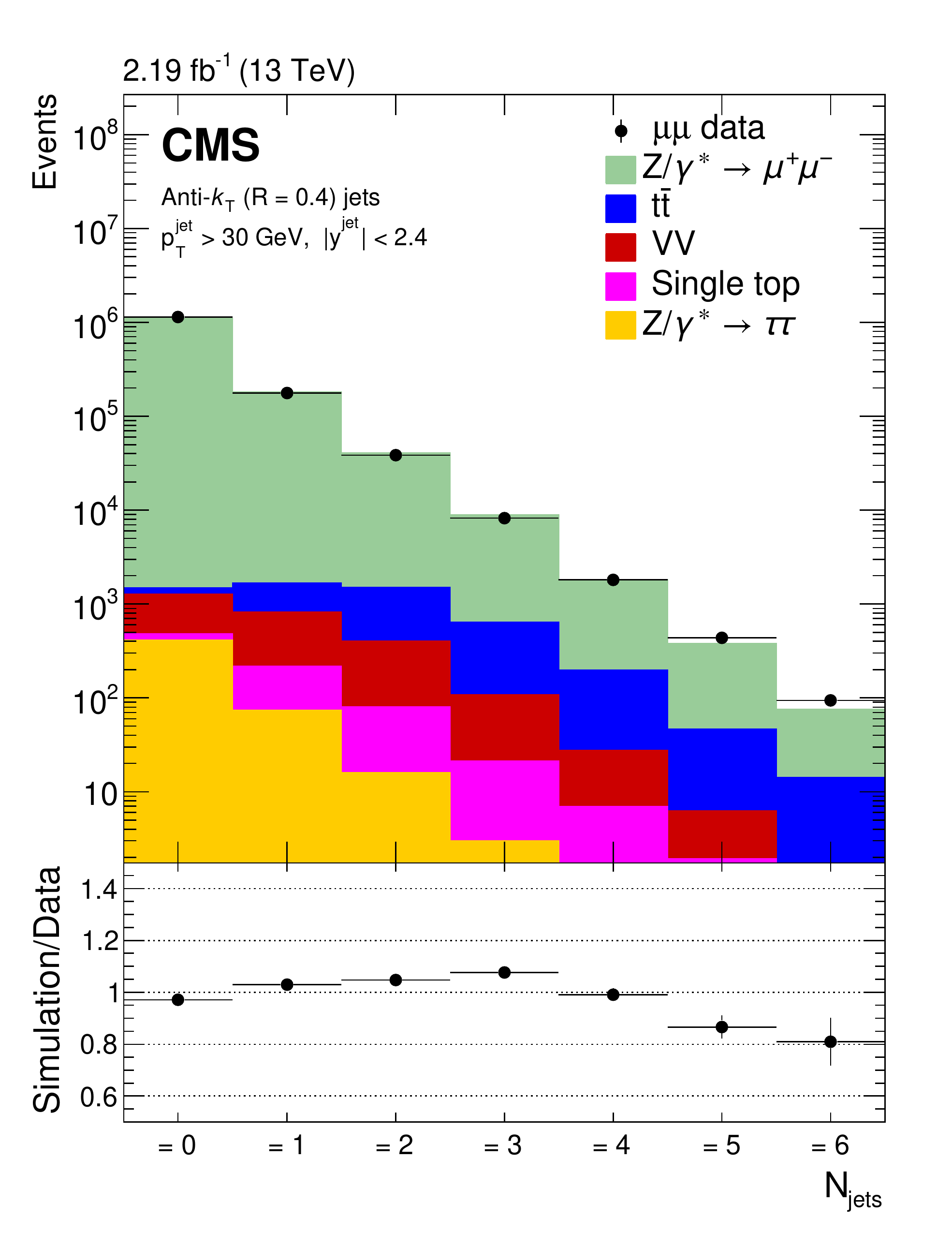}
\caption{\label{fig:CMS_ZWjets_8TeV_Njets_reco} Reconstructed data, simulated signal and background events in the jet multiplicity distributions in \pp collisions at 13 TeV center-of-mass energy at the \LHC for \Wjet events in the muon decay channel of the $W$ boson (top), figure taken from~\cite{Sirunyan:2017wgx}, and for \Zjet events in the muon decay channel of the $Z$ boson (bottom), figure taken from~\cite{Sirunyan:2018cpw}. 
}
\end{figure}
\begin{figure*}[t]
\centering
\begin{minipage}{0.90\textwidth}
    \centering
        \includegraphics[width=\textwidth]{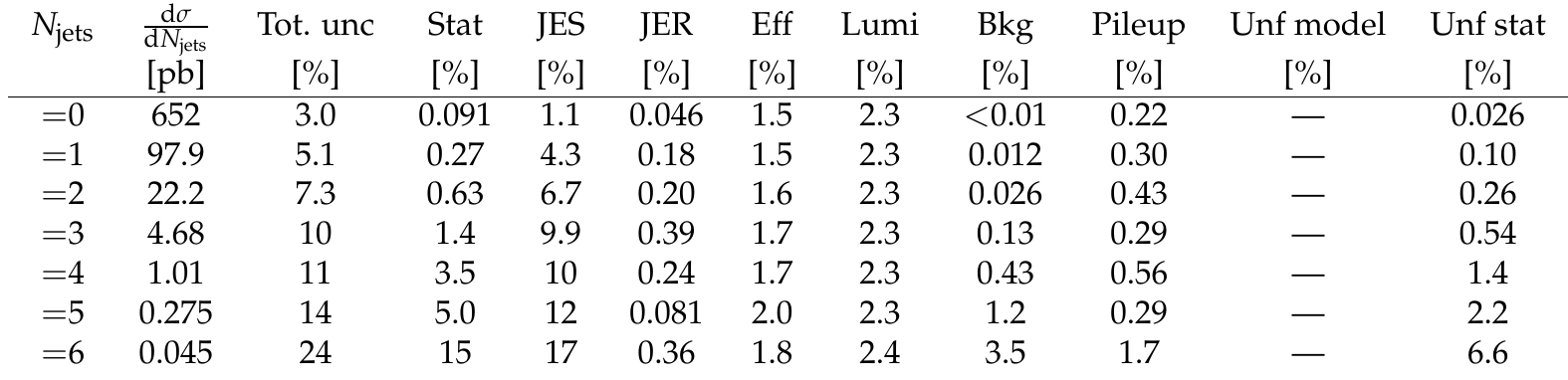}\\
\end{minipage}
\hfill
\begin{minipage}{0.55\textwidth}
    \centering
        \includegraphics[width=1\textwidth]{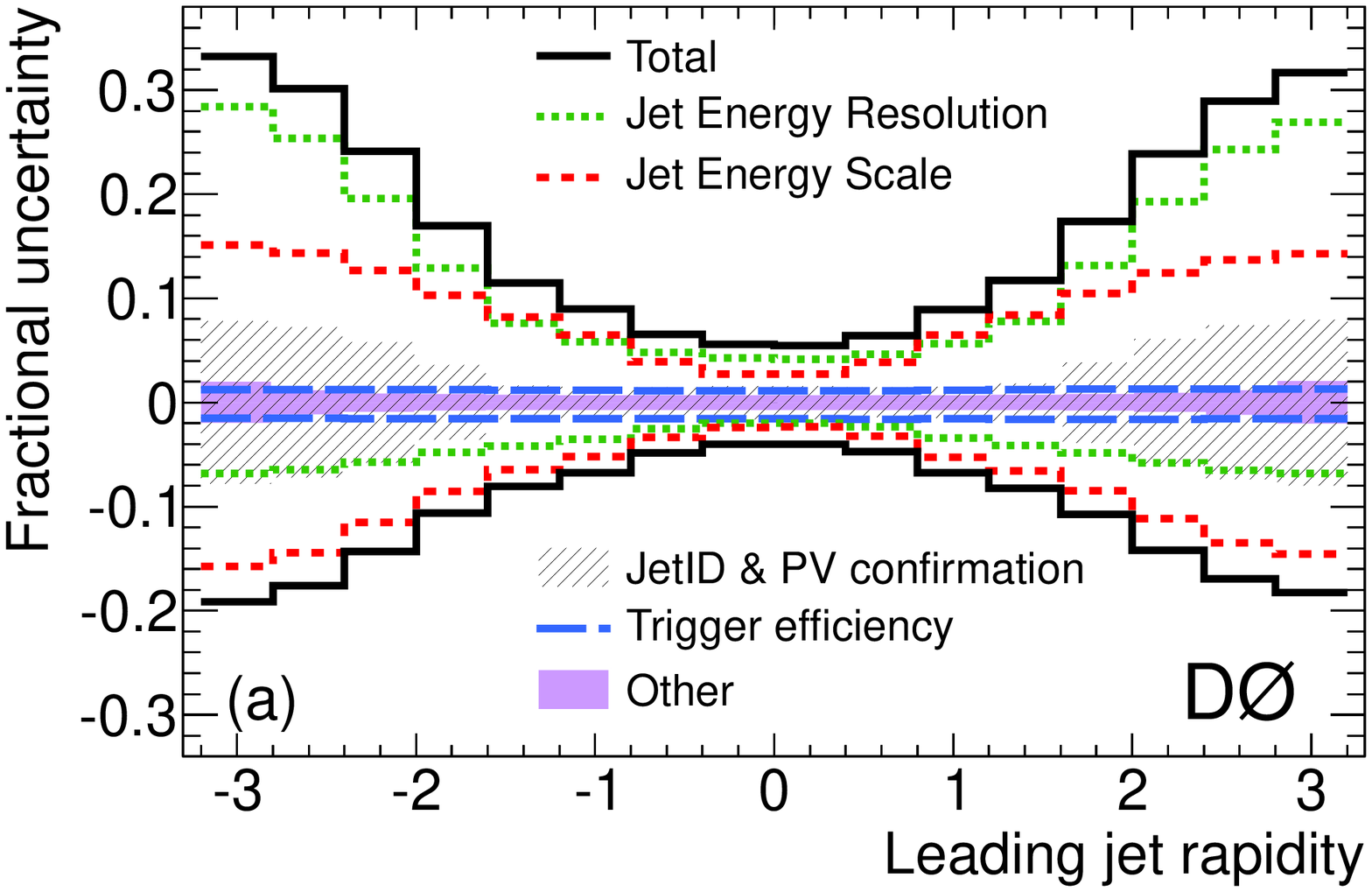}
        \end{minipage}
    \hfill
    \begin{minipage}{0.38\textwidth}
        \centering
        \vspace*{-0.02\textwidth}
        \hspace*{-0.1\textwidth}
        \includegraphics[width=1\textwidth]{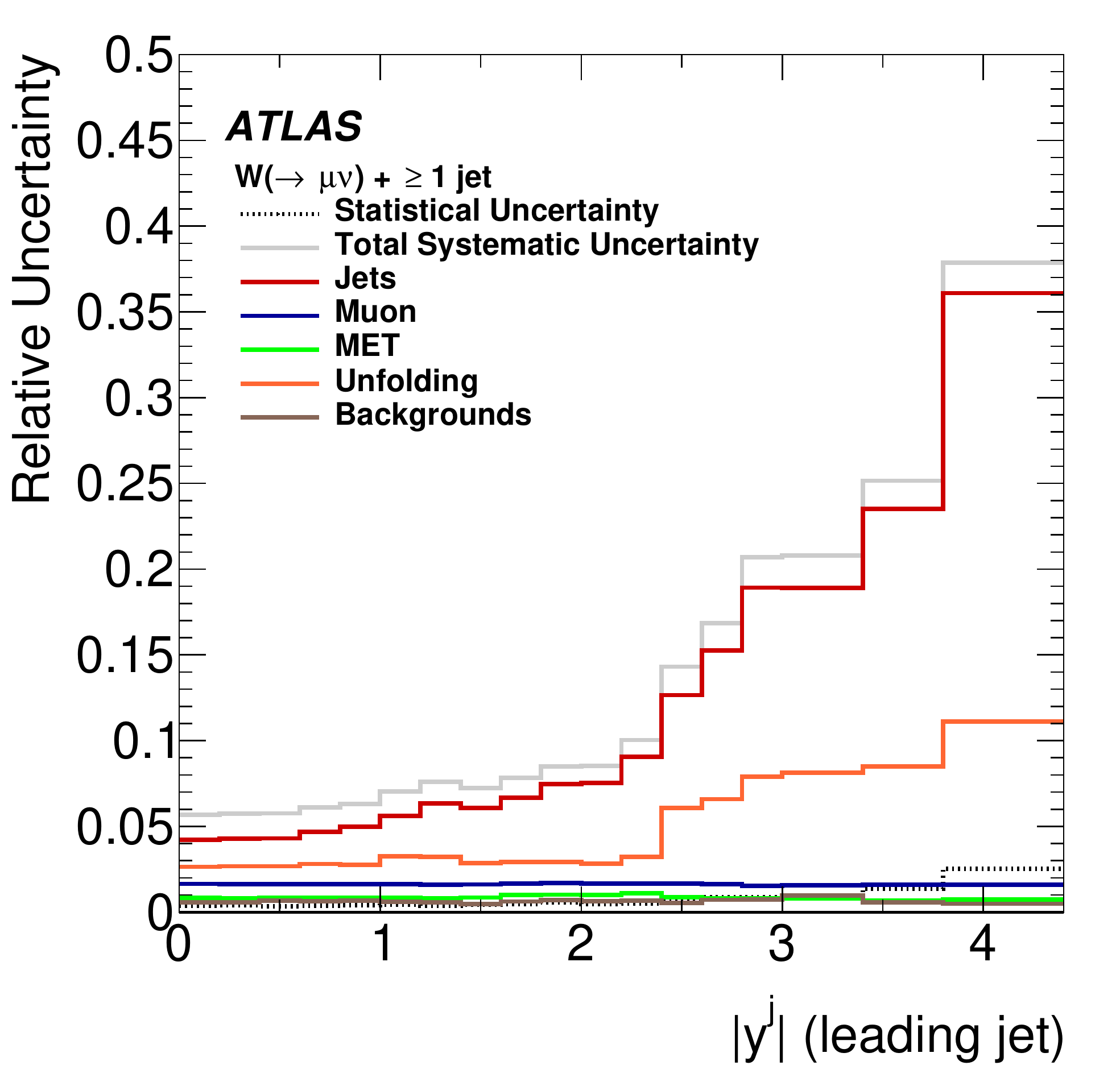}
    \end{minipage}
\caption{\label{fig:Uncertainties_Vjets_13-8TeV} Cross section in exclusive jet multiplicity in \Zjets with 13 TeV \pp collisions at the \LHC for the combination of both decay channels and breakdown of uncertainties (top table), table taken from~\cite{Sirunyan:2018cpw}. Fractional experimental uncertainties in $W + \text{1 jet}$ analyses in 1.96 TeV \ppbar collisions at the \Tevatron (bottom-left plot), figure taken from~\cite{Abazov:2013gpa}, and in 7 TeV \pp collisions at the \LHC  (bottom-right plot) as a function of the leading jet rapidity, figure taken from~\cite{Aad:2014qxa}.  
}
\end{figure*}
\Wjet and \Zjet events have different levels of background contamination. While for both processes at high jet multiplicity the background from $t\bar t$ production is dominant, \Wjet events have a larger background contribution from multi-jet production in which hadronic particles are misidentified as an electron or a muon. The multi-jet background is estimated using data-driven techniques and contributes to $\approx 5-15\%$ of the \Wjet data samples. The background from $t\bar t$ events contributes to about $ 0\%$ (1 jet), $20\%$ ($Z + \text{6 jets}$) and $80\%$ ($W + \text{6 jets}$) and is estimated by MC or with data-driven techniques, and can be suppressed by a $b$-jet veto.
Figure~\ref{fig:CMS_ZWjets_8TeV_Njets_reco} shows the jet multiplicity distributions and the levels of background contamination in \Wjet and \Zjet events in the muon decay channels in the \CMS detector.

\begin{figure}
\centering
\includegraphics[width=0.48\textwidth]{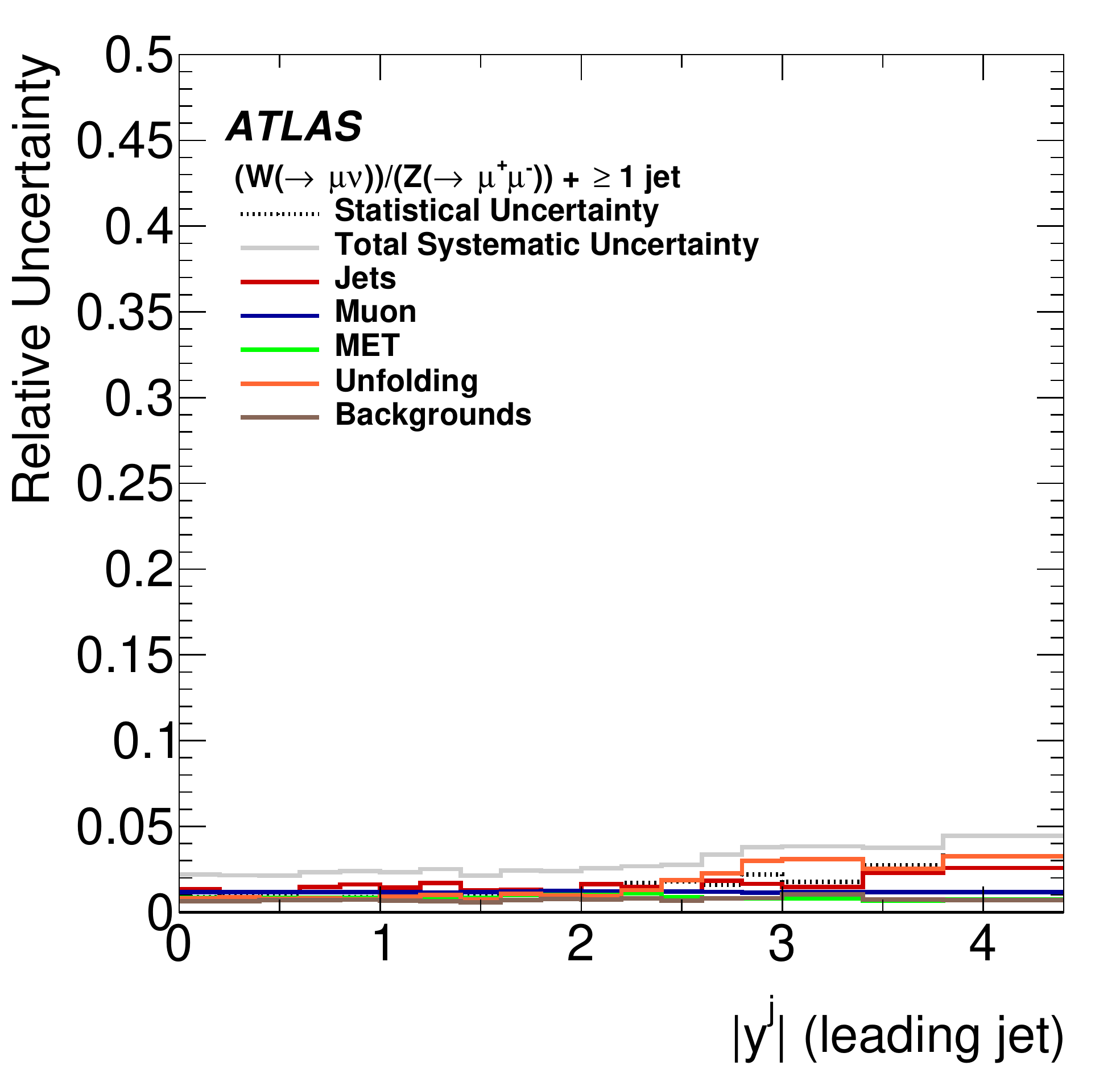}
\caption{\label{fig:ATLAS_Rjets_error} Relative experimental uncertainties on the inclusive ratio $W + \text{1 jet}$ to $Z + \text{1 jet}$  cross section as a function of the leading jet rapidity in 7 TeV \pp collisions at the \LHC. Figure taken from~\cite{Aad:2014rta}.
}
\end{figure}

Experimental uncertainties are dominated at low jet \pT or low $N_{\rm{jets}}$ by  systematics associated to the energy scale and resolution of the jets. At high jet \pT or high $N_{\rm jets}$, unfolding and statistical uncertainties become important. At high $N_{\rm jets}$ the uncertainties on the backgrounds dominate in \Wjet analyses.  
Figure~\ref{fig:Uncertainties_Vjets_13-8TeV} shows the level of experimental uncertainties in the \Zjet production cross section measurement as a function of the jet multiplicity, and the level of jet energy calibration uncertainty in $W + \ge \text{ 1 jet}$ events as a function of the leading jet rapidity. The levels of experimental uncertainties in \WZjets are similar in analyses at 7, 8 and 13 TeV center-of mass energies at the \LHC and comparable with \Tevatron experiments.


Measurements of ratios of differential cross sections allow for partial cancellations of uncertainties (both experimental and theoretical), for example in the ratio of \Wjet to \Zjet differential cross sections, known as $R_{\rm jets}$. The comparison of the experimental uncertainties in Fig.~\ref{fig:Uncertainties_Vjets_13-8TeV} (bottom-right) for \Wjets and in Fig.~\ref{fig:ATLAS_Rjets_error} shows that in $R_{\rm jets}$ the experimental systematics, and especially those associated to jets, cancel from about $40\%$  to about $5\%$ at high jet rapidity. This large cancellation of uncertainties allows for an accurate test of SM predictions at the percent level in a wide region of phase space.

\subsubsection{Cross sections and jet rates}
\label{sec:VLF:exp:vjets}

Measurements of \ensuremath{V+\text{light-jet}}\xspace  production cross sections as functions of the jet multiplicity, jet transverse momenta and the jet rates are carried out at hadron colliders, as they provide benchmarks for the understanding of the underlying QCD dynamics and its modeling in MC generators.
This sections starts with the presentation of the measurements of the associated production of jets and a photon, and concludes with the discussion of those of the associated productions of jets and a massive vector boson, i.e., $W$ or $Z$.

\paragraph{Photon\texorpdfstring{$+\text{jets}$}{+jets} cross section measurements}
\mbox{}\newline

\begin{figure}
\includegraphics[width=0.45\textwidth]{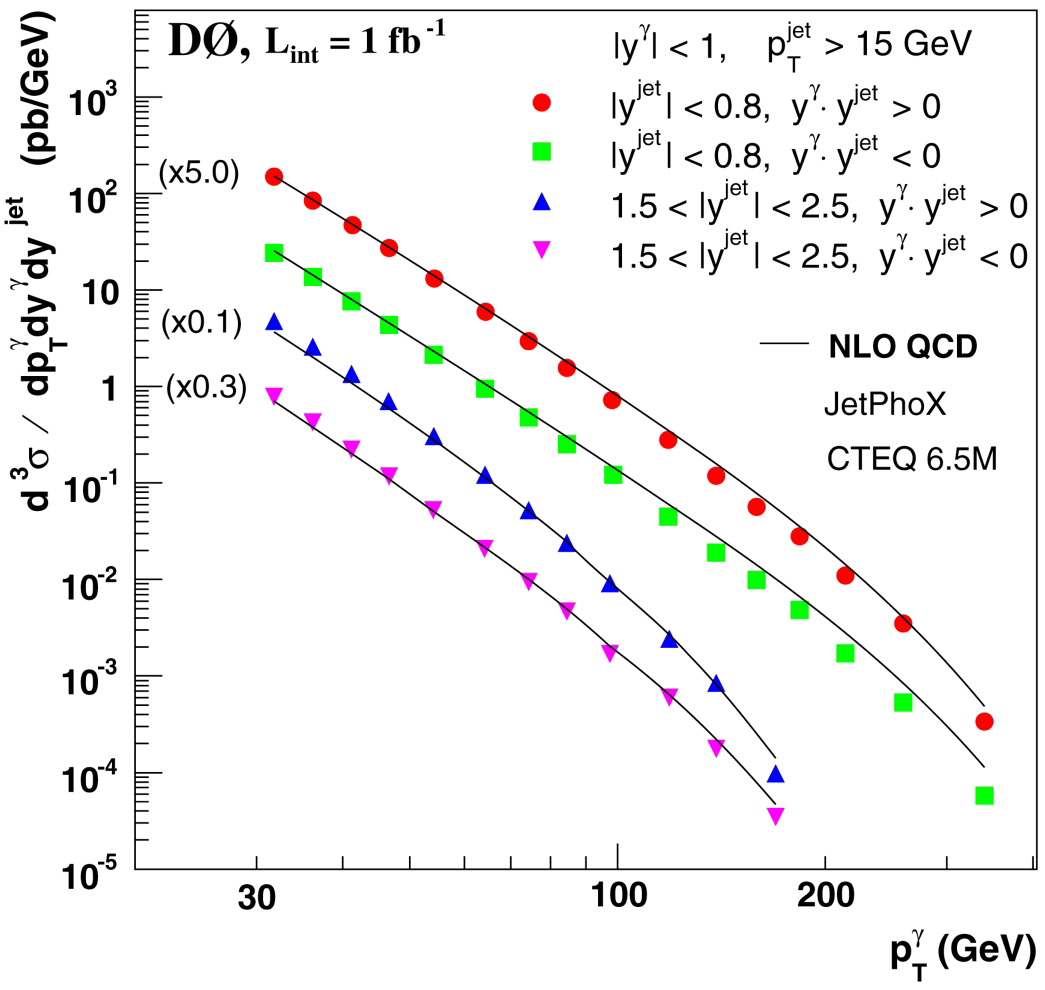}
\includegraphics[width=0.47\textwidth]{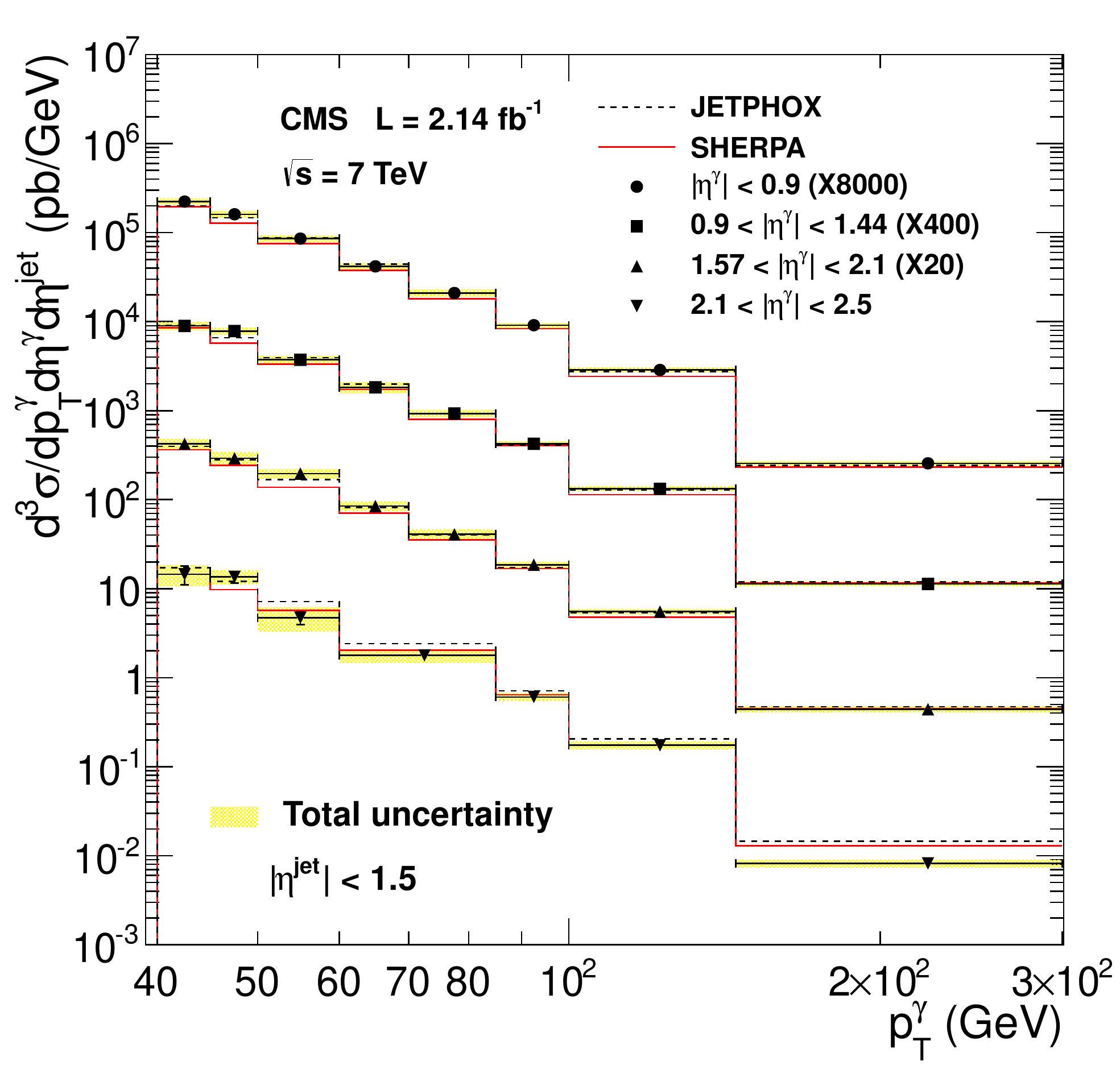}
\caption{\label{fig:D0_CMS_Photonjet_ET} Triple differential cross sections as a function of $\pT^{\gamma}$ in four photon rapidity intervals in \ppbar collisions at 1.96 TeV center-of-mass energy at the \Tevatron, figure taken from~\cite{Abazov:2008er} (top), and in four different ranges of $\eta^{\gamma}$ for $\pT^{\rm jet}$ > 30 GeV and $|\eta^{\rm jet}|$ < 1.5 in \pp collisions with 7 TeV center-of-mass energy at the \LHC (bottom), figure taken from~\cite{Chatrchyan:2013mwa}.  The measured cross sections are compared with the NLO pQCD calculation by \Jetphox and with the LO MC \Sherpa simulation (bottom). 
}
\end{figure}

The processes of \gammajet production have the largest cross sections of all \Vjet processes and they approach the \WZjet cross sections at high photon transverse momentum, i.e., in the regime where weak boson mass effects play a lesser role. 
Figure~\ref{fig:D0_CMS_Photonjet_ET} presents the triple differential cross-section as a function of the photon transverse momentum, and the $\gamma$ and jet pseudo-rapidities, at the \Tevatron in \ppbar collisions at 1.96 TeV center-of-mass energy and at the \LHC in \pp collisions at 7 TeV center-of-mass energy. 
The \DO measurement is carried out in two regions of the jet rapidity and in event configurations in which the jet and photon have either the same or opposite signs in rapidity, while the \CMS analysis is in two regions of the jet pseudo-rapidity and in four regions of the photon pseudo-rapidity. 
In both cases the measurements span several orders of magnitude in the production cross sections. The predictions include the NLO pQCD calculation implemented in \Jetphox and, in the \CMS analysis, the tree-level matrix elements with up to 3 parton-jets matched to parton showering in the \Sherpa generator. The predictions are overall consistent with data but unable to describe the cross section variations across the entire measured range of phase space. \Tevatron results are an important benchmark for theoretical calculations and have served as a stepping stone for more accurate modeling of such processes for the \LHC. However, despite this progress in theoretical predictions, some data-theory discrepancies are observed at the \LHC in specific kinematic regions of \gammajets. For example, in regions with large photon $\eta$ and \pT in $\gamma$ + 1-jet events, the ratios between data and NLO pQCD predictions by \Jetphox and \Sherpa are in the range of 50\% to 70\% ~\cite{Chatrchyan:2013mwa}, and in the photon \pT region greater than 750 GeV in $\gamma$ + 2-jet events~\cite{Aaboud:2016sdm} the discrepancy between data and the NLO calculation reaches about 2$\sigma$. 

\begin{figure}
\centering
\includegraphics[width=0.50\textwidth]{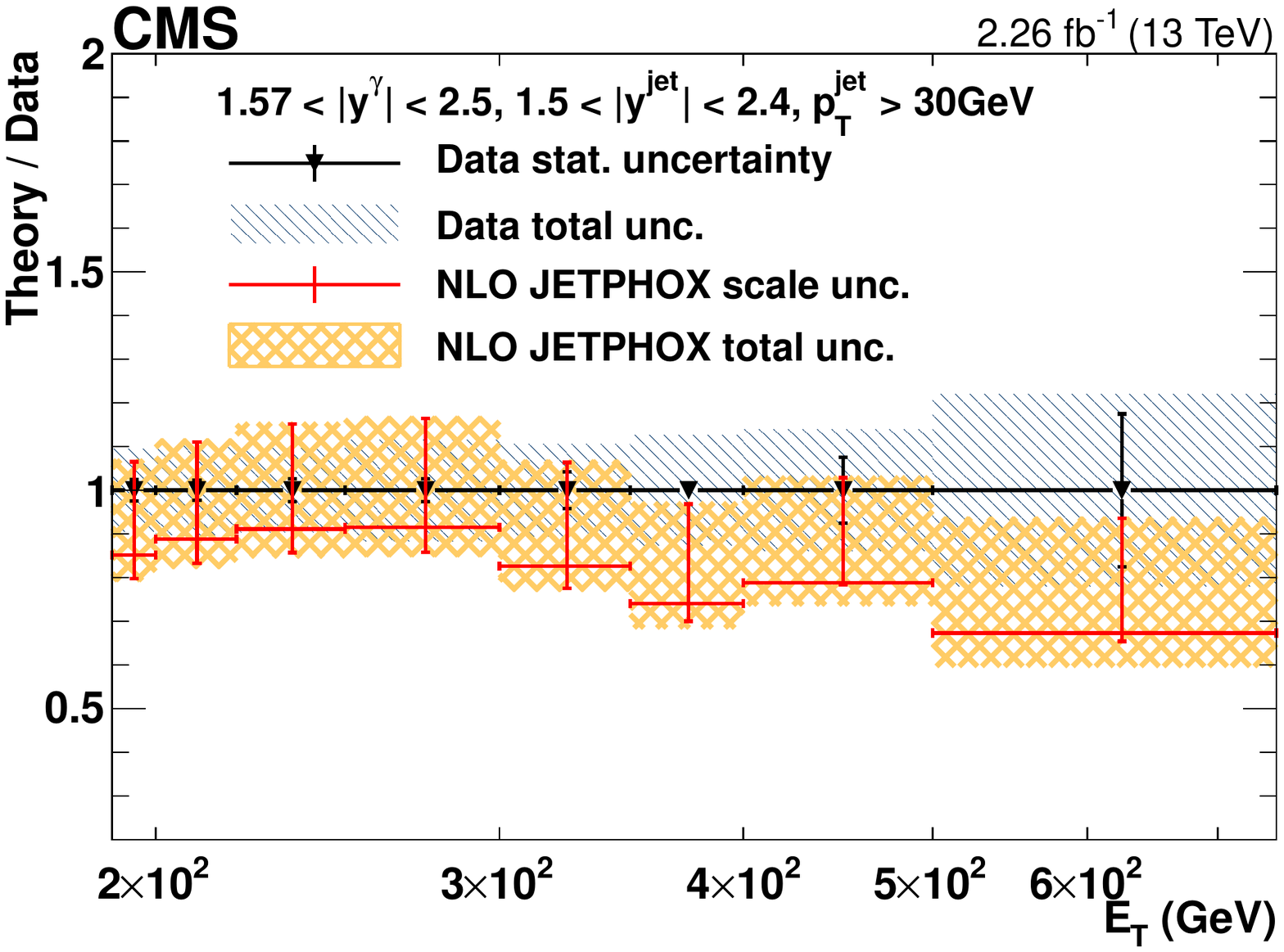}
\includegraphics[width=0.50\textwidth]{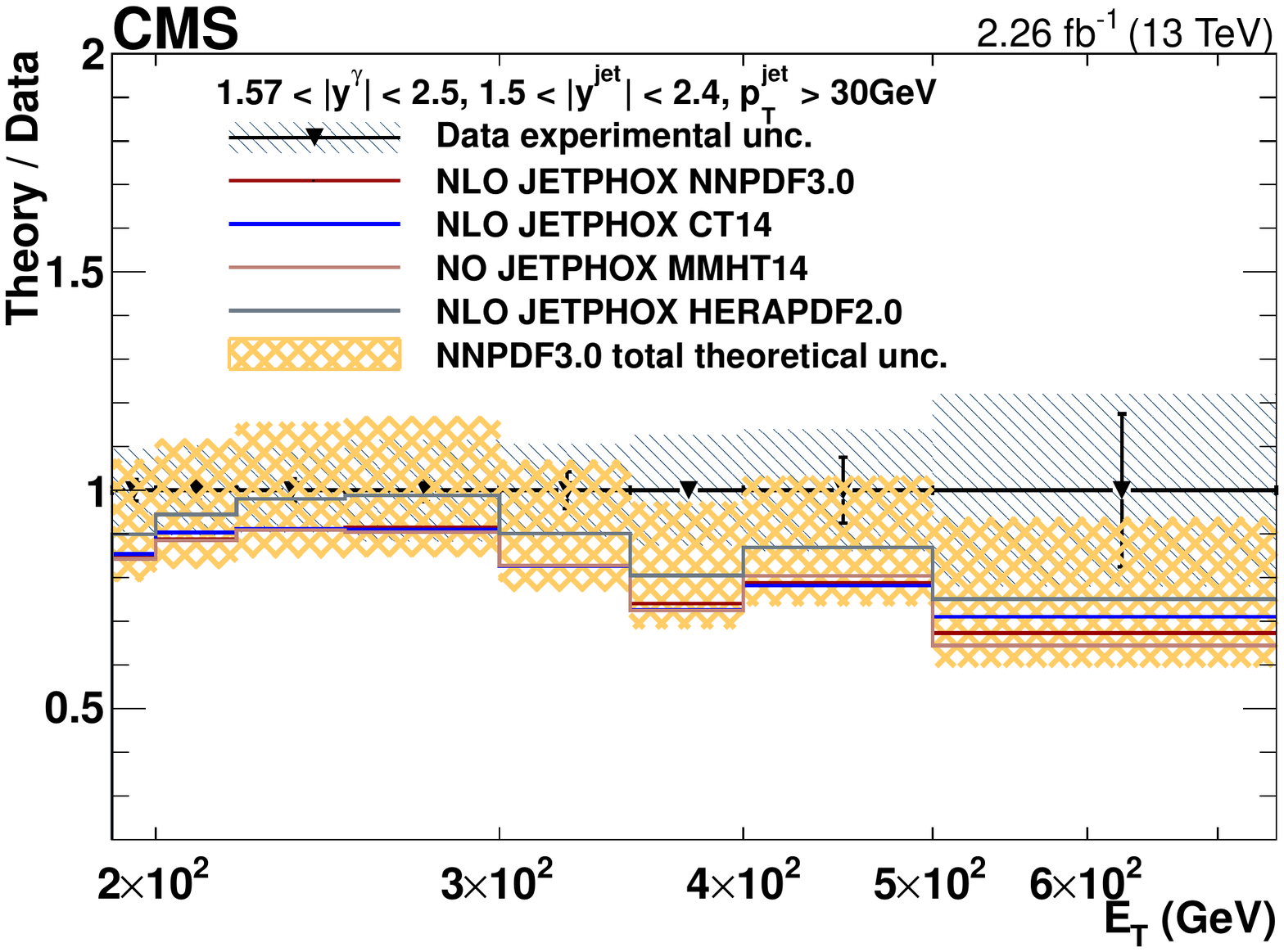}
\caption{\label{fig:CMS_Photonjet_13TeV_ET} Ratios of NLO (\Jetphox) predictions to data as a function of the photon transverse energy in \gammajet events in the forward photon and jet rapidity regions in \pp collisions at the \LHC with 13 TeV center-of-mass energy. The plot on the top shows the contribution of the scale uncertainty to the total theoretical uncertainty in the NLO \Jetphox calculation, compared to the experimental uncertainties. The plot on the bottom shows NLO \Jetphox calculation with various NLO PDF sets, compared to the experimental uncertainties. Figures taken from~\cite{Sirunyan:2628267}.
}
\end{figure}

In addition to testing QCD calculations and MC predictions, the measurements  of  photon $+$ jet  production  can be used to constrain the parton density functions (PDFs). The results in Fig.~\ref{fig:CMS_Photonjet_13TeV_ET} are shown as an example of ratios between theory and \LHC data for the measurement of the differential cross section for \gammajet production as a function of the photon \ET. The measurement is carried out in two photon and two jet rapidity regions, while the figure shows the results in the forward photon and forward jet bin. The measurement shows good agreement between \LHC data and NLO predictions in pQCD from \Jetphox. As Fig.~\ref{fig:CMS_Photonjet_13TeV_ET} (top) shows, 
 the experimental and theoretical uncertainties are comparable, and the theoretical jet scale uncertainty dominates the theoretical uncertainties in the NLO pQCD approximation. 
Figure~\ref{fig:CMS_Photonjet_13TeV_ET} (bottom) shows the NLO prediction by \Jetphox using various NLO PDF sets. Although the differences between the studied PDF sets are small and subleading with respect to the scale uncertainty estimated in the NLO pQCD approximation, new calculations at higher orders in QCD with smaller scale uncertainties, i.e., next-to-next-to-leading order, are available~\cite{Campbell:2017dqk, Campbell:2016lzl,Chen:2019zmr} 
and motivate  the  use  of such measurements at the \LHC to improve the gluon and other PDFs~\cite{Campbell:2018wfu}, especially in the kinematic regions where the experimental uncertainties are smaller or comparable to theoretical uncertainties, e.g., in  low to middle range in photon \ET.

For other inclusive photon and photon $+$ light-jet measurements carried out at the \Tevatron and at the \LHC the reader is referred to the analyses reported in Refs.~\cite{Aad:2010sp,Aad:2011tw,Aad:2013zba,Aad:2016xcr,Aaboud:2017cbm,Aad:2019eqv,Aaboud:2019vpz,Chatrchyan:2011ue,Khachatryan:2010fm,Chatrchyan:2013oda, Aad:2013gaa,ATLAS:2012ar,Abazov:2011rd,Abazov:2009gc,Sirunyan:2019uya,Aad:2019cpw}.


\paragraph{\texorpdfstring{\WZjets}{W/Z+jet} cross sections measurements}
\mbox{}\newline


\begin{figure}
\includegraphics[width=0.46\textwidth]{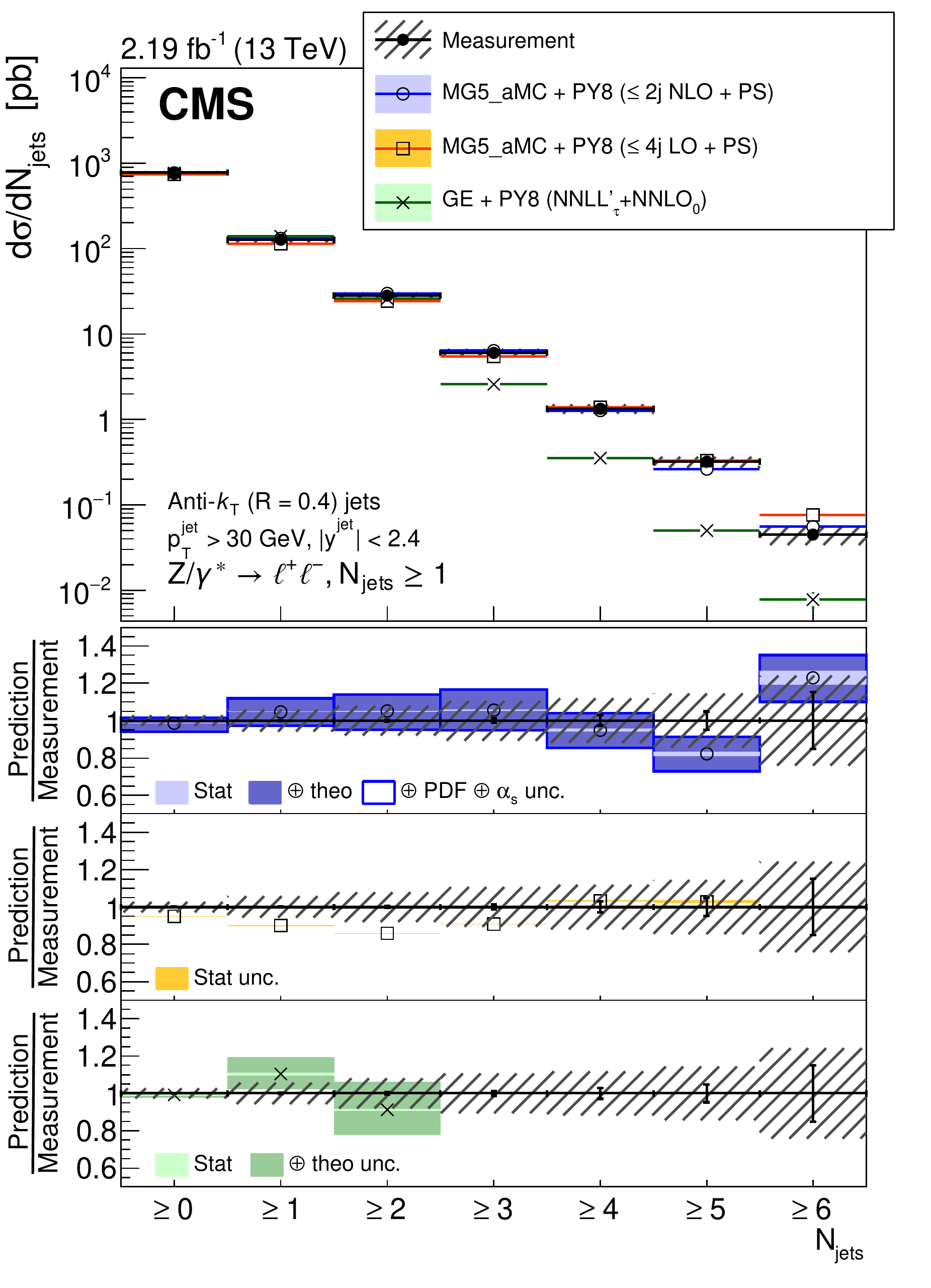}
\caption{\label{fig:CMS_Zjet_13TeV_Njets} Cross section as a function of the inclusive jet multiplicity for \Zjet events at 13 TeV center-of-mass energy in \pp collisions at the \LHC. Figure taken from ~\cite{Sirunyan:2018cpw}. 
}
\end{figure}
\begin{figure*}
\centering
 \begin{minipage}{0.44\textwidth}
    \centering
        \includegraphics[width=\textwidth]{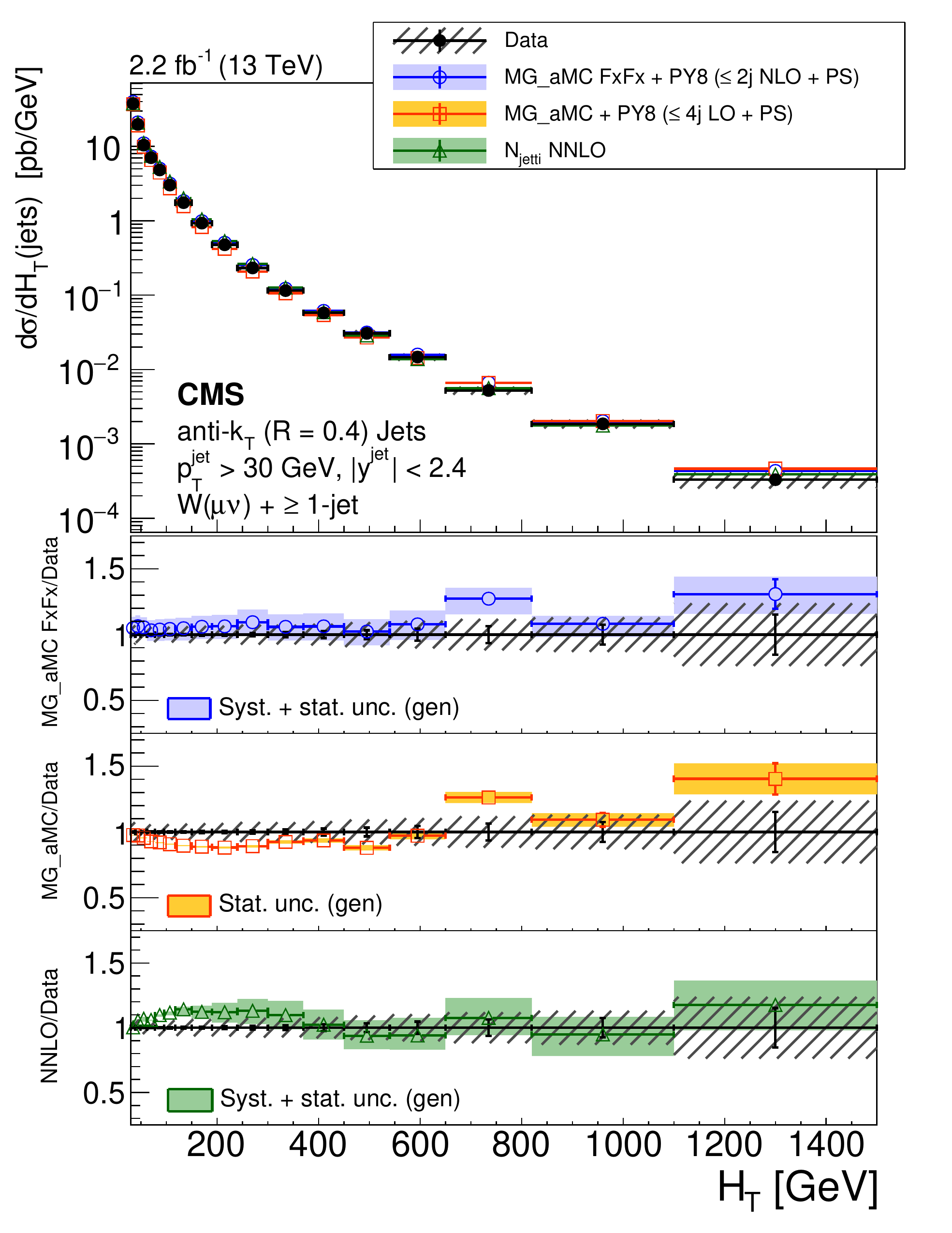}
        \end{minipage}
 \begin{minipage}{0.55\textwidth}
        \centering
        \includegraphics[width=\textwidth]{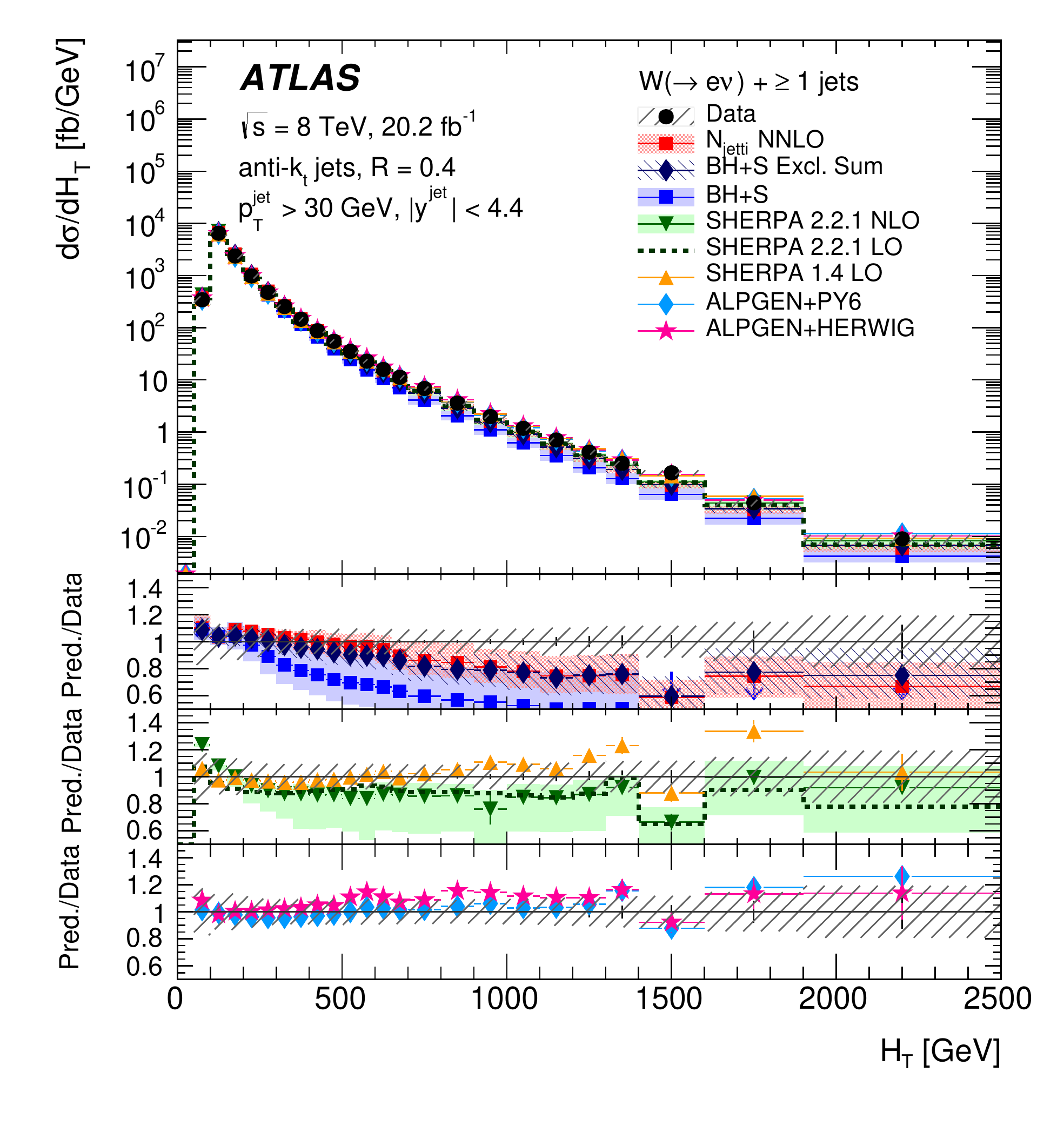}
\end{minipage}
\caption{\label{fig:ATLAS_WZjet_13-8TeV_NjetsHT} Differential cross section for the production of a $W$ boson with at least one jet as a function of \HT at 13 TeV (left), figure taken from ~\cite{Sirunyan:2017wgx}, and at 8 TeV (right), figure taken from ~\cite{Aaboud:2017soa}, in \pp collisions at the \LHC. 
}
\end{figure*}

A typical measurement of \WZjet processes is the production cross section (multiplied by the leptonic branching ratio) as a function of jet multiplicity, as shown in Fig.~\ref{fig:CMS_Zjet_13TeV_Njets} for \Zjet production. 
Such a measurement is important to assess the accuracy of SM predictions that are used to estimate the \WZjet yield in searches of new physics signatures. 

The production of \WZjets is a multi-scale process and various observables can be used to define the scale of the process, depending on the kinematic configuration. A common variable used to set the scale in \WZjet processes is \HT, which is also used to discriminate new physics from SM background, for example in supersymmetry searches. This variable has, however, different definitions: in \ATLAS is defined as the scalar sum of the transverse momenta of leptons (including neutrinos) and jets in the event, while in \CMS as the scalar \pT sum of the jets only.   Figure~\ref{fig:ATLAS_WZjet_13-8TeV_NjetsHT} shows two examples of the differential cross section for $W ~+ \ge 1 $ jet production as a function of \HT by the \CMS and \ATLAS experiments. The measurements in Figures~\ref{fig:CMS_Zjet_13TeV_Njets} and~\ref{fig:ATLAS_WZjet_13-8TeV_NjetsHT} show excellent agreement with theoretical predictions over 4 orders of magnitude in cross-section. The multitude of models that are compared to data show the variety of theoretical approaches that can be validated with such measurements. These results show that MC simulations with multi-parton calculations in the matrix element matched to  the parton shower are in agreement with the data up to very high jet multiplicity and in a broad range in energy scale. The high experimental precision also exposes discrepancies between measurements and predictions. In Figure~\ref{fig:ATLAS_WZjet_13-8TeV_NjetsHT} (right) discrepancies with pQCD NLO calculations (\BlackHat) are visible at large \HT values, when the jets are measured in a broad rapidity region ($|y^{\rm{jet}}|<4.4$).
The accuracy of the calculation improves when higher-order pQCD corrections are included, i.e., in $N_{\rm jetti}$ NNLO~\cite{Boughezal:2015dva,Ridder:2015dxa} or with the {\it exclusive sum} approach~\cite{AlcarazMaestre:2012vp} in \BlackHat~\cite{Berger:2010vm,Berger:2010zx,Bern:2013gka} in which NLO information from higher multiplicity processes are included to the standard fixed-order prediction, cf.\ Sec.\ \ref{sec:VLF:theory}.
%
\begin{figure*}
\centering
\includegraphics[width=0.62\textwidth]{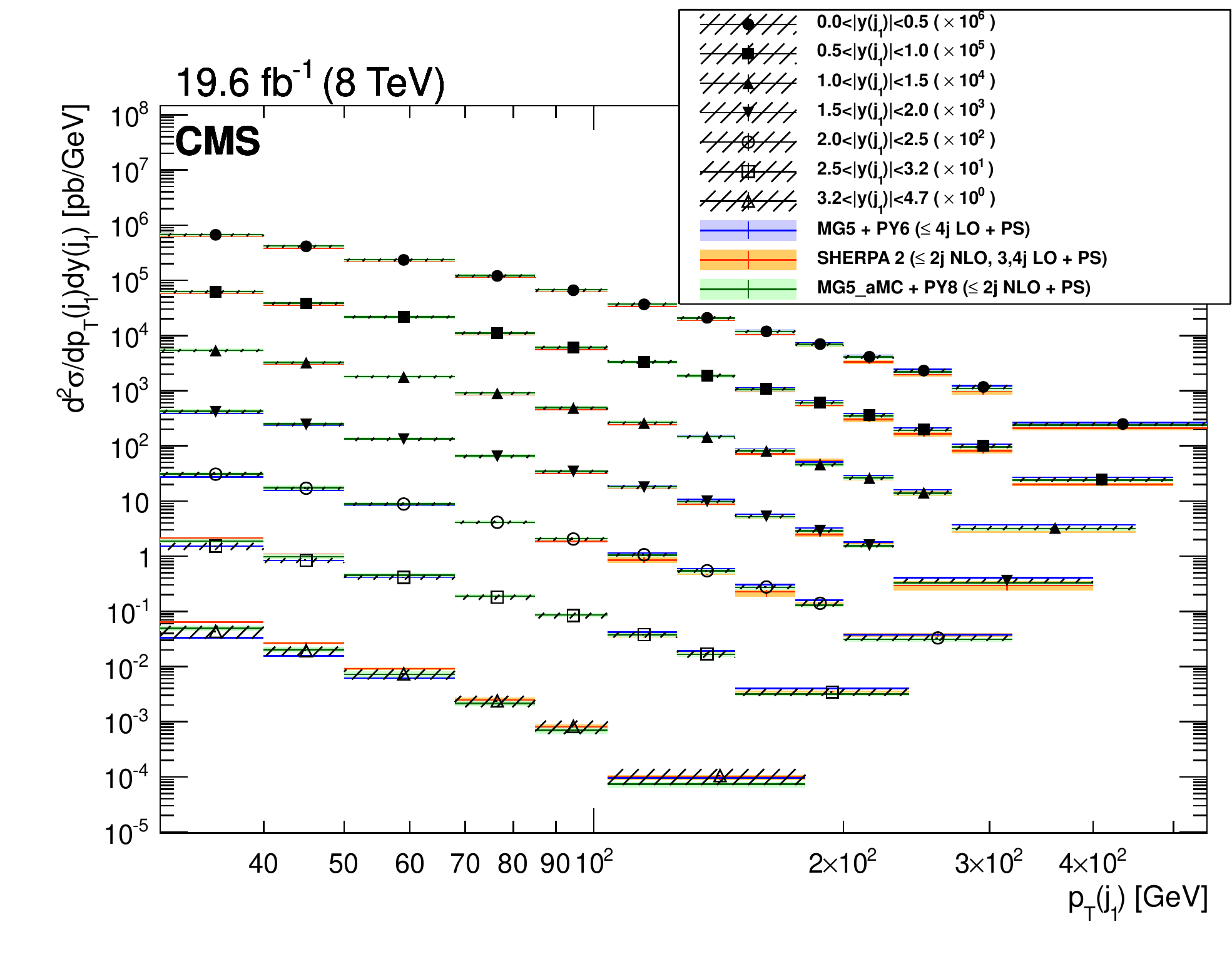}\\
\includegraphics[width=0.47\textwidth]{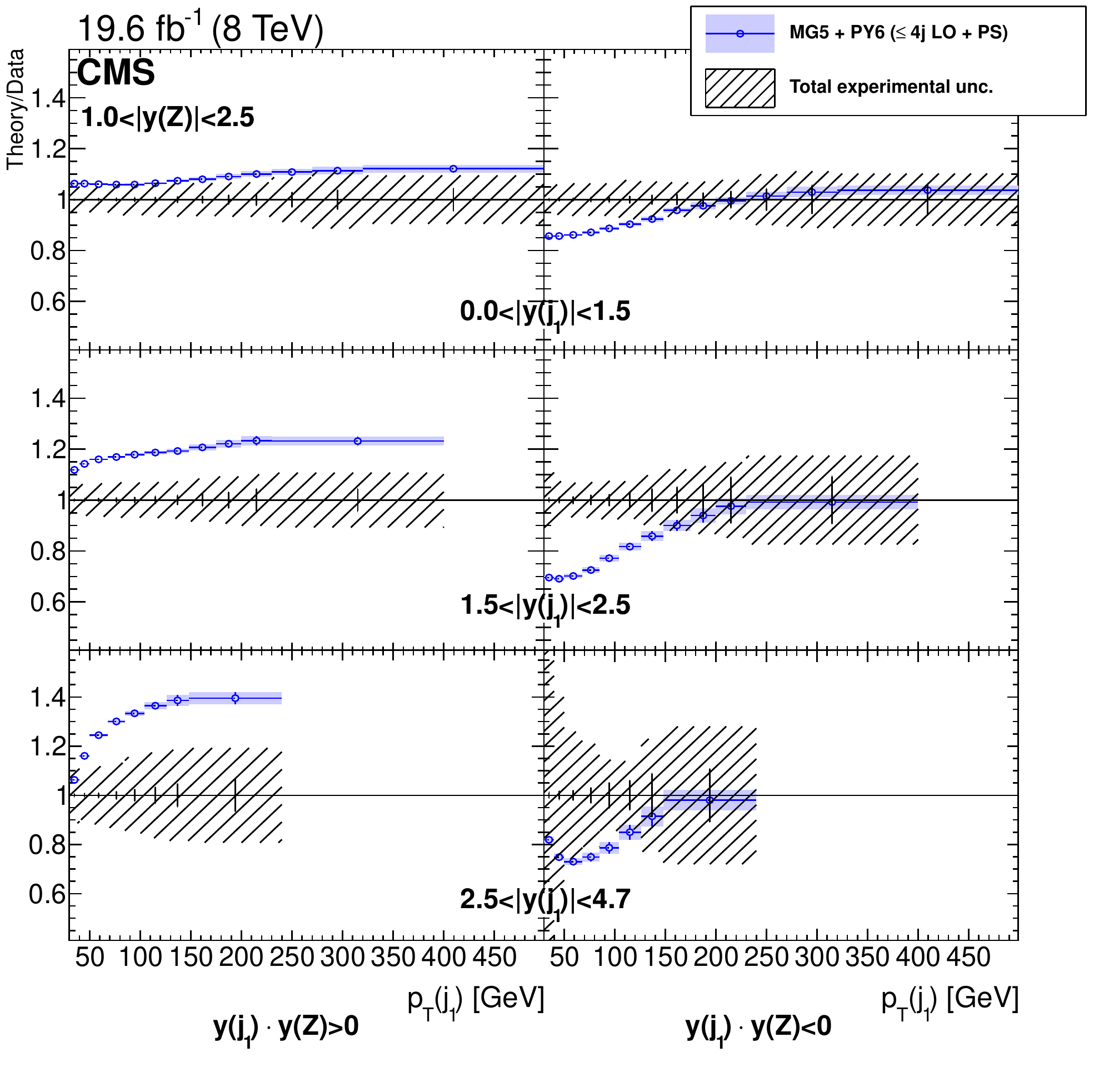}
\includegraphics[width=0.47\textwidth]{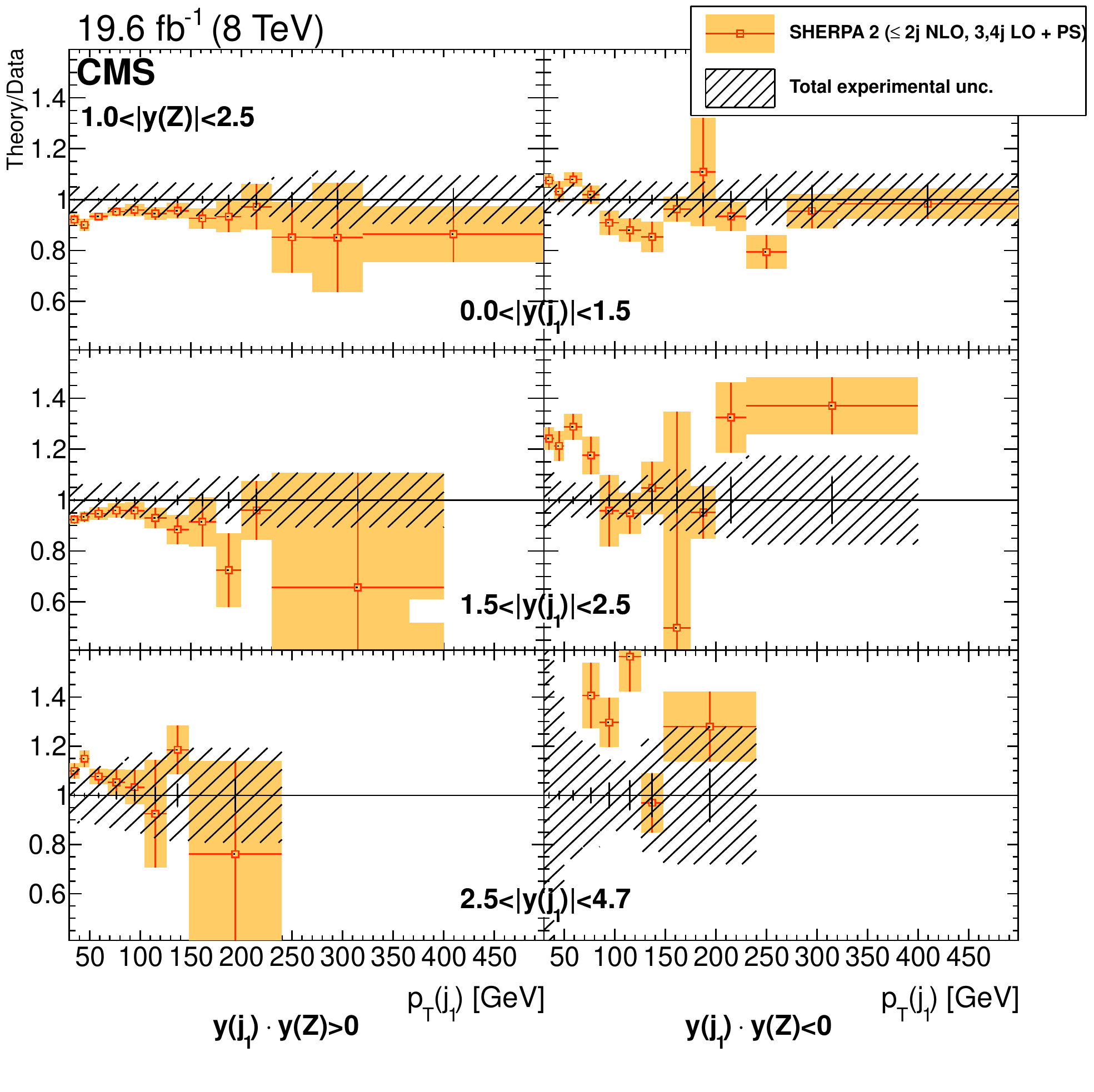}
\caption{\label{fig:CMS_Zjet_8TeV_doublediff} Differential cross section (top) and theory-to-data ratio (bottom) for \Zjet production as a function of the leading jet transverse momentum and rapidity in 8 TeV \pp collisions at the \LHC. Figures taken from~\cite{Khachatryan:2016crw}.
}
\end{figure*}
\par
\begin{figure*}
\centering
\includegraphics[width=0.80\textwidth]{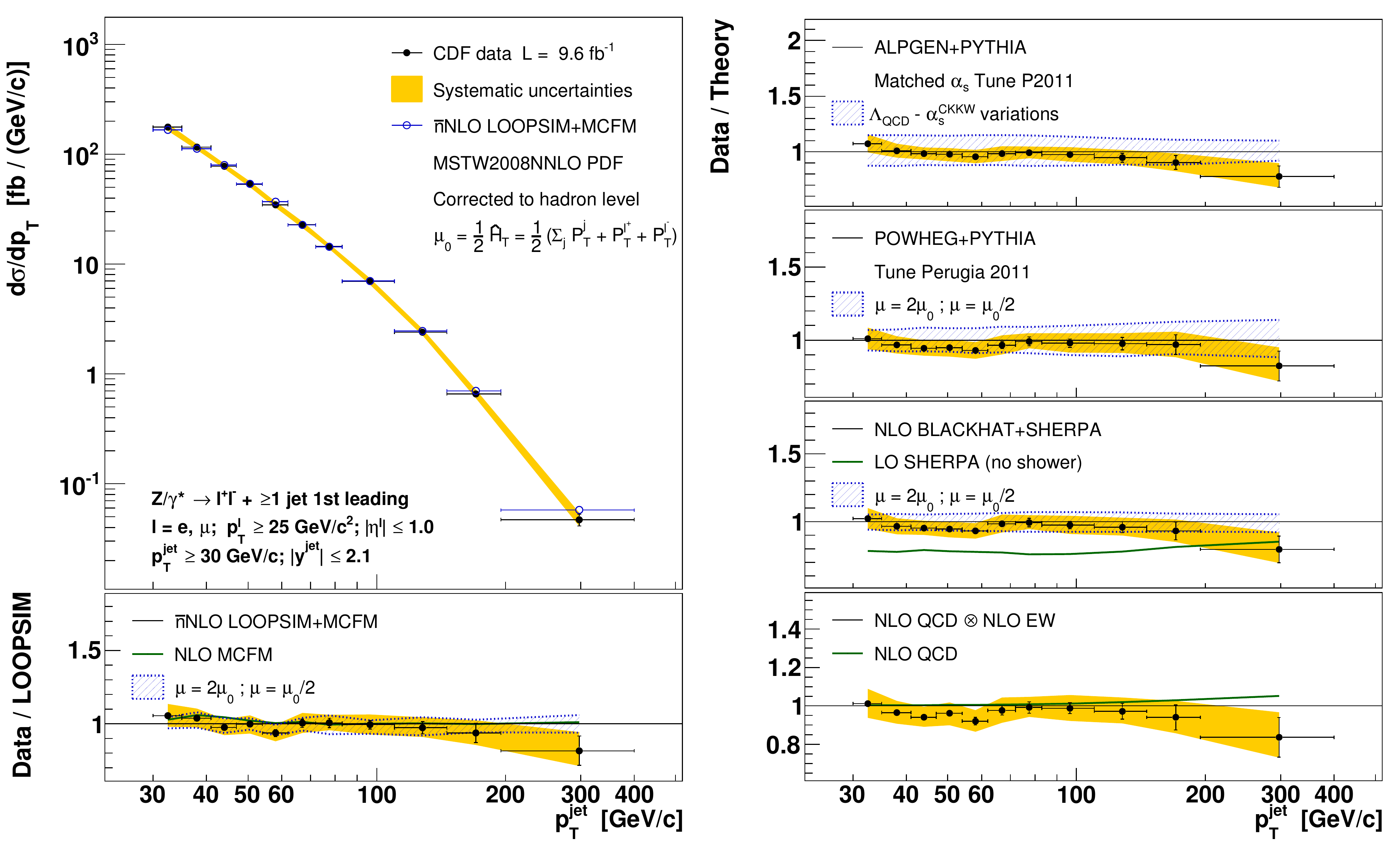}
\caption{\label{fig:CDS_WZjet_8TeV_ptN} Differential cross section and data-to-theory ratios as a function of the leading-jet \pT for $Z + \ge \text{1 jet}$ events in 1.96 TeV \ppbar collisions at the \Tevatron. Figure taken from~\cite{Aaltonen:2014vma}. 
}
\end{figure*}

\begin{figure}
\includegraphics[width=0.42\textwidth]{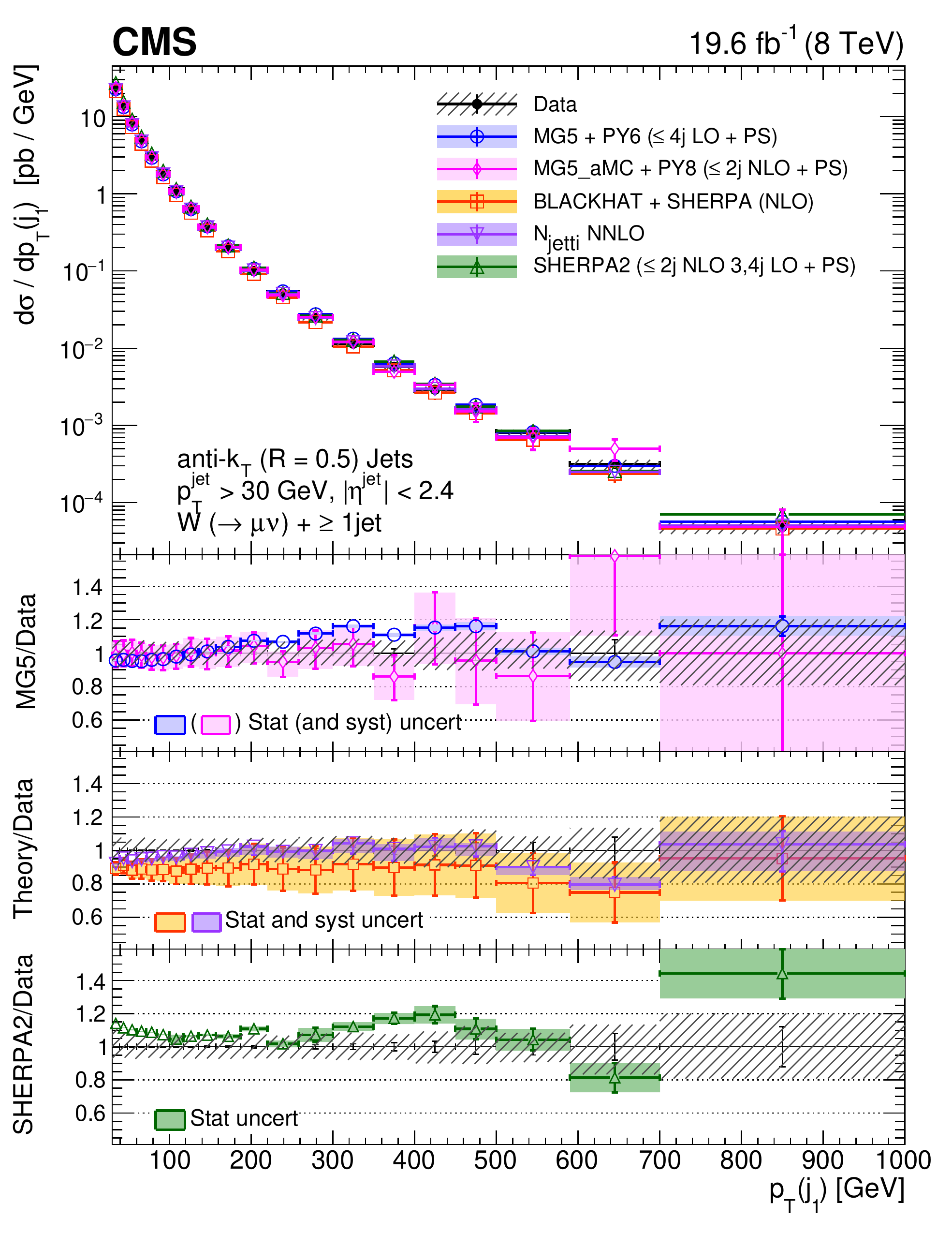}
\caption{\label{fig:CMS_WZjet_8TeV_ptN} Differential cross section as a function of the leading-jet \pT in $W + \ge \text{1 jet}$ events with 8 TeV \pp collisions at the \LHC, figure taken from~\cite{Khachatryan:2016fue}. 
}
\end{figure}

 The large statistics of \WZjet events allows to carry out double differential measurements, similarly to what is typically done in inclusive jet or \gammajet measurements, see two examples in Refs.~\cite{Aad:2019hga,Khachatryan:2016fue} by the \ATLAS and \CMS experiments.  
Figure~\ref{fig:CMS_Zjet_8TeV_doublediff} shows an example of a double differential cross section as a function of the leading jet \pT and rapidity that is performed in a broad region of phase space up to jet $\pT=500$ GeV and $|y| = 4.7$. 
Such measurements are expected to provide valuable input to the understanding of parton density functions in addition to QCD dynamics. However, in several regions of phase space the precision of experimental results is higher than current prediction-to-prediction differences, and discrepancies between data and theoretical predictions can be up to $40\%$ 
i.e., larger than the effects from PDFs, and thus the potential PDF sensitivity of the data cannot be exploited.
The MC simulation that includes NLO QCD corrections provides a more accurate normalization and better modeling of the shapes of the distributions than LO QCD predictions. 
Similar conclusions can be reached from several other results on \Zjet and \Wjet processes: the inclusion of higher-order QCD corrections in fixed-order calculations and in MC simulations generally provides predictions that more accurately describe the data and are also more precise. Figures~\ref{fig:CDS_WZjet_8TeV_ptN} and~\ref{fig:CMS_WZjet_8TeV_ptN} show examples of such an effect. Better agreement with data and smaller uncertainties are found in predictions that include higher-order corrections, such as in the NLO \MGaMC MC prediction with respect to the LO \Madgraph, the ${\rm \bar{n}}$NLO approximation with \LOOPSIM+ \MCFM~\cite{Rubin:2010xp} and the NNLO
$N_{\rm jetti}$ calculation, for the distribution of the leading jet \pT at the \Tevatron, up to 400 GeV, and 
at the \LHC, up to 1 TeV.
Such an effect is corroborated by other studies, such as those in Refs.~\cite{Sirunyan:2018cpw,Sirunyan:2017wgx,Aad:2019hga}, where data are compared to NNLO predictions. 
These experimental results also show that such processes can test theoretical calculations in a broad region of phase space with great precision, i.e., with an uncertainty as low as few percent in the $W + \ge \text{1 jet}$ events.

\begin{figure*}
\centering
\includegraphics[width=0.36\textwidth]{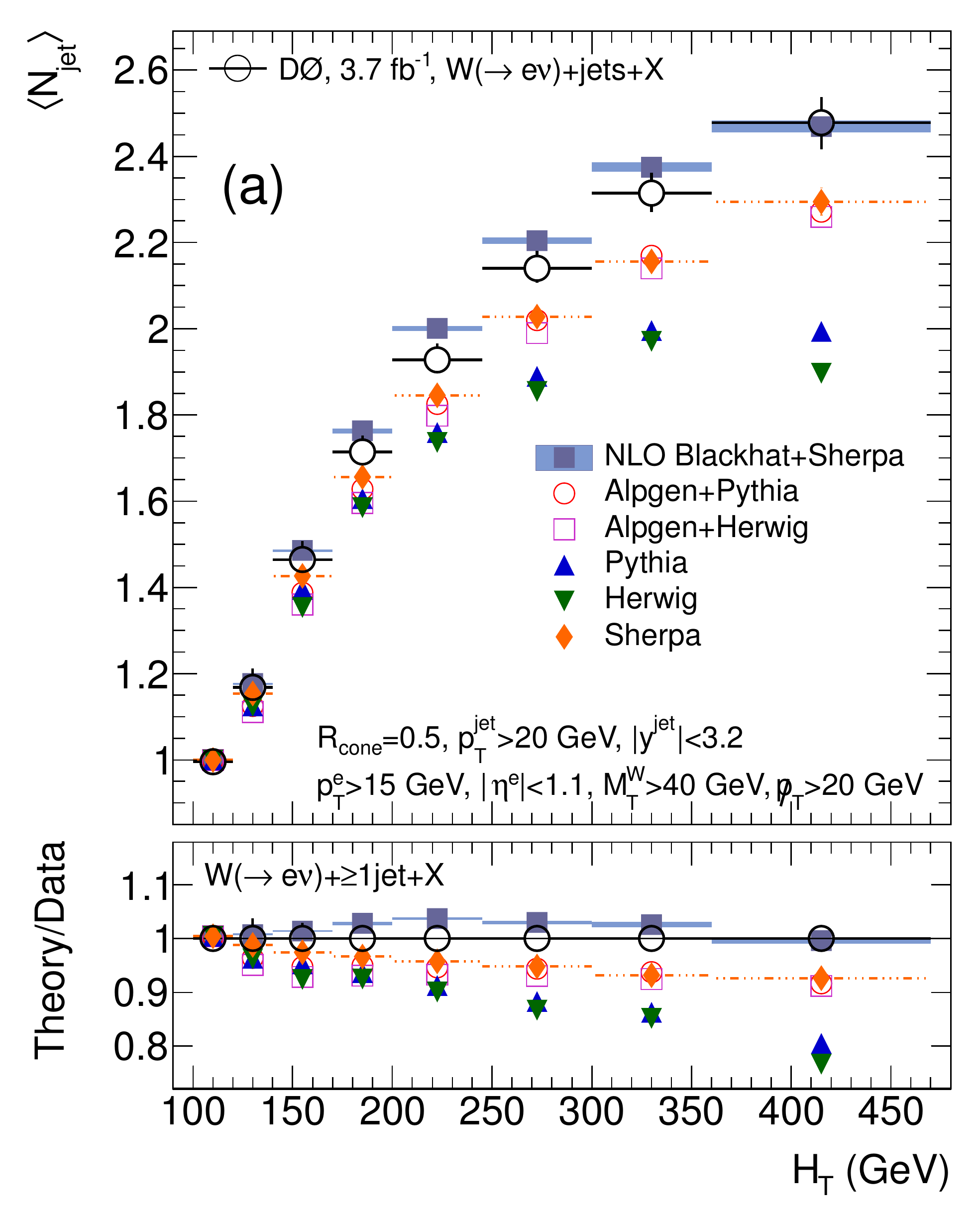}
\includegraphics[width=0.46\textwidth]{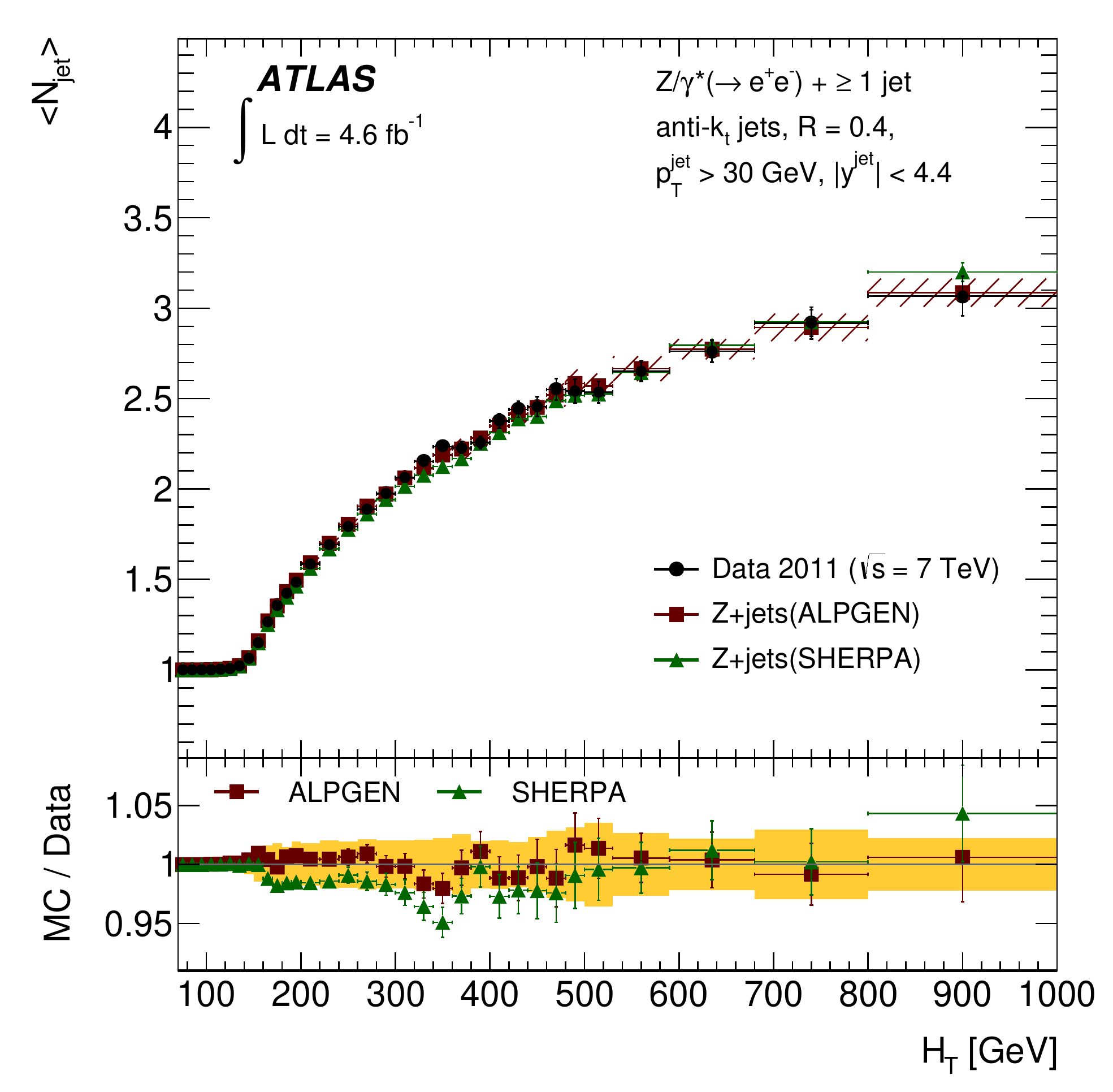}
\caption{\label{fig:ATLAS_Zjet_7TeV_Njets-HT} Average number of jets $\langle N_{\rm jet}\rangle$ as a function of \HT in \Wjet events  in 1.96 TeV \ppbar collisions at the \Tevatron (left), figure taken from~\cite{Abazov:2013gpa}, and in \Zjet events with 7 TeV \pp collisions at the LHC (right), figure taken from~\cite{Aad:2013ysa}. 
}
\end{figure*}

\begin{figure}
\includegraphics[width=0.46\textwidth]{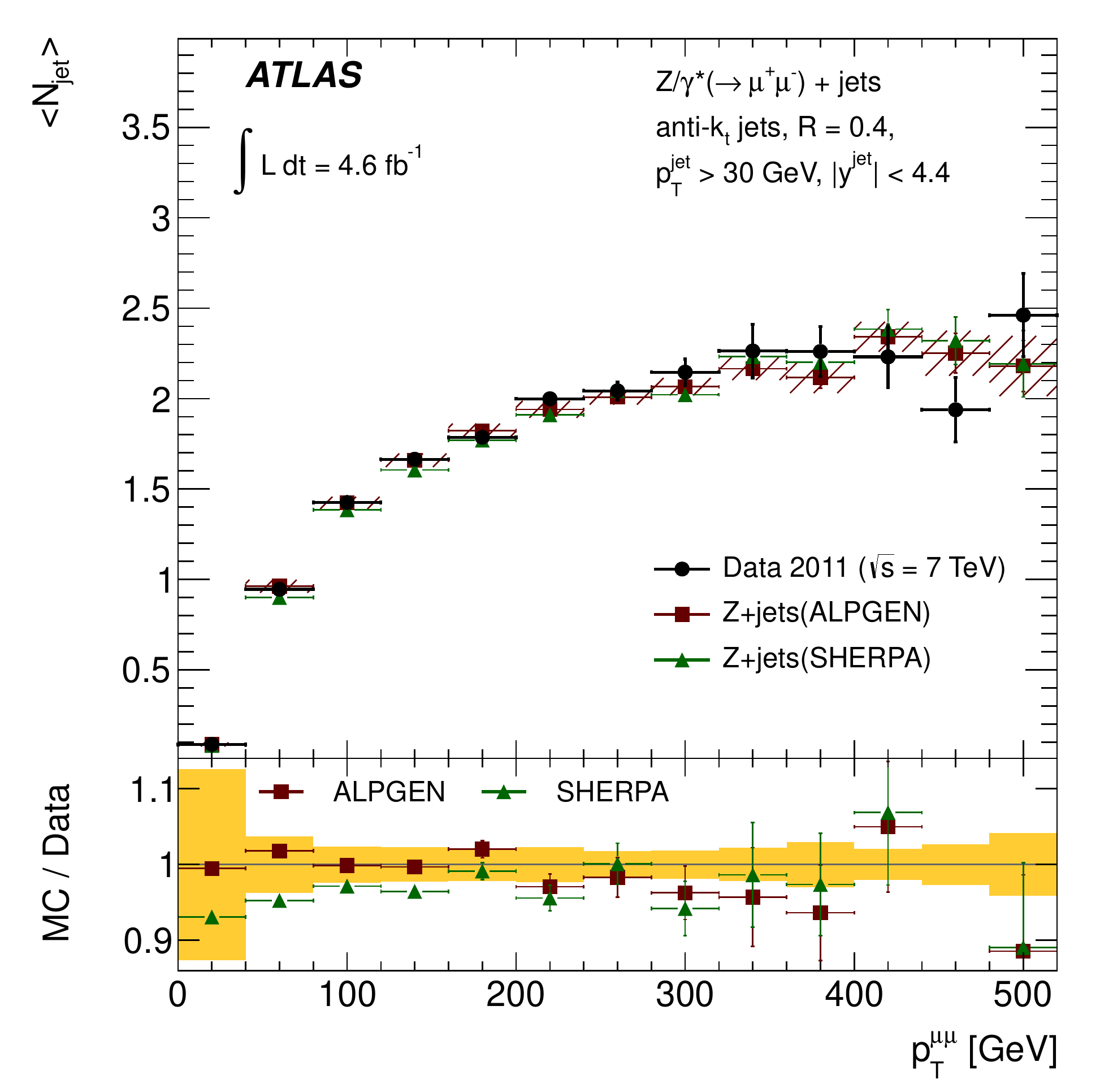}
\caption{\label{fig:ATLAS_Zjet_7TeV_Njets-pT} Average number of jets $\langle N_{\rm jet}\rangle$ as a function of the  $Z$ boson \pT in \Zjet events with 7 TeV \pp collisions at the LHC, figure taken from~\cite{Aad:2013ysa}. 
}
\end{figure}

 The jet multiplicity in \WZjet events is correlated to the energy scale of the process.  Figure~\ref{fig:ATLAS_Zjet_7TeV_Njets-HT} shows the correlation between the average jet multiplicity ($\langle N_{\rm jet}\rangle$) and \HT in \Wjet and \Zjet events at the \Tevatron and at the \LHC, and a similar correlation is demonstrated between $\langle N_{\rm jet}\rangle$ and the $Z$ boson \pT in \Zjets events in Fig.~\ref{fig:ATLAS_Zjet_7TeV_Njets-pT}. At $\HT \approx 300$ GeV in $W$ events or $\pT^Z\approx300$ GeV in $Z$ events the average jet multiplicity is about 2, while at $\HT=1000$ GeV the average jet multiplicity reaches 3. 
 In the \Tevatron study in Figure~\ref{fig:ATLAS_Zjet_7TeV_Njets-HT} (left) a fixed-order NLO calculation is used to computed the mean number of jets in an inclusive $W$ + $n$-jet sample by using the following prescription to improve the description beyond the NLO approximation: $\langle N_{\rm{jets}}\rangle = n + ( d \sigma^\text{NLO}_{n+1}+ d \sigma^\text{LO}_{n+2}) / d\sigma^\text{NLO}$~\cite{AlcarazMaestre:2012vp}.
 Such a calculation describes well this effect while the MC simulations underestimate the effect of the correlation. At the \LHC a good agreement between data and simulation is found. 
%

Multi-differential cross section measurements of the vector boson production allow detailed studies of QCD dynamics. In the absence of QED corrections, the five-dimensional differential cross-section $\done\sigma/(\done\pT^Z\,\done y^Z \,\done m^Z \,\done\cos\theta \,\done\phi)$ that describes the kinematics of the two leptons from  the $Z$ boson  decay  can  be  decomposed  into  a  sum  of  nine  harmonic  polynomials $P_i(\cos\theta,\phi)$ and eight dimensionless angular coefficients $A_i=A_i(\pT^Z,y^Z,m^Z)$ ($i=0..7$), which represent the ratios of helicity cross-sections with respect to the unpolarised one ($\sigma^{U+L}$) \cite{Mirkes:1992hu,Mirkes:1994eb} \footnote{In the presence of QED corrections, the expansion in terms of sperical 
 harmonics does not terminate after $l=2$, but instead turns into an infinite sum.},
\begin{equation}
  \label{eq:vlf:exp:angular_coeff}
  \begin{split}
    \lefteqn{\done\sigma/(\done\pT^Z\,\done y^Z\,\done m^Z\,\done\cos\theta\,\done\phi)}\\
    =&\;3\, \done\sigma^{U+L}/(16\pi\,\done\pT^Z\,\done y^Z\,\done m^Z)\\
     &\times
      \left\{
        (1+\cos^2\theta)
        +\tfrac{1}{2}\,A_0\,(1-3\cos^2\theta)\right.\\
     &\hspace*{15pt}\left.
        +A_1\sin 2\theta \cos\phi
        +\tfrac{1}{2}\,A_2\,\sin^2\theta\cos 2\phi\right.\\
     &\hspace*{15pt}\left.
        +A_3\sin\theta\cos\phi
        +A_4\cos\theta\right.\\
     &\hspace*{15pt}\left.
        +A_5\sin^2\theta\sin 2\phi
        +A_6\sin 2\theta\sin \phi\right.\\
     &\hspace*{15pt}\left.
        +A_7\sin\theta\sin\phi
        \vphantom{\cos^2}
      \right\}\;.
  \end{split}\nnb
\end{equation}
In this formulation the dependence on the QCD dynamics from the $Z$ boson production mechanism, i.e., $\pT^Z$, $y^Z$, and $m^Z$, is entirely provided by the $A_i$ coefficients and $\sigma^{U+L}$. It is clear, however, that in order to access all eight coefficients, the full dependence on $\theta$ and $\phi$ has to be analyzed. In particular, for $Z$ production at LO in QCD, i.e., $\order(\alphaS^0)$, only $A_4$ is non-zero, while at NLO, i.e., $\order(\alphaS)$, also the $A_{0-3}$ receive non-zero contributions due to the spin-1 nature of the additional gluon. The final coefficients, $A_{5-7}$, receive contributions starting at NNLO QCD, i.e., $\order(\alphaS^2)$, arising through the effective $ggZ$ interaction \cite{Hagiwara:1991xy}, and are thus comparatively small.

\begin{figure}[ht]
\centering
\hspace*{-0.05\textwidth}
\includegraphics[width=0.50\textwidth]{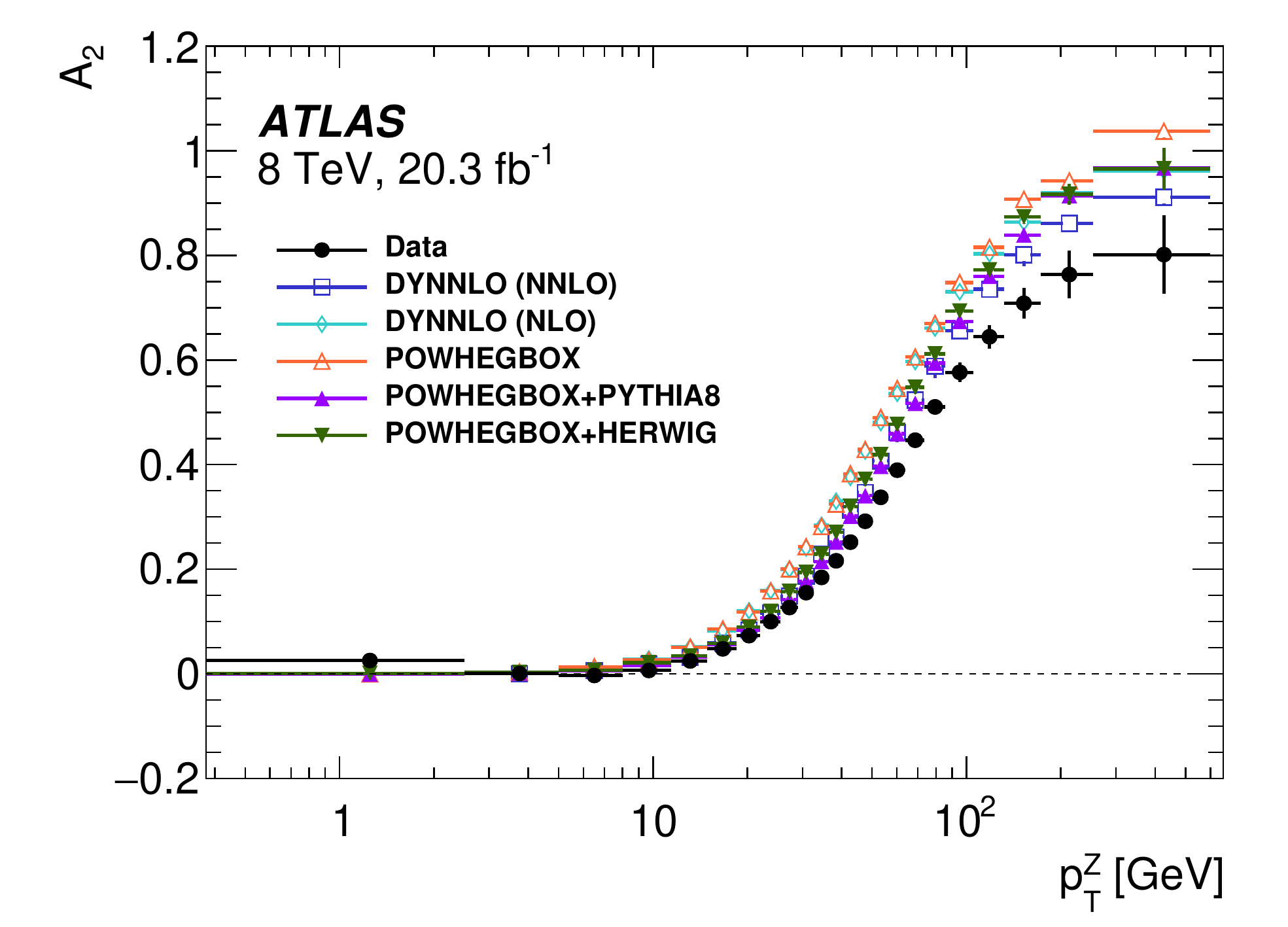}
\caption{\label{fig:ATLAS_8TeV_Zpol_A2} Distribution of the angular coefficient $A_2$ as a function of $p^{Z}_{\rm T}$, integrated over $y^Z$, in data, measured in 8 TeV \pp collisions at the \LHC, compared to the predictions at NLO, i.e., $\order(\alphaS)$, and NNLO, i.e., $\order(\alphaS^2)$, in pQCD, as well as to those from NLO calculations with two different parton-shower models. Figure taken from~\cite{Aad:2016izn}.
}
\end{figure}

\begin{figure}[ht]
\centering
\includegraphics[width=0.45\textwidth,height=0.38\textwidth]{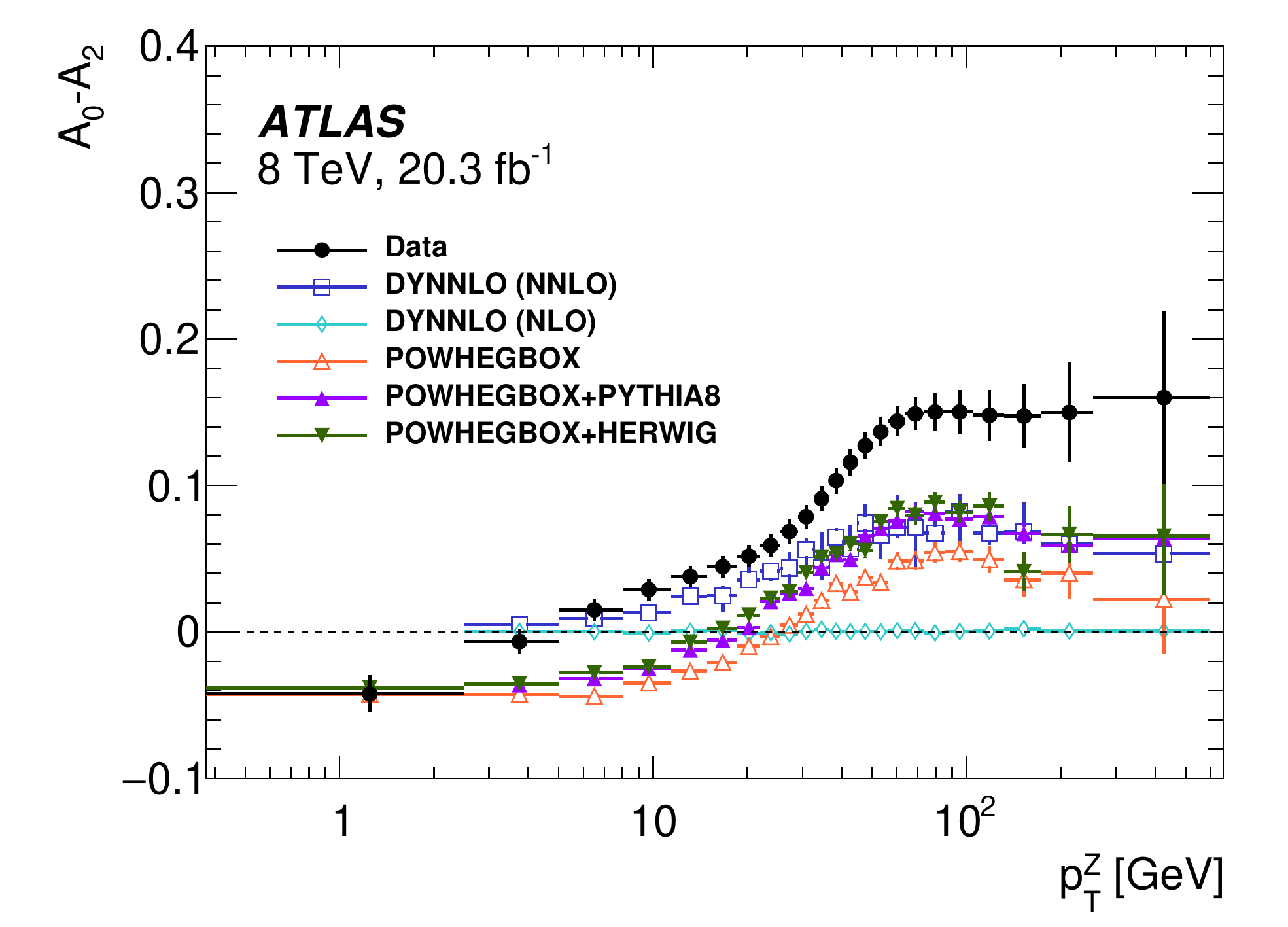}
\includegraphics[width=0.44\textwidth,height=0.40\textwidth]{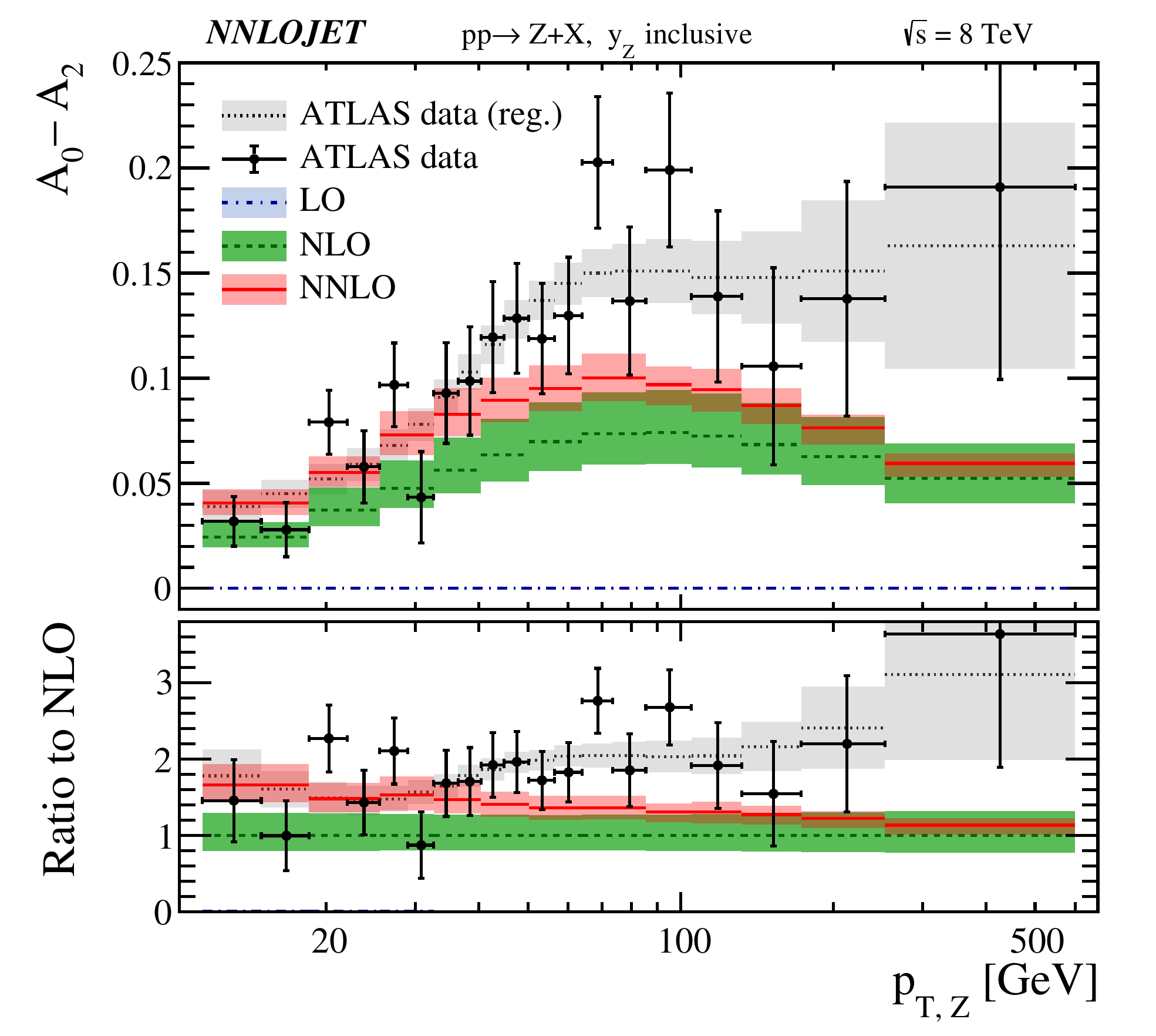}
\caption{\label{fig:ATLAS_8TeV_Zpol_A0mA2} Distribution of the angular coefficient $A_0-A_2$ as a function of $p^{Z}_{\rm T}$, integrated over $y^{Z}$, in data, measured in 8 TeV \pp collisions at the \LHC, compared to the predictions at NLO, i.e., $\order(\alphaS)$, and NNLO, i.e., $\order(\alphaS^2)$, in pQCD, as well as to those from NLO calculations with two different parton-shower models (top), figure taken from~\cite{Aad:2016izn}.  Theoretical predictions with associated uncertainties at $\order(\alphaS)$ (denoted here as LO), $\order(\alphaS^2)$ (denoted here as NLO), and $\order(\alphaS^3)$ (denoted here as NNLO) are compared to data (bottom), figure taken from~\cite{Gauld:2017tww}.
}
\end{figure}

The \CDF collaboration at the \Tevatron carried out a measurement of some of the $A_i$ angular coefficients in \ppbar collision data at a center-of-mass energy of 1.96 TeV~\cite{Aaltonen:2011nr}, and the average value of the $A_4$ coefficient was used to indirectly measure the weak mixing angle $\sin^2\theta_W$~\cite{Aaltonen:2013wcp}.
The \ATLAS and \CMS collaborations at the \LHC measured the angular coefficients for $W$ boson polarization at 7 TeV~\cite{Aad2012_Wpol,Chatrchyan:2011ig}
, and more recently for the $Z$ boson at 8 TeV~\cite{Khachatryan:2015paa,Aad:2016izn}. From the $Z$ boson angular coefficient $A_4$ measured at the \LHC, the $\sin^2\theta_W$ parameter is also extracted~\cite{ATLAS-CONF-2018-037}. Other measurements of $\sin^2\theta_W$ at the \LHC are included in Refs.~\cite{Chatrchyan:2011ya,Aad:2015uau,Aaij:2015lka, Sirunyan:2018swq}, and a legacy combination of \Tevatron measurements can be found in Ref.~\cite{Aaltonen:2018dxj} (see references therein for individual \Tevatron measurements).
As illustrated in Fig.~\ref{fig:ATLAS_Zjet_7TeV_Njets-pT}, for high values of the $Z$ boson \pT the measurements become sensitive to the production of the $Z$ boson in association with jets. Although NLO and NNLO are in general agreement with the $A_i$ distributions as functions of the $Z$ boson \pT in data, the $A_2$ coefficient, which is among the most sensitive coefficients to higher-order corrections, increases less steeply in the data than in the calculations as the $Z$ boson \pT increases, see Fig.~\ref{fig:ATLAS_8TeV_Zpol_A2}. The difference between $A_0$ and $A_2$ coefficients, i.e., $A_0-A_2$, is particularly interesting since it is zero if calculated at NLO in pQCD, the so-called Lam-Tung relation \cite{Lam:1978pu,Lam:1980uc}, and becomes positive at NNLO in pQCD. As Fig.~\ref{fig:ATLAS_8TeV_Zpol_A0mA2} (top) shows, the measured values of $A_0-A_2$ increase for increasing values of $p^{Z}_{\rm T}$, up to about 0.15, while significant deviations are observed in the comparison with MC predictions that include NLO pQCD calculations matched to the parton shower. While \ATLAS and \CMS experimental measurements are consistent, the \CMS measurement is not sufficiently precise to show significant disagreement between data and predictions.
These measurements prompted a dedicated study~\cite{Gauld:2017tww} of the $A_i$ coefficients in $Z$ boson events that calculated $\order(\alphaS^3)$ pQCD corrections and their uncertainties on $A_0-A_2$. As seen in Fig.~\ref{fig:ATLAS_8TeV_Zpol_A0mA2} (bottom), the $\order(\alphaS^3)$ corrections are large and lead to a significant improvement in the agreement with the data, however a tendency of underestimating the data is visible at high $Z$ boson \pT. It must be noticed that in Fig.~\ref{fig:ATLAS_8TeV_Zpol_A0mA2} (bottom), differently from the results shown in Fig.~\ref{fig:ATLAS_8TeV_Zpol_A2} and Fig.~\ref{fig:ATLAS_8TeV_Zpol_A0mA2} (top),  the $\order(\alphaS^2)$ and $\order(\alphaS^3)$ calculations are denoted as NLO and NNLO, respetively. 

\begin{figure}[th]
\centering
\includegraphics[width=0.40\textwidth]{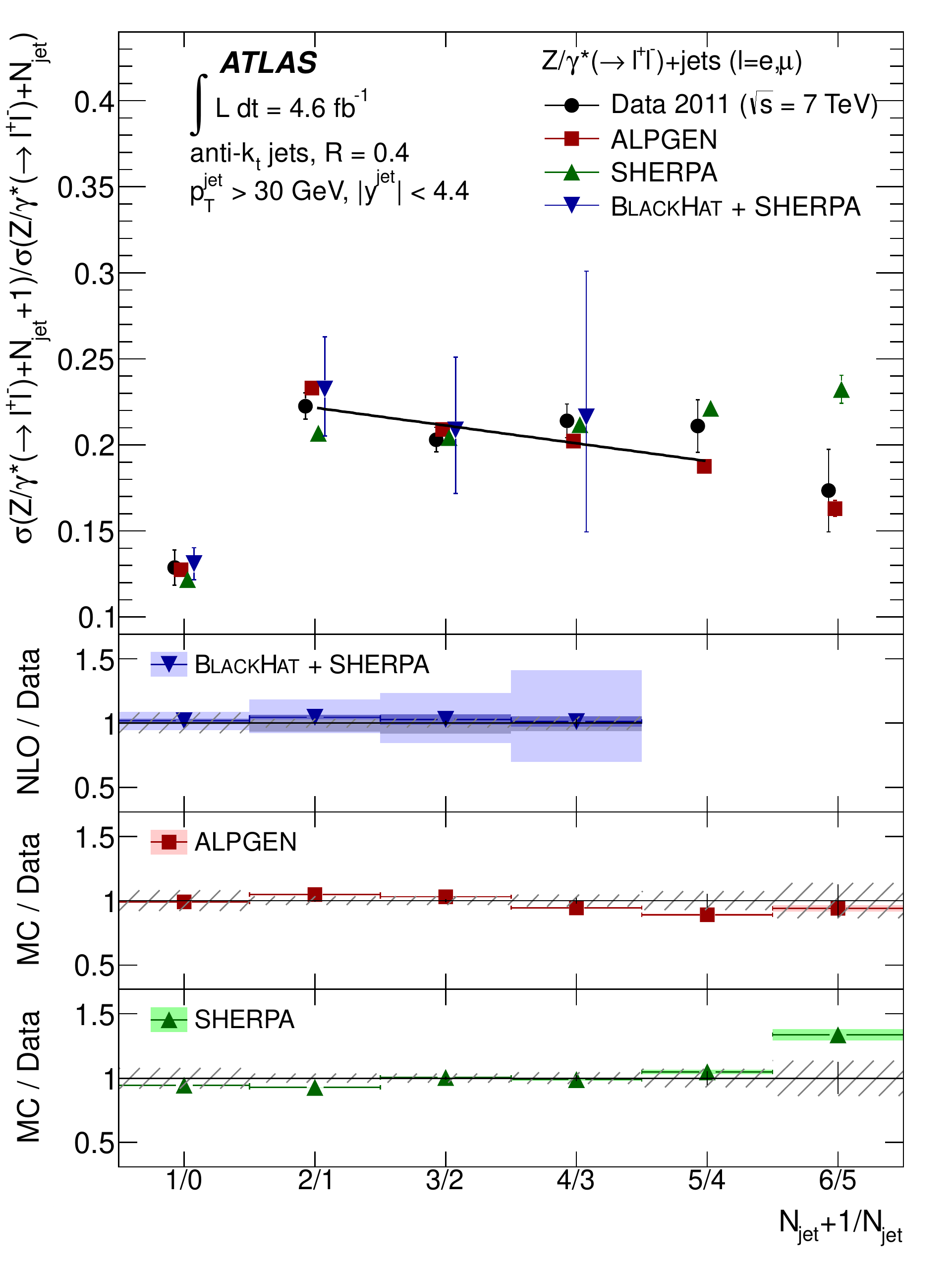}
\includegraphics[width=0.40\textwidth]{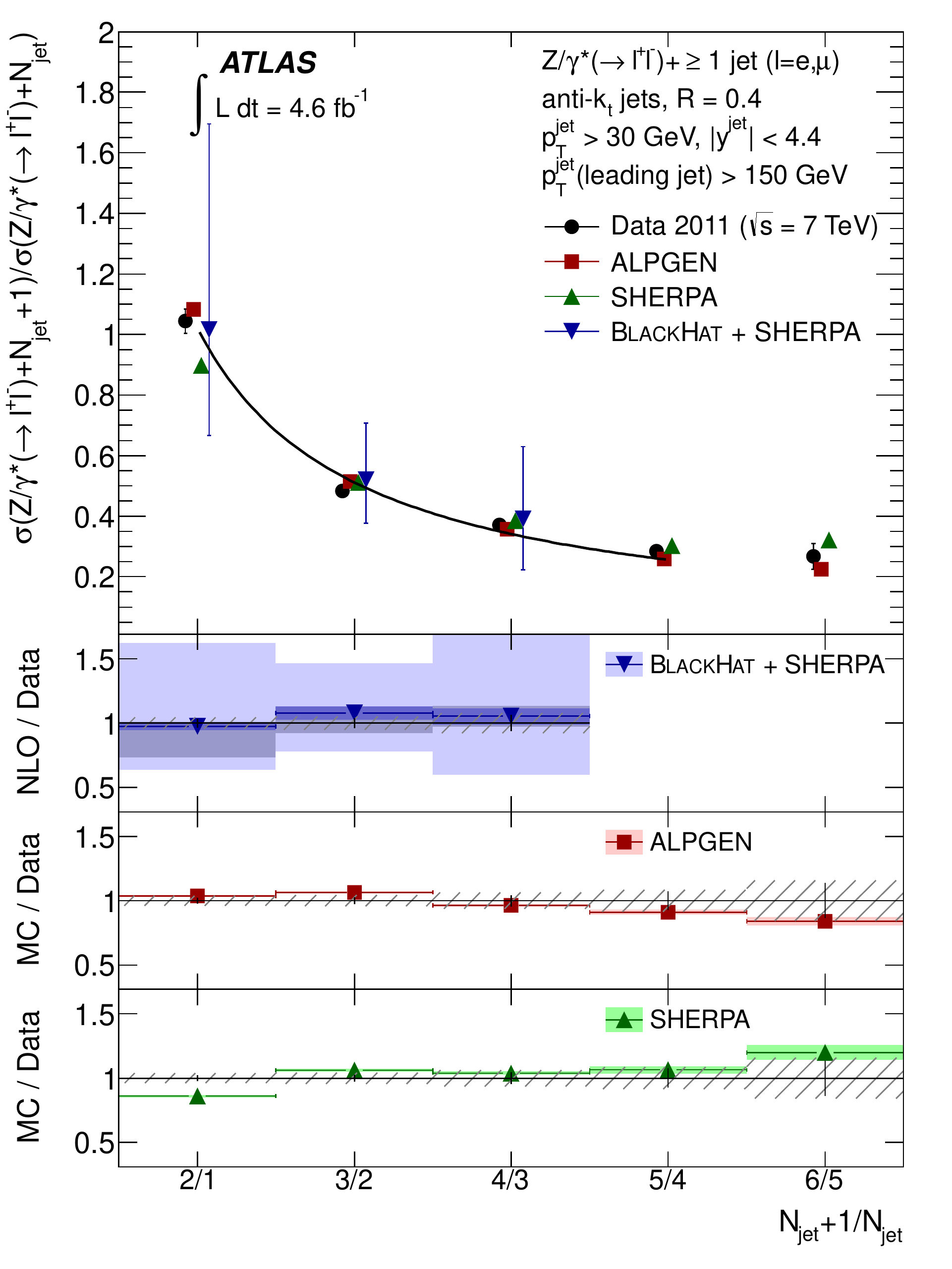}
\caption{\label{fig:ATLAS_Zjets_7TeV_scaling} Ratio of \Zjet cross sections for successive exclusive jet multiplicities, $N_{\rm jet}$, in events selected with symmetric jet selection, i.e., $\pT>30$ GeV for all jets in the event (top) and  in events with at least one jet with $\pT > 150$ GeV (bottom), with 7 TeV \pp collisions at the \LHC. The figures include comparisons with fixed-order calculation, MC simulations as well as a linear fit (top) and a Poisson fit (bottom) to the data. Figures taken from~\cite{Aad:2013ysa}.
}
\end{figure}
\begin{figure*}
\centering
\includegraphics[width=0.47\textwidth]{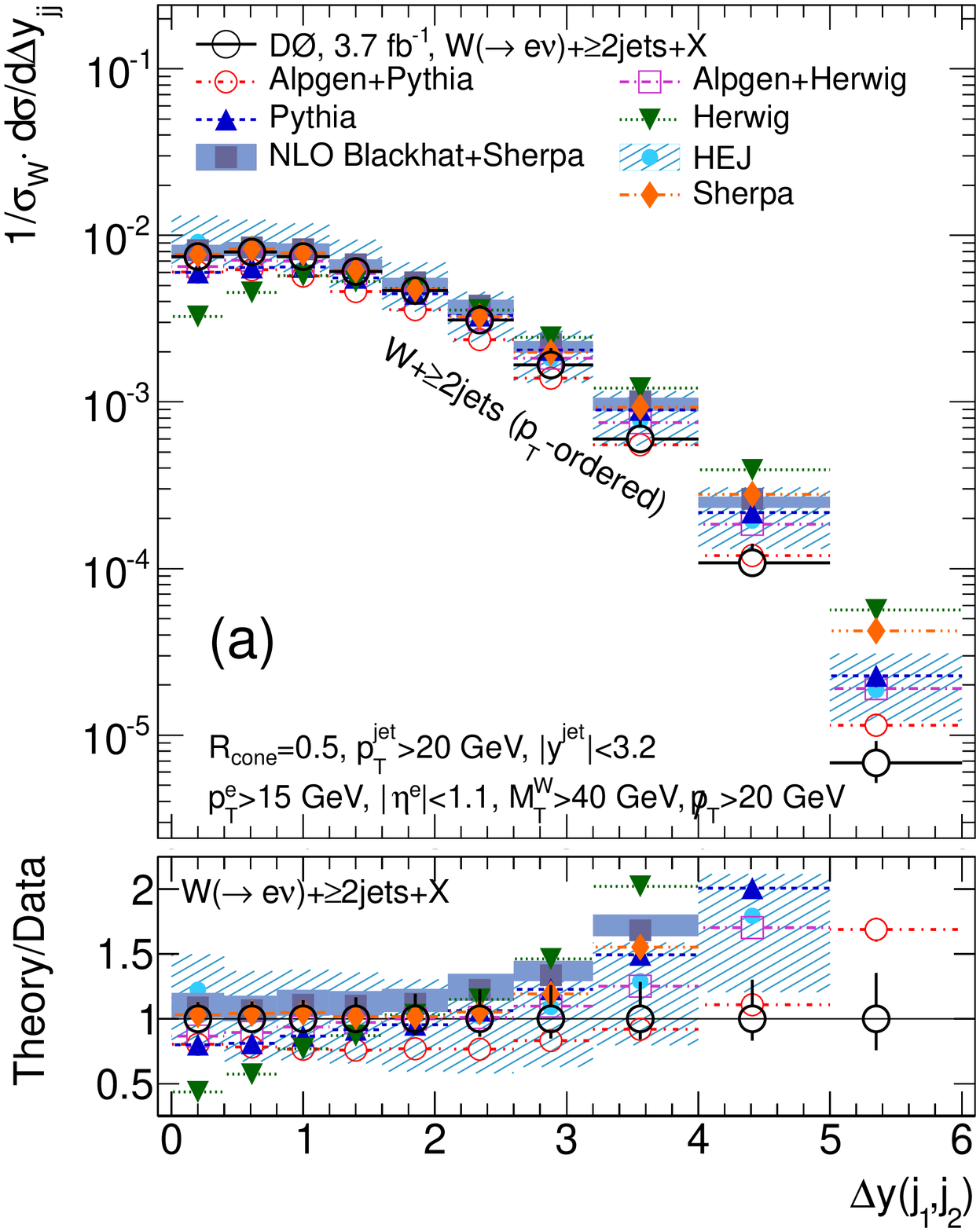}
\includegraphics[width=0.39\textwidth]{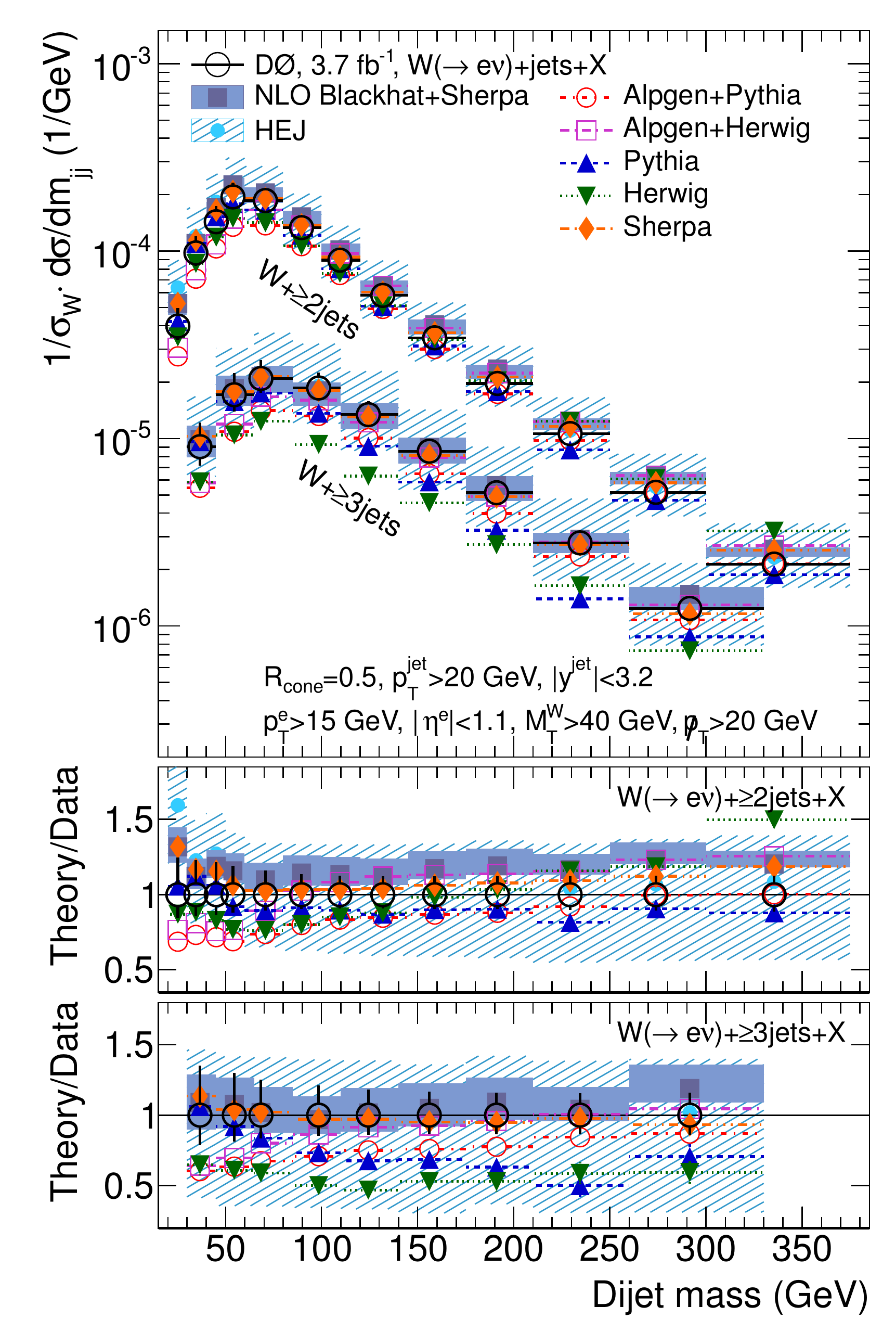}
\caption{\label{fig:Tevatron_Wjet_dymjj} Cross section for the production of $W + \ge$ 2 jets as a function of the difference in the rapidity (left) and the dijet invariant mass (right) between the two leading jets, at 1.96 TeV \ppbar collisions at the \Tevatron. The figures include comparisons with a fixed-order calculation and MC simulations. Figures taken from~\cite{Abazov:2013gpa}.
}
\end{figure*}
\begin{figure*}
\centering
\includegraphics[width=0.76\textwidth]{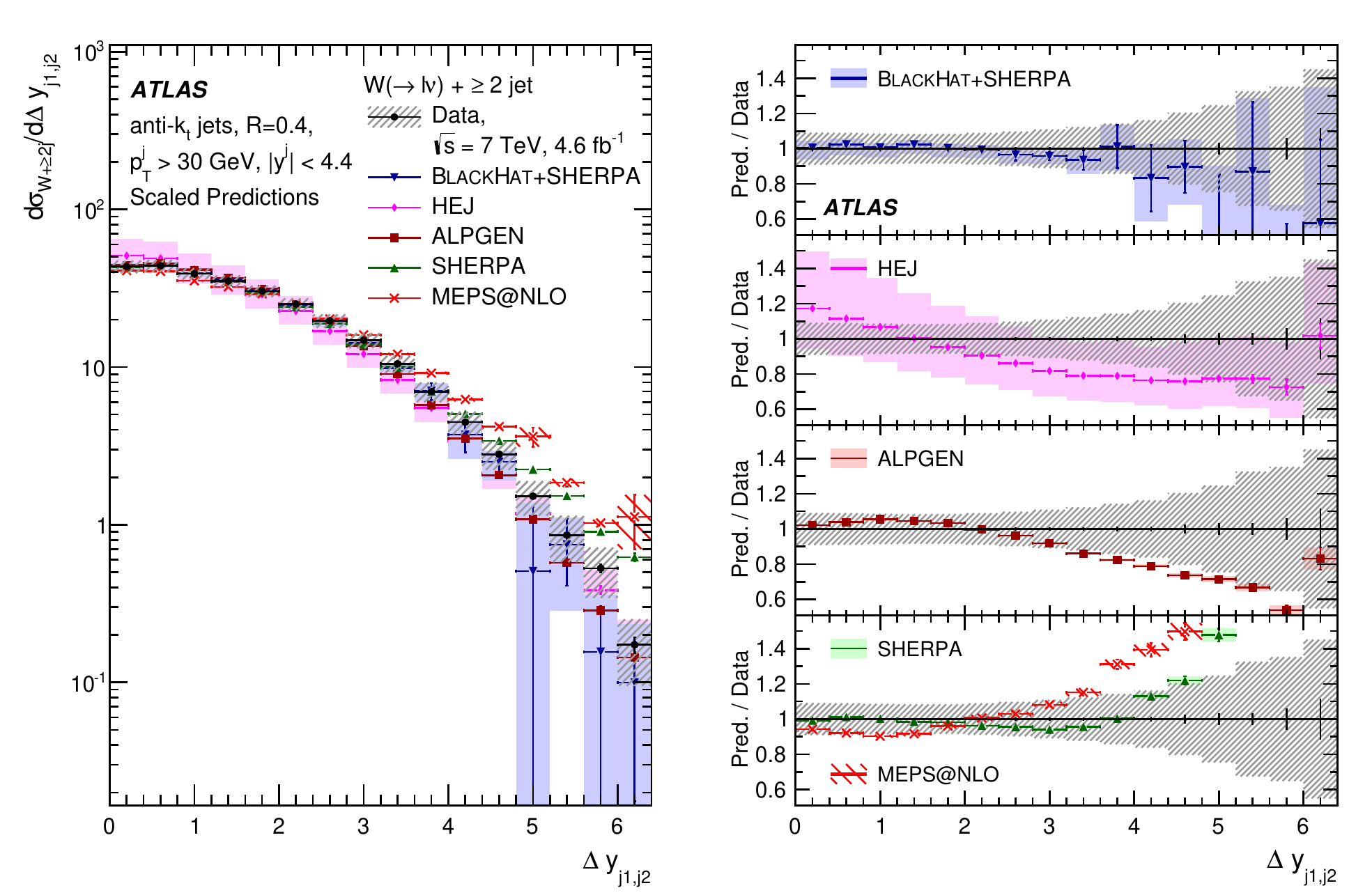}
\includegraphics[width=0.76\textwidth]{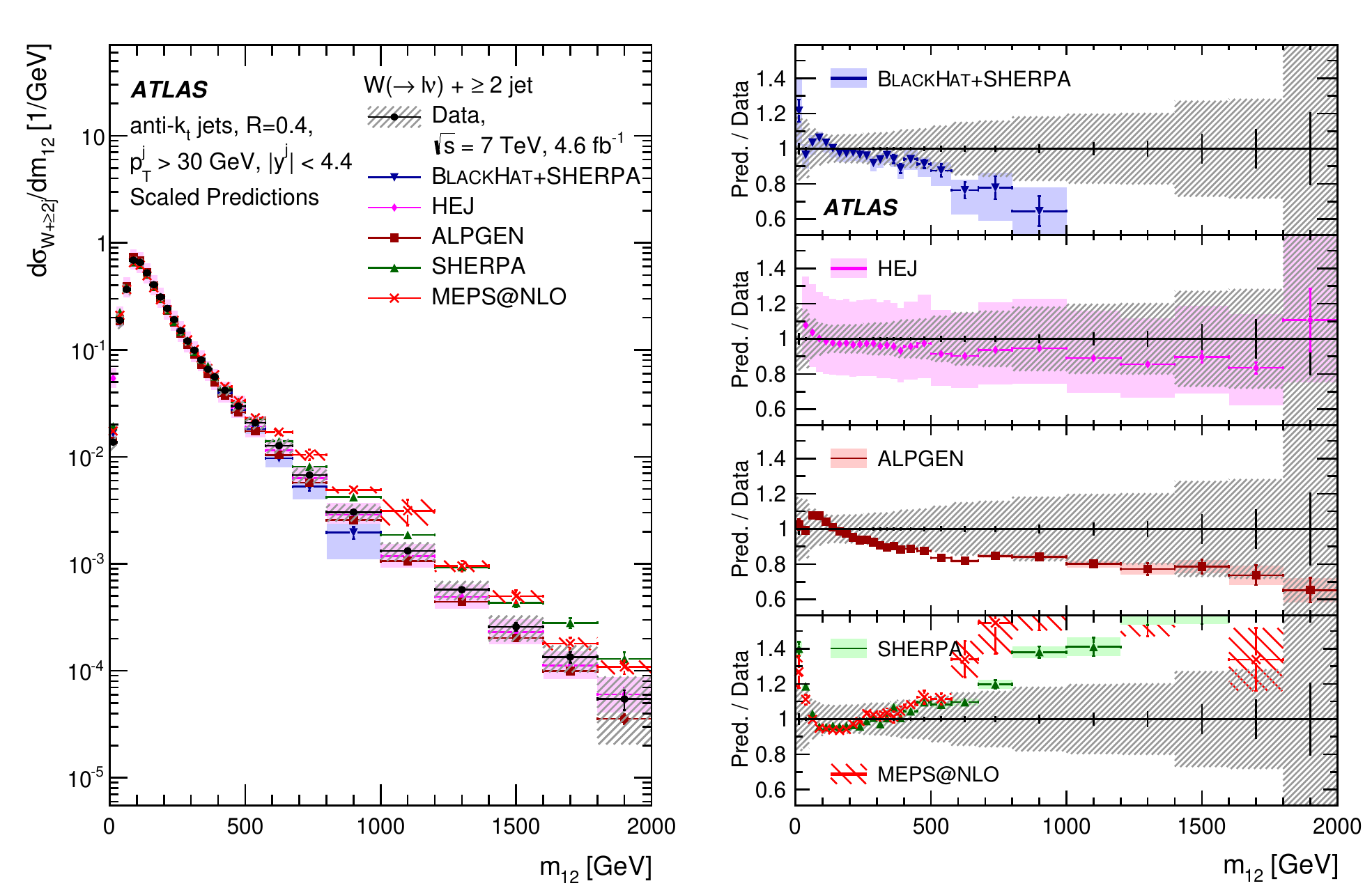}
\caption{\label{fig:ATLAS_Wjet_7TeV_dymjj} Cross section for the production of $W + \ge$ 2 jets as a function of the difference in the rapidity (top) and the dijet invariant mass (bottom) between the two leading jets, at 7 TeV \pp collisions at the \LHC. The figures include comparisons with a fixed-order calculation and MC simulations. Figures taken from~\cite{Aad:2014qxa}.
}
\end{figure*}
The studies of QCD scaling properties are useful for a better understanding of  QCD dynamics and in analyses that employ jet vetoes to separate signal processes from \WZjets backgrounds~\cite{Gerwick:2012hq,Berends:1989cf}. Figure~\ref{fig:ATLAS_Zjets_7TeV_scaling} from Ref.~\cite{Aad:2013ysa} reports a study of two different types of scaling in \Zjets in the exclusive jet multiplicity ratios $R_{(n+1)/n}  =  N_{Z+(n+1)} / N_{Z+n}$. When a symmetric selection of the jet transverse momenta is applied, i.e., $\pT>30$ GeV for all jets, the so-called "Staircase scaling" is seen, whereas when an asymmetric selection of the jet transverse momenta is applied, i.e., $\pT({\rm leading})>150$ GeV and $\pT>30$ GeV for all other jets, a falling distribution is seen, i.e., the so-called "Poisson scaling". The scaling properties measured in data are well reproduced by the theory.
The staircase scaling is a property of non-abelian theories with $R_{(n+1)/n} = R = e^{-b}$, as  $\sigma_n = \sigma_0 e^{-bn}$, and occurs in events with democratic jet selection and no major scale separations. The first bin of the distribution in Fig.~\ref{fig:ATLAS_Zjets_7TeV_scaling} (top), i.e., $R_{1/0}$, is suppressed by PDF effects by about $60\%$.
The Poisson scaling (already known from final-state-radiation QED at $e^+e^-$ colliders) occurs in events that feature large differences between the scale $Q$ of the process and the radiation cut-off scale $Q_0$. For $Q \gg Q_0$ each emission is independent from the previous one (the primary emission is typically off the hard parton leg), while for $Q\approx Q_0$ the emissions are correlated (for secondary emissions from secondary quark lines). In this configuration the ratio $R_{(n+1)/n} = \langle n\rangle/n+1$ follows a Poissonian distribution with $P_n=\tfrac{1}{n!}\,\langle n\rangle^n e^{-\langle n\rangle}$ and occurs in abelian theories too.
Asymptotically for large $N_{\rm jets}$ the staircase approximation dominates as can be seen in Fig.~\ref{fig:ATLAS_Zjets_7TeV_scaling} (bottom).

For other cross section measurements of \WZjet processes at the \Tevatron Run-2 and at the \LHC, the reader is referred to the publications in Refs.~\cite{Aaltonen:2007ip,Aaltonen:2007ae,Abazov:2011rf,Abazov:2008ez,Abazov:2009av,Aad:2010ab,Aad:2012en,Aad:2011qv,Sirunyan:2018cpw,Sirunyan:2017wgx,Khachatryan:2014zya,Khachatryan:2014uva,Chatrchyan:2011ne,Aaij:2019ctd,Aad:2019hga,Aaboud:2017hbk,AbellanBeteta:2016ugk}.

\subsubsection{Event properties}
\label{sec:VLF:exp:prop}


The measurements of angular distributions provide important tests of the modeling of QCD in the theory, as these measurements are sensitive to the parton emission at small and large angles. Hard emissions at large angles are typically calculated by matrix elements, while unresolved soft or collinear radiation is typically modeled in MC generators by the parton shower.
Measurements of the angular ($\Delta\phi$(j$_1$,j$_2$)) or rapidity ($\Delta y $(j$_1$,j$_2$)) separation between the two associated leading jets or their invariant mass ($m_{jj}$) distribution are important for studies of vector boson fusion or scattering to disentangle the electroweak from the QCD production mechanisms, see Sec.~\ref{sec:VBF}.
Figures~\ref{fig:Tevatron_Wjet_dymjj} and~\ref{fig:ATLAS_Wjet_7TeV_dymjj} show selected measurements of $\Delta y $(j$_1$,j$_2$) and $m_{jj}$ at the \Tevatron and at the \LHC, respectively, in events with a $W$ boson produced in association with at least two jets selected in a broad kinematic region, i.e., jet $\pT >$ 30(20) GeV and $|y| <$ 4.4(3.2) at the \LHC (\Tevatron). The fixed-order NLO calculation (\BlackHat) is in good agreement with data on $\Delta y$(j$_1$,j$_2$), especially at the \LHC. 
A similar level of discrepancy is seen at the \Tevatron and at the \LHC  for \Sherpa, while \Alpgen and \HEJ (based on BFKL-like resummation) MC generators are in better agreement with the data. 
The fixed-order NLO calculation (\BlackHat) is in good agreement with the data in the $m_{jj}$ distribution in the range accessible by the \Tevatron, i.e., up to about 300 GeV.
The \LHC measurement of the $m_{jj}$ distribution extends to 2 TeV, and the fixed-order NLO calculation is compared to \LHC data up to 1 TeV, showing good agreement up to approximately 500 GeV.
The \HEJ simulation is in agreement with data over the whole $m_{jj}$ range at the \Tevatron and at the \LHC, but its associated uncertainties are large.  Significant discrepancies in the high $m_{jj}$ region at the \LHC are visible in LO and NLO multi-leg MC predictions in the \Sherpa and \MEPSatNLO calculations. In such a kinematic region important contributions are expected from the modeling of the beam remnant, underlying event, multi-parton interactions and parton shower.
Similar measurements at the \LHC~\cite{Khachatryan:2016fue,Khachatryan:2016crw,Aaboud:2017hbk} in a restricted phase space, i.e., jet $|y| < 2.4$ and $m_{jj}$ up to 700 GeV -- 1 TeV, are compatible with those presented above, but in such a kinematic region do not show significant data-theory discrepancies. 
In an updated analysis of $W$ + $\ge$ 2-jet events at the \LHC in the same broad phase space of jet $\pT >$ 30 GeV and $|y| <$ 4.4, which extends the reach of the $m_{jj}$ distribution to 3 TeV~\cite{Aaboud:2017soa}, a very good agreement is found between data and  updated theoretical calculations, while the same level of discrepancy is observed with the older LO \Sherpa version (v1.4).
These measurements show that the extension of the kinematic reach of the \LHC can expose theoretical mismodeling and can be used to improve the theoretical predictions.

\begin{figure}
\centering
\hspace*{-0.05\textwidth}
\includegraphics[width=0.43\textwidth]{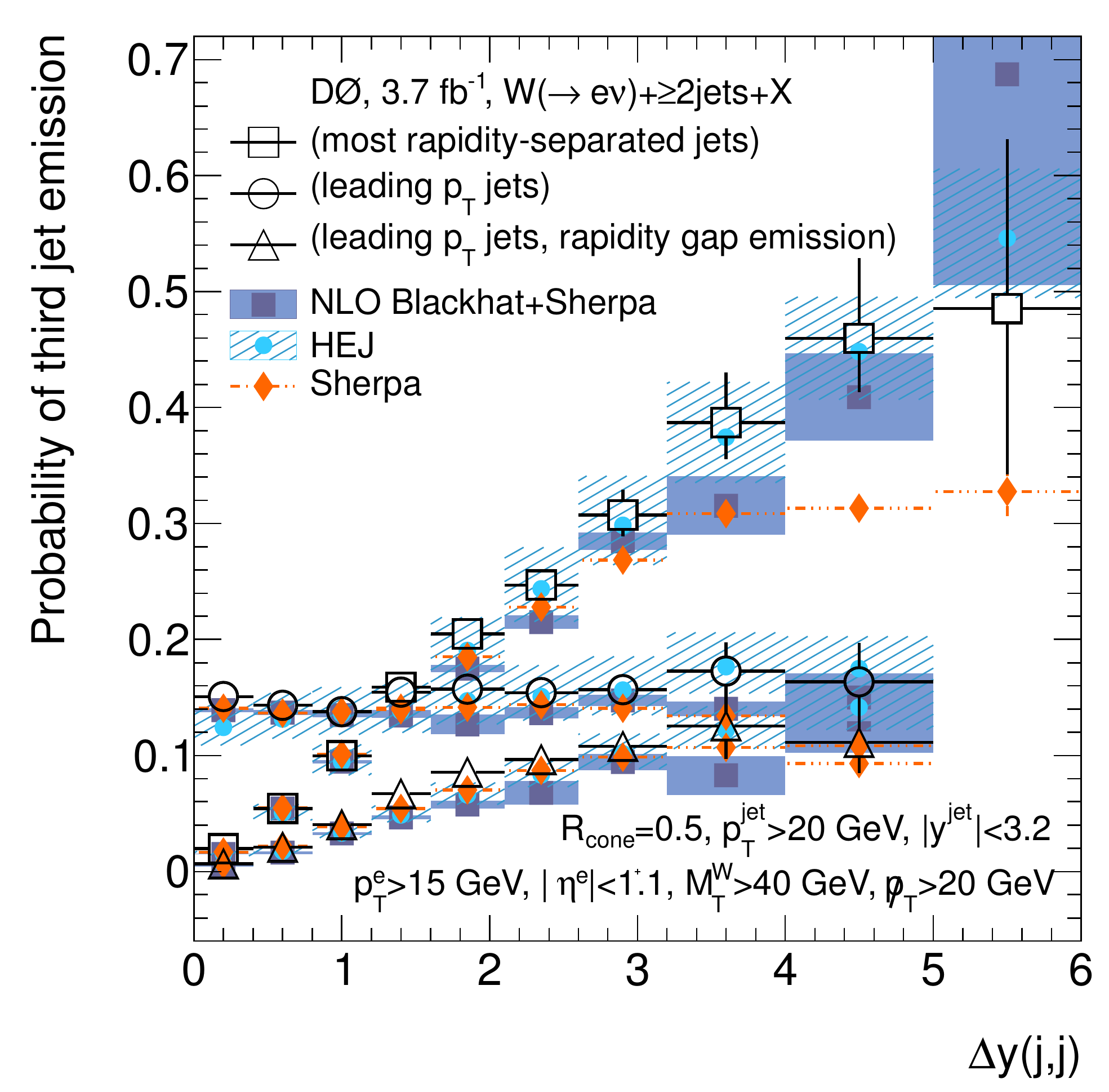}
\caption{\label{fig:D0_196TeV_3rdJetProb} Measurement of the probability for the emission of a third jet in events with $W + \ge$ 2 jets as a function of the dijet rapidity separation of the two tagging jets at 1.96 TeV \ppbar collisions at the \Tevatron, and comparisons with theory predictions. The definitions of jet tagging are the two most-rapidity-separated jets, the two highest-\pT jets, and the two highest-\pT jets with a $3^{rd}$ jet produced in the rapidity gap between them. Figure taken from~\cite{Abazov:2013gpa}. 
}
\end{figure}

The \DO collaboration has also studied in Ref.~\cite{Abazov:2013gpa} the probability of emission of a $3^{rd}$ jet in events with a $W$ and at least two associated jets, as a function of the rapidity separation between the two tagged jets, under various definitions of jet tagging (two most-rapidity-separated jets, the two highest-\pT jets, or the two highest-\pT jets with a $3^{rd}$ jet produced in the rapidity gap between them), see Fig.~\ref{fig:D0_196TeV_3rdJetProb}. Such a measurement provides a laboratory for studies of rapidity gaps, central jet veto and vector boson fusion jet dynamics. They can test the high-\pT and the wide-angle jet production, in a complementary way to studies of dijet events.
The results show that in such configurations there are competing effects of increasing phase space for high-\pT jet emission between jets and decreasing PDFs at large $x$. The BFKL-based resummation calculation in \HEJ generator describes best the data.
%

\begin{figure}
\centering
\begin{minipage}{0.45\textwidth}
    \centering
    \includegraphics[width=\textwidth,height=0.95\textwidth]{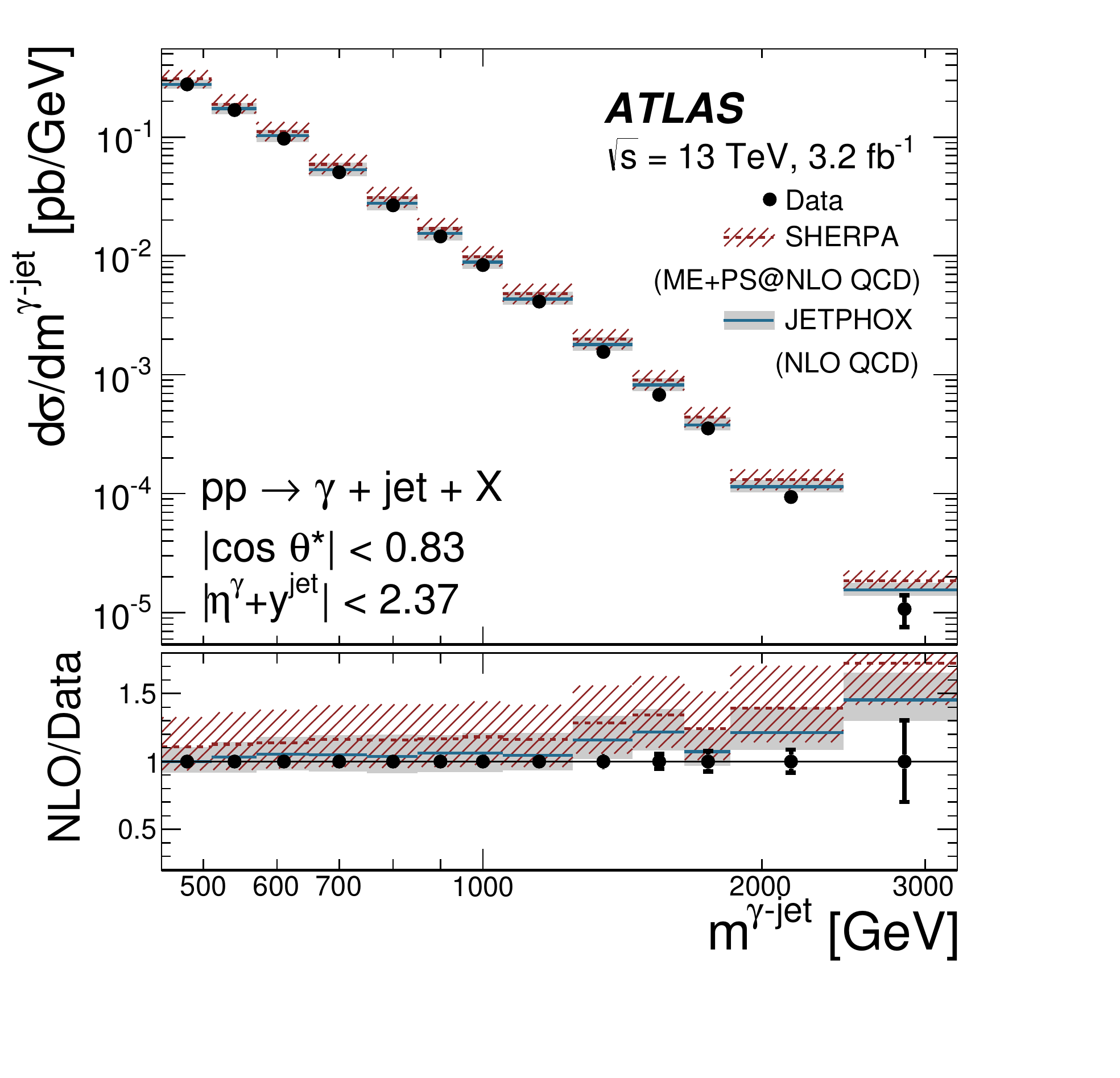}
    \end{minipage}
\hfill
\vspace*{-0.04\textwidth}
\hspace*{-0.04\textwidth}
\begin{minipage}{0.45\textwidth}
    \includegraphics[width=\textwidth,height=1.15\textwidth]{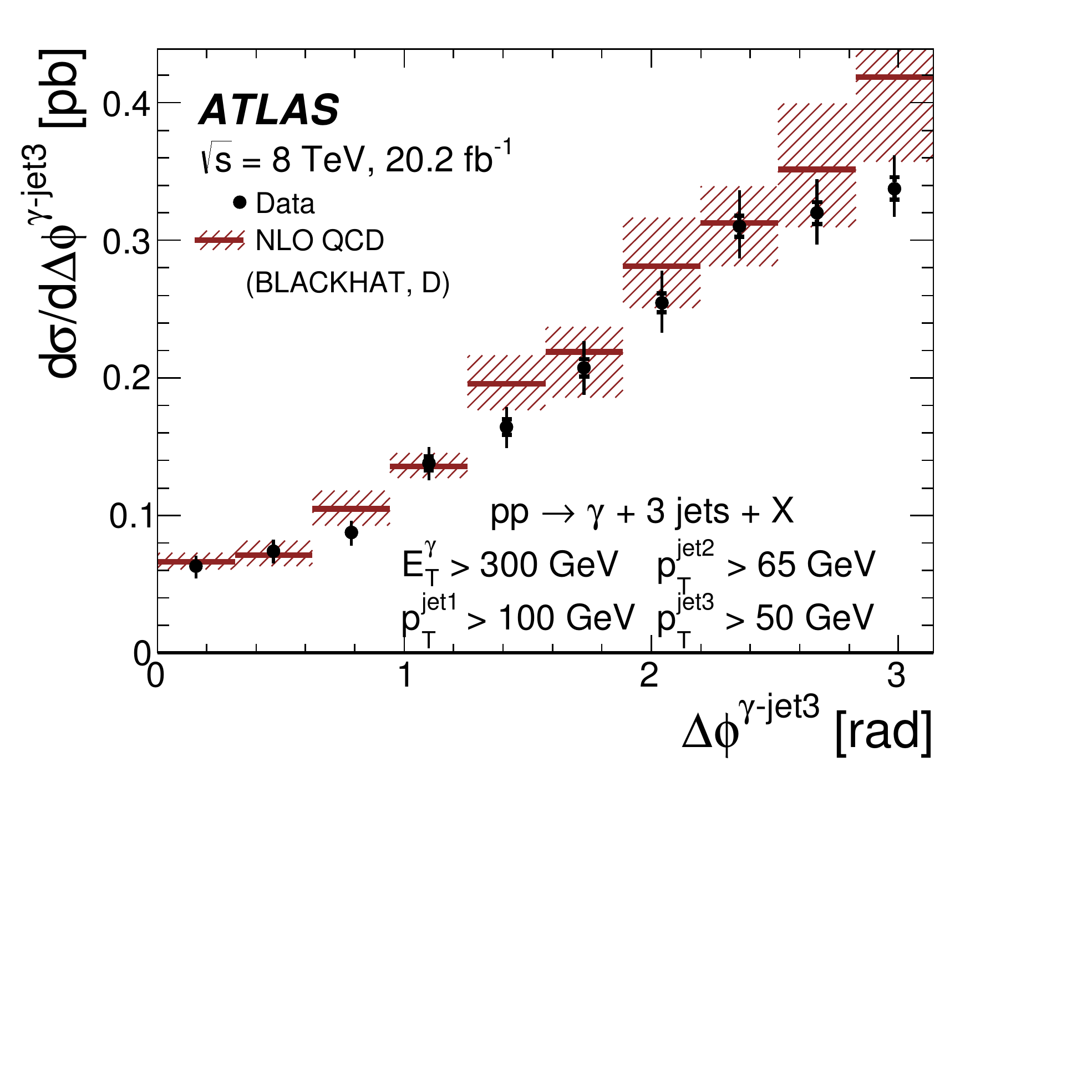}
    \end{minipage}
    \vspace*{-0.15\textwidth}
\caption{\label{fig:ATLAS_Photonjet_7-13TeV_m_deltaPhi} Differential cross sections measured at the \LHC for isolated-photon plus jet production as a function of  $m^{\gamma-{\rm{jet}}}$ at 13 TeV center-of-mass energy (top), figure taken from~\cite{Aaboud:2017kff}, and for isolated-photon plus three-jet production as a function of $\Delta\phi^{\gamma-{\rm{jet3}}}$ for $\ET^{\gamma}>300$ GeV  at 8 TeV center-of-mass energy (bottom), figure taken from~\cite{Aaboud:2016sdm}, and comparisons with theoretical predictions from fixed-order calculations and a MC generator at NLO in pQCD. 
}
\end{figure}

The study of events with a photon and jets is used for searches of new physics signatures, such as heavy resonance states decaying into a photon and a jet. Such new physics processes can produce distinct features in the  \gammajet final state, such as deviation in the invariant mass of the photon and the jet ($m^{\gamma-{\rm{jet}}}$) or angular correlations between the photon and the jet with respect to SM expectations. Figure~\ref{fig:ATLAS_Photonjet_7-13TeV_m_deltaPhi} presents measurements of differential cross sections as a function of $m^{\gamma-{\rm{jet}}}$ in \gammajet events at the \LHC at 13 TeV center-of-mass energy, and as a function of azimuthal angular separation between the photon and the third leading jet in events with a photon and at least 3 jets in \pp collisions at 8 TeV center-of-mass energy.
The differential cross section of $d\sigma/dm^{\gamma-{\rm{jet}}}$ shown in Fig.~\ref{fig:ATLAS_Photonjet_7-13TeV_m_deltaPhi} (top) is monotonically decreasing by more than four orders of magnitude up to the highest measured value of $m^{\gamma-{\rm{jet}}}=3.25$ TeV.
Both NLO QCD predictions (fixed-order \Jetphox and \Sherpa MC generator that includes the matching of NLO matrix-element with parton showering) describe the data within the experimental and theoretical uncertainties. However, in the highest $m^{\gamma-{\rm{jet}}}$ range a trend of the simulation to overestimate the data is seen. 
Figure~\ref{fig:ATLAS_Photonjet_7-13TeV_m_deltaPhi} (bottom) shows that the cross-section $d\sigma/\Delta\phi^{\gamma−{\rm{jet3}}}$ increases as $\Delta\phi^{\gamma−{\rm{jet3}}}$ increases, indicating the preference for back-to-back configuration between the photon and the third leading jet in \gammajet events. 
The fixed-order NLO pQCD prediction by \BlackHat gives an adequate description of the angular correlations and their evolution with energy scale, however shows a tendency to systematically overestimate the data.

A recent analysis of $\gamma$ + 2 jets + X production at 13 TeV at the \LHC with 36.1 $\rm{fb^{-1}}$ of integrated luminosity is carried out in two distinct regions of phase space: one enriched with direct photon production and one with photon fragmentation processes. Experimental cross sections are measured as a function of several observables, including $m^{\rm{jet+jet}}$, $m^{\gamma \rm{+jet+jet}}$ as well as azimuthal and rapidity differences between the photon and leading jet and between the two jets. Good agreement between data and MC predictions with tree-level multi-jet matrix element merged to parton shower or with NLO accuracy in QCD are observed in the sample enriched with direct photon production, whereas discrepancies are observed in the sample enriched with fragmentation processes. The precision of the measurement is significantly better than the differences between the predictions indicating that theoretical uncertainties are much larger than those of experimental nature~\cite{Aad:2019cpw}.

\begin{figure*}
\centering
\includegraphics[width=0.45\textwidth]{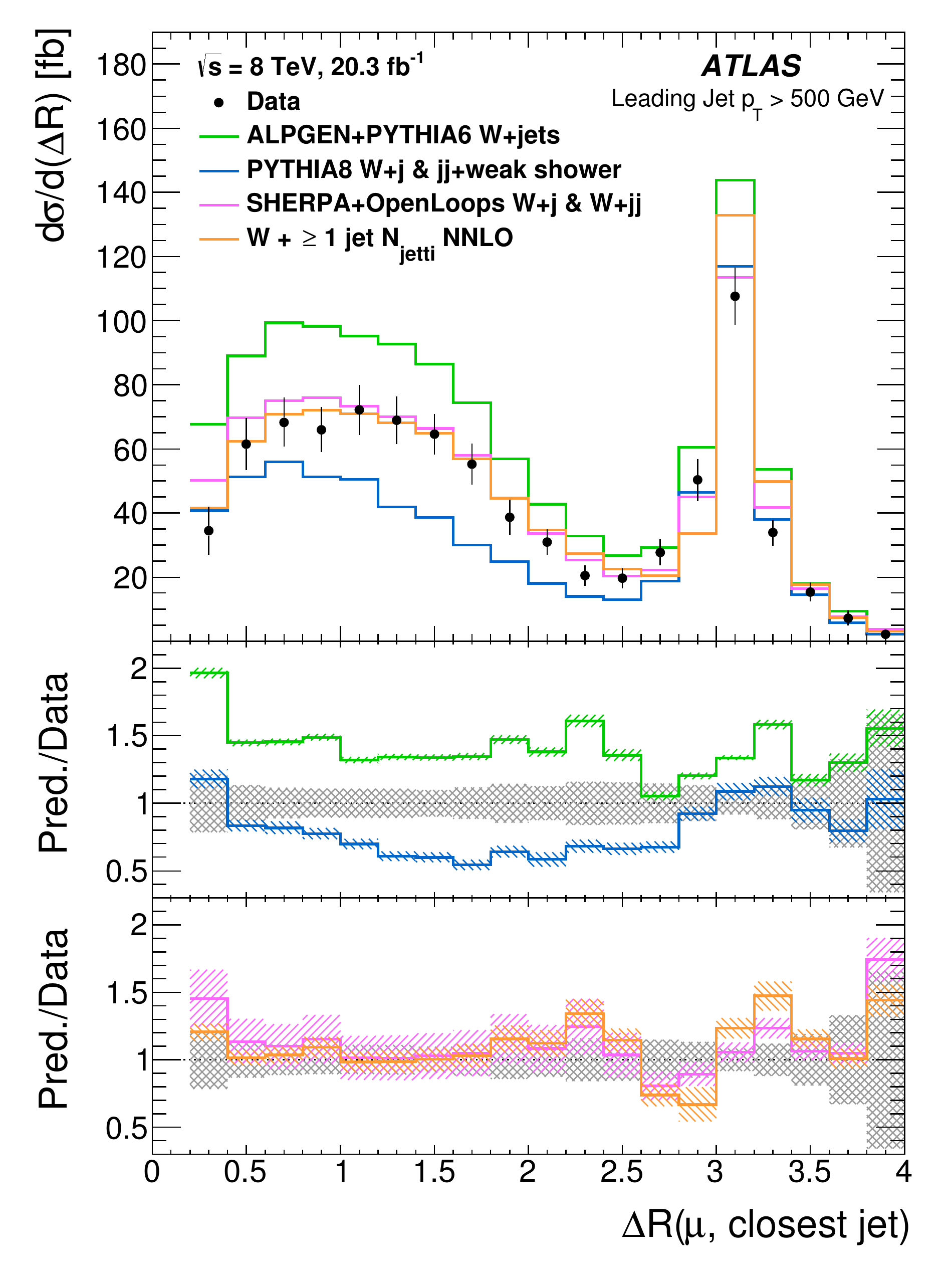}
\includegraphics[width=0.45\textwidth]{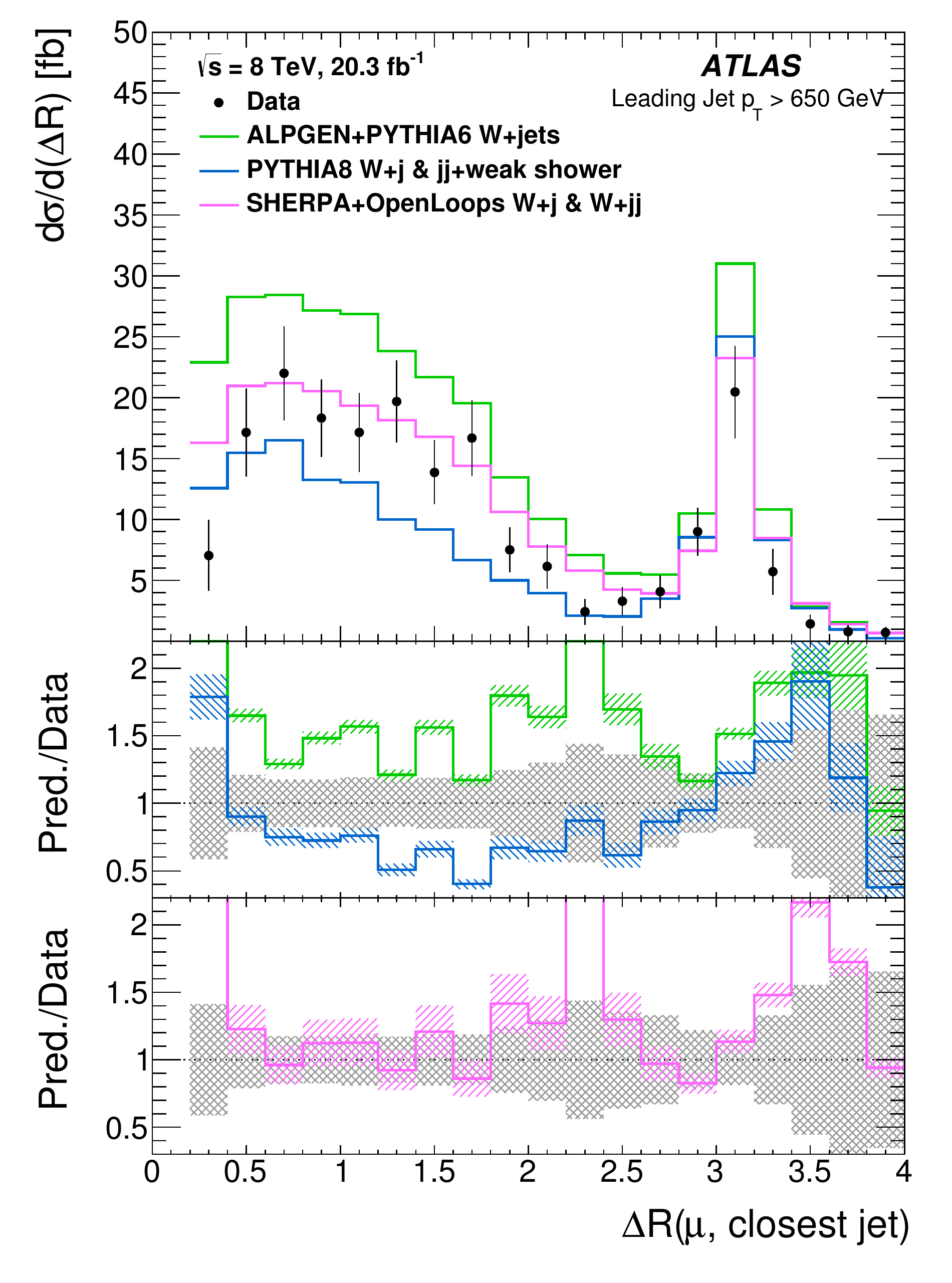}
\caption{\label{fig:ATLAS_Wjets_8TeV_collinear} Differential \Wjet cross section as a function of the angular separation $\Delta R$ between the $W$ decay muon and the closest jet for events with \pT(leading jet) > 500 GeV (left) and \pT(leading jet) > 650 GeV (right), with 8 TeV \pp collisions at the \LHC. Several different theoretical predictions from MC generators are compared to experimental data. Figures taken from~\cite{Aaboud:2016ylh}.
}
\end{figure*}
\begin{figure*}
\centering
\includegraphics[width=0.38\textwidth]{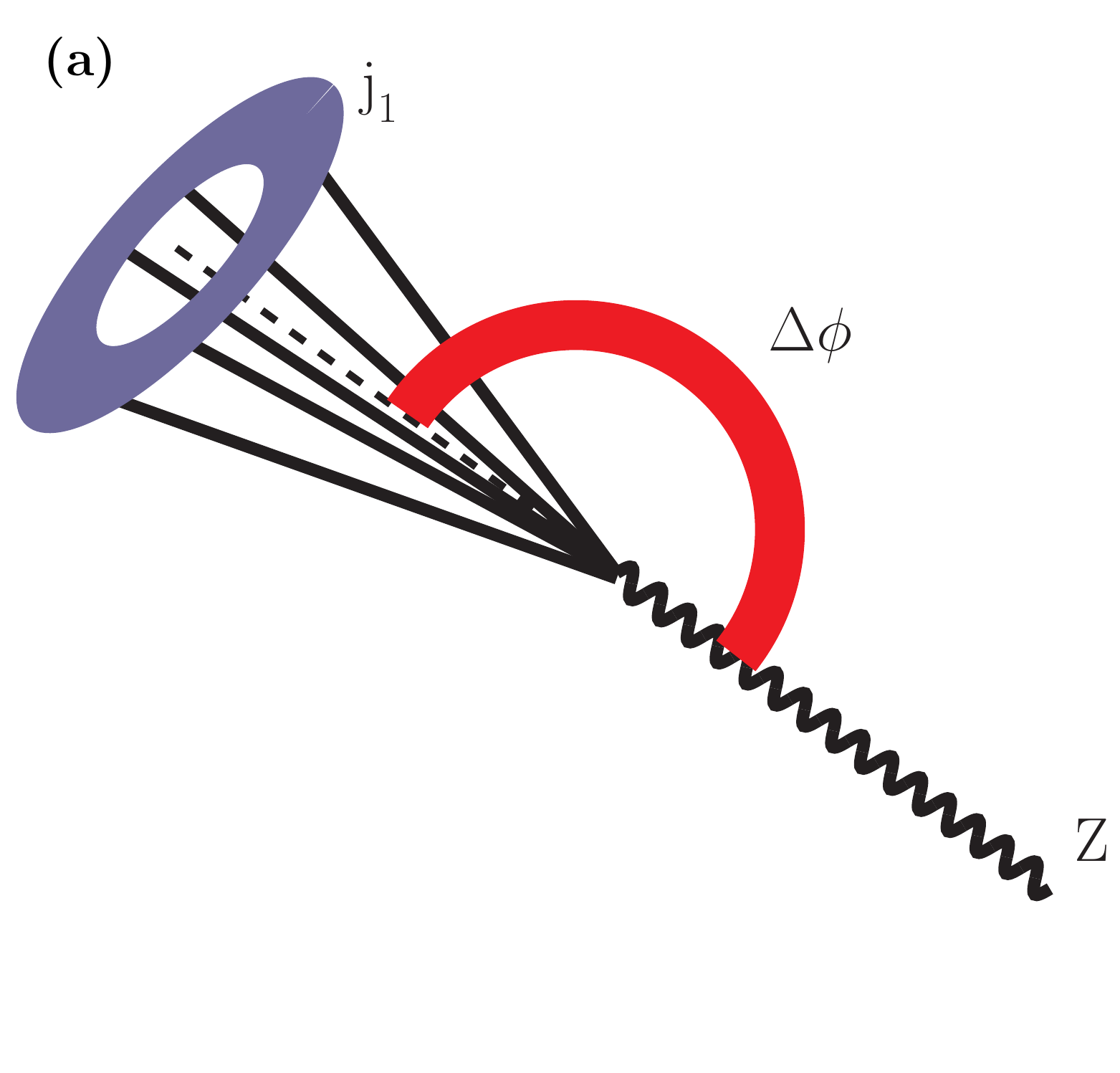}
\hspace*{0.1\textwidth}
\includegraphics[width=0.45\textwidth]{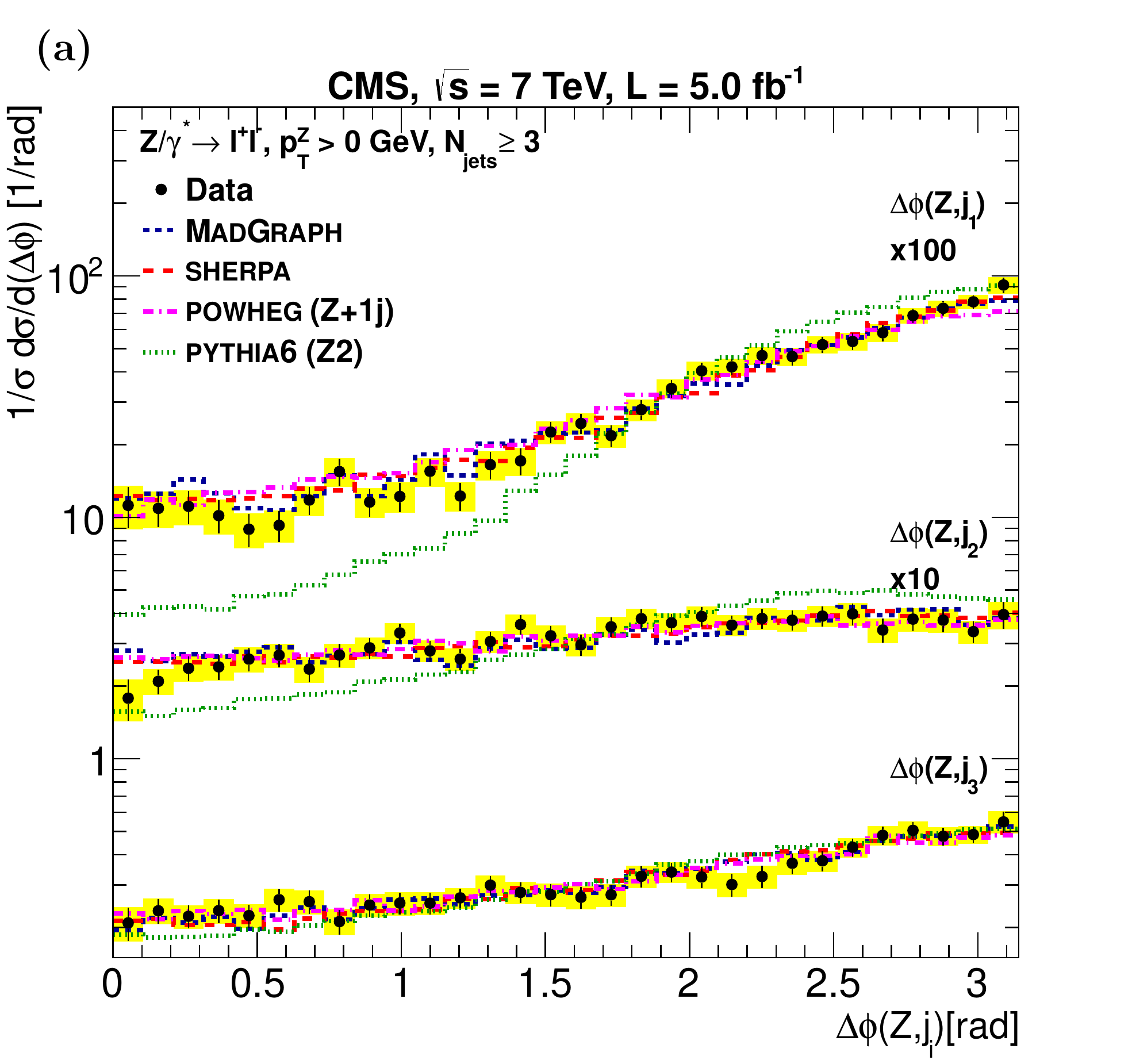}\\
\includegraphics[width=0.38\textwidth]{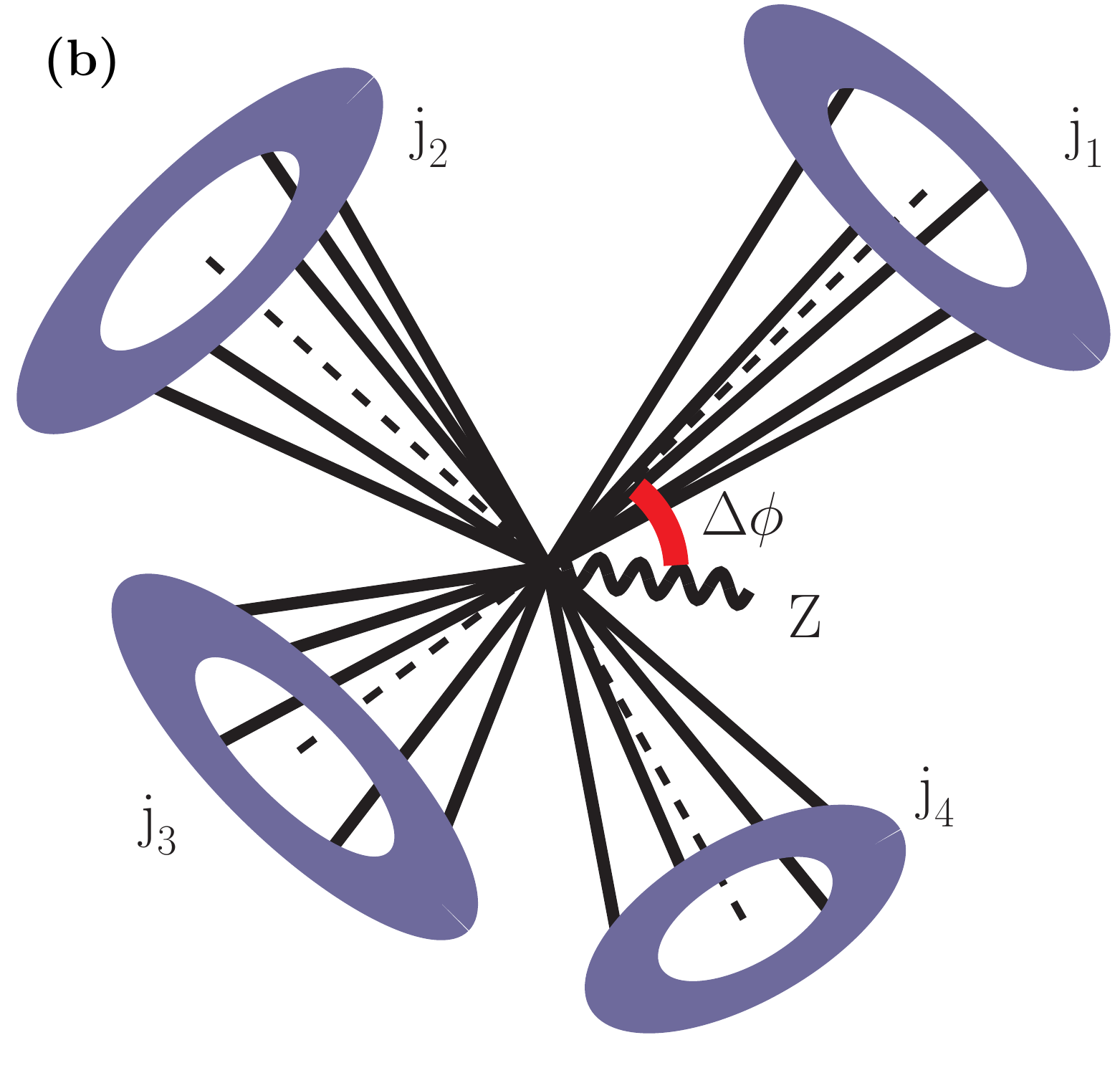}
\hspace*{0.1\textwidth}
\includegraphics[width=0.45\textwidth]{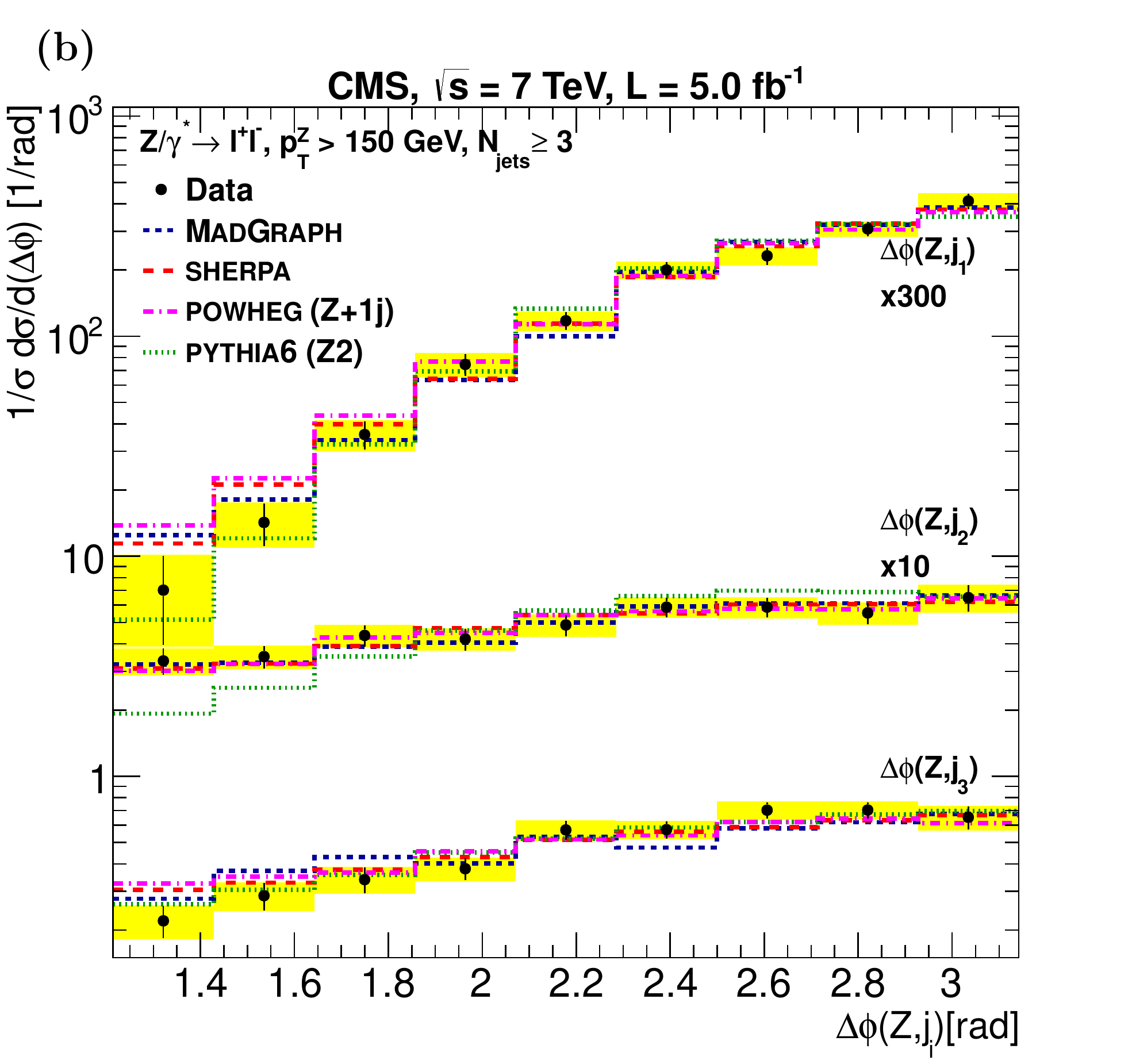}
\caption{\label{fig:CMS_Zjets_8TeV_eventshapes} Topology of \Zjet events for $\Delta\phi(Z,j_1) \rightarrow \pi$ (top-left), and for $\Delta\phi(Z,j_1) \ll \pi$ (bottom-left). Normalized $\Delta\phi(Z,j_i)$ ($i=1,2,3$) distributions for the inclusive $N_{\rm jets}\ge 3$ for $\pT^Z> 0$ GeV (top-right) and for $\pT^Z> 150$ GeV (bottom-right), with 8 TeV \pp collisions at the \LHC, compared to theoretical predictions from MC generators.
Figures taken from~\cite{Chatrchyan:2013tna}.
}
\end{figure*}

The real emission of a vector boson from an initial- or final-state quark has a collinear divergence in the limit of a massless boson. This may be detected as a collinear enhancement in the distribution of the angular separation between the vector boson and the closest jet.
Parton shower algorithms are implemented in MC generators to account for QCD and QED emissions in the soft and collinear approximation and an analogous mechanism occurs for the emission of real weak bosons.
At very high energies the real emission of weak bosons in dijet events can significantly contribute to the inclusive \Wjet measurement. Measurements of \Wjet production at the \LHC are often insensitive to such an effect as they require large separation between the decay charged lepton and any of the jets.
The analysis in Ref.~\cite{Aaboud:2016ylh} studies event configurations in which a muon from a $W$ decay is produced close to a high transverse momentum jet.
Figure~\ref{fig:ATLAS_Wjets_8TeV_collinear}
shows the differential cross section for \Wjet events with at least one jet with $\pT>500$ GeV and higher, and any additional jets with $\pT>100$ GeV, as a function of the $\Delta R$ distance between the $W$ decay muon and the closest jet.
An enhancement of the collinear event fraction is expected for increasing values of the leading jet \pT, as the $W$ emission from the jet is enhanced. This effect is illustrated in Fig~\ref{fig:ATLAS_Wjets_8TeV_collinear}: as the value of leading-jet \pT increases from $\pT>500$ GeV to $\pT>650$ GeV,  the  fraction  of  events in the collinear region at low $\Delta R$ increases with respect to the fraction of events in the back-to-back configuration.  
The \Alpgen MC simulation for \Wjet production overestimates the data, especially in the collinear region. The prediction by \PythiaEight,  which  is  modified  to explicitly include the process of  $W$ boson emission as electroweak final-state radiation in the parton shower of a dijet event, underestimates the data in the collinear region.
The best agreement over the entire distribution is provided by \Sherpa+ \OpenLoops $W+ \text{1-jet}$ and $W+ \text{2-jet}$ calculation that incorporates NLO QCD and NLO EW corrections. In the high-\pT regime the NLO EW corrections have a  significant effect, up to about $20\%$.
The "$W+ \ge 1$ jet $N_{\rm jetti}$ NNLO" prediction, which uses a technique based on $N$-jettiness to split the phase space for the real emission corrections, provides a description very similar to \Sherpa+ \OpenLoops. 
Such a topology will be more accessible and important with Run-2 data at 13 TeV center-of-mass energy and with larger data sets at the \LHC, as well as at higher proton collision energies, for example at future higher-energy proton colliders.

Other studies of correlations between the vector boson and the jets are undertaken as they provide important benchmarks for calculations and for the tuning of MC simulations. One example is the study of the $Z$ boson production in a boosted regime of the $Z$ boson that is important for modeling the background from $Z$ boson decaying into neutrinos in searches of new physics with missing transverse energy in the final state. Figure~\ref{fig:CMS_Zjets_8TeV_eventshapes} shows different levels of azimuthal correlations between the $Z$ boson and the three leading jets ($\Delta\phi(Z,j_i)$) in $Z + \ge \text{3 jets}$ events in two different event configurations, i.e., with $\pT^Z>0$ GeV or $>150$ GeV. Large correlations are visible between the $Z$ and the leading jet, whereas smaller correlations are present between the $Z$ and the subleading jets. In events with a boosted Z the correlation between the $Z$ boson and the leading jet is enhanced.
A good modeling is provided by LO multi-leg (\Sherpa, \Madgraph) and NLO $Z + \text{1 jet}$ (\Powheg) generators. The \PythiaSix prediction, which relies on the parton shower simulation for parton emission shows better modeling in the small $\Delta \phi$ region in the high $Z$ \pT regime, where the soft and collinear approximation of the parton shower is most applicable.

\begin{figure}[tb]
\centering
\hspace*{-0.06\textwidth}
\includegraphics[width=0.60\textwidth]{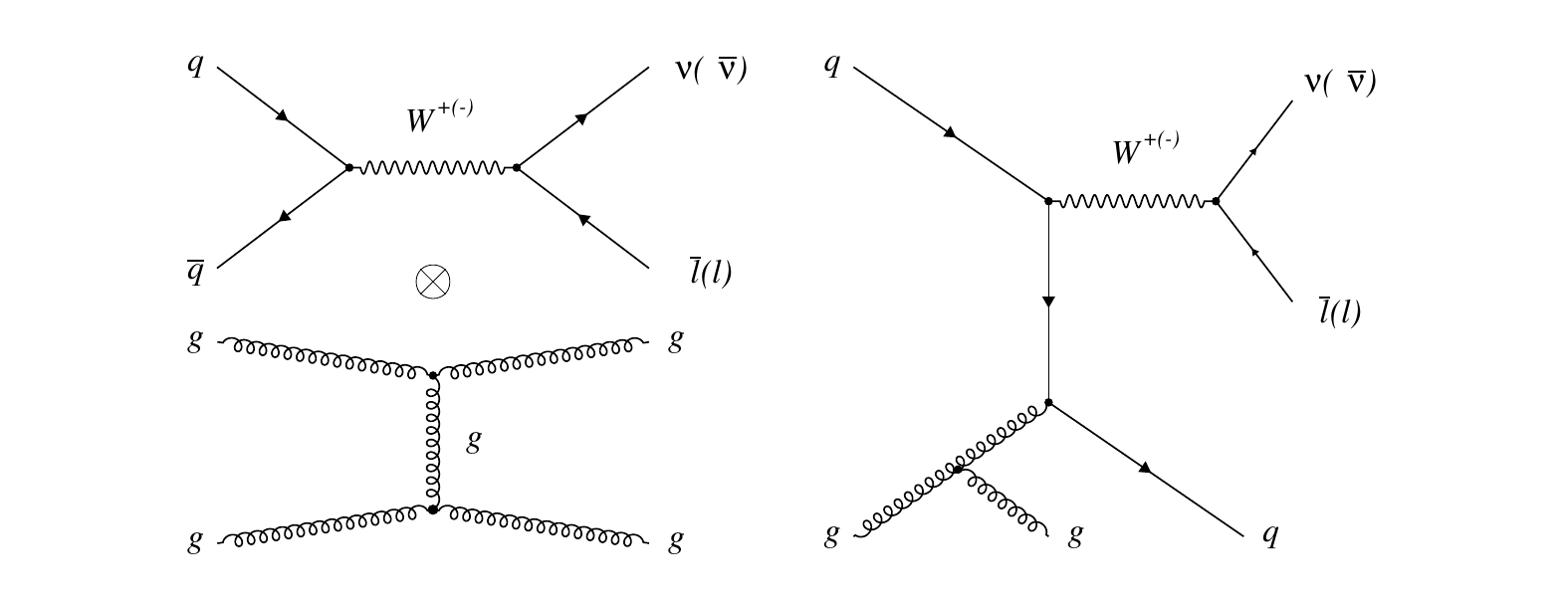}
\includegraphics[width=0.40\textwidth]{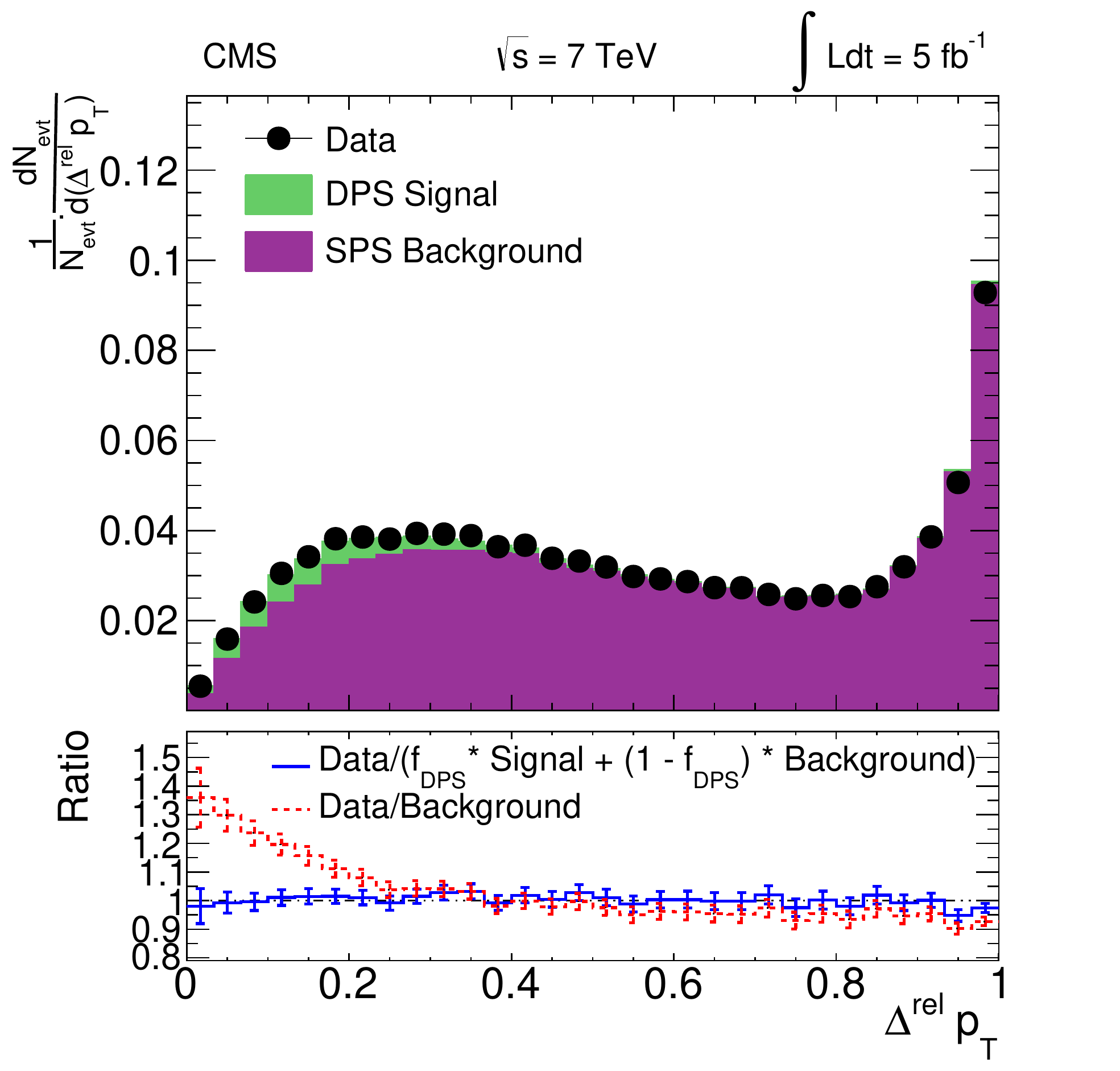}
\caption{\label{fig:CMS_WjetsDPI_8TeV_dpi} Feynman diagrams for $W + \text{2-jet}$ production from DPS  (top-left) and SPS (top-right); fit results for the DPS-sensitive observable $\Delta \pT^\text{rel}$ at 7 TeV \pp collisions at the \LHC (bottom). Figures taken from~\cite{Chatrchyan:2013xxa}.
}
\end{figure}

Multi-Parton Interactions (MPI) are a necessary ingredient of simulations for the description of particle multiplicities and energy flow, and may contaminate event samples for precision measurements (e.g., for Higgs boson properties) and new physics searches. The greater the $\sqrt{s}$ (thus the lower the parton momentum fraction $x$), 
 the bigger the impact of MPI at high \pT. Therefore, the MPI contribution is generally more significant at the \LHC than at the \Tevatron. 
The impact of MPI in physics processes is difficult to measure as it co-exists with initial and final state radiation, beam remnants and the hard interaction. Experimentally the MPI contribution must also be disentangled from pileup interactions. 
Double parton scattering (DPS) is a specific case of MPI and its production cross section is typically parameterised as $\sigma_\text{DPS}=\frac{\sigma_A\cdot\sigma_B}{\sigma_\text{eff}}$, where $\sigma_A$ and $\sigma_B$ are the parton level cross sections of the two underlying processes, assumed to be independent, while $\sigma_\text{eff}$ is an effective area parameter and is assumed to be independent of phase space and process.
These assumptions are tested by measuring $\sigma_\text{eff}$ in several processes and at different energy scales. 
The DPS contribution to the inclusive $W$ production is studied in $W + \text{2-jet}$ events in Refs.~\cite{Aad:2013bjm, Chatrchyan:2013xxa}. Figure~\ref{fig:CMS_WjetsDPI_8TeV_dpi} (top) shows examples of the two contributions to the $W + \text{2-jet}$ event sample: DPS (top left) and single parton scattering (SPS) (top right). The fraction of DPS events in $W + \text{2-jet}$ data and $\sigma_\text{eff}$ are extracted from a fit of DPS and SPS templates to the normalized transverse momentum balance $\Delta \pT^\text{rel}=\frac{|\vec{p}_{\rm T}^{~j1}+\vec{p}_{\rm T}^{~j2}|}{|\vec{p}_{\rm T}^{~j1}|+|\vec{p}_{\rm T}^{~j2}|}$. Figure~\ref{fig:CMS_WjetsDPI_8TeV_dpi} shows the template fit results compared to data. The values of $\sigma_\text{eff}$ measured at 7 TeV by the \ATLAS and the \CMS experiments are $15\pm 3 ~{\rm (stat.)} ^{+5}_{-3} ~{\rm (syst.)}$ and $20.7 \pm 0.8 ~{\rm (stat.)} \pm 6.6 ~{\rm (syst.)}$ mb, respectively.
In order to test the energy dependence of $\sigma_\text{eff}$ it is important to repeat such measurements at experiments with greater center-of-mass energies. Moreover, higher $\sqrt{s}$ in future measurements implies a larger phase space available for DPS and thus a greater need for more precise DPS measurements.

For other measurements of \Vjet properties, the reader is referred to the following Refs.~\cite{Abazov:2009pp,Aaboud:2017hbk,Aad:2013ysa,Aaboud:2017hox,Aad:2013ueu,Chatrchyan:2011ig,Aaij:2013nxa}.

\subsubsection{Cross section ratios}
\label{sec:VLF:exp:ratios}

\begin{figure}
\centering
\includegraphics[width=0.46\textwidth]{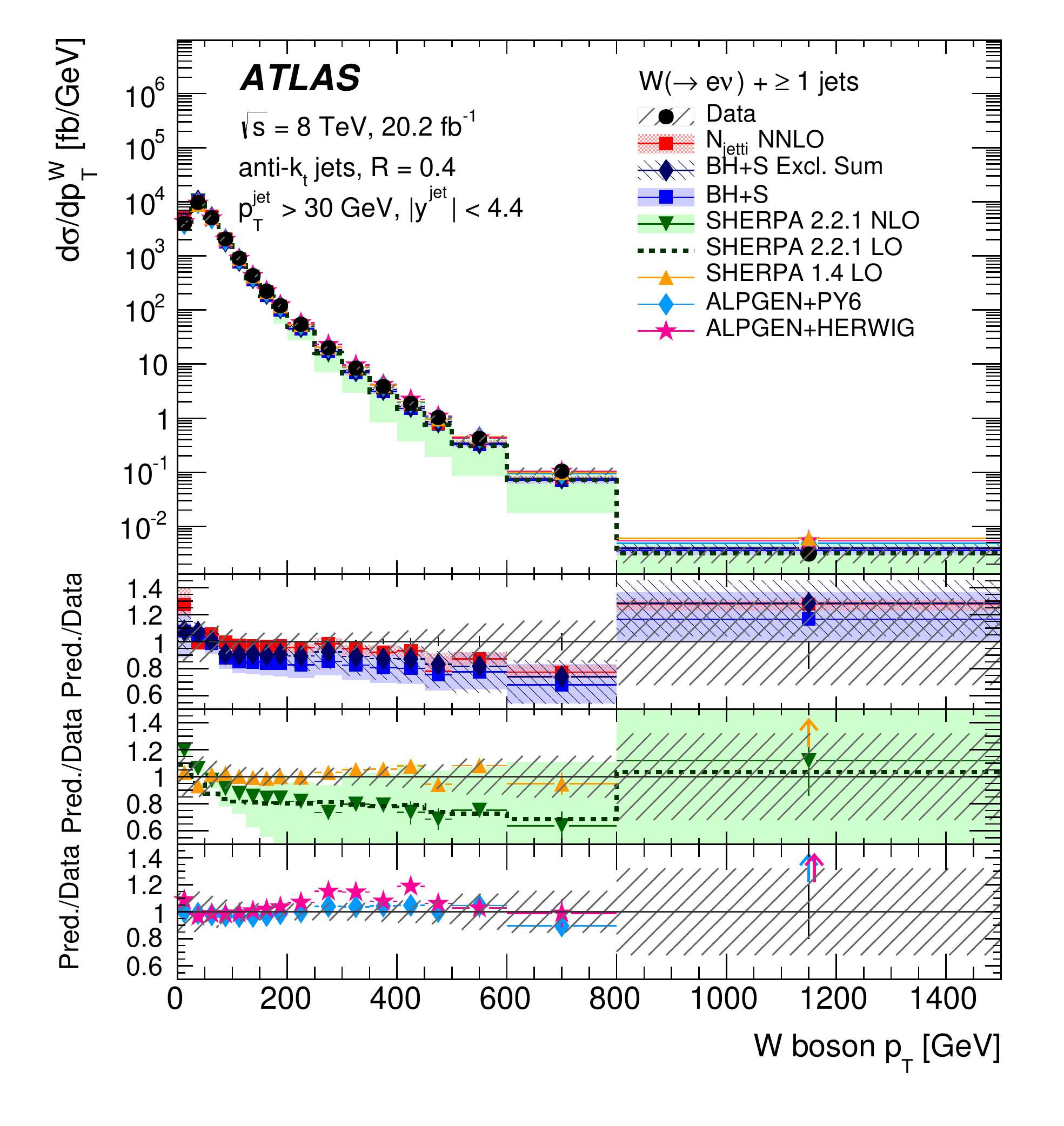}
\caption{\label{fig:ATLAS_Wjets_8TeV} Differential cross section for the production of a $W$ boson as a function of the $W$ \pT for events with $N_{\rm jets} \ge 1$ in \pp collisions at 8 TeV center-of-mass energy at the \LHC. Experimental results are compared to several theoretical predictions calculated at different orders and with different approximations in pQCD, and with different parton shower implementations. Figure taken from~\cite{Aaboud:2017soa}.
}
\end{figure} 
\begin{figure*}
\centering
\includegraphics[width=0.46\textwidth]{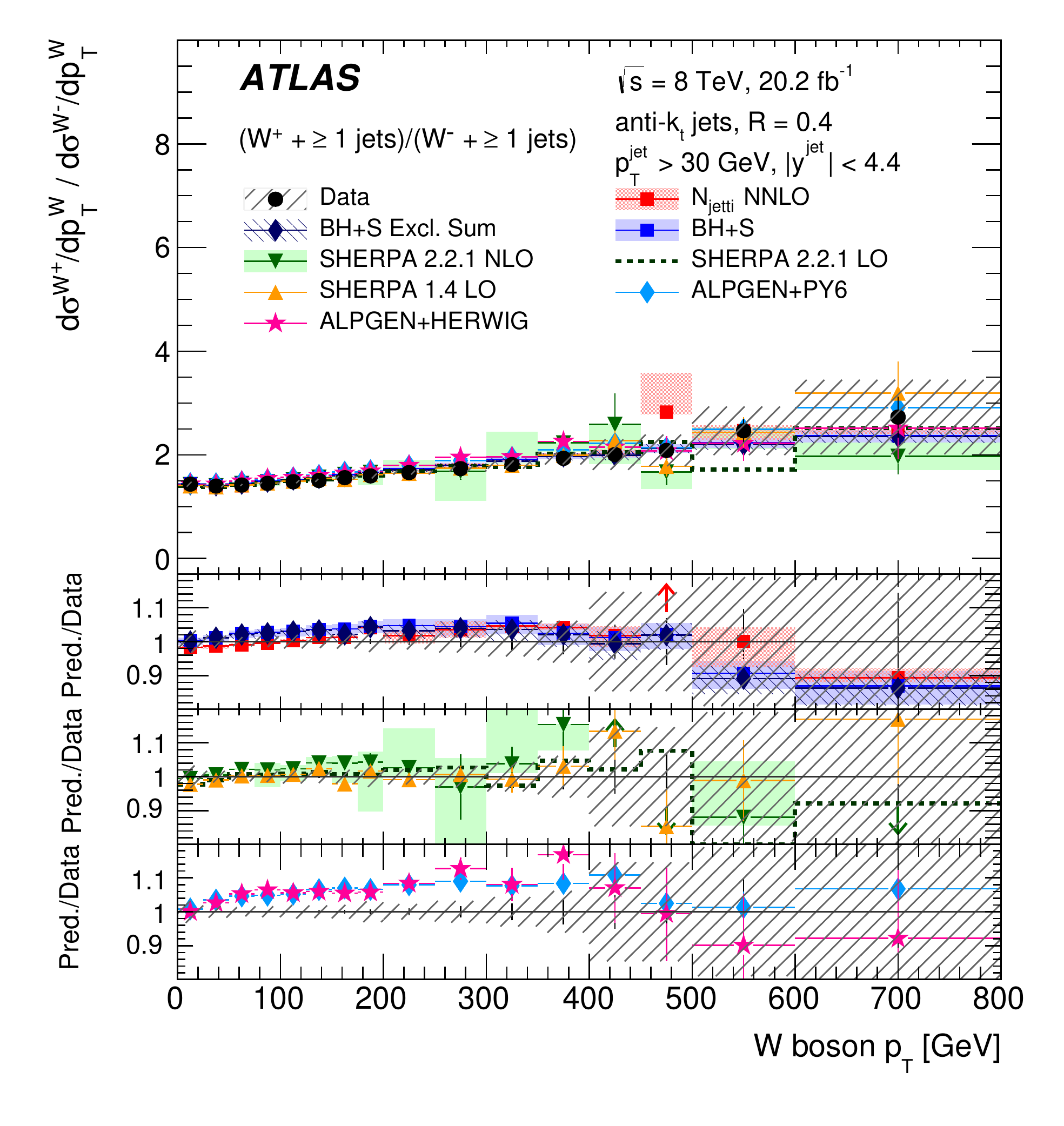}
\includegraphics[width=0.46\textwidth]{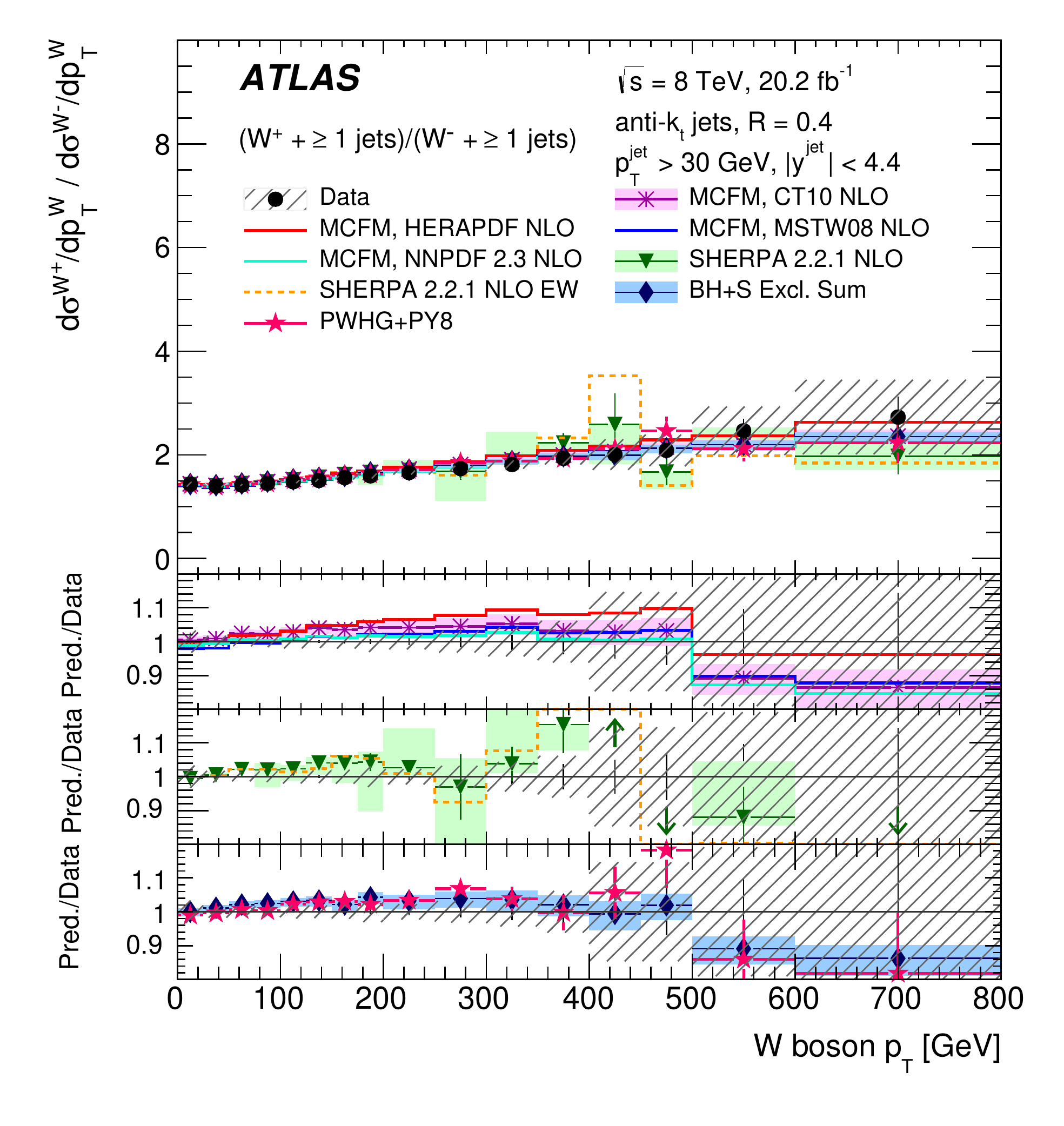}
\caption{\label{fig:ATLAS_Wjets_8TeV_chargeratio} Differential cross section for the $W^+$/$W^-$ ratio as a function of the $W$ \pT for events with $N_{\rm jets} \ge 1$ in \pp collisions at 8 TeV center-of-mass energy at the \LHC. Experimental results are compared to several theoretical predictions calculated at different orders, with different approximations in pQCD and with different parton shower implementations as well as to NLO \MCFM predictions with four different PDF sets. Figures taken from~\cite{Aaboud:2017soa}.
}
\end{figure*}
\begin{figure*}
\centering
\includegraphics[width=0.45\textwidth]{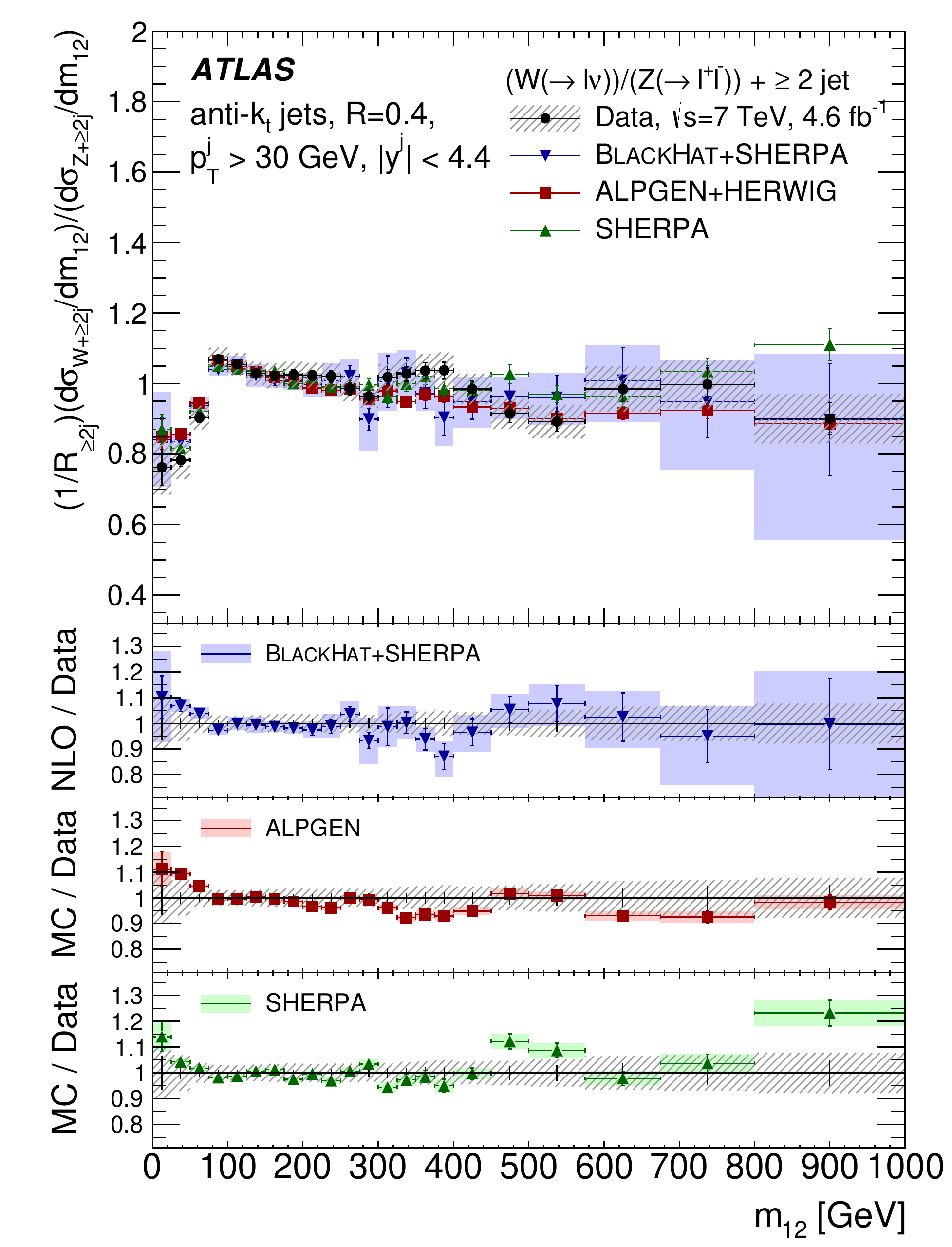}
\includegraphics[width=0.45\textwidth]{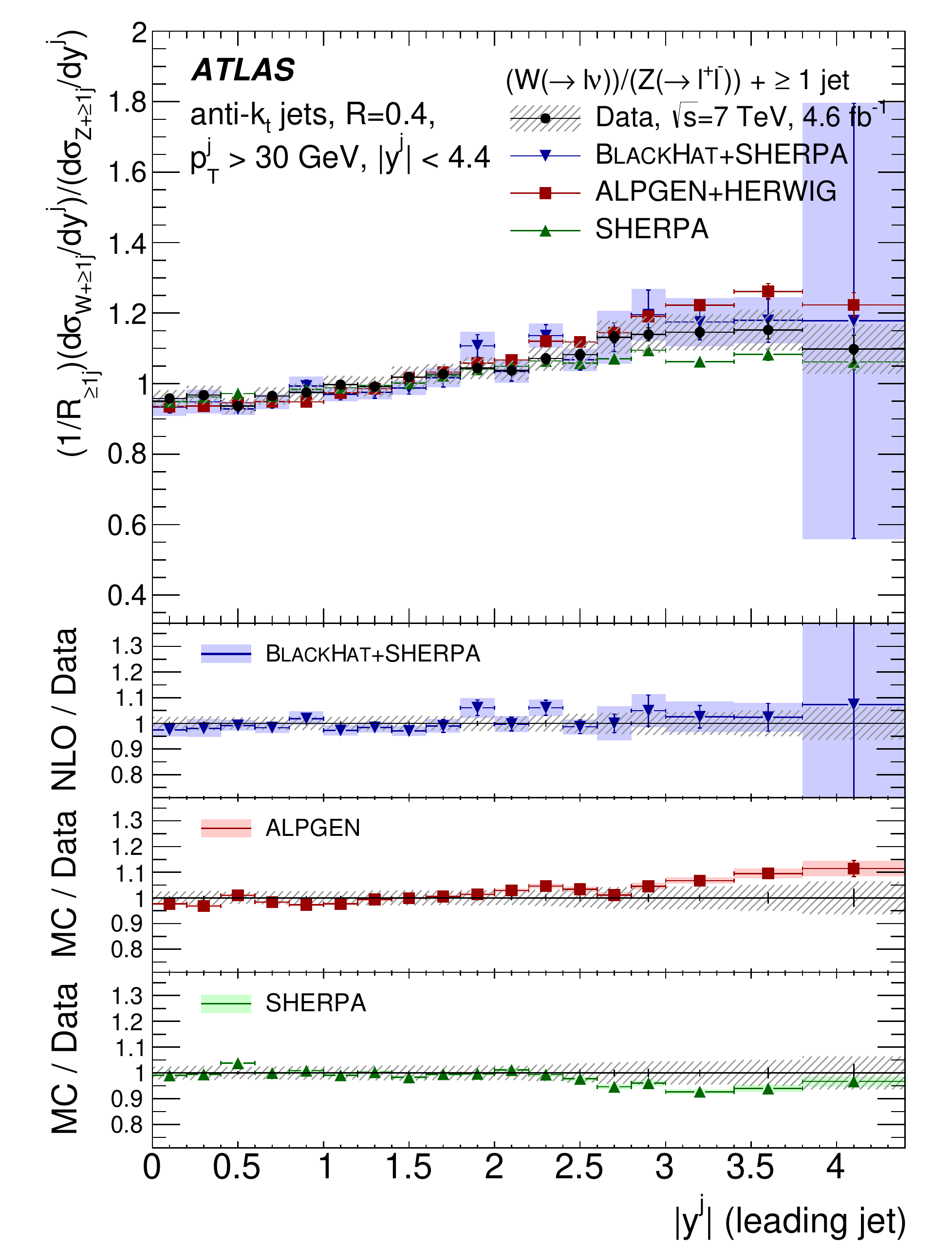}
\caption{\label{fig:ATLAS_Rjets_mjj_y} The normalized ratio of \Wjet and \Zjet production cross sections, $R_{\rm jets}$, as a function of the dijet invariant mass, $m_{12}$, for $N_{\rm jets} ≥ 2$ (left) and the leading-jet absolute rapidity, $|y^j|$, for $N_{\rm jets} ≥ 1$ (right) with 7 TeV \pp collisions at the \LHC. Experimental results are compared to NLO calculations in pQCD by \BlackHat as well as to MC generators. Figures taken from~\cite{Aad:2014rta}.
}
\end{figure*}
Measurements of production cross sections of individual \Vjet processes are limited in high-statistic regions of phase space by systematic uncertainties that are common between processes, while measurements  of ratios of different \Vjet processes exploit partial cancellations of experimental uncertainties as well as theoretical effects that are common between the two processes in the ratio.
Such ratios can provide high-precision tests of the Standard Model, as they are sensitive to non-universal corrections in QCD and electroweak calculations as well as PDFs. 
In this section several ratios will be presented. In the ratio of $W^+$ to $W^−$ productions, pQCD and electroweak effects cancel to large extent, making this measurement particularly sensitive to PDFs, and specifically to the ratio of up-quark to down-quark at high Bjorken-$x$. In the $W$ to $Z$ boson ratio, effects from non-perturbative QCD processes largely cancel at high energy scales, whereas other effects do not cancel, such as boson mass effects at low energy scales, quark–gluon and quark–antiquark contributions to \Vjet productions, and non-universal electroweak corrections. Such a ratio is therefore useful for validating theoretical predictions used to estimate \Wjet or \Zjet backgrounds in searches for new physics.
Similarly, in the ratio $Z$ to $\gamma$ bosons,
 mass effects cancel at high energy scales, whereas higher-order QCD and electroweak corrections can have large contributions, thus making such a ratio a precise test of higher-order effects in perturbative calculations.

 In the analysis presented in Ref.~\cite{Aaboud:2017soa} measurements are carried out for $W^{\pm}$ production as well as for $W^+$ and $W^−$ productions and the cross-section ratio of $W^+/W^−$ in events with the $W$ boson produced association with jets, as a function of a number of variables  that are sensitive to higher-order terms and to the PDFs. 
 In the $W^+/W^−$ ratio in $W$ events with at least one associated jet, many of the experimental and theoretical uncertainties cancel out, making it a more precise test of the theoretical predictions, especially in a kinematic regime with $x$ values higher (approx. up to $x=$0.1–-0.3) than what is typically accessible in measurements of inclusive $W$ production at \ATLAS and \CMS ($10^{−4}< x < 10^{−1}$). 
 Figure~\ref{fig:ATLAS_Wjets_8TeV} shows the differential cross section as a function of the \pT of the $W$ boson for events with $W^{\pm} + \ge \text{1 jet}$ production. Good overall agreement is found between the data and most of LO, NLO and NNLO calculations. 
 Variations in the modeling of different \Sherpa generator versions are seen, whereas different parton shower models interfaced to the \Alpgen generator show little impact, with \Pythia providing a slightly better description of the data.  
In the $W^+/W^−$ cross-section ratio in Fig.~\ref{fig:ATLAS_Wjets_8TeV_chargeratio}, differences due to QCD and electroweak higher-order effects cancel out to a large extent. In the ratio, the experimental precision is greatly improved and most predictions show a trend to overestimate the data.  
 The data is also compared to different PDF sets with a common calculation by the \MCFM program. Sensitivity to PDFs is visible in the variation of agreement between data and the different PDF sets, especially in the region of \pT $\approx$ 200 -- 400 GeV, where experimental uncertainties are in the $2\%$ -- $6\%$ range. In this region the predictions from different PDF sets may differ by about $2\%$ to $5\%$ and in some cases differ from data up to 2–3 standard deviations.

The production mechanisms of \Wjets and \Zjets are very similar once the kinematic effect of the different boson masses and the leptonic branching ratios are taken into account. In ratios of differential cross sections in \Wjet and \Zjet events ($R_{\rm jets}$), the experimental uncertainty cancel significantly, as can be seen in Fig.~\ref{fig:ATLAS_Rjets_error}.
Theoretical uncertainties, if treated as correlated between the two types of processes, can be significantly reduced too: QCD scale variations, estimated at NLO in pQCD, and PDF uncertainties overall account for a $2-4\%$ level uncertainty in $R_{\rm jets}$ with at least one jet in the final state with jet $\pT \simeq 800$ GeV, to be compared to the $20\%$ level of uncertainties in events with a $W$ boson with at least one associated jet. Such a reduction in theoretical uncertainties is also visible in the phenomenological study presented in Fig.~\ref{fig:VLF:theory:pT-ratio} in Sec.~\ref{sec:VLF:theory:ho}.
Figure~\ref{fig:ATLAS_Rjets_mjj_y} shows that the level of mismodeling of the MC simulations that is seen in the cross section measurement as a function of $m_{jj}$ for \Wjets in Fig.~\ref{fig:ATLAS_Wjet_7TeV_dymjj} is largely reduced in $R_{\rm jets}$. This effect points towards an underlying cause for the mismodeling in MC generators that has the same effect in both processes. 
The low part of the $m_{jj}$ distribution in \Wjet or \Zjet events has sensitivity to jet kinematics and non-perturbative effects in soft QCD radiation that differ  between $W$ and $Z$ events and do not cancel in the $R_{\rm jets}$ ratio, as can be seen in the $R_{\rm jets}$ values lower than 1 for $m_{jj} < 100$ GeV in Fig.~\ref{fig:ATLAS_Rjets_mjj_y} (left).
The agreement between predictions and data in the region of high rapidity of the \pT-leading jet, see Fig.~\ref{fig:ATLAS_Rjets_mjj_y} (right), can be affected by the modeling of the parton shower and PDF. 
Such a ratio measurement is not only important for a better understanding of the theoretical modeling of \WZjet processes, but also for the estimation of backgrounds on searches for new physics. For example the calculation of such a ratio is used as {\it transfer factor} to estimate the $Z(\rightarrow\nu\nu)+\text{jet}$ background yield in a search signal region by extrapolating the measurement of the $W(\rightarrow l\nu)+\text{jet}$ yield from a data control region, see examples in Refs.~\cite{Aad2014,Chatrchyan2012,Aad:2014nra,Aad:2015txa,Khachatryan:2016whc,Sirunyan:2018owy}.

\begin{figure*}
\centering
    \includegraphics[width=0.40\textwidth,height=0.45\textwidth]{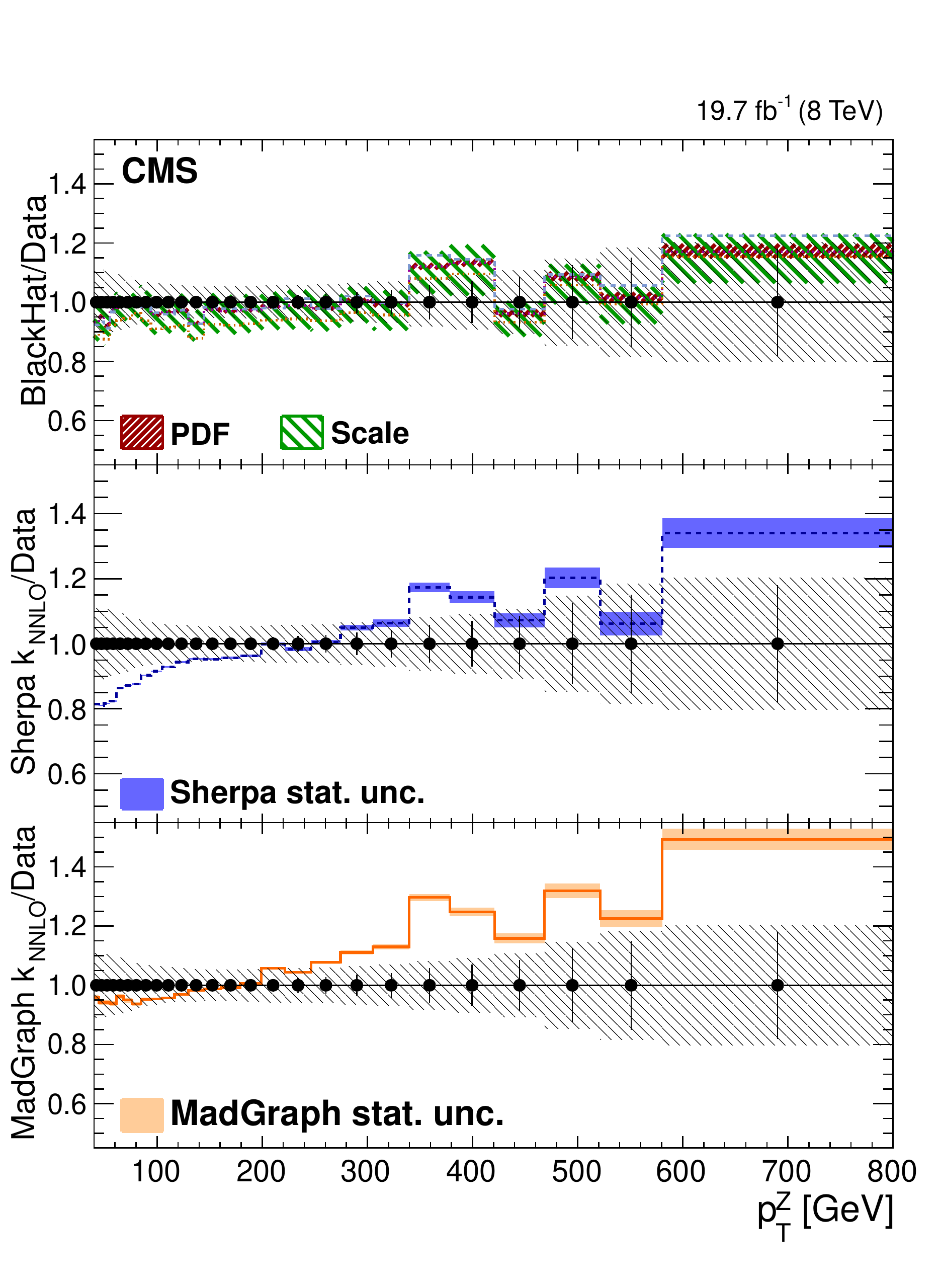}
    \hspace*{0.02\textwidth}
    \includegraphics[width=0.40\textwidth,height=0.47\textwidth]{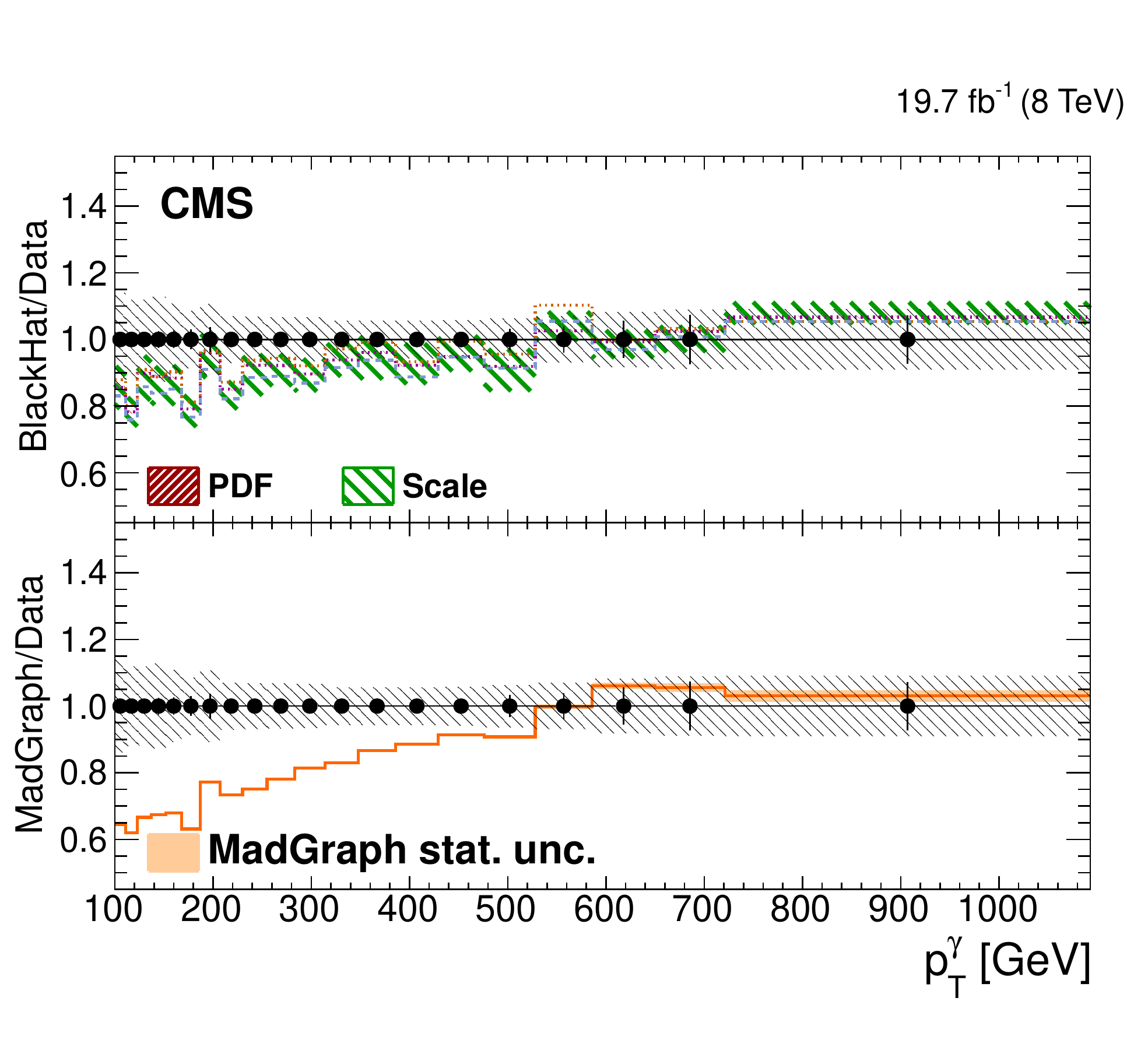}
    \hspace*{-0.03\textwidth}
    \includegraphics[width=0.44\textwidth]{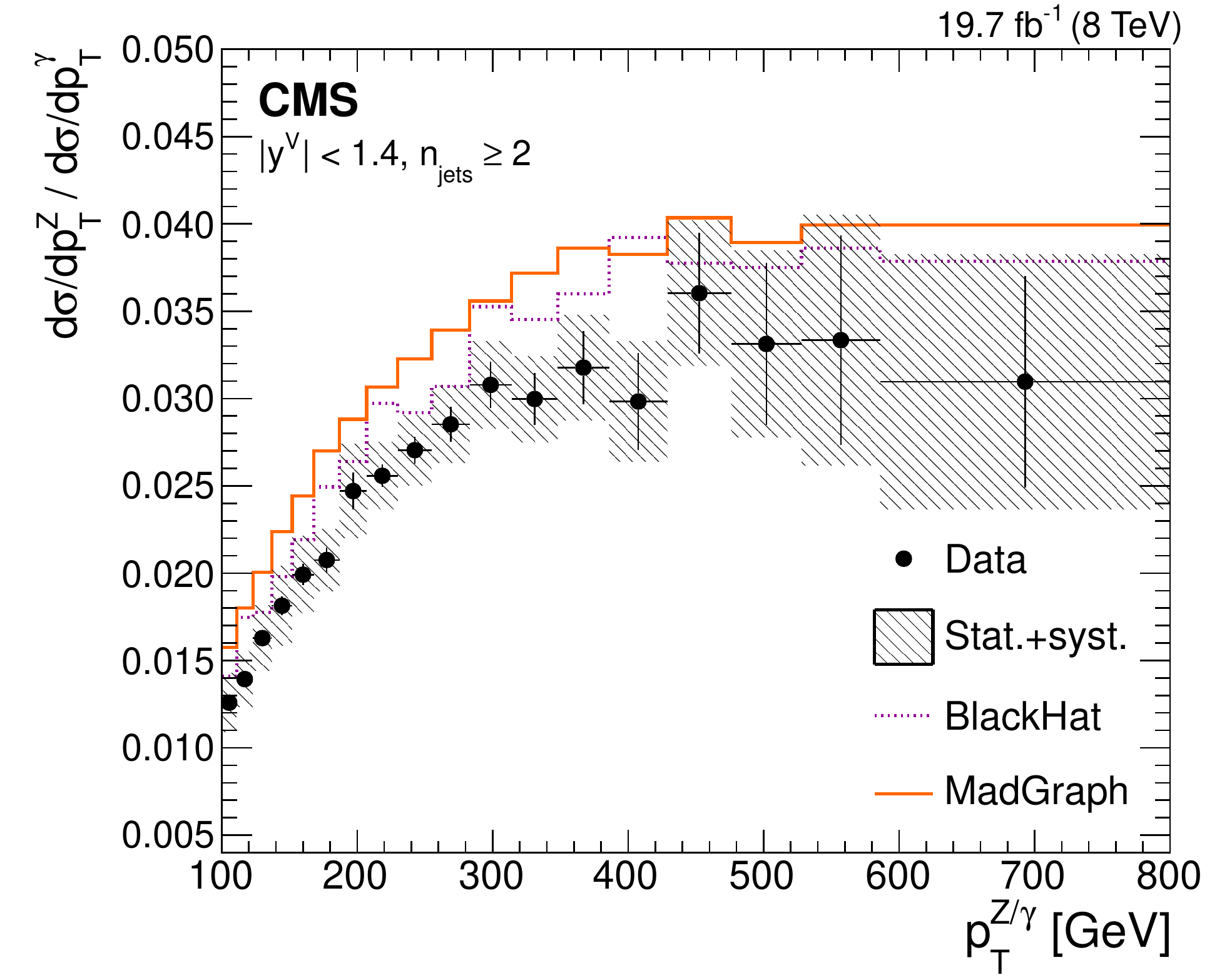}
    \hspace*{-0.03\textwidth}
    \includegraphics[width=0.45\textwidth,height=0.38\textwidth]{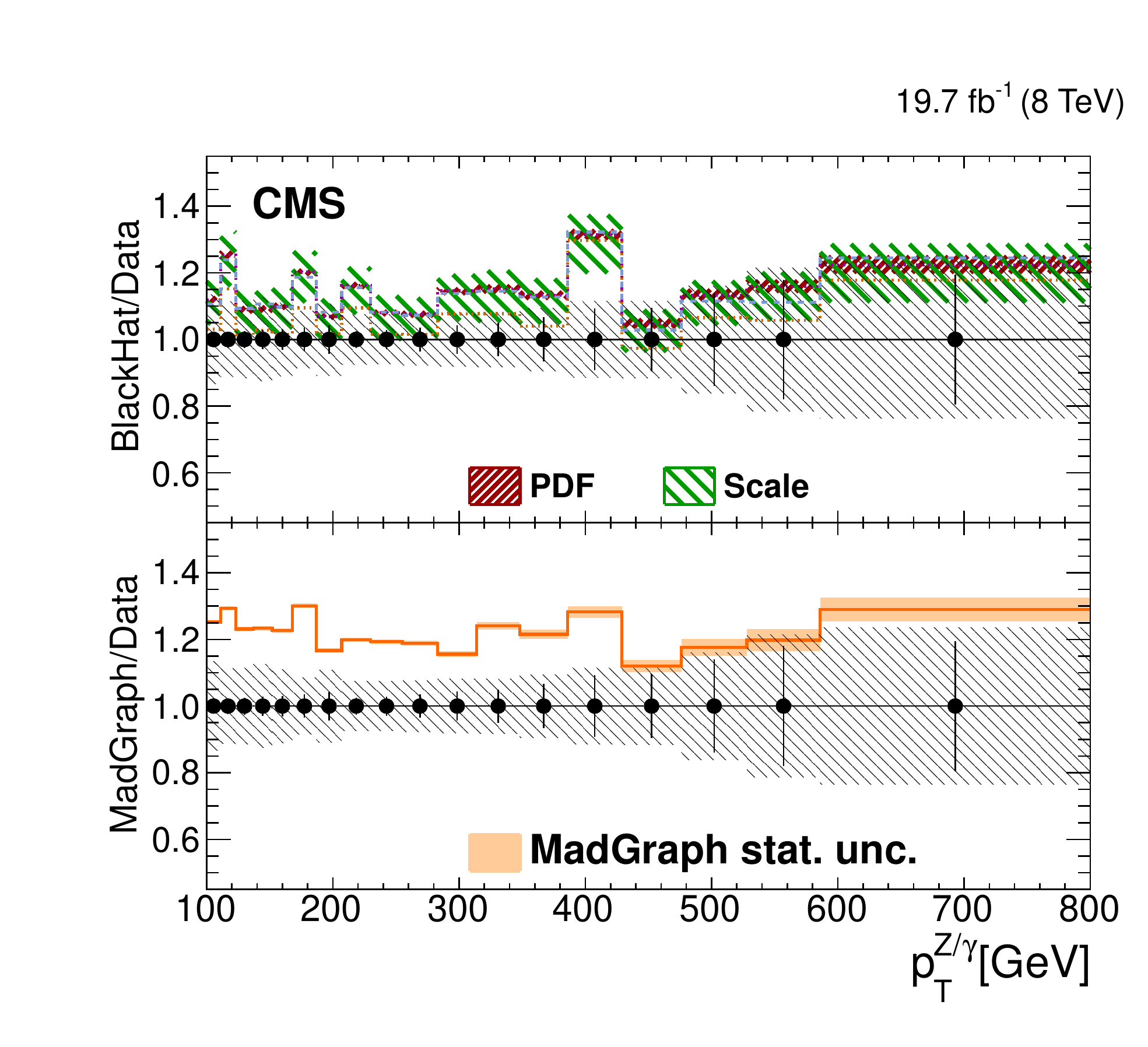}
\caption{\label{fig:CMS_Zgammajets_pT} Prediction-to-data ratios of differential cross sections for $Z$ and central ($|y^{\gamma}|<1.4$) $\gamma$ productions as a function of the boson \pT for inclusive \Zjet and \gammajet processes with a $N_{\rm jets}\ge 2$ selection, with 8 TeV \pp collisions at the \LHC (top). Differential cross section ratio of \Zjets over \gammajets and its prediction-to-data ratios as a function of the boson transverse-momentum for central bosons ($|y^V|<1.4$) in the $N_{\rm jets}\ge 2$ sample (bottom). Experimental results are compared to NLO pQCD calculation by \BlackHat and to the \Madgraph MC generator. Figures taken from~\cite{Khachatryan:2015ira}.
}
\end{figure*}

The ratio of \gammajets and \Zjets is also of great interest, especially in the high-\pT region of the vector bosons, where the $Z$ boson mass effects play a less significant role than in the low-\pT region. This ratio can test the impact of QCD and electroweak higher order corrections with greater experimental accuracy, thanks to cancellation of experimental systematic uncertainties such as jet energy calibration and luminosity, see Fig.~\ref{fig:CMS_Zgammajets_pT}. 
The NLO pQCD calculation by \BlackHat describes well the $Z$-boson \pT spectrum but it tends to underestimate the low part of the $\gamma$ \pT spectrum in events with at least two associated jets. In the LO multi-leg MC generators (\Madgraph+ \PythiaSix and \Sherpa) a similar systematic trend of mismodeling the $Z$ \pT in $Z+ \ge  \text{2 jets}$ is seen, as well as a significant bias at modeling the shape of the photon \pT distribution (see \Madgraph+ \PythiaSix prediction).  
The \pT distribution of the $Z/\gamma$ ratio flattens at high boson \pT values, i.e., greater than 350 GeV. 
The boson \pT shape mismodeling observed in the  individual $Z$ and $\gamma$ production events largely cancels out in the $Z/\gamma$ ratio and a residual over-estimation of the ratio by a flat $20\%$ is observed in QCD LO multi-leg MC generators (see \Madgraph+ \PythiaSix prediction). A less significant systematic mismodeling is also visible in the NLO fixed-order calculation by \BlackHat.  

\begin{figure*}
\centering
\hspace*{-0.08\textwidth}
\includegraphics[width=0.34\textwidth]{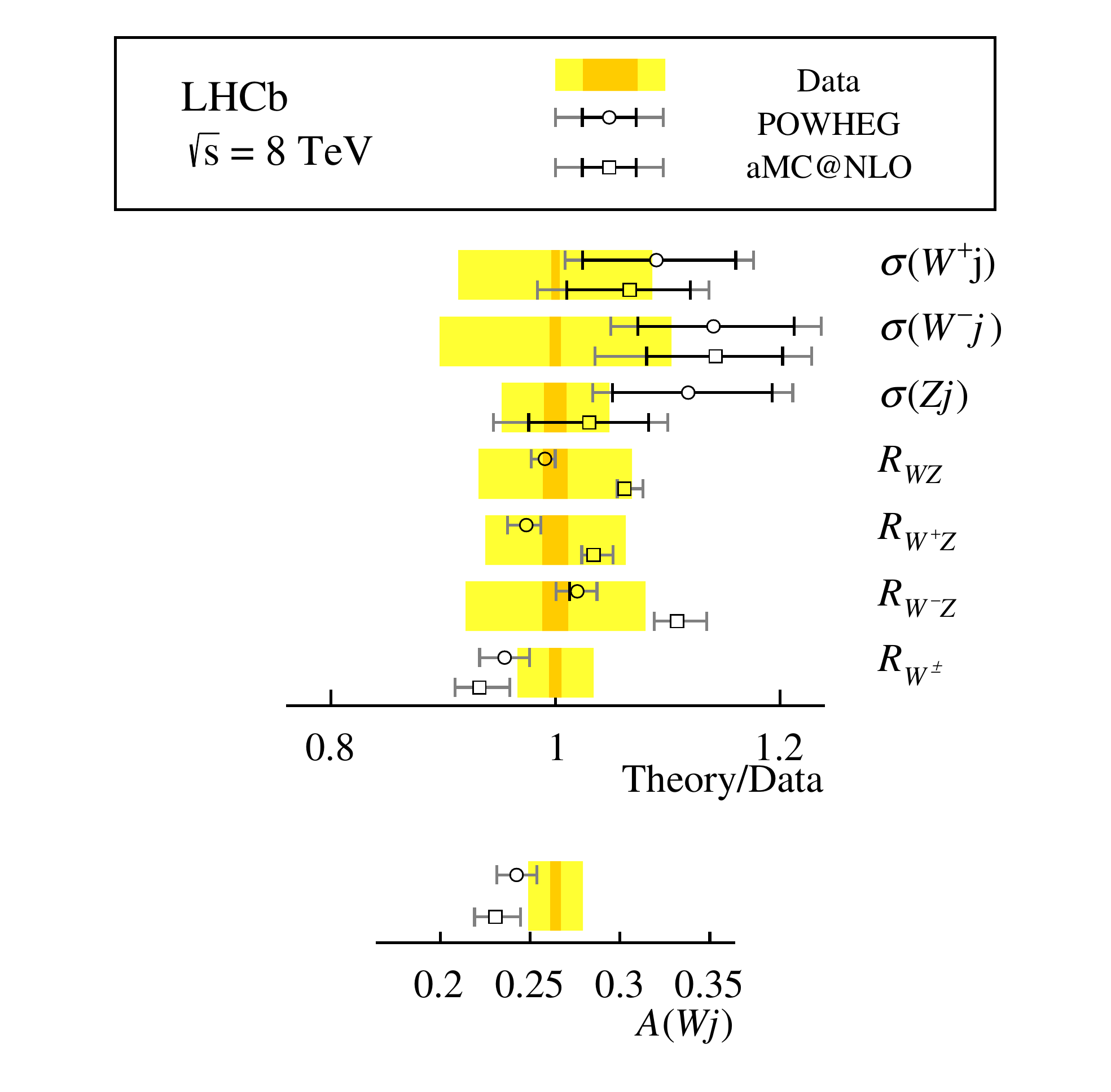}
\hspace*{-0.02\textwidth}
\includegraphics[width=0.35\textwidth]{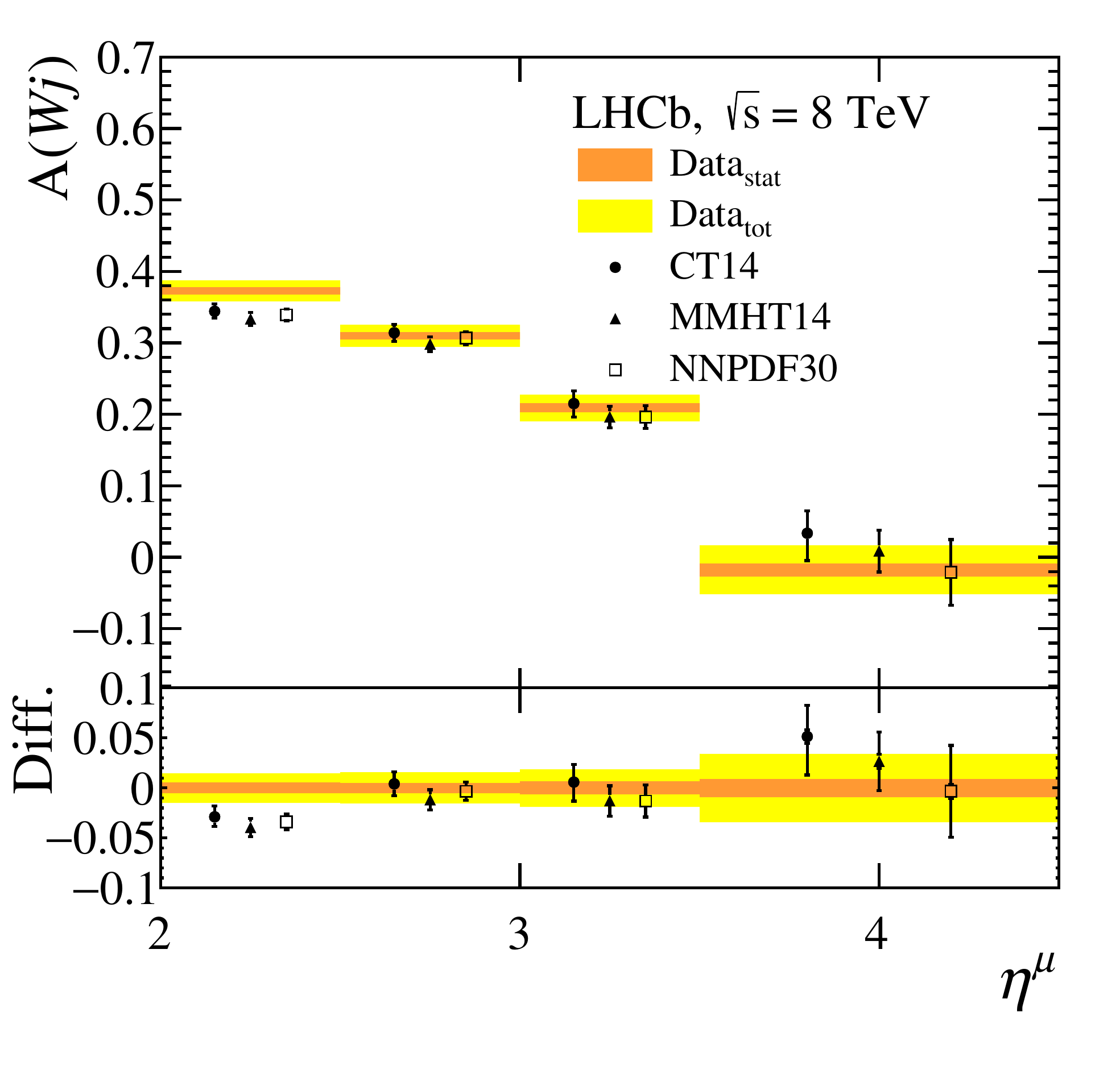}
\includegraphics[width=0.35\textwidth]{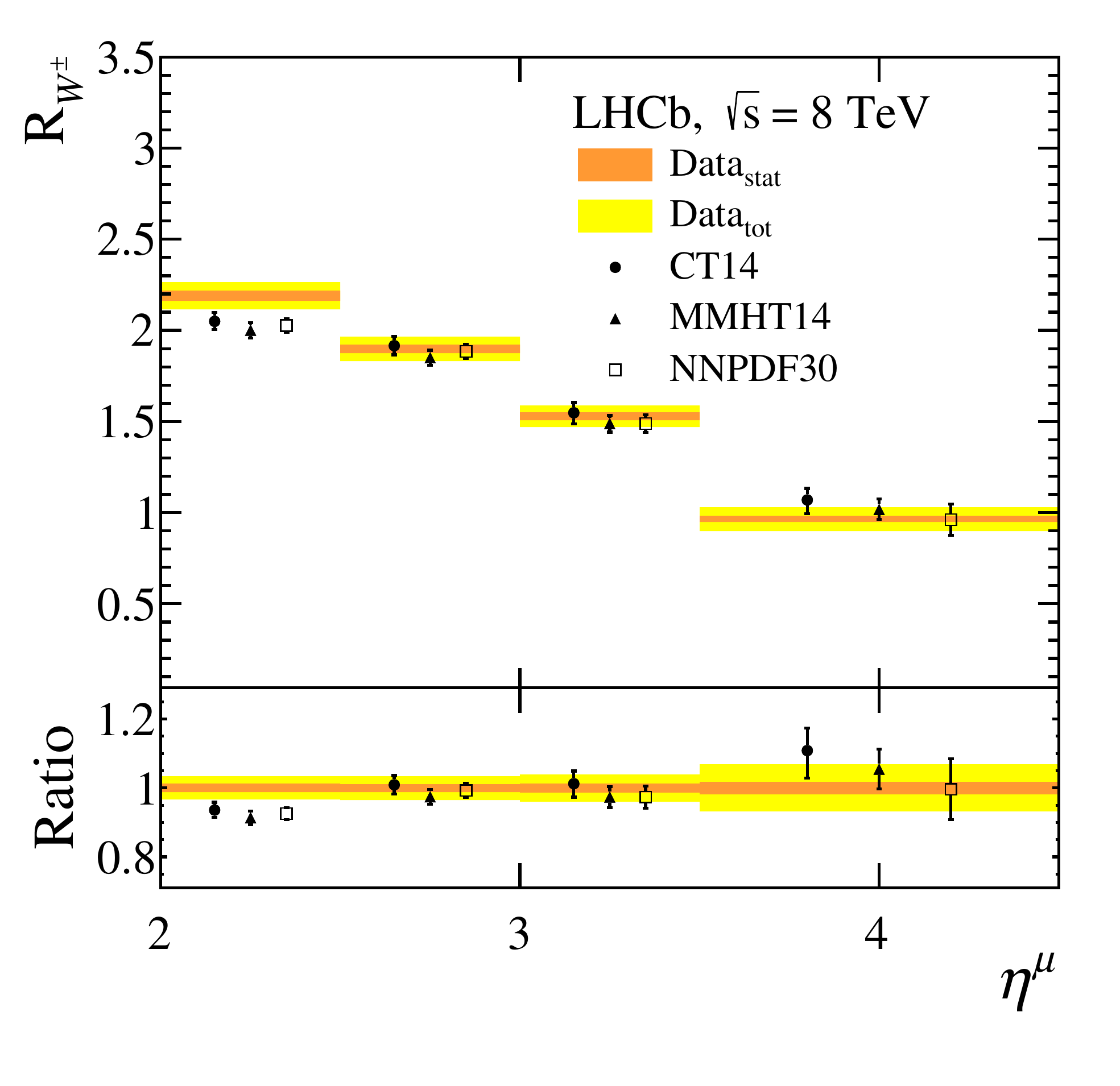}
\caption{\label{fig:LHCb_WWjets_8TeV_lepteta} Summary of the \WZjet measurements performed in a fiducial region with the LHCb experiment at 8 TeV \pp collisions at the \LHC (left), compared to predictions from two MC generators. The cross sections for $W^+ + \ge  \text{1 jet}$, $W^-+ \ge \text{ 1 jet}$ and $Z ~+ \ge \text{1 jet}$ are denoted as $\sigma(W^+j)$,  $\sigma(W^-j)$, and $\sigma(Zj)$, respectively, while the ratios $\sigma(Wj)/\sigma(Zj)$, $\sigma(W^+j)/\sigma(Zj)$, $\sigma(W^-j)/\sigma(Zj)$ and $\sigma(W^+j)/\sigma(W^-j)$ are denoted as $R_{WZ}$, $R_{W^{+}Z}$, $R_{W^{-}Z}$ and $R_{W^{\pm}}$, respectively. The asymmetry of $\sigma(W^+j)$ and $\sigma(W^-j)$ is denoted as $A(Wj)$.
The asymmetry $A(Wj)$ (middle) and  the  ratio $R_{W^{\pm}}$ (right) as a function of the muon pseudorapidity $\eta^\mu$, compared to NLO calculations performed with the \FEWZ calculation and three different PDF sets. Figures taken from~\cite{AbellanBeteta:2016ugk}. 
}
\end{figure*}

The production of $W$ and $Z$ bosons in association with jets is studied in the forward region of proton-proton collisions with the \LHCb experiment. Such measurements provide additional tests of the SM in a region of phase space not directly accessible by \ATLAS and \CMS at the \LHC and provide additional constraints on PDF in a different range of Bjorken-$x$. 
As shown in Fig.~\ref{fig:LHCb_WWjets_8TeV_lepteta}, in the \LHCb experiment the charged leptons from the weak boson decay are reconstructed in the forward pseudo-rapidity region of $2.0 < \eta^{\rm{l}}  < 4.5$, while jets in the region $2.2< \eta^{\rm{jet}}<4.2$ with the anti-$k_t$ algorithm with distance parameter $R=0.5$ and $\pT^{\rm{jet}}>20$ GeV~\cite{AbellanBeteta:2016ugk}.
Cross sections and their respective ratios are measured for $W^+ + \ge  \text{1 jet}$, $W^-+ \ge \text{ 1 jet}$ and $Z ~+ \ge \text{1 jet}$. 
In addition, the asymmetry of $W^++ \ge \text{ 1 jet}$ and $W^- + \ge \text{ 1 jet}$ production and the asymmetry as a function of the charged lepton $\eta$ are measured. Due to the cancellation of scale uncertainties, the ratios as a function of the charged lepton $\eta$ are expected to provide sensitivity to the PDFs. Figure~\ref{fig:LHCb_WWjets_8TeV_lepteta} shows the broad range of measurements that are carried out in this analysis and the extensive comparisons with predictions from different MC generators and PDFs.  Overall a good agreement is seen between data and predictions, however slightly larger values of the asymmetry and the ratio of $W^++ \ge \text{ 1 jet}$ to $W^- + \ge \text{ 1 jet}$ cross sections are seen in data than in NLO QCD predictions in the first bin of the charged lepton $\eta$.

\section{Electroweak production of a vector boson and two jets}
\label{sec:VBF}
\subsection{Theoretical predictions}
\label{sec:VBF:theory}

The production of a single vector boson in vector boson fusion constitutes an 
experimental signature of special interest 
because of its sensitivity to the self-interactions 
of the electroweak gauge bosons.
It presents a prime testbed for searches for new physics 
signals that are connected to the electroweak symmetry breaking.

\begin{figure*}[t!] 
  \centering
  \begin{minipage}{0.24\textwidth}
    \centering
    \includegraphics[width=\textwidth]{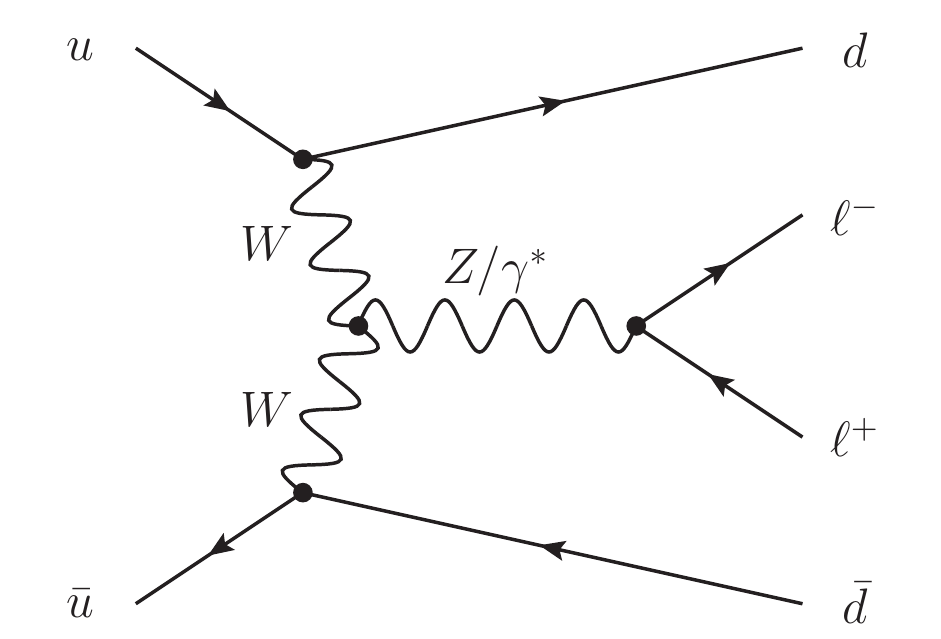}\\
    (a)
  \end{minipage}
  \hfil
  \begin{minipage}{0.24\textwidth}
    \centering
    \includegraphics[width=\textwidth]{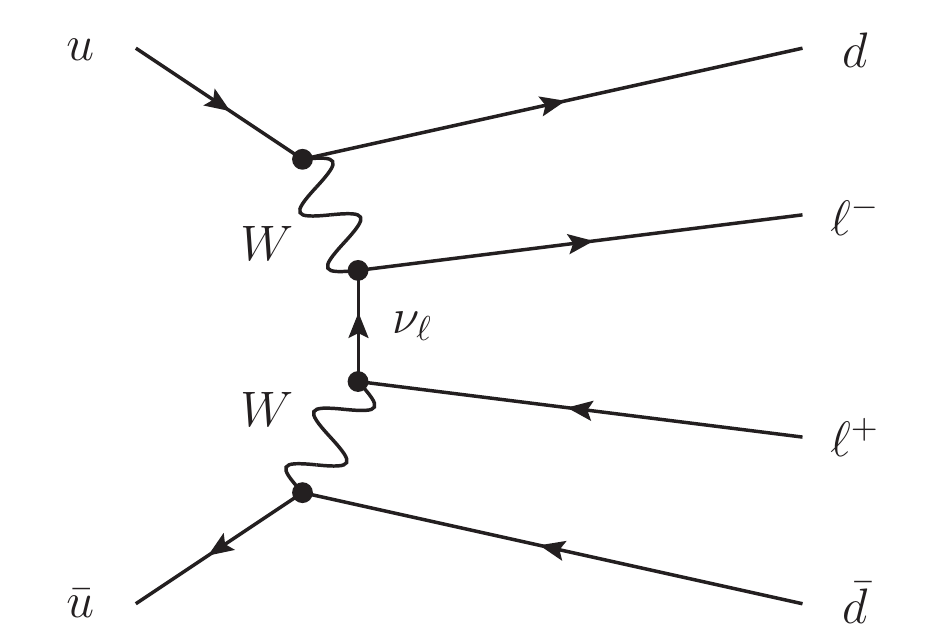}\\
    (b)
  \end{minipage}
  \hfil
  \begin{minipage}{0.24\textwidth}
    \centering
    \includegraphics[width=\textwidth]{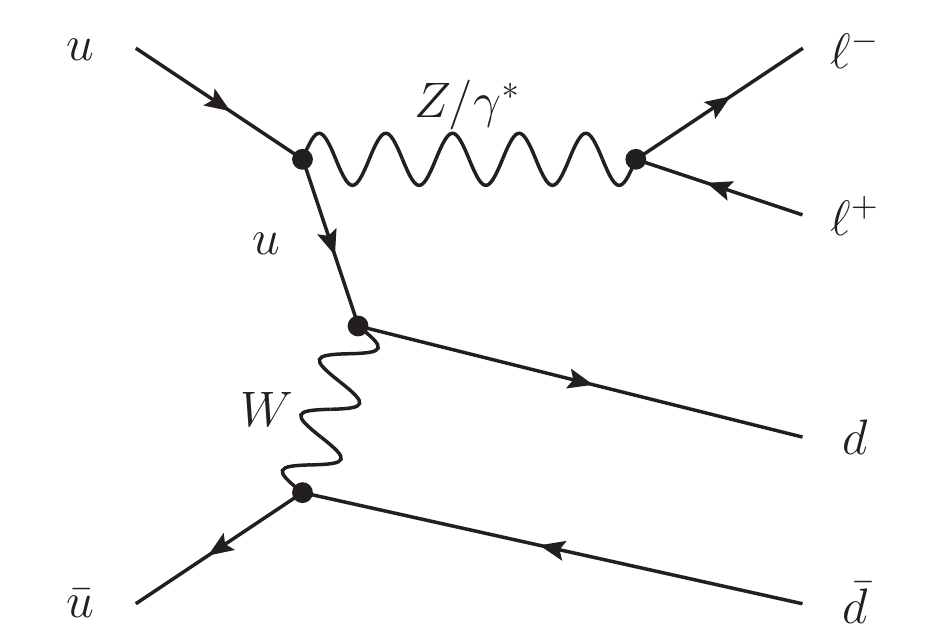}\\
    (c)
  \end{minipage}
  \hfil
  \begin{minipage}{0.24\textwidth}
    \centering
    \includegraphics[width=\textwidth]{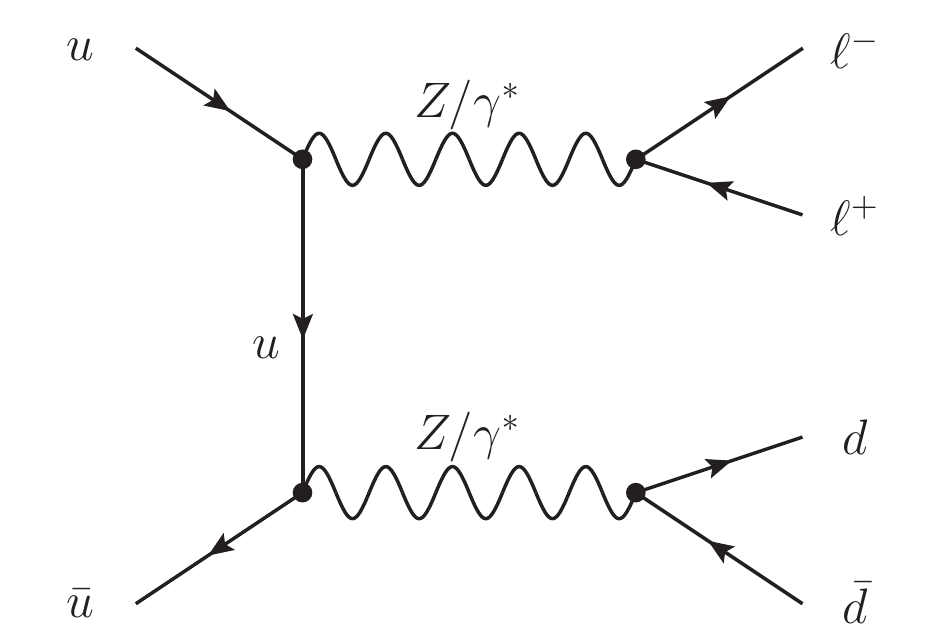}\\
    (d)
  \end{minipage}
  \caption{\label{fig:VBF:theory:diagrams}
    Representative Feynman diagrams for the production of two charged leptons in association with 
    two jets at $\order(\alpha^4)$:
    vector boson fusion (a), 
    multi-peripheral (b),
    bremsstrahlung-like (c),
    and semileptonic diboson production (d).
  }
\end{figure*}

The electroweak production of a single vector boson proceeds 
at $\order(\alpha^4)$ at leading order and contains multiple distinct topologies. 
Of particular interest are: a) the classic vector boson fusion topologies, 
b) the closely related multi-peripheral topologies, 
c) bremsstrahlung-like electroweak boson emission off electroweak quark scattering topologies, 
and d) semileptonic diboson production topologies ($s$-channel). 
They are depicted in Fig.\ \ref{fig:VBF:theory:diagrams}. 
Although not all topologies exist for all external flavor configurations, 
the different diagrams of Fig.\ \ref{fig:VBF:theory:diagrams} interfere and cannot be separated. 
Nonetheless, in different regions of the phase space, different topologies 
will dominate and suitable approximations can be constructed. 
Besides the diboson region, in which both the invariant mass of the 
lepton pair and the invariant mass of the two final state jets are close to 
the nominal $W$ and $Z$ boson masses, 
the vector boson fusion region is of particular interest. 
This region is characterised by a large invariant mass of the two final-state jets and their large separation in rapidity, typically $m_{jj}>600\,\text{GeV}$ 
and $\Delta y>4.5$.
Here, a subclass of the diagrams in Fig.\ \ref{fig:VBF:theory:diagrams}(a) 
dominate the cross section.\footnote{
  Typically, to maximise data statistics, experimental measurements 
  apply much looser cuts. 
  The suitability of the VBF-approximation in such a phase space 
  must be confirmed if theory predictions calculated in this approximation 
  are to be tested against data.
}

The main irreducible background in experimental measurements of the 
electroweak production of a single vector boson in association with at 
least two jets is its QCD production channel, proceeding at 
$\order(\alphaS^2\alpha^2)$ at LO.\footnote{
  This distinction between QCD and EW production channels is tied to 
  a leading order interpretation of the process where their interference 
  is small in the VBF phase space region. It breaks down in other 
  regions or at higher-orders.
}
As the electroweak production mode is characterised by a color-neutral 
$t$-channel exchange, it exhibits reduced hadronic activity in the 
central region between the leading jets. 
Such a suppression does not exist in its QCD production mode,
and a veto on 
central or, in fact, any additional jet activity can further enhance 
the sought after signal. 
Such jet vetoes, however, are typically badly described by fixed-order 
perturbation theory due to the emergence of logarithms of the ratio of the 
hard scale and the jet veto scale and the best available description for this 
observable is offered by conventional parton showers.

\subsubsection{Higher order computations}
\label{sec:VBF:theory:fixedorder}

\begin{figure*}[t!]
  \centering
  \includegraphics[width=0.47\textwidth,height=0.54\textwidth]{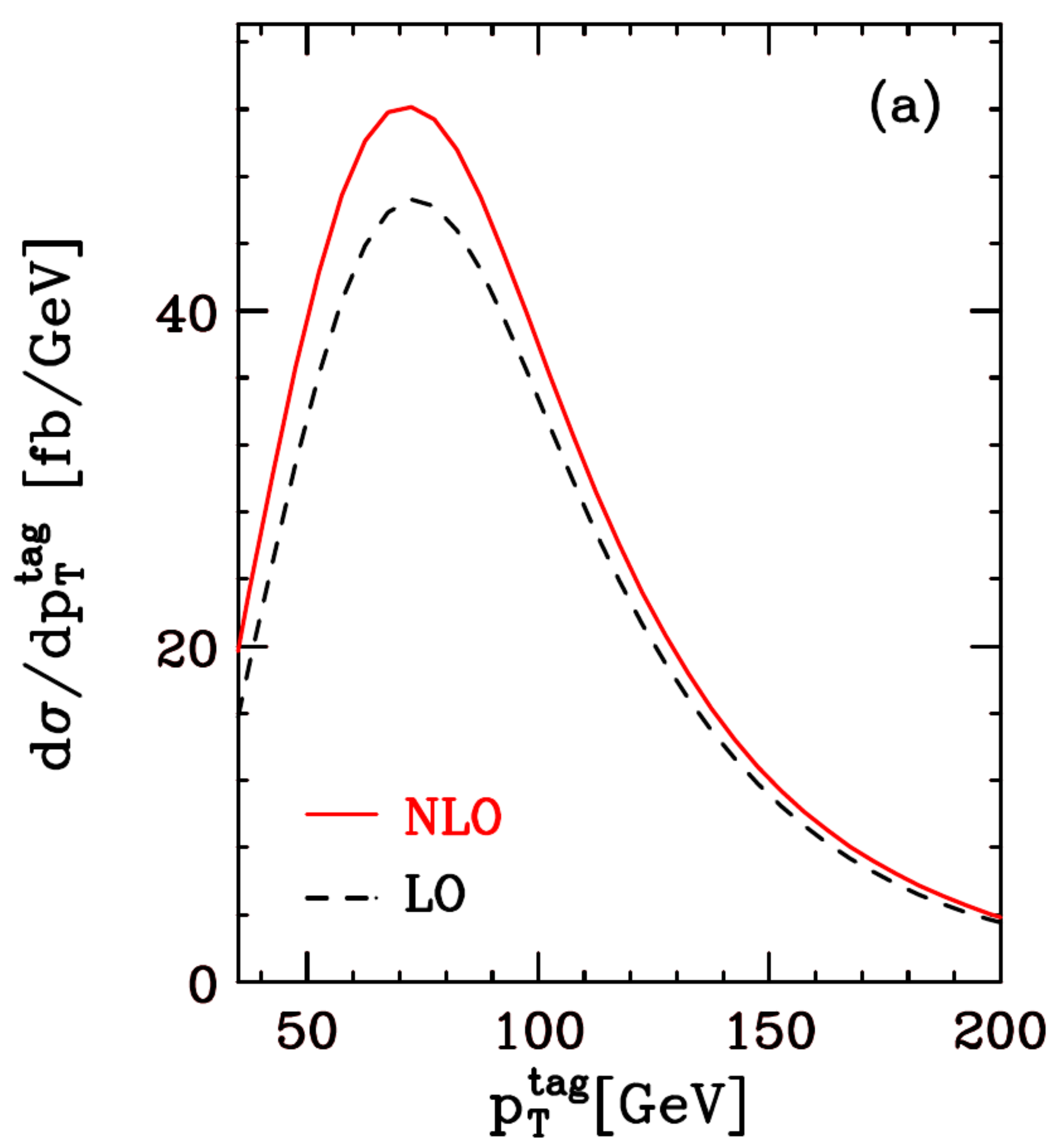}
  \hfill
  \includegraphics[width=0.47\textwidth]{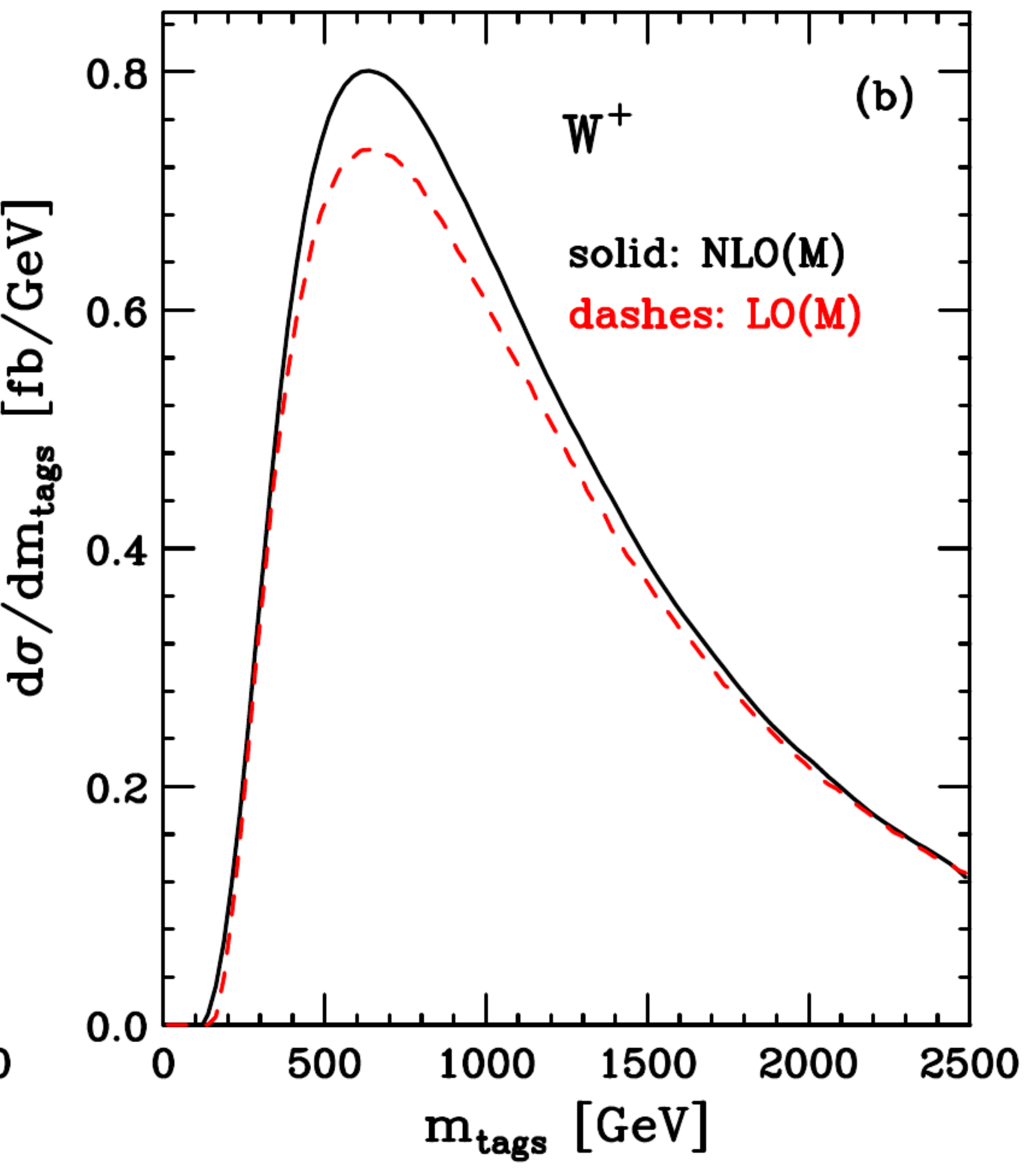}
  \caption{\label{fig:VBF:theory:nlo}
    Leading tagging jet transverse momentum in prompt photon production
    in association with at least two jets 
    through vector boson fusion calculated at LO and NLO QCD accuracy 
    using \Vbfnlo (left), 
    figure taken from \cite{Jager:2010aj}.
    Invariant mass of the tagging jet pair in the production of a charged lepton and a neutrino  
    in association with at least two jets 
    through vector boson fusion calculated at LO and NLO QCD accuracy 
    using \Vbfnlo (right), 
    figure taken from \cite{Oleari:2003tc}.  
  }
\end{figure*}

All higher-order calculations to date have been performed in the 
above-introduced VBF approximation wherein not only $s$-channel 
contributions are neglected but also $t$-/$u$-channel interferences 
are not taken into account. 
This simplifies the calculation immensely in two ways. 
Firstly, it separates the two quark 
lines in color space, thus effectively rendering the calculation 
a ``double-DIS'' one.
And secondly, it facilitates the calculation of its 
NLO QCD correction by removing all components at 
$\order(\alphaS\alpha^4)$ which possess EW divergences 
with respect to the $\order(\alphaS\alpha^3)$ Born process.
The production processes of all EW vector bosons, $W$, $Z$ and $\gamma$, 
in this approximation are implemented in the \Vbfnlo library \cite{Baglio:2014uba,
  Oleari:2003tc,Jager:2010aj}.
Figure\ \ref{fig:VBF:theory:nlo} displays the results for  
$W$ and photon production in vector boson fusion, respectively. 
As can be seen, the NLO QCD 
corrections are generally small. 
Complete EW corrections are not known and will have to be calculated 
for the full $\order(\alpha^4)$ process.

\subsubsection{Monte Carlo event generators}
\label{sec:VBF:theory:MCs}

\begin{figure*}[t!]
  \centering
  \includegraphics[width=0.47\textwidth,height=0.47\textwidth]{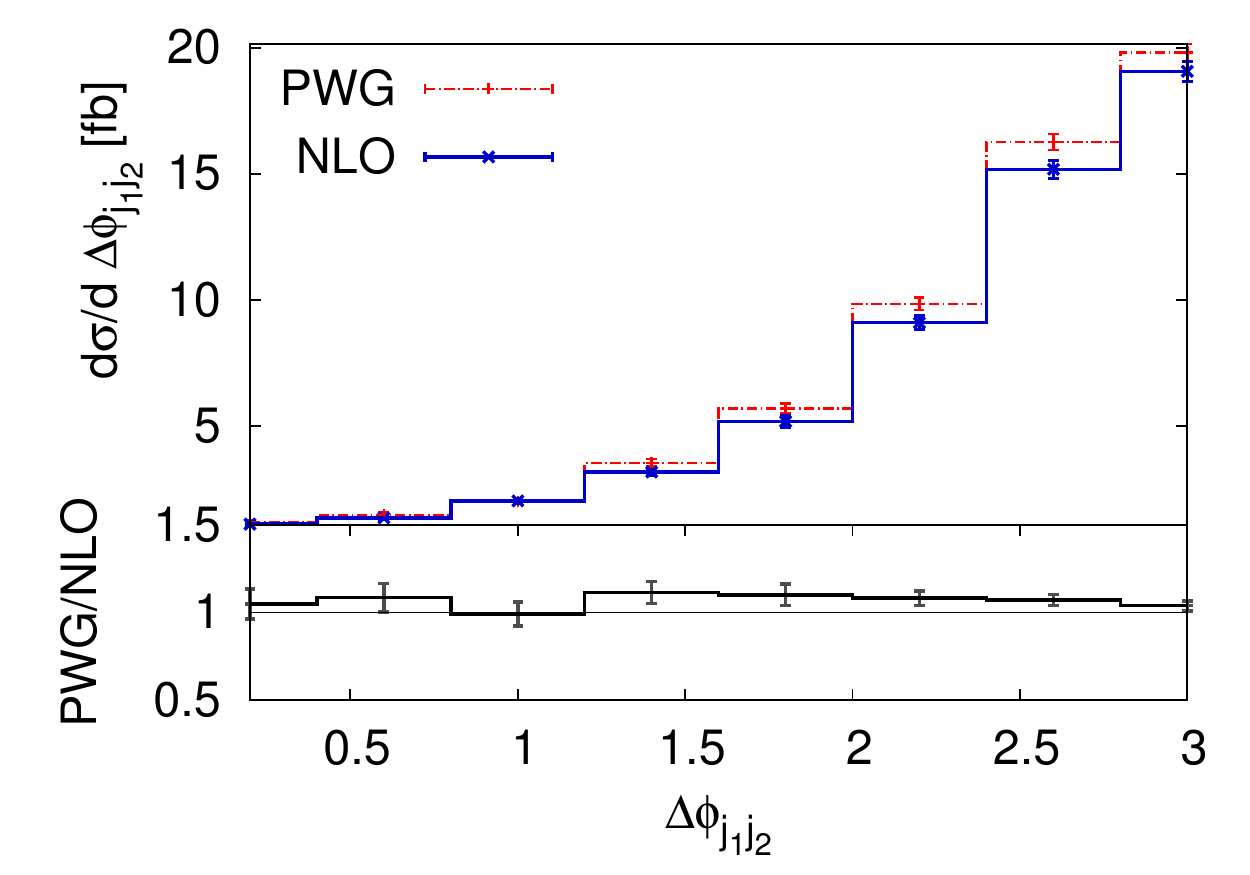}
  \hfill
  \includegraphics[height=0.50\textwidth,angle=270,origin=c]{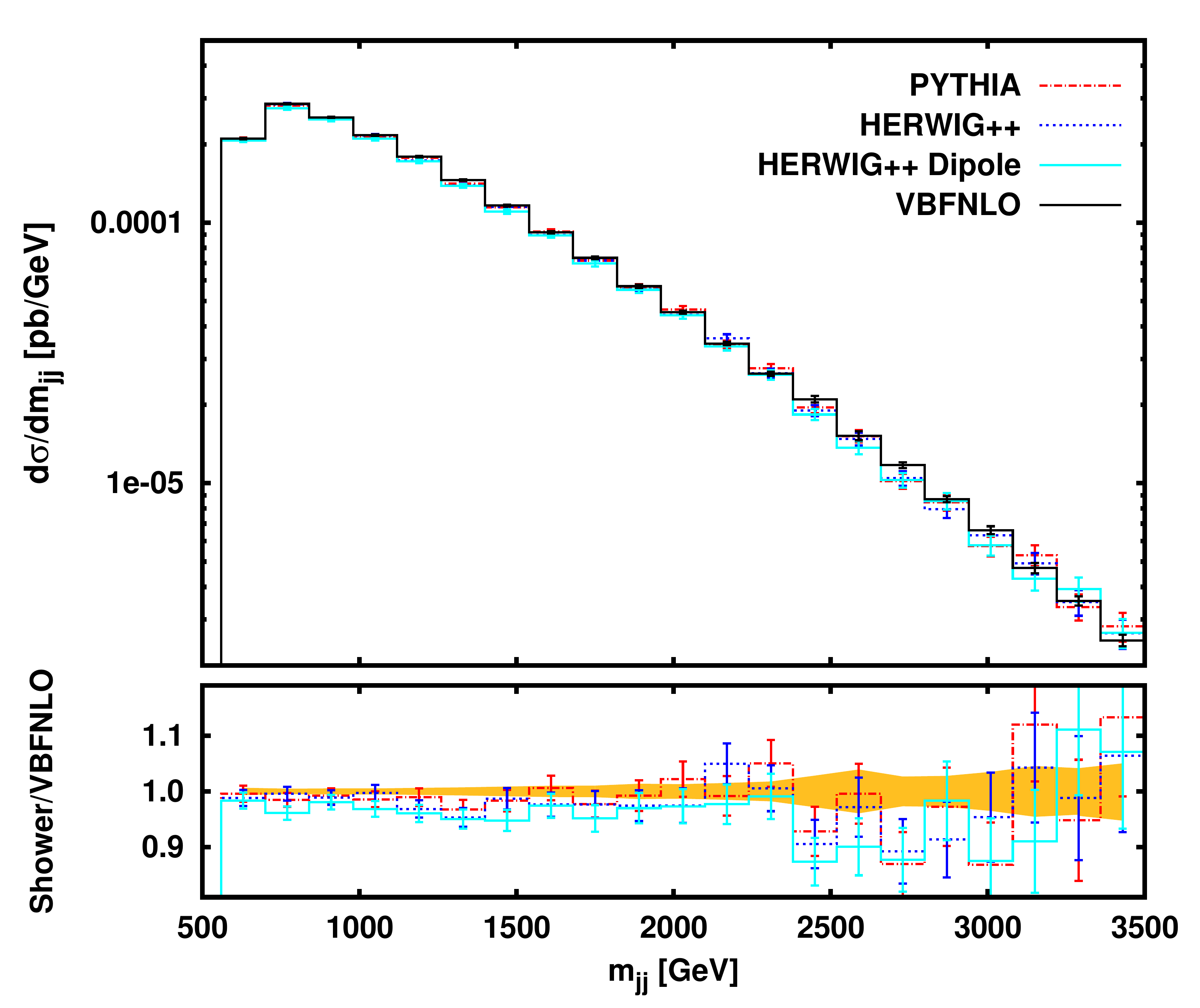}
  \caption{\label{fig:VBF:theory:mc}
    Azimuthal separation of the tagging jets in the production of a pair of charged leptons 
    in association with at least two jets through vector boson fusion 
    calculated at NLO QCD with \Vbfnlo and at NLO QCD accuracy matched to the parton shower 
    using \Powheg (left), 
    figure taken from \cite{Jager:2012xk}.
    Invariant mass of the tagging-jet system in the pair production of a charged lepton and neutrino  
    in association with at least two jets through vector boson fusion
    calculated at NLO QCD with \Vbfnlo and at NLO QCD accuracy matched to the parton shower 
    using \Powheg, compared to predictions using \Pythia and \Herwigpp (right), 
    figure taken from \cite{Schissler:2013nga}.
  }
\end{figure*}

The above-mentioned fixed-order NLO QCD calculations have been matched to 
parton showers. 
Explicit implementations exist in the \Powheg \cite{Jager:2012xk,
  Schissler:2013nga} generator, but are also available in the 
automated \NLOPS tools.
Figure\ \ref{fig:VBF:theory:mc} displays results for $Z$ and 
$W$ production in vector boson fusion, well reproducing the 
fixed-order results. 
Further, multi-jet-merged calculations exist at LO accuracy.

One key aspect in the selection of VBF-type events is the 
aforementioned rapidity gap. 
Therefore, a good description of the radiation pattern of the 
third jet, and any further higher-order radiation, is mandatory. 
Care must be taken to ensure the initial color and starting scale 
assignment in the parton showers is correct in order to 
preserve the unique rapidity gap structure and not spuriously 
fill it with additional radiation. 
The supplementation with LO matrix elements in the above matching 
helps in controlling the associated uncertainties on the level 
of a few percent, but higher accuracy would be desirable.
\subsection{Experimental results}
\label{sec:VBF:exp}
\begin{table*}
    \centering
    \begin{tabular}{c|p{4cm}|p{4cm}|p{4cm}}
       $m_{jj}$ cut  &  $\sqrt{s}=$7~TeV  & $\sqrt{s}=$8~TeV  & $\sqrt{s}=$13~TeV   \\ \hline
        120~GeV  & $154\pm58$~fb~\cite{Chatrchyan:2013jya}  & $174\pm43$~fb~\cite{Khachatryan:2014dea} & $534\pm60$~fb~\cite{Sirunyan:2017jej} \\
        250~GeV &   & $54.7\pm11.2$~fb~\cite{Aad:2014dta} & $119\pm26$~fb~\cite{Aaboud:2017emo}  \\
         1 TeV &   & $10.7\pm 2.1$~fb~\cite{Aad:2014dta} &  $37.4\pm 6.5$~fb~\cite{Aad:2020sle} \\
    \end{tabular}
    \caption{Summary of VBF \Z production cross sections measured at the \LHC in the $\ell\ell$jj final state with different $m_{jj}$ definitions and different proton collision energies. All quoted cross sections are for a single lepton flavor.}
    \label{tab:vbfz}
\end{table*}
\begin{table*}
    \centering
    \begin{tabular}{c|p{4cm}|p{4cm}|p{4cm}}
       $m_{jj}$ cut  &  $\sqrt{s}=$7~TeV  & $\sqrt{s}=$8~TeV  & $\sqrt{s}=$13~TeV   \\ \hline
        120~GeV &  &  & $6.23\pm 0.62$~pb~\cite{Sirunyan:2019dyi} \\
        500~GeV & $2.76\pm0.67$~pb~\cite{Aaboud:2017fye}  & $2.89\pm0.51$~pb~\cite{Aaboud:2017fye} & \\
         1 TeV  &   & $0.42\pm 0.10$~pb~\cite{Khachatryan:2016qkk} &  \\
    \end{tabular}
    \caption{Summary of VBF \W production cross sections measured at the \LHC in the $\ell\nu$jj final state with different $m_{jj}$ definitions and different proton collision energies. All quoted cross sections are for a single lepton flavor.}
    \label{tab:vbfw}
\end{table*}

Initial measurements of the electroweak production of a vector boson and two jets were performed by the \CMS collaboration in the final state with two charged leptons and two jets  (VBF \Z channel) with 7~TeV proton collision data~\cite{Chatrchyan:2013jya}. 
The precision obtained in this first measurement was around 30\%, mostly limited by systematic uncertainties on the jet energy scale and the background modeling.  

Improved measurements have been obtained with 8~TeV data, by 
both \ATLAS~\cite{Aad:2014dta} and \CMS~\cite{Khachatryan:2014dea},
with precisions around 20\%, and signal significances just above 5 standard deviations. 
Measurements of the VBF \Z process with 13~TeV have also been performed by \ATLAS~\cite{Aaboud:2017emo} with 2015 data,  by \CMS~\cite{Sirunyan:2017jej} with 2016 data, and by  
\ATLAS~\cite{Aad:2020sle} with the full Run 2 data.
These \ATLAS and \CMS measurements reach an overall precision of 20\% and 10\%, respectively.

It should be noted that the measurements provided by \ATLAS and \CMS are quite different and complementary. 
Since the first measurement, \CMS has defined the VBF \Z signal in an inclusive phase space in the four-fermion final state $\ell\ell$jj, with $m_{\ell\ell}>50$~GeV and $m_{jj}>120$~GeV, in which all pure EW diagrams of order $\alpha_{\rm EW}^4$ contribute to the signal definition, whereas 
\ATLAS has performed measurements with the signal defined in higher dijet mass  fiducial phase space regions. 
The \ATLAS signal definition is at particle level, where the dijet invariant mass condition is implemented on the two \pT-leading jets after clustering the final state particles, and the simulation setup includes NLO QCD corrections, implemented with \Powheg~\cite{Oleari:2003tc,Jager:2012xk}, and does not include $s$-channel diboson contributions. 
\begin{figure*}[htb] 
\centering
\includegraphics[width=0.42\textwidth,height=0.45\textwidth]{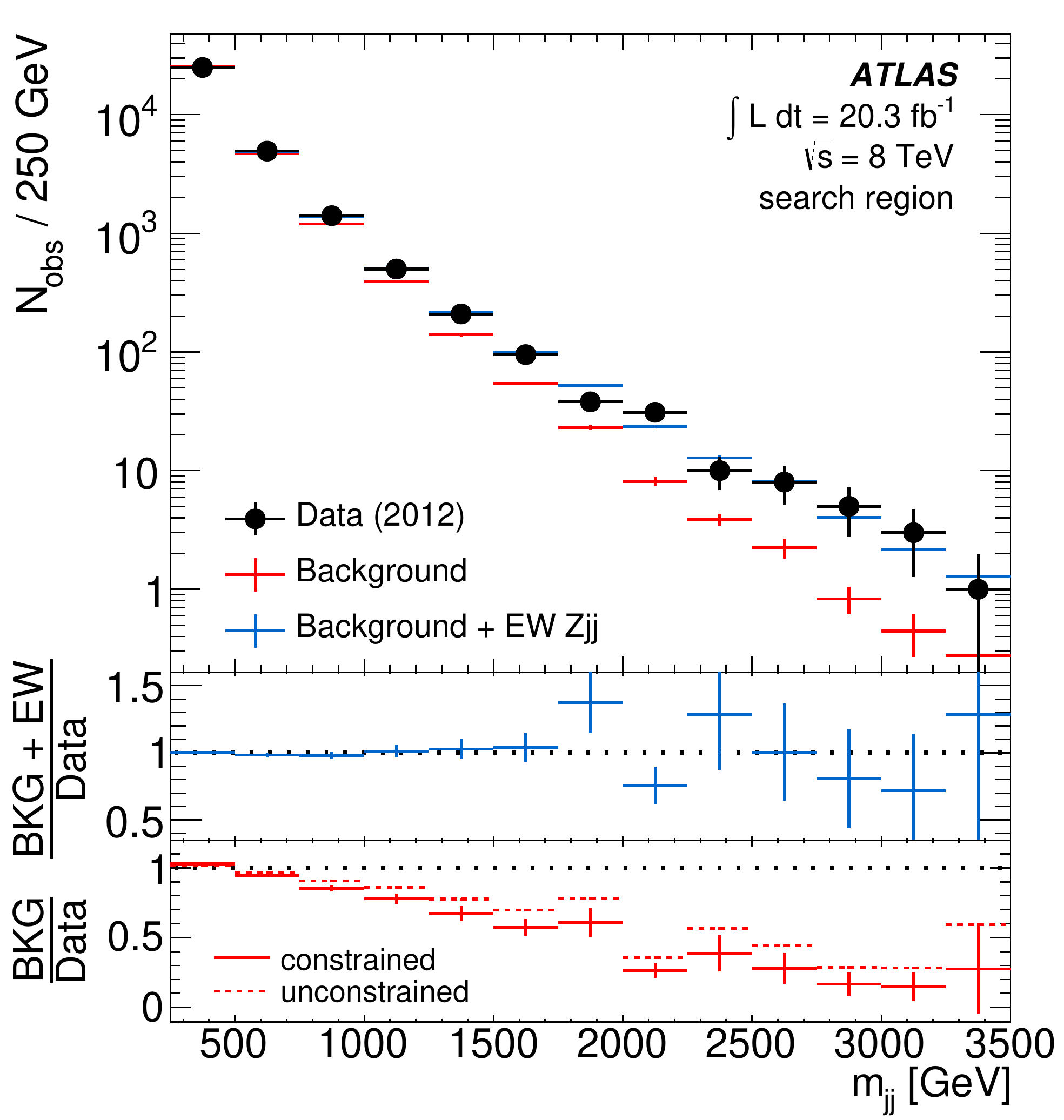}
\includegraphics[width=0.57\textwidth,height=0.45\textwidth]{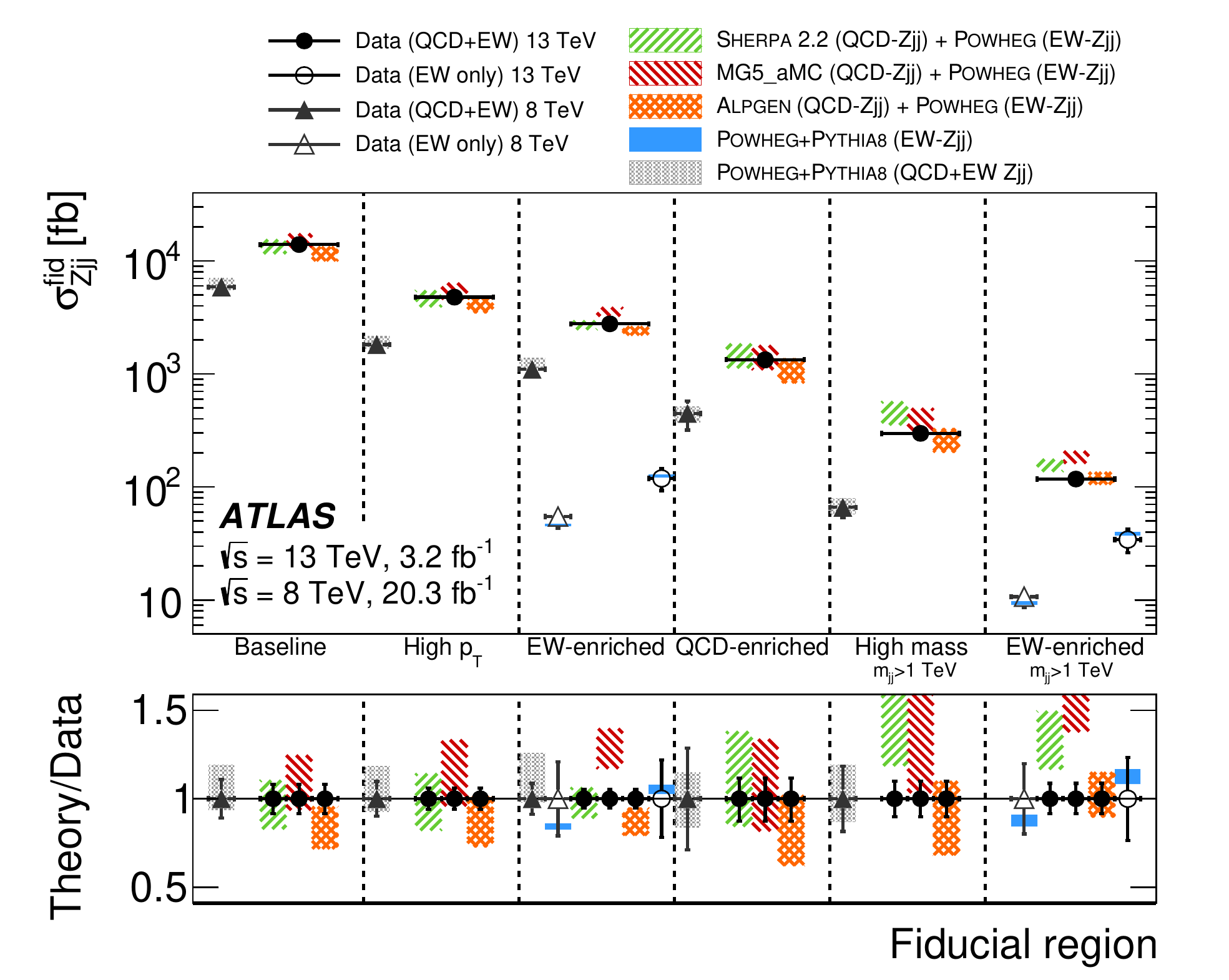}
\caption{
  The dijet invariant mass distribution in the search region. The signal and (constrained) background templates are scaled to match the number of events as a result of a fit to data. The lowest panel shows the ratio of constrained and unconstrained background templates to the data (left), figure taken from~\cite{Aad:2014dta}.
 Cross-sections measurements
in different fiducial regions at 8 and 13 TeV, compared to theoretical predictions (shaded/hatched bands). The bottom panel shows the ratio of the various theory predictions to data as shaded bands. Relative uncertainties in the measured data are represented by an error bar centered at unity (right), figure taken from~\cite{Aaboud:2017emo}. 
\label{fig:VBFZ}}
\end{figure*}
Figure~\ref{fig:VBFZ} (left) shows the dijet invariant mass distribution that is used by \ATLAS to extract the signal contribution in the high dijet mass tail with 8~TeV data~\cite{Aad:2014dta}.
 Measurements by \CMS are  extracted by fitting the dijet invariant mass but also making use of more sophisticated multivariate discriminants with different event observables. Among the \CMS multivariate inputs is an internal jet composition discriminator used to separate features of quark- and gluon-initiated jets, applied to the two VBF tagging jets~\cite{CMS:2013kfa,CMS:2017wyc}. 
The \ATLAS measurements include several additional fiducial regions where inclusive cross sections are also measured. Figure~\ref{fig:VBFZ} (right) shows a summary of such inclusive measurements with 13~TeV data~\cite{Aaboud:2017emo}.
The most recent \ATLAS results~\cite{Aad:2020sle} focus on differential cross-section measurements, both for the electroweak signal component, 
and inclusively for the signal and background production, for different observables.
Table~\ref{tab:vbfz} shows a summary of  
inclusive VBF \Z cross sections measured to date at the LHC.

\begin{figure*}[htb] 
\centering
\includegraphics[width=0.46\textwidth,height=0.45\textwidth]{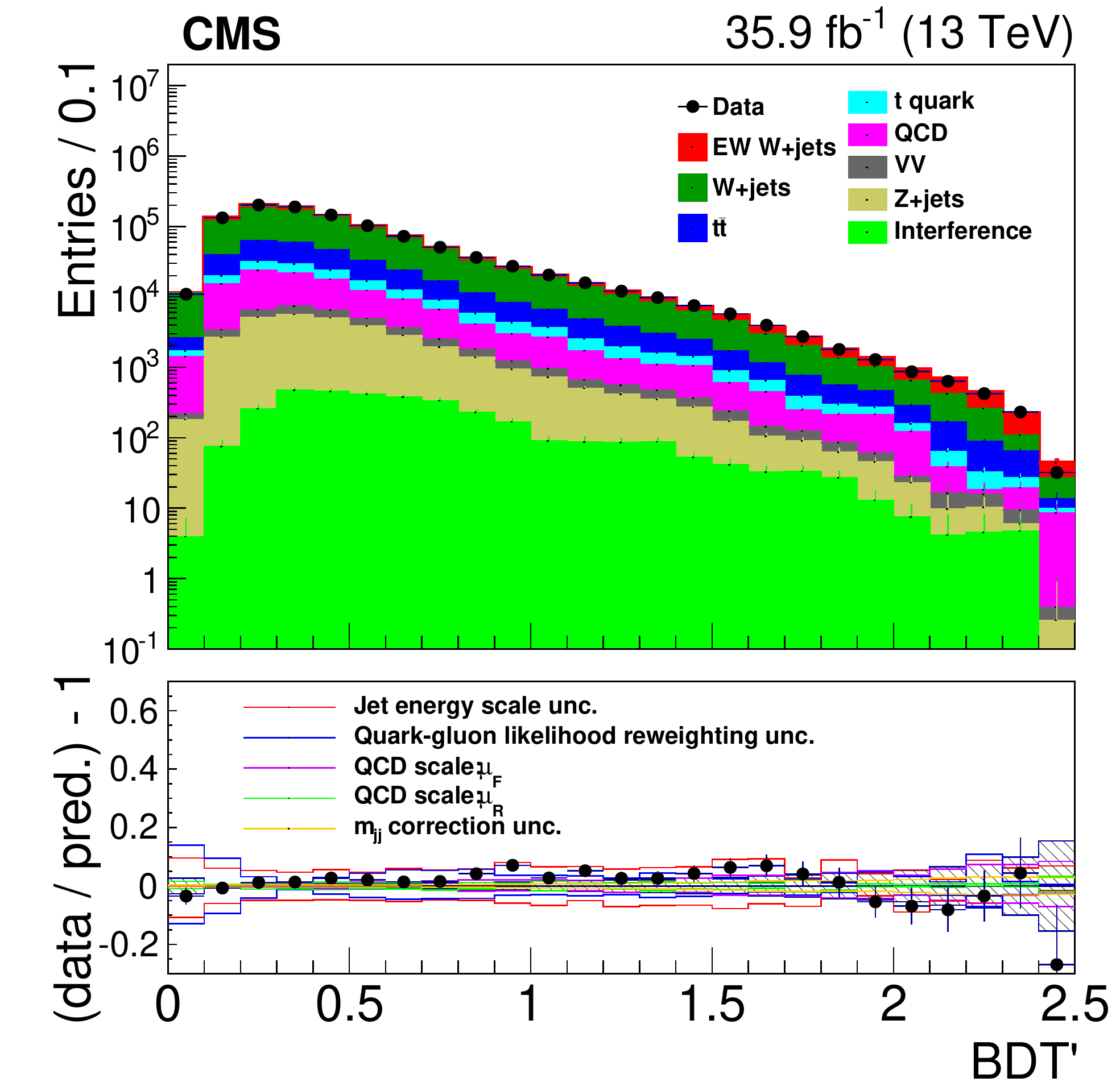}
\includegraphics[width=0.53\textwidth,height=0.45\textwidth]{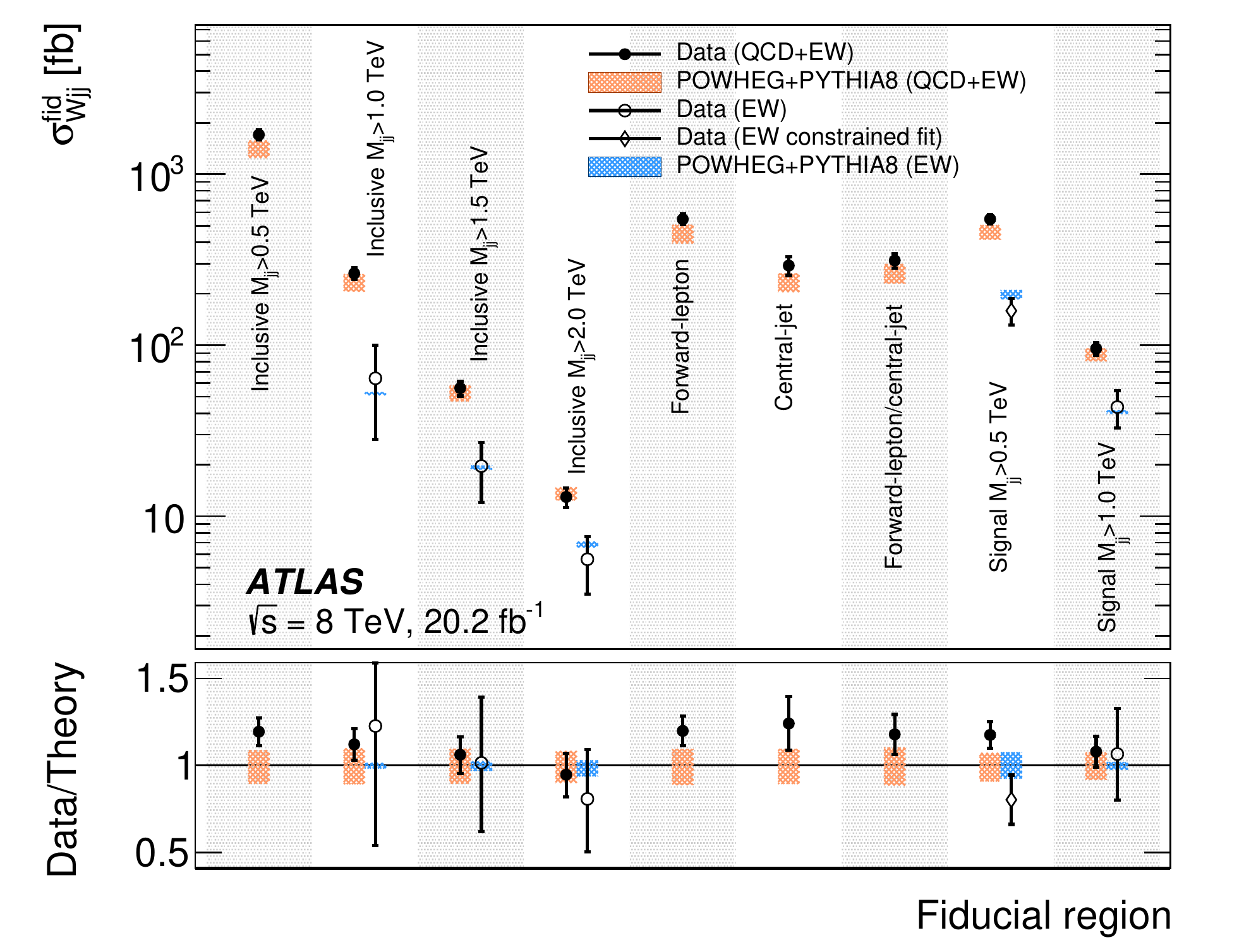}
\caption{
Data and pre-fit MC simulation of the  multivariate output distribution  
used to extract the signal in the electron channel at 13 TeV (left), figure taken from~\cite{Sirunyan:2019dyi}. 
Cross sections for \W plus two jets inclusive and EW signal productions in different  particle-level fiducial regions at 8 TeV (right), figure taken from~\cite{Aaboud:2017fye}.
\label{fig:VBFW}}
\end{figure*}

Analogous measurements have been performed in the single charged lepton plus dijet final state (VBF \W channel) by \CMS with 8~TeV collision data~~\cite{Khachatryan:2016qkk}, by \ATLAS with 7~TeV and 8~TeV collision data~\cite{Aaboud:2017fye}, and by \CMS with 13~TeV collision data~\cite{Sirunyan:2019dyi}.  
A variety of signal definitions has also been chosen for the VBF \W channel. The \CMS collaboration has used four-fermion LO definitions with $m_{jj}>120$~GeV  (as for VBF \Z) and with $m_{jj}>1$~TeV, while the \ATLAS collaboration makes use of NLO signal modeling with  $m_{jj}>0.5, 1, 2$~TeV cuts defined at particle level after parton showering and jet clustering. 
Figure~\ref{fig:VBFW} shows the multivariate output distribution used to measure the inclusive cross section at 13 TeV in the electron channel, and different particle-level fiducial cross sections performed at 8 TeV, respectively in \CMS and \ATLAS. Table~\ref{tab:vbfw} shows a summary of  
inclusive VBF $W$ cross sections measured to date at the \LHC.

\begin{figure*}[htb] 
\centering
\includegraphics[width=0.46\textwidth]{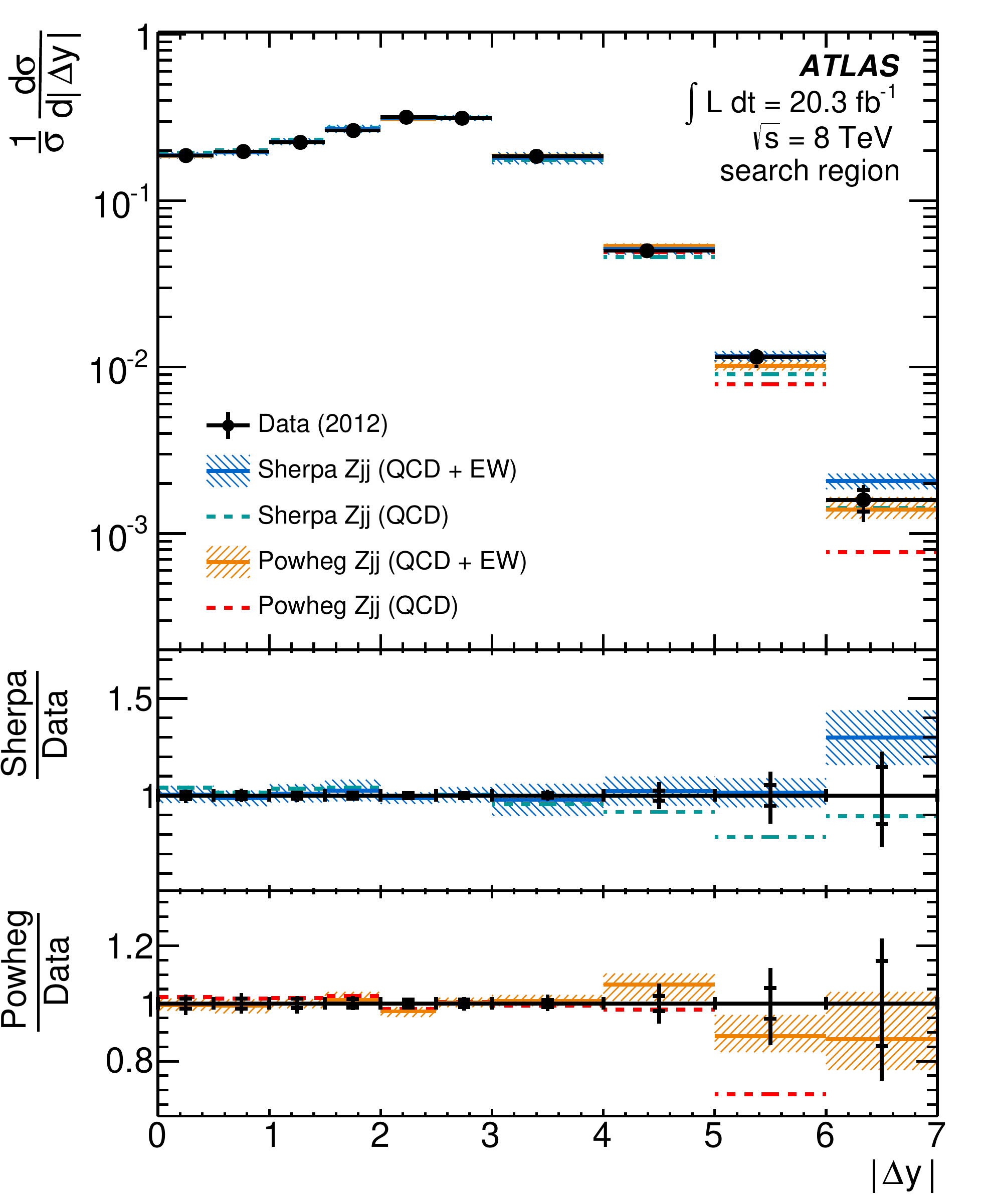}
\includegraphics[width=0.46\textwidth]{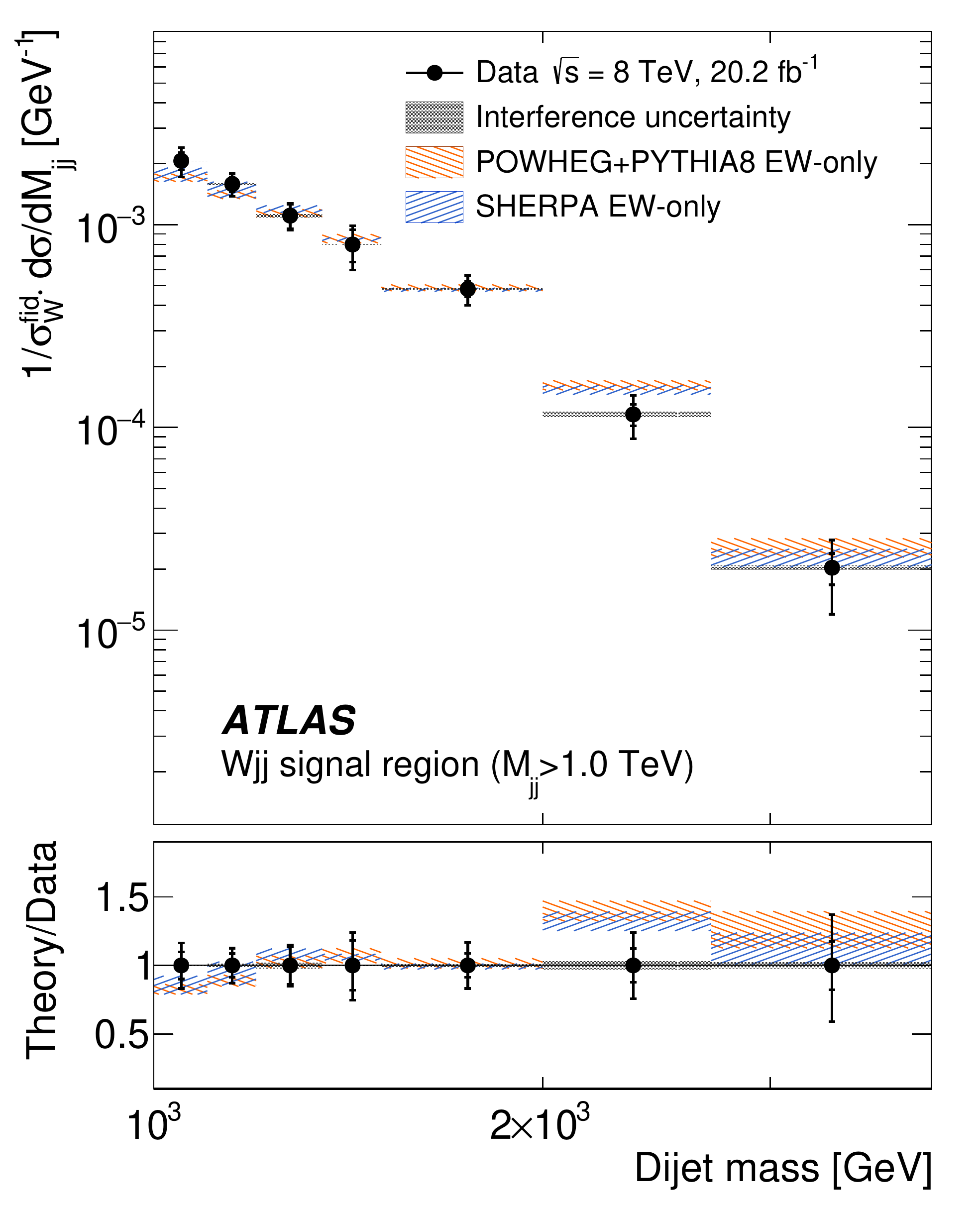}
\caption{
 Unfolded normalised differential cross section distribution at 8 TeV as a function of the rapidity separation between the leading jets in the search region. Particle-level predictions are shown for  strong and electroweak $Z$jj production (left), figure taken from~\cite{Aad:2014dta}.
 Unfolded normalized differential EW $W$jj production cross sections at 8 TeV as a function of the dijet invariant mass for the signal fiducial region (right), figure taken from~\cite{Aaboud:2017fye}. Both statistical (inner bar) and total (outer bar) measurement uncertainties are shown, as well as ratios of the theoretical predictions to the data.
\label{fig:VBFVunf}}
\end{figure*}

Both the VBF \Z and \W measurements by \ATLAS at 8~TeV~\cite{Aad:2014dta,Aaboud:2017fye} include a large number of differential distributions unfolded to particle level for both inclusive and signal contributions in different fiducial regions,
as shown in Fig.~\ref{fig:VBFVunf}.

Interference effects between signal and background sources have been evaluated in the range from 2\%--12\%  of the total signal, depending on the channel and the selected phase space, and are generally positive. Results by \CMS include a full simulation of  interference contributions that are implemented in the cross section extraction fits.

\begin{figure*}[htb] 
\centering
\includegraphics[width=0.42\textwidth]{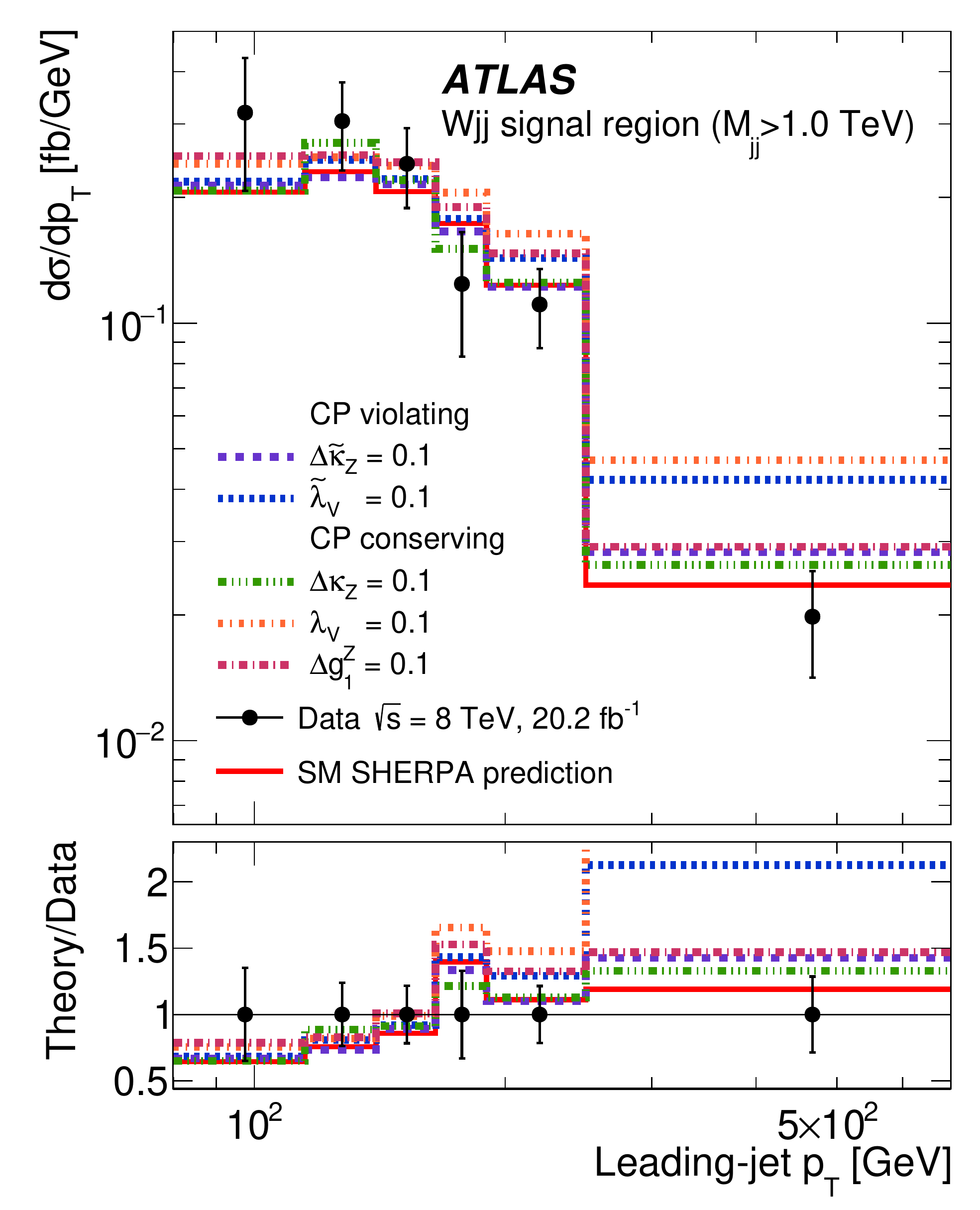}
\includegraphics[width=0.55\textwidth]{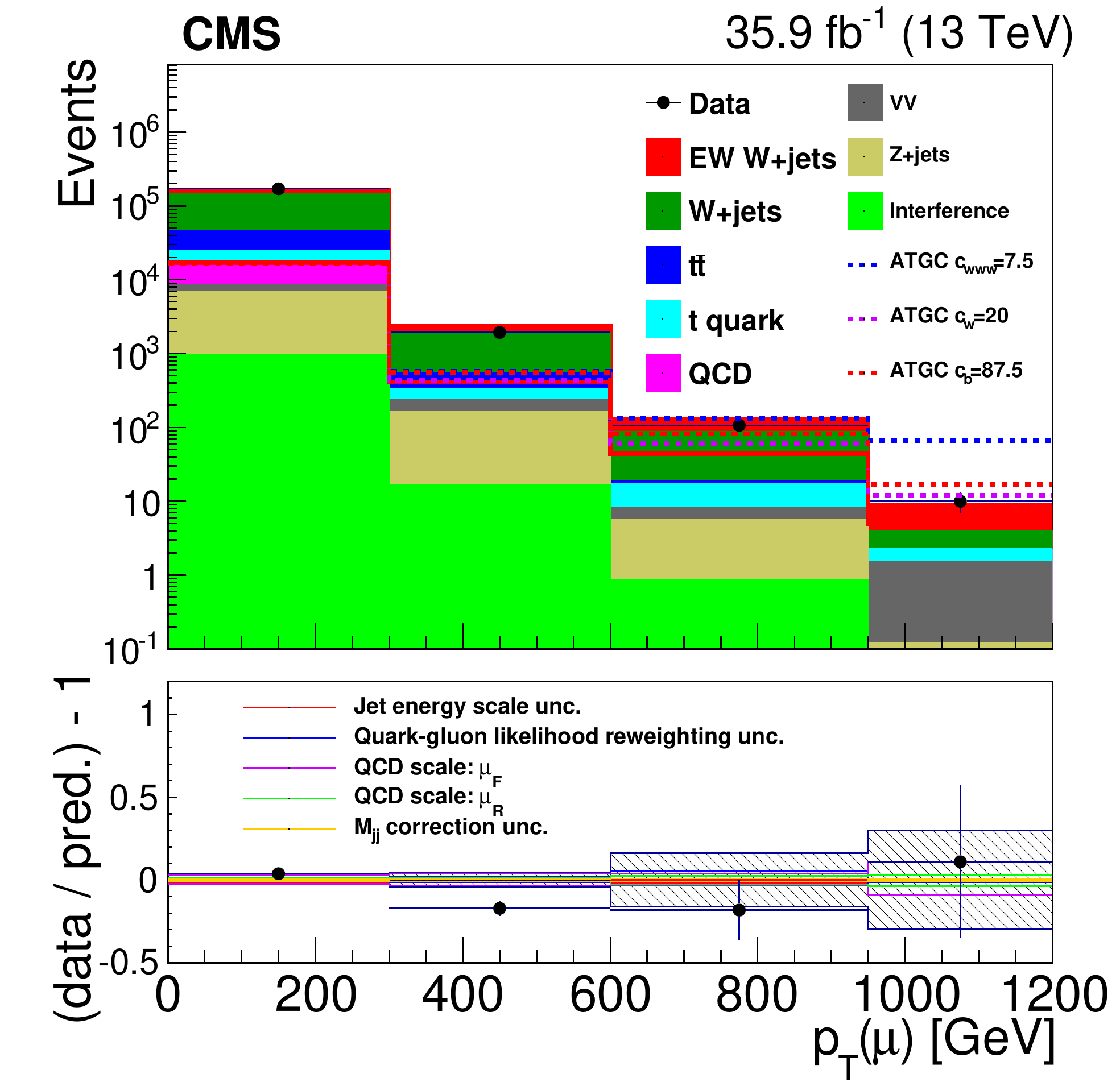}
\caption{
 Leading jet \pT for the SM (solid line) and with anomalous TGC parameter deviations of  0.1 from the SM values (dashed lines) compared to unfolded differential VBF $W$  cross-section distribution measured in 8 TeV data in the $m_{jj}>1$~TeV signal region (left), figure taken from~\cite{Aaboud:2017fye}.
 Muon \pT distribution in 13 TeV data and SM backgrounds, and various scenarios for anomalous TGCs.  The lower panel shows the ratio between data and prediction minus one with the statistical uncertainty from simulation (gray hatched band) as well as the leading systematic uncertainties (right), figure take from~\cite{Sirunyan:2019dyi}.
\label{fig:VBFtgc}}
\end{figure*}

The structure of the $WWZ$ and $WW\gamma$ triple gauge couplings (TGCs) can be explored with VBF \Z and \W measurements, and anomalous contributions to the TGCs have been searched for by both \ATLAS and \CMS in the context of the LEP effective Lagrangian approach ~\cite{Hagiwara:1986vm} and effective field theory operators in the HISZ basis~\cite{Hagiwara:1993ck}.
Limits on anomalous coupling parameters have been extracted by \ATLAS fitting alternatively the dijet invariant mass and the leading jet \pT, in fiducial signal regions~\cite{Aad:2014dta,Aaboud:2017fye}. Limits have also  been extracted by the \CMS collaboration, fitting the \pT distributions of the two charged leptons or the single charged lepton in a more inclusive $Vjj$ phase space~\cite{Sirunyan:2017jej,Sirunyan:2019dyi}, and resulted to be more stringent because of the larger $\sqrt{s}$ of the analyzed data set, and of a larger acceptance in the high \pT tails where anomalous TGC effects are generally expected. 
Examples of kinematic distributions used to fit anomalous TGC contributions are shown in Fig~\ref{fig:VBFtgc}.
The reported TGC sensitivities are  comparable or even more stringent than some obtained from diboson channels with the same data luminosity, revealing an unexpected high sensitivity  of the TGC studies from VBF $V$ measurements. 

The full Run 2  \ATLAS results~\cite{Aad:2020sle} focus on extracting limits on the interference between the Standard Model and dimension-six scattering amplitudes.

\begin{figure*}[htb] 
\centering
\includegraphics[width=0.49\textwidth]{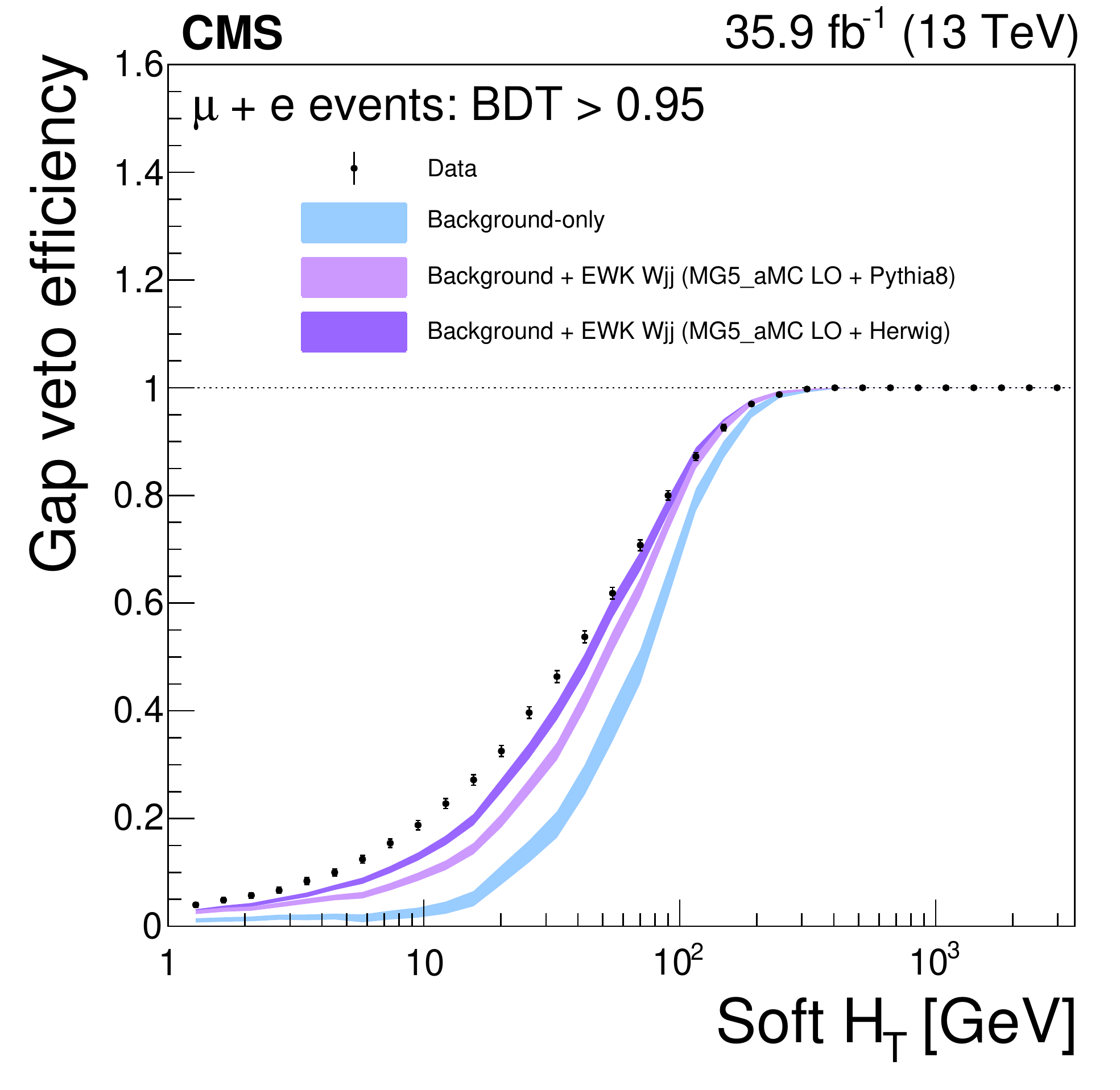}
\includegraphics[width=0.49\textwidth]{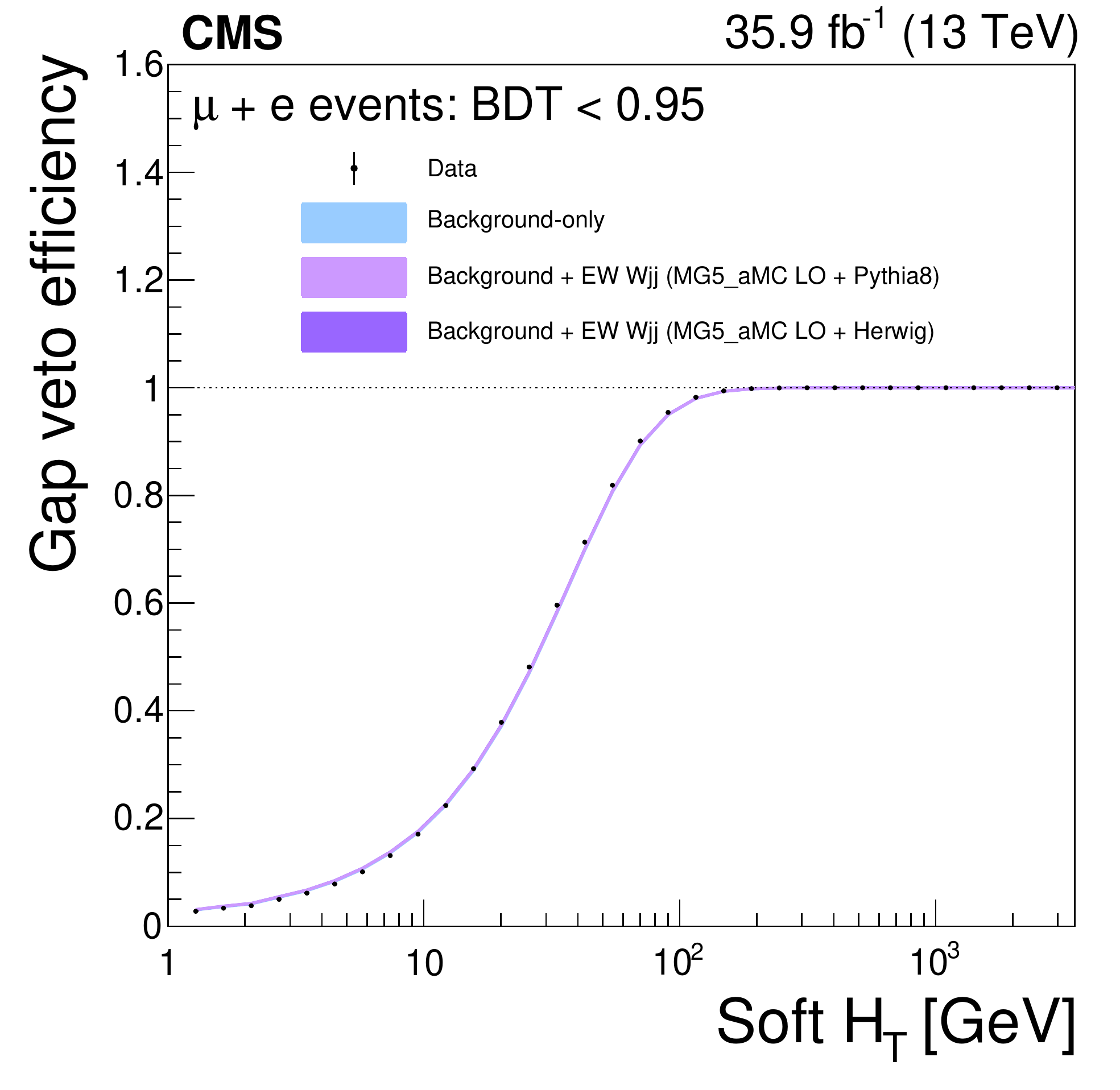}
\caption{
Veto efficiencies of hadronic gap activity at 13 TeV, evaluated with charged particles in signal-enriched (left) and background-enriched (right) regions. The data are compared with the background-only prediction as well as background+signal with \Pythia~\cite{Sjostrand:2014zea} or \Herwig~\cite{Bahr:2008pv} parton showering. Figures taken from~\cite{Sirunyan:2019dyi}.
\label{fig:VBFgap}}
\end{figure*}

A number of QCD studies of hadronic activity in the selected 
$V$ plus two jets events have been carried out by \CMS and \ATLAS with the 7~TeV and 8~TeV data~\cite{Chatrchyan:2013jya,Khachatryan:2014dea,Aad:2014dta,Aaboud:2017fye}.
Inclusive studies of "radiation patterns" have been performed following the prescriptions and suggestions in Ref.~\cite{Binoth:2010nha},
where  model dependencies are estimated by comparing different generators. The results~\cite{Chatrchyan:2013jya} show a good agreement between data and the predictions by \Madgraph interfaced to \Pythia parton shower (ME--PS) for all 
 chosen observables that are sensitive to hadronic activity. 
Dedicated studies, restricted to the additional hadronic activity in the expected rapidity gap between the two tagging jets, have also been performed. They are particularly interesting when making use of the larger 13~TeV data set~\cite{Sirunyan:2017jej,Sirunyan:2019dyi}. 
The hadronic activity in the rapidity gap is measured in signal-enriched regions that have similar signal and background yields, using as observables the standard reconstructed jets or jets reconstructed by clustering tracks ("soft track jets") from charged particles. The latter are used as they can be effectively cleaned from pileup contributions allowing precise low-\pT measurements~\cite{CMS:2009sva,CMS:2010rua}. 
Monte~Carlo based studies of the additional jet activity in VBF $W$ and $Z$ channels revealed interesting differences in the prediction of different parton shower setups~\cite{Schissler:2013nga}. 

Figure~\ref{fig:VBFgap} shows the gap veto efficiency for 
the "soft" \HT observable, i.e., the scalar sum of track jets \pT in the rapidity gap region, in signal and background enriched samples. 
In the background dominated sample the agreement of the data with 
the predictions is very good. 
The data in the signal region clearly disfavor the background-only predictions and are in reasonable agreement with the presence of the signal with the \Herwigpp PS predictions for gap activities above 20 GeV, while the signal with \Pythia PS seems to generally overestimate the gap activity. In the events with very low gap activity, in particular below 10 GeV, as measured with the soft track jets, the data indicates gap activities also below the \Herwigpp PS predictions.


\section{Associated production of a vector boson and heavy-flavor jets}
\label{sec:VHF}
\subsection{Theoretical predictions}
\label{sec:VHF:theory}

The third important class of vector boson production processes 
is the production in association with heavy quarks, namely $b$- 
and $c$-quarks. 
The characterizing feature of these quark flavors is their relatively large mass, in 
comparison with the proton, in combination with a life-time long 
enough to form hadrons that decay after 
macroscopic path lengths. 
Indeed, this feature, leading to the presence of differentiable 
secondary decay vertices, is used in most tagging algorithms that identify the presence of heavy-flavor hadrons.
The top quark associated production features very different 
dynamics and will not be discussed in this review.

In the following section, the features and availability of calculations 
for this process class will be reviewed, while a comprehensive review 
of the calculation techniques can be found in  \cite{Cordero:2015sba}.

\subsubsection{Higher order calculations and flavor schemes}
\label{sec:VHF:theory:ho}

\begin{figure*}
  \centering
  \begin{minipage}{0.22\textwidth}
    \centering
    \includegraphics[height=1.4\textwidth]{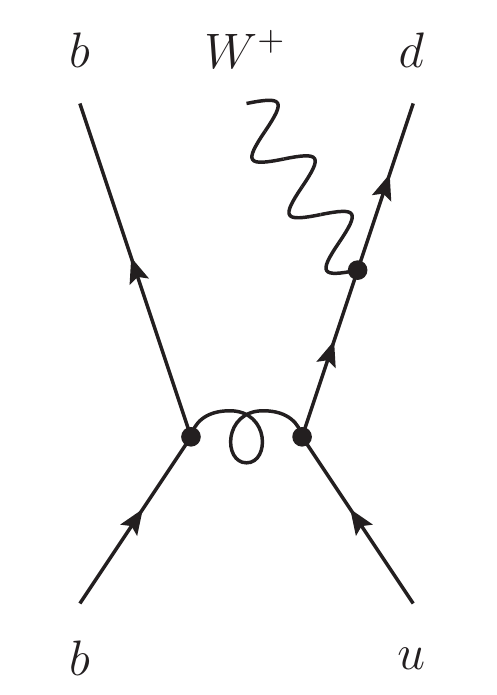}\\
    (a)
  \end{minipage}
  \hfill
  \begin{minipage}{0.44\textwidth}
    \centering
    \includegraphics[height=0.7\textwidth]{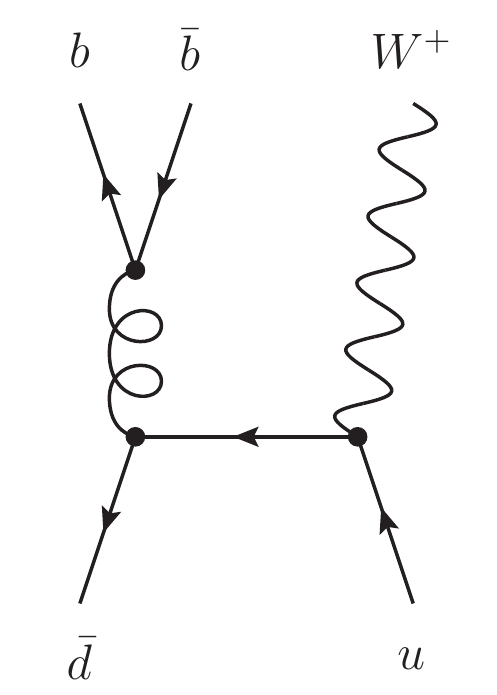}
    \hspace*{-0.05\textwidth}
    \includegraphics[height=0.7\textwidth]{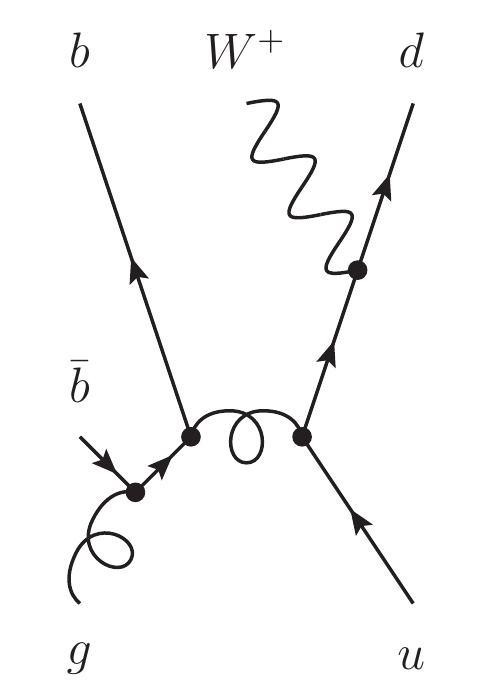}\\
    (b)
  \end{minipage}
  \hfill
  \begin{minipage}{0.22\textwidth}
    \centering
    \includegraphics[height=1.4\textwidth]{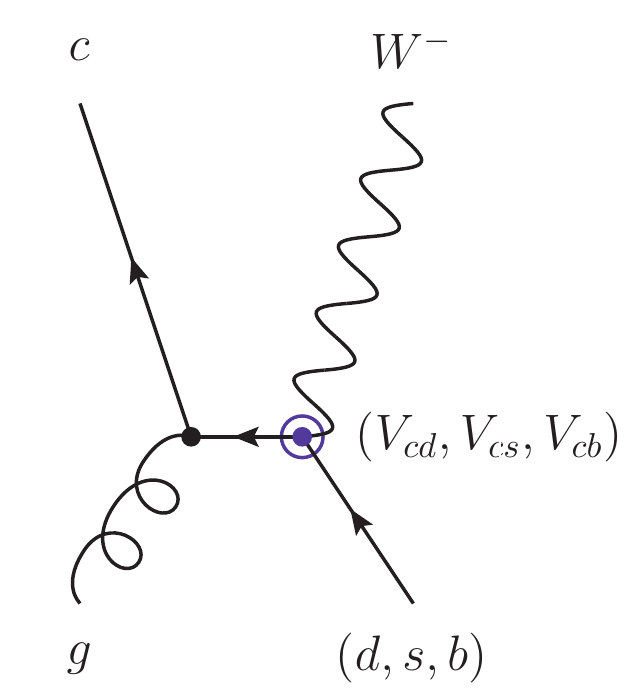}\\
    (c)
  \end{minipage}
  \caption{\label{fig:VHF_W_diagrams}
    Representative Feynman diagrams for the production of a $W$ boson and (at least) 
    one $b$ quark at LO in the $n_f=5$ scheme (a), the $n_f=4$ scheme (b) and 
    additional contributions in the production of (at least) one $c$ quark in 
    association with the $W$ boson (c).
  }
\end{figure*}

\begin{figure*}
  \centering
  \begin{minipage}{0.22\textwidth}
    \centering
    \includegraphics[height=1.4\textwidth]{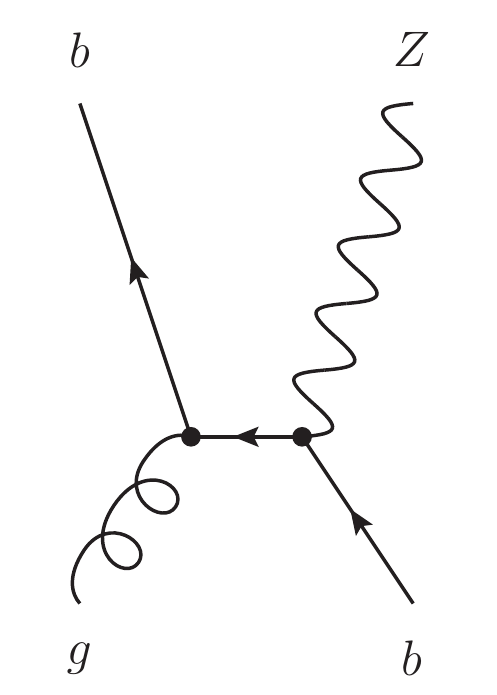}\\
    (a)
  \end{minipage}
  \hspace*{0.11\textwidth}
  \begin{minipage}{0.44\textwidth}
    \centering
    \includegraphics[height=0.7\textwidth]{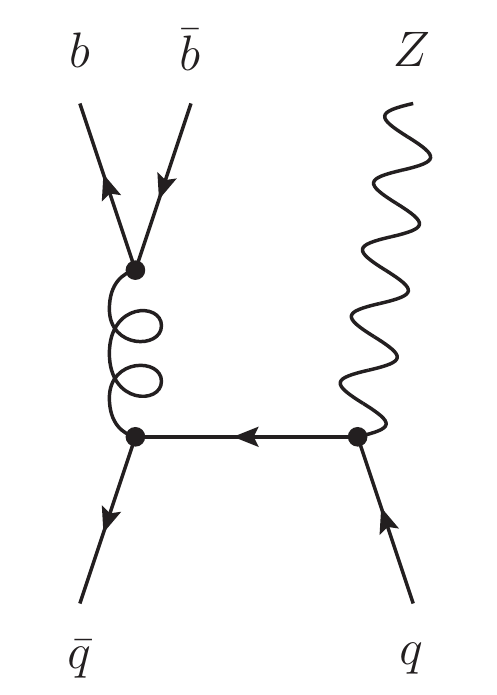}
    \hspace*{-0.05\textwidth}
    \includegraphics[height=0.7\textwidth]{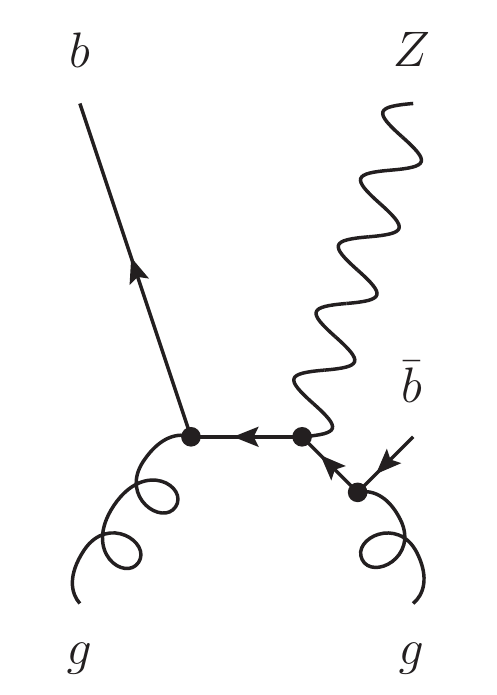}\\
    (b)
  \end{minipage}
  \caption{\label{fig:VHF_Z_diagrams}
    Representative Feynman diagrams for the production of a $Z$ boson and (at least) 
    one $b$ quark at LO in the $n_f=5$ scheme (a), the $n_f=4$ scheme (b).
  }
\end{figure*}

\begin{figure*}
  \centering
  \includegraphics[width=0.47\textwidth,height=0.44\textwidth]{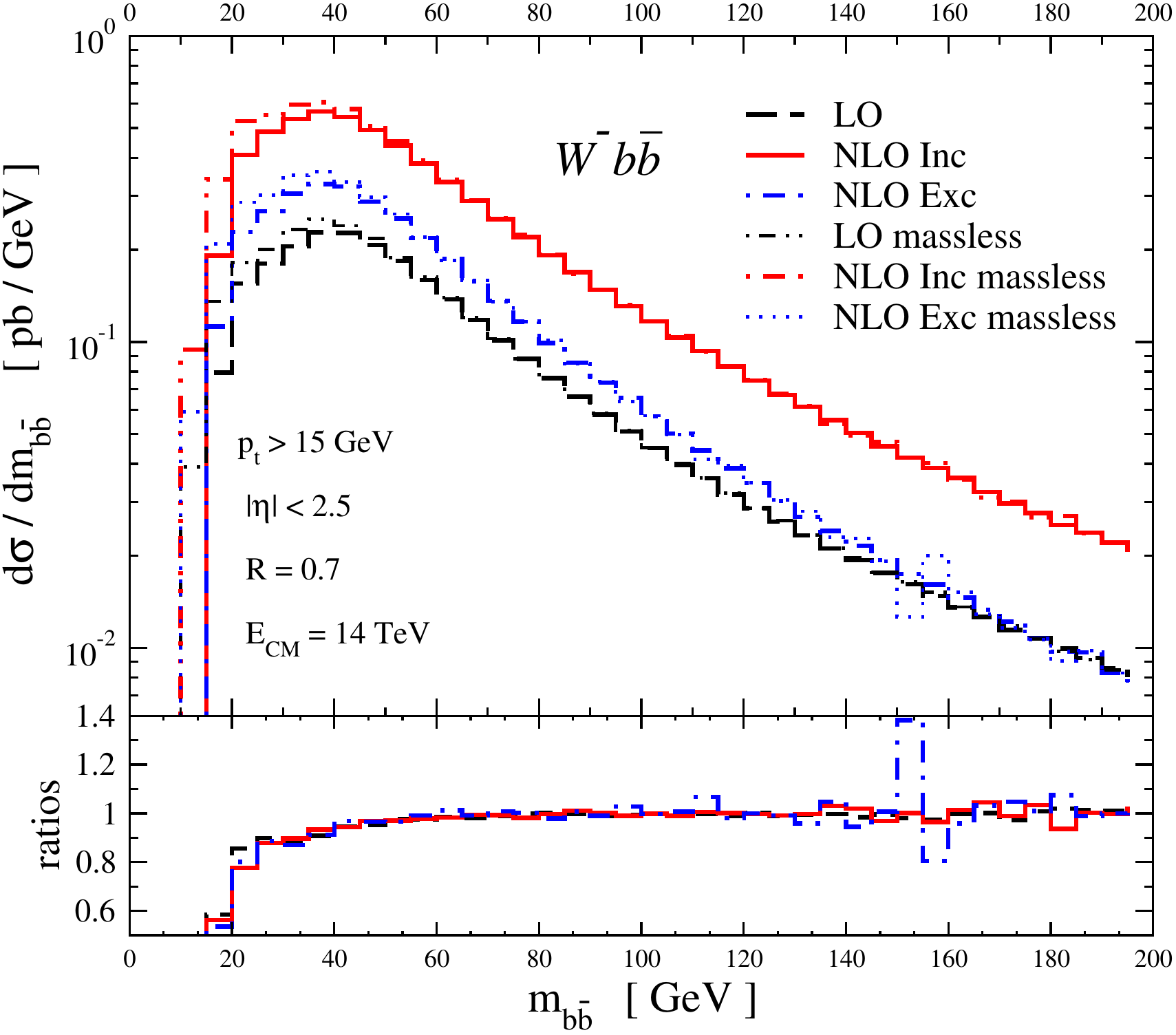}
  \hfill
  \includegraphics[width=0.47\textwidth,height=0.43\textwidth]{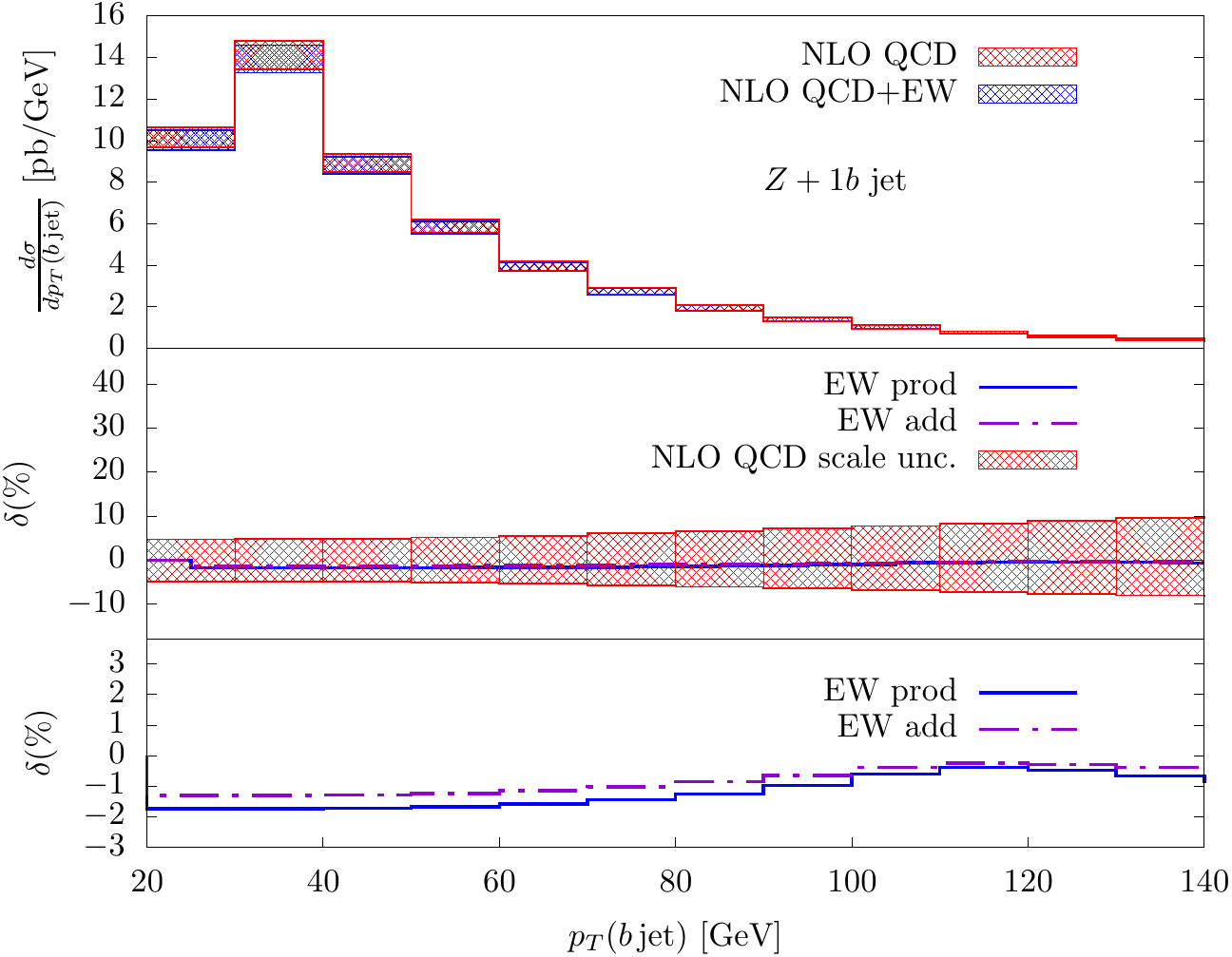}
  \caption{\label{fig:VHF:theroy:nlo}
    Di-$b$-jet invariant mass in the pair production of a charged lepton and a neutrino 
                   in association with at least two $b$-jets 
                   calculated at LO and NLO QCD in both the $\nf=4$ and $\nf=5$ massless
                   quark flavor schemes (left), 
                   figure taken from \cite{Cordero:2009kv}.  
    Leading $b$-jet transverse momentum in the pair 
                    production of two charged leptons in association with at least one $b$-jet 
                    calculated at NLO QCD with and without NLO EW corrections in the $\nf=5$ 
                    massless quark flavor scheme (right), 
                    figure taken from \cite{Figueroa:2018chn}.
  }
\end{figure*}

Heavy-quark processes in general can be calculated in at least two different 
approaches. 
For definiteness, when $b$-quark associated production is considered 
either only the $dusc$-quarks are considered 
massless and the full mass dependence of the $b$-quark is retained
(the four massless flavor scheme, $n_f=4$, also referred to as 4F) 
or all five light quark flavors are considered massless (the five 
massless flavor scheme, $n_f=5$, also referred to as 5F). 
While the former correctly describes all effects that are due to the 
$b$-mass, the latter allows for the $b$-quark to be extracted directly 
from the proton, resumming its contribution to the proton's parton 
density. 
By consistency, the $b$-quark is thus also only included in the running 
of the strong coupling in the $n_f=5$ scheme. 
The choice of scheme, thus, has a non-negligible effect on the value 
of the strong coupling constant on scales beyond the $b$-quark mass. 
In particular, the value of $\alpha_s(m_Z)$ differs in both schemes.\footnote{
  A possible remedy is explored in Ref. \cite{Bertone:2015gba} where 
  \emph{doped} PDFs are introduced, running $\alpha_s$ in $n_f=5$ and 
  the evolve the PDFs in $n_f=4$, which however has not seen a wide-spread 
  use so far.
} Thus, ideally, one would like to have a calculation with both the finite-mass 
and the resummation effects accounted for. 
In consequence to these considerations, methods have been formulated 
that combine both ansatzes, like the FONLL method \cite{Cacciari:1998it,Forte:2010ta}.
The diagrams contributing in the respective cases are shown in Figures 
\ref{fig:VHF_W_diagrams} and \ref{fig:VHF_Z_diagrams} for $W$ and $Z$ 
associated heavy flavor production, respectively.

The generalisation for the different mass-dependence treatments of charm 
quarks is mostly straight-forward \cite{Ball:2015tna}. 
Because the charm quark's isospin partner, the strange quark, is  
not mass suppressed, in contrast to the bottom's isospin partner, 
the top quark, and the respective inter-generational mixing matrix elements 
$V_{cd}$ are sizeable in comparison to $V_{ub}$ and $V_{cb}$, 
additional topologies contribute in $W$ associated charm production 
that are strongly suppressed in $W$ associated bottom production, 
cf.\ Figure \ref{fig:VHF_W_diagrams}(c).

One particular aspect of all parton level calculations that has 
to be kept in mind is that no 
flavor-jet related observable can be defined in complete analogy 
to the experimental definitions. 
This roots in the fact that the identification of heavy flavors 
in an experimental setting relies on the properties of heavy-flavor 
hadrons, in particular their finite lifetime allowing for 
a measurable spatial separation of production and decay vertices. 
A parton level heavy-flavor tag, on the other hand, can only 
involve the partonic jet constituents. 
The requirement for infrared safety then generally necessitates 
a signed counting of heavy flavor quanta to guarantee that a 
collinear $g\to q\bar{q}$ splitting does not alter the jet flavor 
tag.
One useful example here is the flavor-\kT 
algorithm \cite{Banfi:2006hf}. 
Its anti-\kT relative, however, is not infrared safe starting at NNLO as through 
its cone-like structure, soft wide-angle $g\to q\bar{q}$ splittings 
carry the possibility of losing one heavy quark thereby again altering 
the jet flavor tag \cite{Gauld:2020deh}.
This is 
problematic in the limit that the splitting gluon itself comes from a 
soft $q\to qg$ splitting, rendering the cancellation of its infrared 
divergence incomplete \cite{Banfi:2006hf}.

Four and five flavor calculations thus exist for $W/Z/\gamma+b$, 
$W/Z/\gamma+b\bar{b}$ and $W/Z/\gamma+b\bar{b}+\text{jet}$ production 
at NLO QCD \cite{Campbell:2005zv,Campbell:2006cu,Cordero:2009kv,
  Stavreva:2009vi,Hartanto:2013aha} 
and NLO EW \cite{Figueroa:2018chn} accuracy. 
Fig.\ \ref{fig:VHF:theroy:nlo} shows selected calculations for the 
$W+b\bar{b}$ and $Z+b$ process.
A $Z/\gamma+b$ calculation at NNLO QCD accuracy in the $n_f=5$ scheme 
became available recently \cite{Gauld:2020deh}, accounting for $b$-quark 
mass effects at NLO QCD accuracy with the aforementioned FONLL method 
and is expected to impact heavy flavour PDF extractions in particular.

\subsubsection{Monte Carlo event generators}
\label{sec:VHF:theory:mc}
\begin{figure*}
  \centering
  \includegraphics[width=0.47\textwidth]{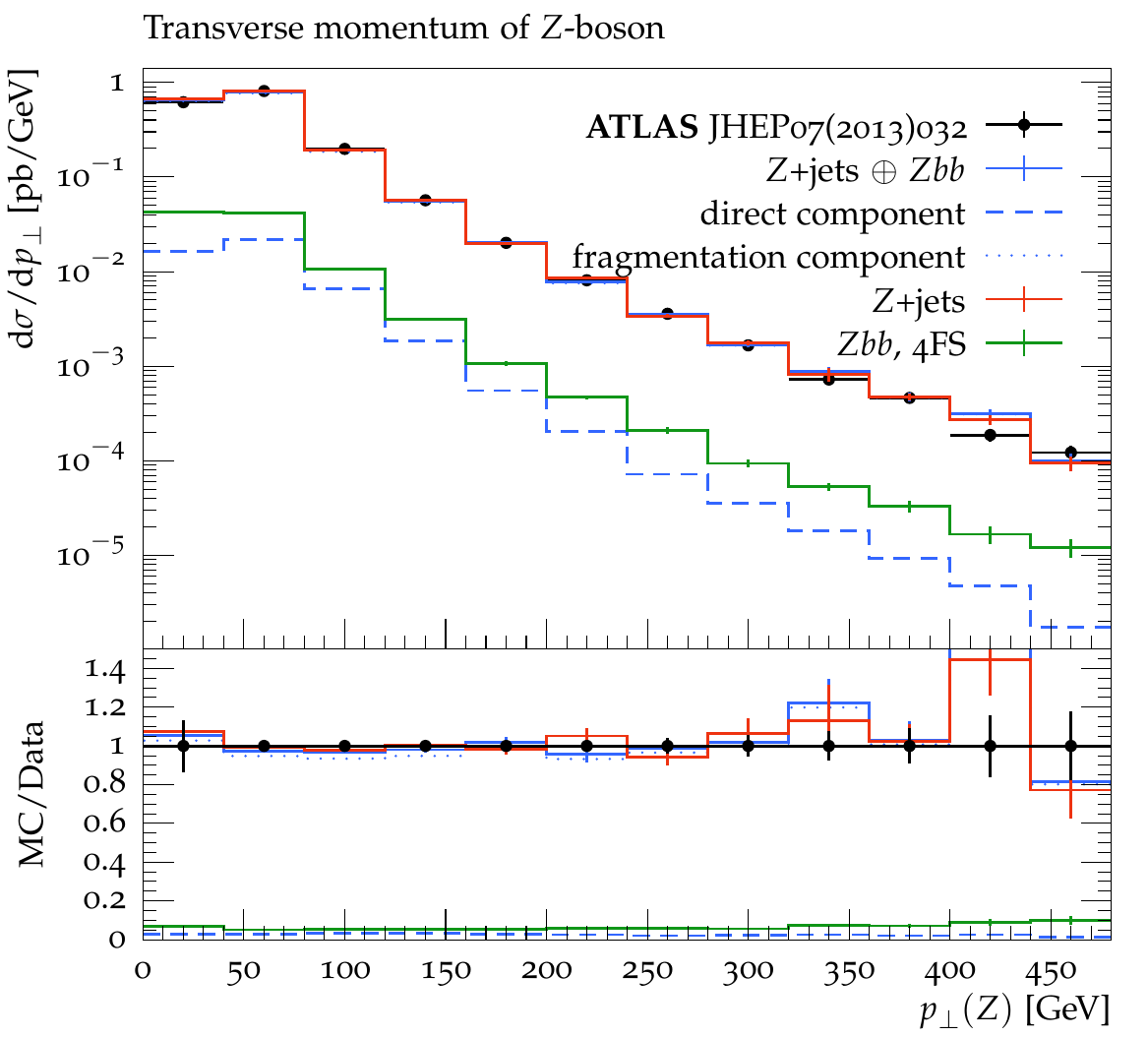}
  \hfill
  \includegraphics[width=0.47\textwidth]{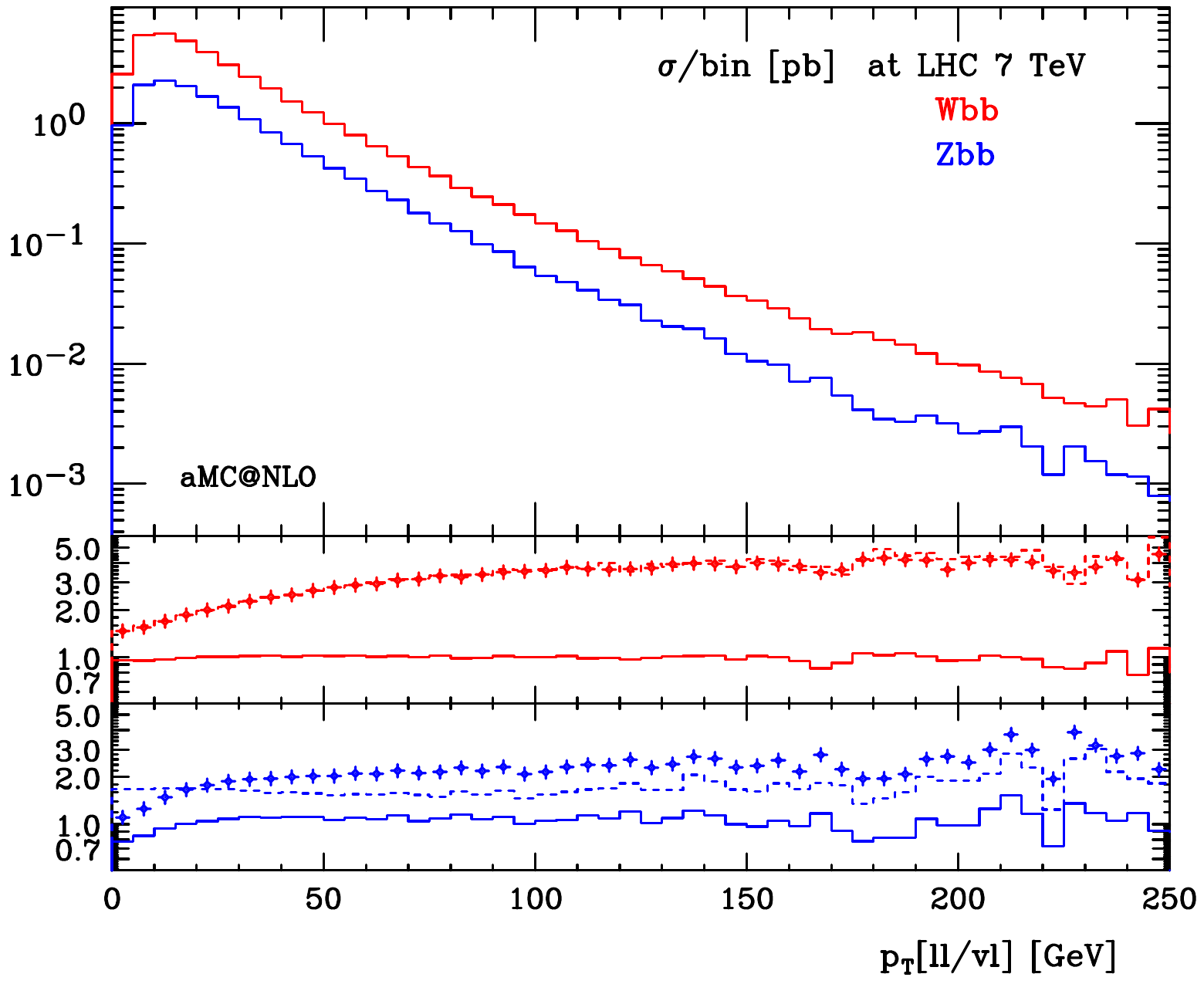}
  \caption{\label{fig:VHF:theory:mc} Reconstructed $Z$ boson transverse momentum in the pair production of two charged leptons in association with at least one $b$ jet calculated using the fusing method to combine $n_f=4$ and $n_f=5$ schemes in the \MEPSatNLO method in \Sherpa (left), figure taken from \cite{Hoche:2019ncc}. Reconstructed vector boson transverse momentum in the production of a pair of charged leptons, and a charged lepton and neutrino in association with a pair of $b$ jets calculated at NLO QCD matched to the parton shower in \aMCatNLO (right). The insets show the ratios of the \aMCatNLO result over the corresponding NLO (solid), \aMCatLO (dashed) and LO (crosses) results.  Figure taken from~\cite{Frederix:2011qg}.}
\end{figure*}
Along with the automation of matching NLO QCD calculations to parton 
showers, the availability of precision Monte Carlo event generation 
grew for this process class. 
Thus, the existing NLO QCD matched results produced by the 
\MCatNLO \cite{Frederix:2011qg} or \Powheg \cite{Oleari:2011ey} 
generators represent the state-of-the art for any fixed flavor number scheme. 
Recently, multi-jet merged predictions at NLO accuracy became available 
in the \MEPSatNLO method combining the $n_f=4$ and $n_f=5$ scheme 
\cite{Hoche:2019ncc}.
Typically, in the multi-jet merging approach problems arise with double counting of 
contributions already present in the general \Vjets multi-jet merged 
calculations in the massless limit. 
Therefore, various strategies have been devised to address this issue. 
Besides a more phenomenological and not theoretically rigorous 
approach, known as Heavy Flavor Overlap Removal \cite{Mangano:2001xp}, 
applied so far 
to LO-accurate simulations only, more rigorous approaches use schemes 
equivalent to the FONLL approach \cite{Hoche:2019ncc}. 
Figure \ref{fig:VHF:theory:mc} shows the result of both approaches.

\subsection{Experimental results}
\label{sec:VHF:exp}

Processes involving vector bosons in association with bottom or 
charm quarks
provide stringent tests of QCD predictions and are the largest backgrounds in studies of the Higgs boson decaying to two $b$ quarks, in measurements of the properties of the productions of single or pairs of top quarks, and in numerous searches for physics beyond the SM.

\subsubsection{Heavy-flavor identification in jets}
\label{sec:VHF:exp:tag}
\begin{figure*}[htb] 
\centering
\includegraphics[width=0.45\textwidth,height=0.40\textwidth]{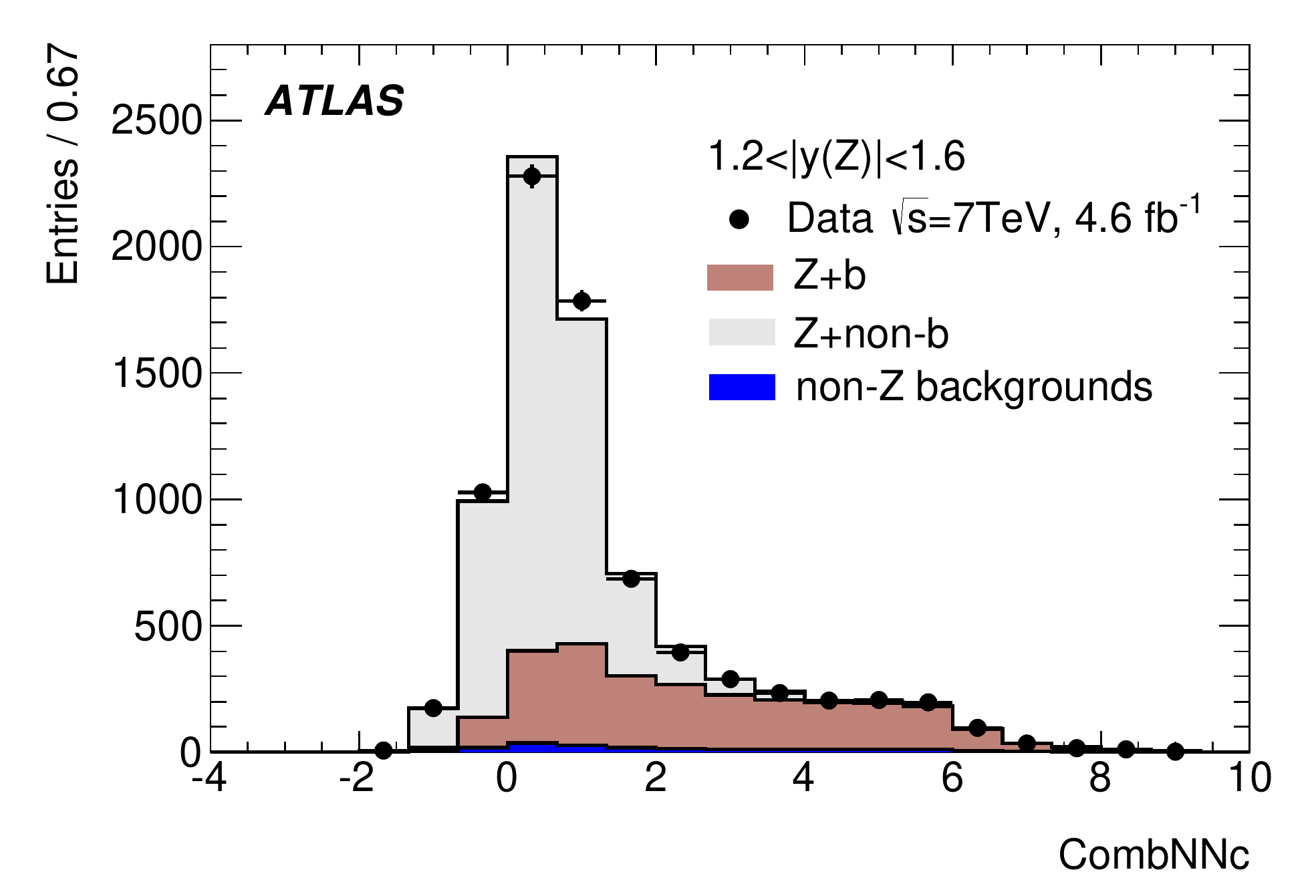}
\includegraphics[width=0.4\textwidth]{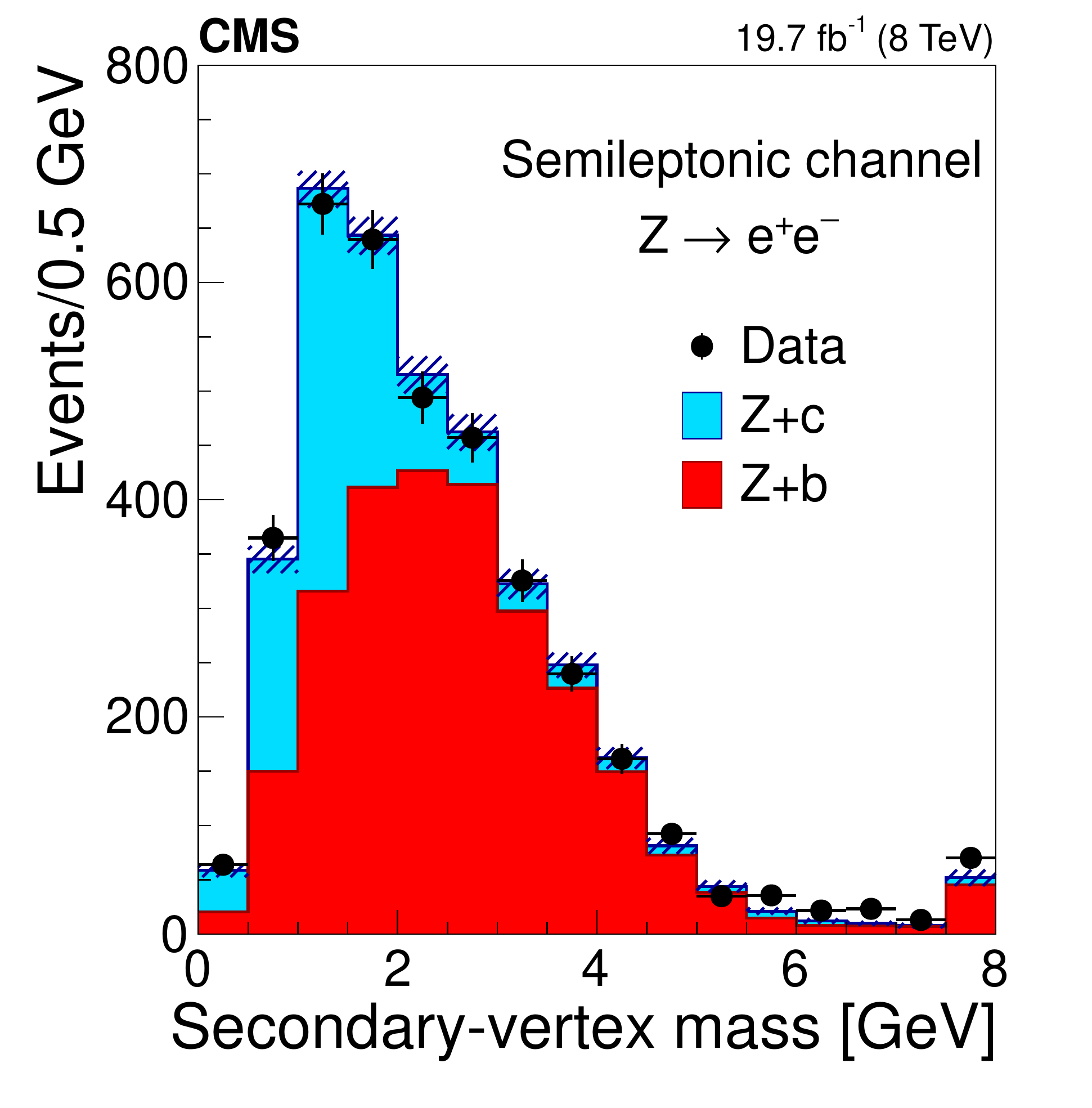}
\caption{Combined neural network output distribution providing separation between jet flavors in \Zjet events at 7 TeV(left), figure taken from~\cite{Aad:2014dvb}. Secondary-vertex mass distributions for jets associated to a \Z boson decaying to electrons, after background subtraction, at 8 TeV (right), figure taken from~\cite{Sirunyan:2017pob}. 
\label{fig:VHF:tags}}
\end{figure*}

The identification of jets originating from $b$ or $c$ quarks (heavy-flavor jets) is of primary importance for many measurements and searches with proton collision data.  Detectors with precise charged-particle tracking as well as electron and muon identification are well suited to identify heavy-flavor jets,
exploiting mainly the presence of displaced tracks from which a secondary vertex (SV) may be reconstructed. 
Figure~\ref{fig:VHF:tags} shows examples of distributions of combined multivariate algorithms and reconstructed secondary vertex mass used to identify and separate heavy-flavor jets in association with vector bosons.
The heavy-flavor identifications algorithms and their performances have been described in detail for \LHC 7 and 8 TeV proton collision data~\cite{Aad:2015ydr,Chatrchyan:2012jua}, and for 13 TeV collision data~\cite{Aaboud:2018xwy,Sirunyan:2017ezt}.

In simulated events different procedures can be applied to assign a flavor to a jet. A simple parton-level angular association was mostly used by \Tevatron and \LHC Run 1 data. A particle-level definition commonly employed for \LHC Run 2 data makes use of a
ghost association~\cite{Cacciari:2007fd} of heavy-flavor hadrons to generator-level particle jets. In all definitions, precedence is first given to the $b$-quark flavor, and then to c-quark flavor.

Both parton- and particle-based heavy-flavor definitions have been used to define the $V$ plus heavy-flavor jet measurements described in the following.

\subsubsection{\texorpdfstring{$V+b$}{V+b}-quark productions}
\label{sec:VHF:exp:Vb}

%

Studies of the production of prompt photons in association with $b$ quarks have been performed
with \Tevatron data and compared to various QCD  predictions. 
Measurements by \DO~\cite{Abazov:2009de,Abazov:2012ea,D0:2012gw}
were performed  differentially in the photon $\pT$,
and in the photon and jet rapidity for both 
$\gamma$+$b$ and $\gamma$+$c$ productions,
separating the jet flavors with a combined displaced track jet probability.
Similar results have been produced by \CDF~\cite{Aaltonen:2013ama,Aaltonen:2009wc}, alternatively making use of the invariant mass of reconstructed secondary vertices to separate jet flavors. 
All results showed a need for higher-order perturbative QCD corrections beyond NLO, in the larger $\pT>70$~GeV regions. 

\begin{figure*}[htb] 
\centering
\includegraphics[width=0.58\textwidth]{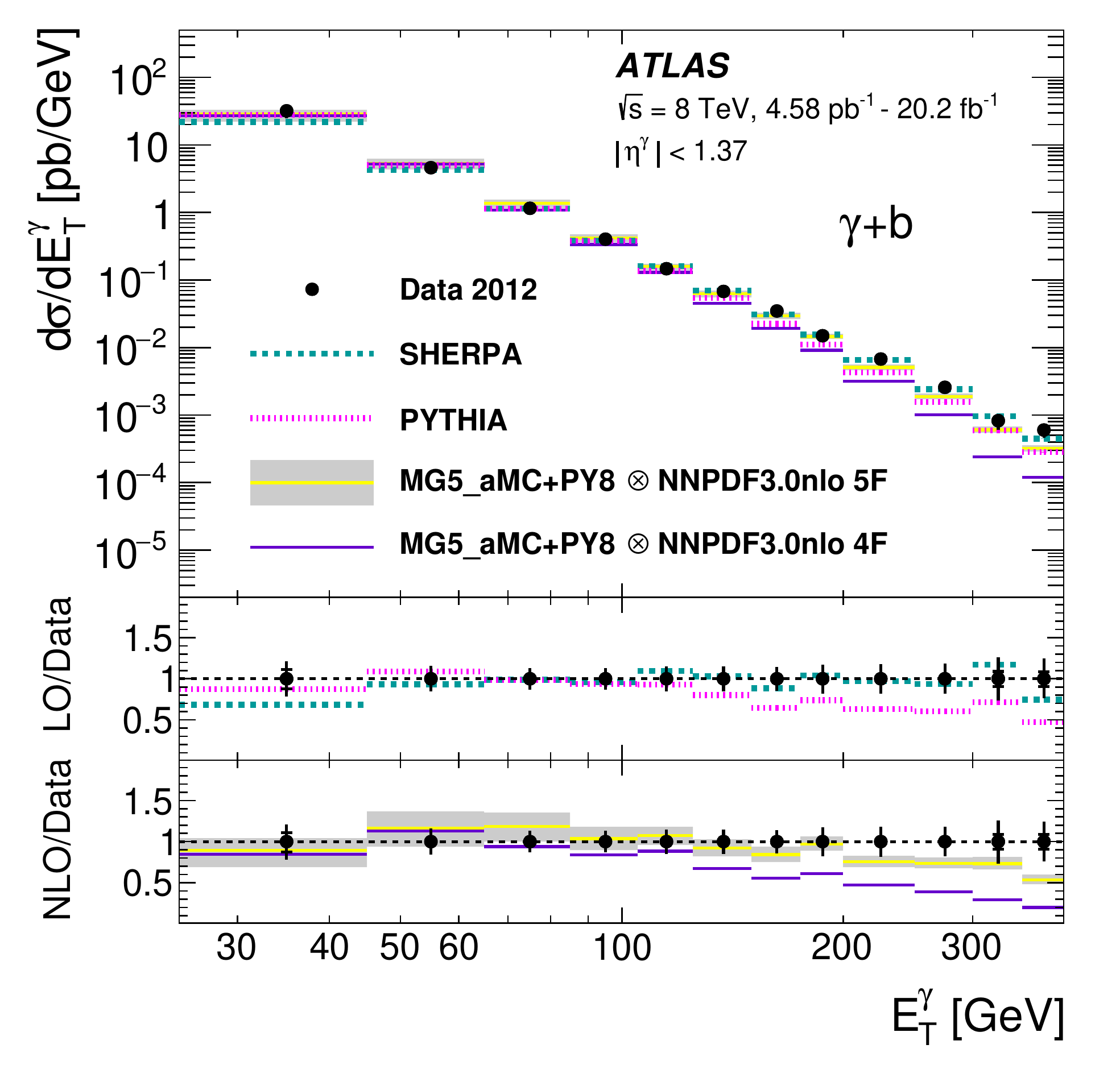}
\includegraphics[width=0.4\textwidth]{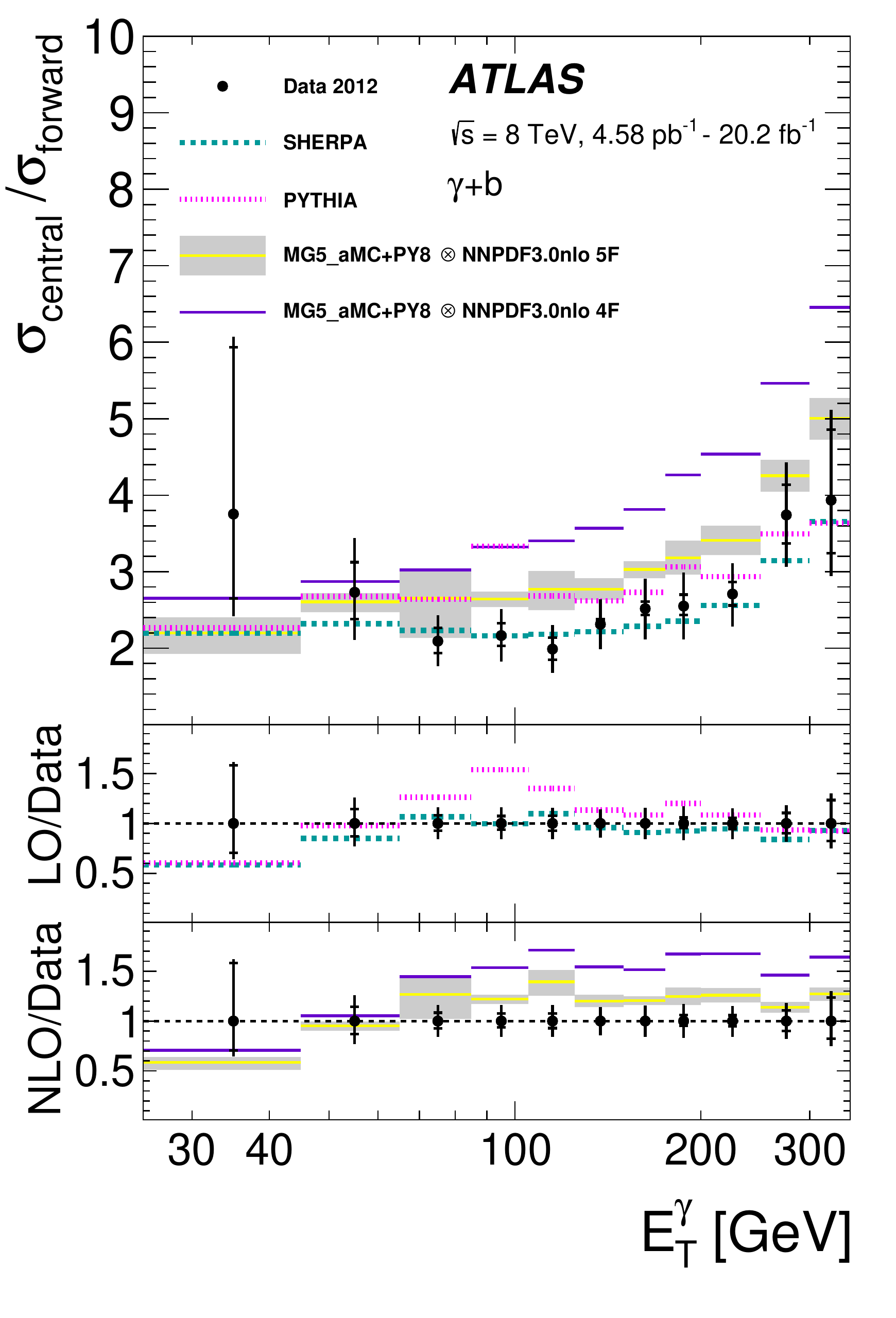}
\caption{
Measured $\gamma+b$  cross sections at 8 TeV. Differential cross sections as a function of the photon transverse energy in the central region ($|\eta^\gamma|<1.37$) (left).  Cross-section ratios of the central region, $|\eta^\gamma|<1.37$, to the forward region, $1.56<|\eta^\gamma|<2.37$, as a function of the photon transverse energy (right). Figures taken from~\cite{Aaboud:2017skj}.
\label{fig:VHF:gb}}
\end{figure*}

The \ATLAS collaboration has measured isolated-photon plus heavy-flavor jet production in 8 TeV proton collisions~\cite{Aaboud:2017skj}. 
Results are provided differentially in the transverse energy of the photon and in two photon pseudorapidity regions, and compared to LO
and NLO QCD calculation with 5F and 4F schemes, 
as shown in Fig~\ref{fig:VHF:gb}.
The NLO predictions underestimate the data in the kinematic region with 
$\ET^\gamma \geq 125$~GeV with the 4F scheme, and in the kinematic region with 
$\ET^\gamma \geq 200$~GeV with the 5F scheme. 
The 4F predictions for the cross-section ratios  overestimate the data for $\ET^\gamma \geq 65$~GeV.
The best description of the data is provided by \Sherpa predictions, which include up to three additional partons and are computed in the  5F scheme.

The first measurement of the associated production of a \Z boson with a $b$-jet was performed by \dzero~\cite{Abazov:2004zd} indicating a ratio to light jets around 2\%, in agreement with existing NLO QCD predictions. Similar results where derived also by \cdf~\cite{Abulencia:2006ce}, including a fiducial \Z $+ b$ cross section with a total uncertainty around 40\%.

A subsequent \cdf analysis with a larger data sample reported similar results for the fractions of associated $b$-jets, with improved precision, and differential distributions in jet \ET , jet $\eta$, \Z boson transverse momentum, number of jets, and number of $b$-jets~\cite{Aaltonen:2008mt}. Results were consistent with predictions from LO Monte Carlo generators and NLO QCD calculations  within uncertainties.
The invariant mass distribution of the tracks forming the secondary vertex was used to extract the $b$-jet fractions. 

More recent measurements by \dzero determined the $b$-jet ratios to light jets with a precision around 10\% using a peculiar technique that combines the properties of the tracks associated to the jet~\cite{Abazov:2010ix}. 
A more recent  \dzero publication reported the fractions of $b$-to-light jet associated production as a function of the \Z boson transverse momentum, jet transverse momentum, jet pseudorapidity, and the azimuthal angle between the \Z boson and the jet~\cite{Abazov:2013uza}. Existing predictions from  Monte Carlo event generators did not provide a consistent description of all the examined variables.

In the meantime first measurements of $\Z+ b$ productions with \LHC data were performed by \ATLAS~\cite{Aad:2011jn} with 7~TeV proton collision data, reporting both a fiducial cross section 
and the ratio to the inclusive \Z cross section in the same fiducial region, both with a precision around 30\%. 
Similar measurements were then performed by \CMS~\cite{Chatrchyan:2012vr} with a  larger data sample, 
allowing the precision to improve to better than 20\%. 
The measured cross sections and the kinematic distributions of the $b$-jet and charged leptons were found to be in reasonable agreement with existing predictions.  

\begin{figure*}[htb] 
\centering
\includegraphics[width=0.48\textwidth]{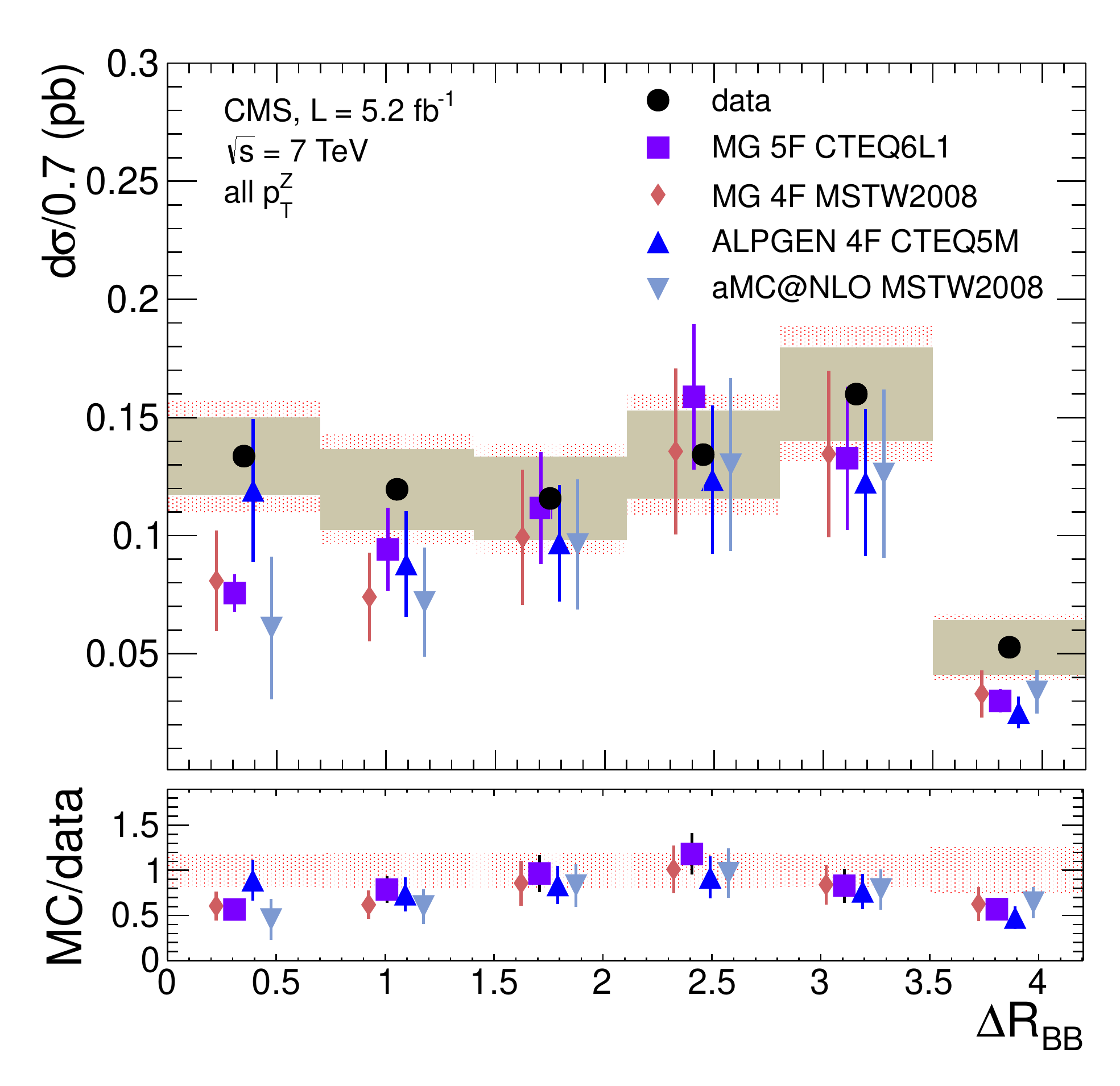}
\includegraphics[width=0.48\textwidth]{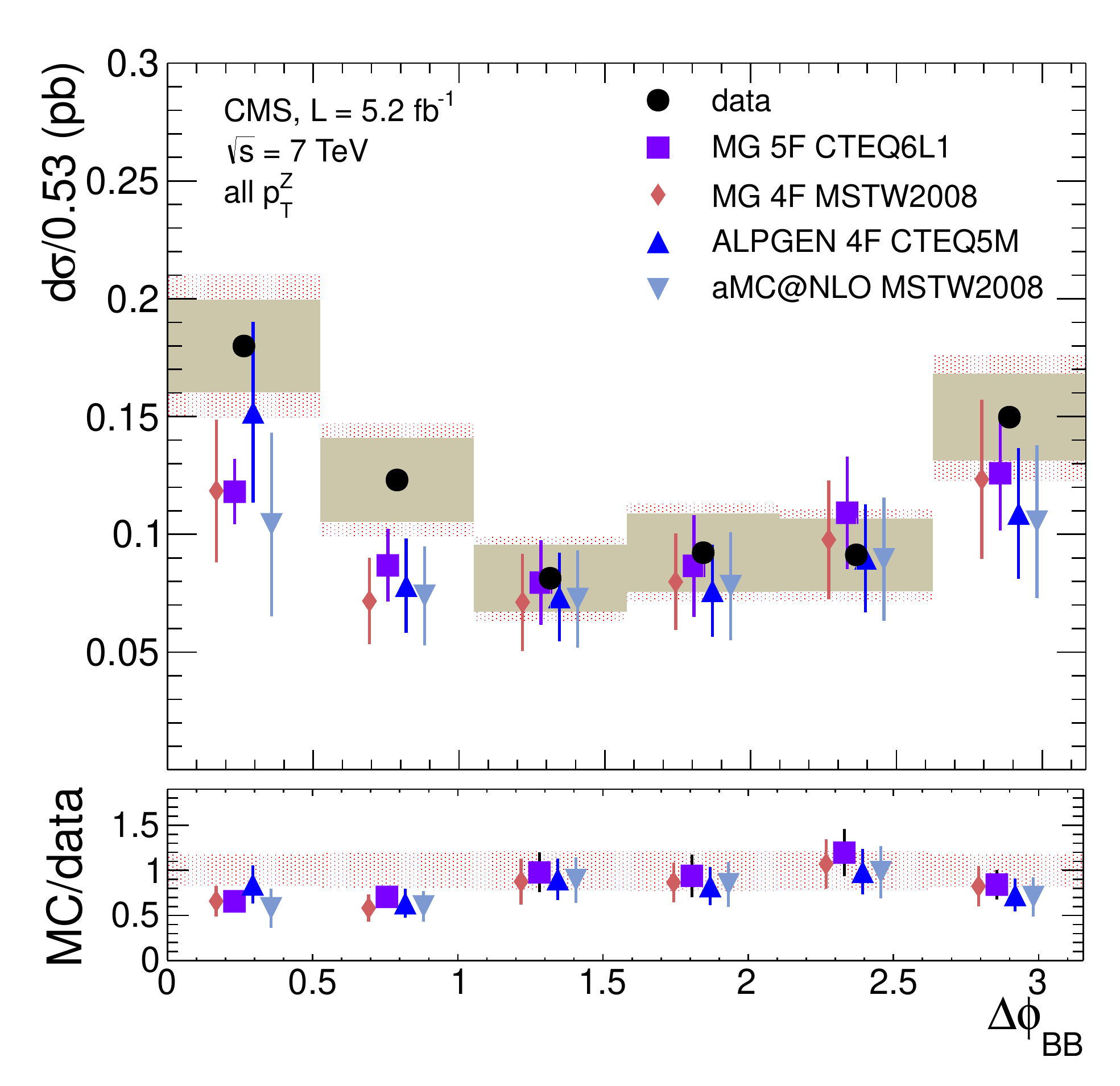}
\caption{
Measured $Z+BB$ differential cross sections at 7 TeV as a function of $\Delta R (BB)$ (left) and $\Delta\phi (BB) $ (right). Measurements are compared to the hadron-level predictions by \Madgraph in the four- and five-flavor schemes. Figures taken from~\cite{Chatrchyan:2013zja}.  
\label{fig:VHF:ZBB}}
\end{figure*}

Dedicated measurements of the production of two $b$-hadrons ($B$) together with a \Z boson were performed with 7~TeV data by \CMS with particular focus on the angular correlations between the $b$-hadrons and the \Z boson~\cite{Chatrchyan:2013zja}.
The $b$-hadrons are identified by means of displaced secondary vertices, without the use of reconstructed jets, permitting the study of $b$-hadron pair production with an angular separation smaller than the jet radius. 
The results shown in Fig.~\ref{fig:VHF:ZBB} indicate that the 5F description may not be well suited to describe the collinear production of $b$-hadrons.

Other measurements of total cross sections, separately for a \Z boson produced with exactly one $b$-jet and with at least two $b$-jets have been produced by \CMS~\cite{Chatrchyan:2014dha}. For those results data favor the predictions in the five-flavor scheme, where $b$-quarks are assumed massless, while predictions in the four-flavor scheme show a clear disagreement in the $\Z+1 ~b$-jet final state.

The \LHCb collaboration has produced complementary measurements
of  $\Z+ b$-jet cross section in the forward  pseudorapidity range 2.0$<\eta$ 4.5 and with jet \pT above 10 or 20 GeV~\cite{Aaij:2014gta}. The results yield a 25-30\% precision and are in reasonable agreement 
with both massless and massive bottom-quark calculations.

\begin{figure*}[htb] 
\centering
\includegraphics[width=0.44\textwidth,height=0.45\textwidth]{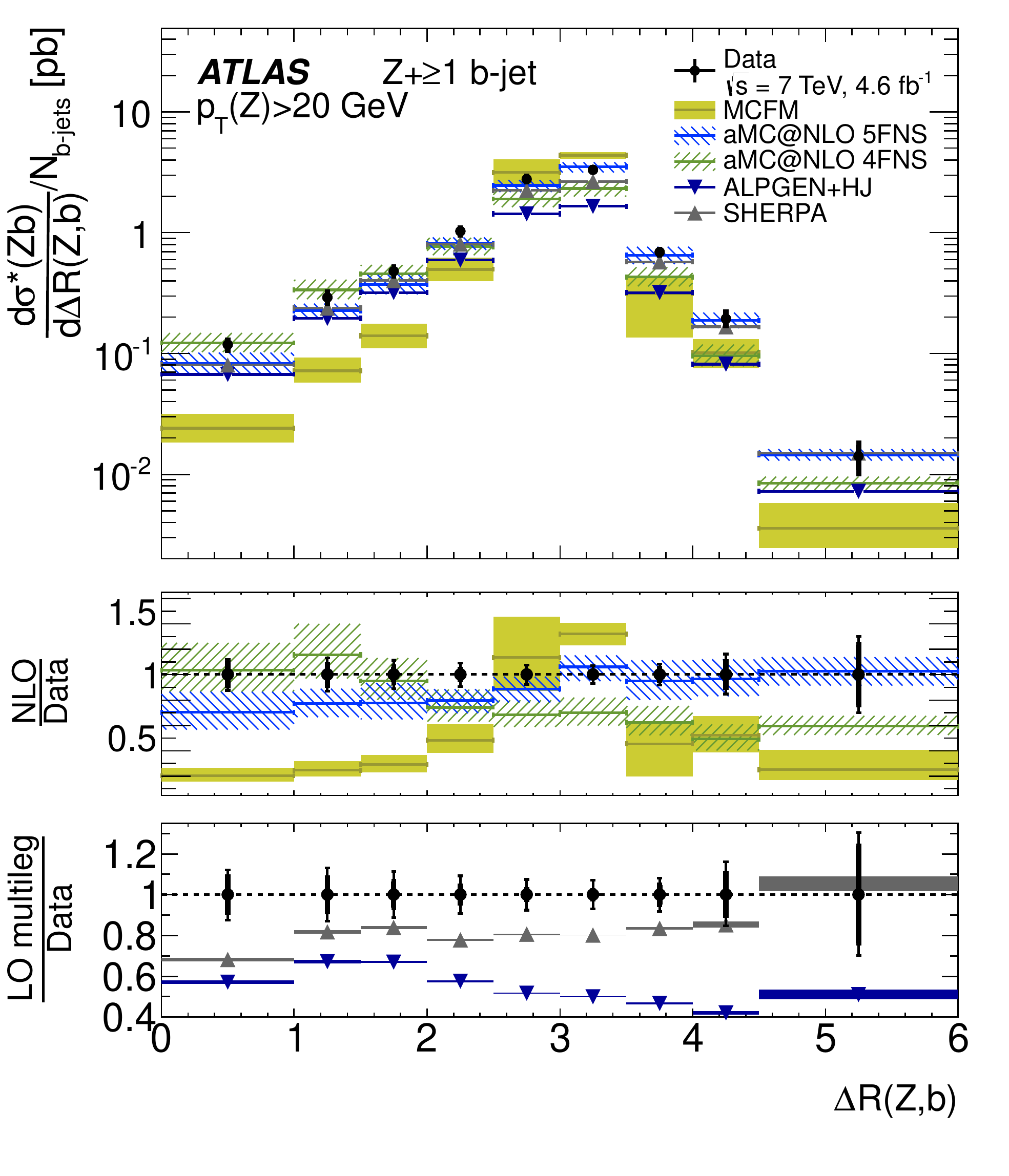}
\includegraphics[width=0.49\textwidth]{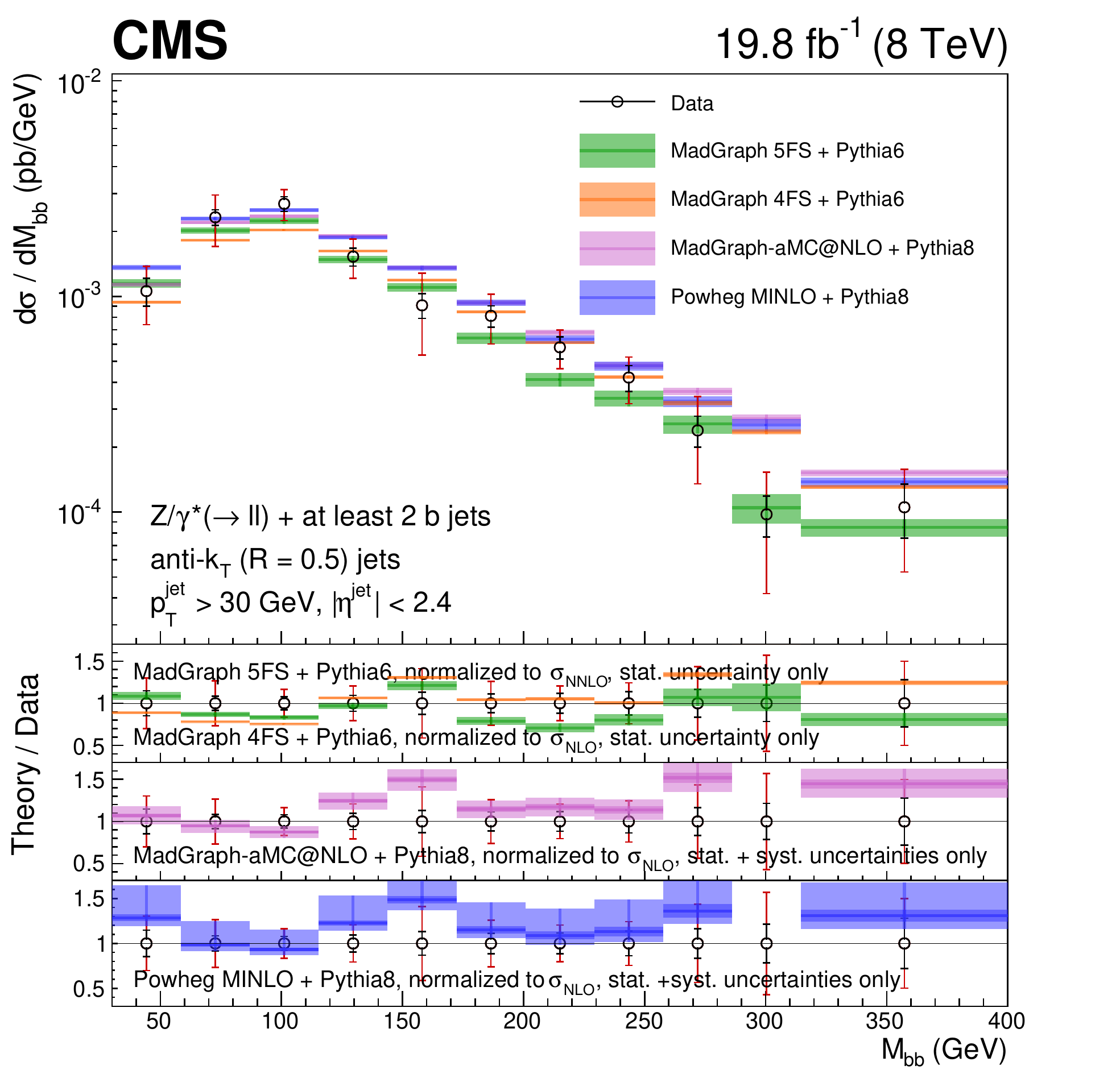}
\caption{
Measured $\Z + b$ differential cross sections at 7 TeV (left) and 8 TeV (right). The \Z plus $b$-jet cross section as a function of $\Delta R(Z,b)$ (left), figure taken from~\cite{Aad:2014dvb}. The \Z plus 2 $b$-jets cross section as a function of the invariant mass of the $b$-jet pair (right), figure taken from~\cite{Khachatryan:2016iob}.  
\label{fig:VHF:ZB}}
\end{figure*}

Further differential measurements of $\Z+b$-jet productions have been performed by \ATLAS with 7 TeV data~\cite{Aad:2014dvb}, and by \CMS with 8 TeV data~\cite{Khachatryan:2016iob}. 
The \ATLAS total cross section results are generally in good agreement with predictions from \MCFM. Predictions obtained using \MGaMC with a 4F scheme underestimate the $\Z+ 1 ~b$ cross sections, while predictions with the 5F scheme seem to underestimate the $\Z + 2 ~b$ yields. 
Interesting disagreements between predictions and data are also reported in the differential distributions, as for example the angular separation between the \Z boson and the $b$-jet shown in Fig.~\ref{fig:VHF:ZB}, where missing higher order QCD corrections in the predictions might explain the discrepancies. 

 The \CMS results for $\Z + b$ productions with 8 TeV collision data have also been compared with a variety of predictions, yielding fair agreement with the data results.  Predictions with 4F scheme seem to underestimate the total $\Z + 1 ~b$ cross sections and fail to describe simultaneously both the low- and high-\pT $b$-jet regions.
In the case of \Z boson in association with two $b$-jets, the data distributions are generally well reproduced by the predictions, as for example the dijet mass shown in Fig.~\ref{fig:VHF:ZB}.

\begin{figure*}[htb] 
\centering
\includegraphics[width=0.48\textwidth]{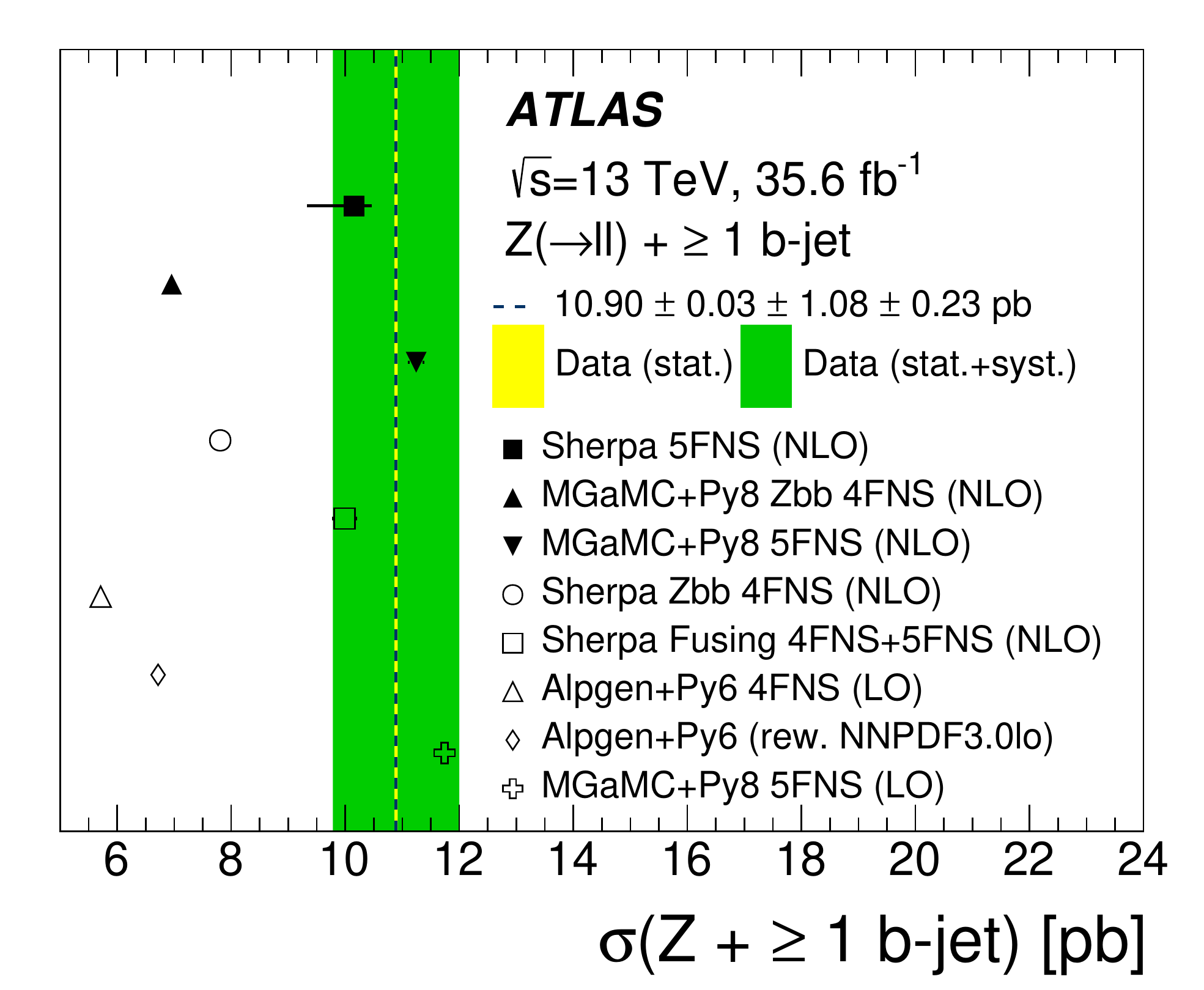}
\includegraphics[width=0.48\textwidth]{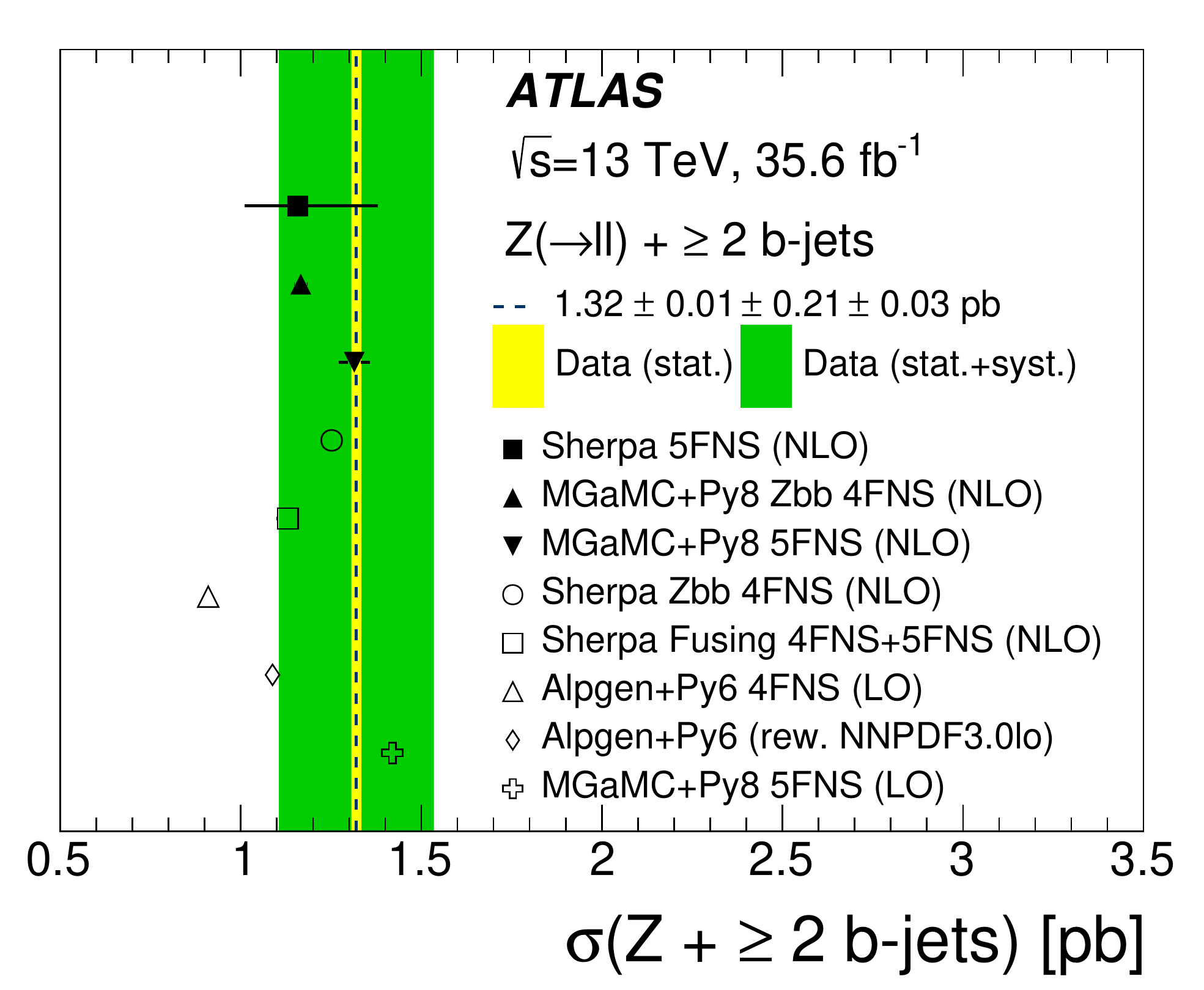}
\caption{Measured cross-sections for $\Z ~+ \ge 1 ~b$-jet (left) and $\Z ~+ \ge 2 ~b$-jets (right). The data are compared to different predictions in the 4F and 5F approximations. The yellow band corresponds to the statistical uncertainty of the data, and the green band to statistical and systematic uncertainties of the data, added in quadrature. The error bars on the \Sherpa 5F (NLO) predictions correspond to the statistical and theoretical uncertainties added in quadrature. Only statistical uncertainties are shown for the other predictions. Figures taken from~\cite{Aad:2020gfi}.
\label{fig:VHF:ZB2}}
\end{figure*}

Measurements of $\Z + b$ and $\Z + bb$ productions with 13 TeV collision data have been performed by \ATLAS~\cite{Aad:2020gfi}.
A summary of the total measured cross sections and comparisons with different predictions is shown in Fig.~\ref{fig:VHF:ZB2}. 
It can be seen that the 5F scheme predictions at NLO accuracy agree better with data than 4F scheme ones, and that the 4F predictions underestimate data in events with at least one $b$-jet.

Early measurements of $W + b$ rates by \cdf~\cite{Aaltonen:2009qi}
with \Tevatron data revealed some excess over the existing predictions with LO multijet-merged~\cite{Mangano:2001xp} and NLO accurate calculations~\cite{Campbell:2006cu,FebresCordero:2006nvf,Campbell:2008hh} that were not confirmed by subsequent similar \DO measurements~\cite{D0:2012qt}.

\begin{figure*}[htb] 
\centering
\includegraphics[width=0.55\textwidth]{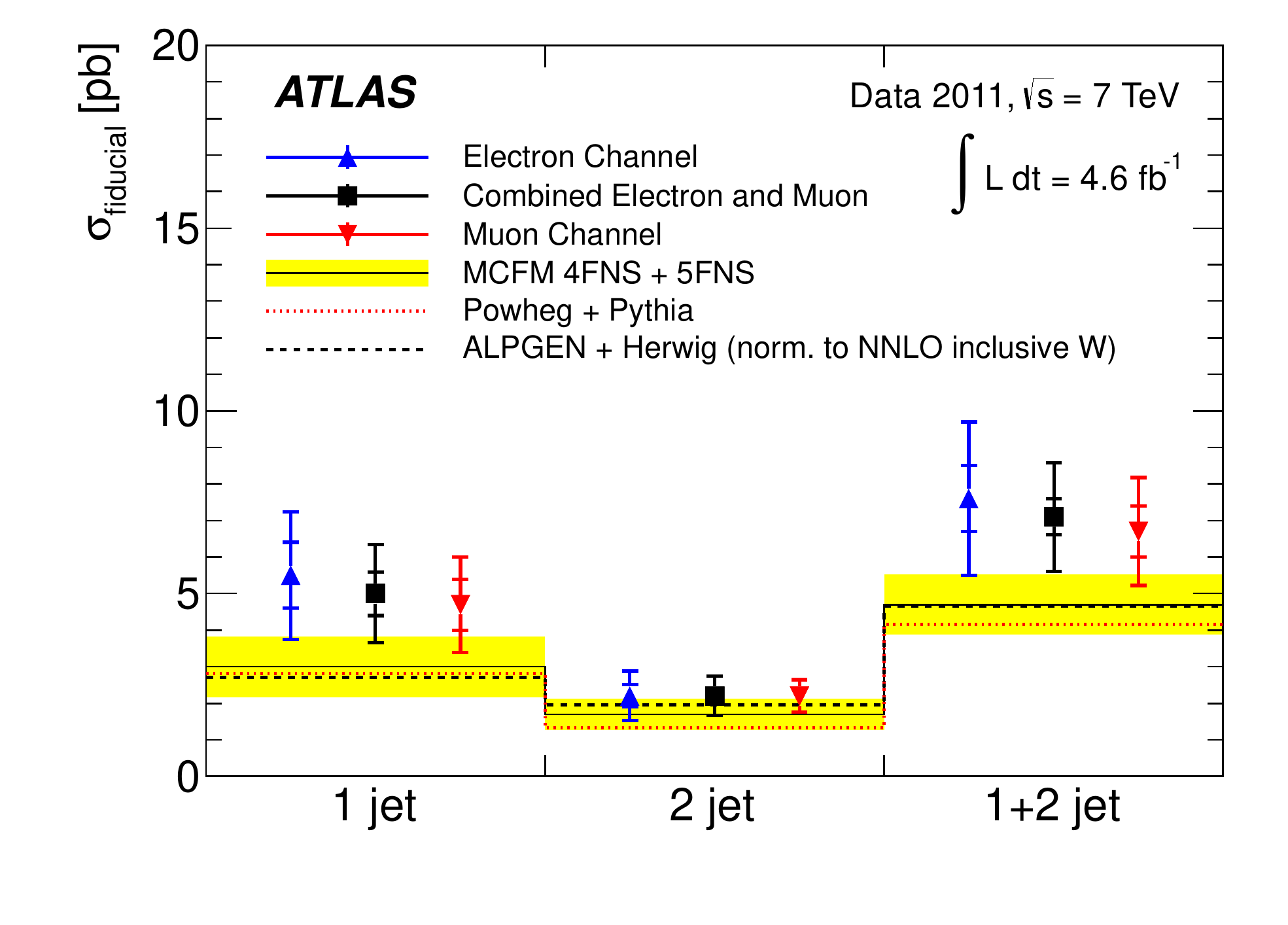}
\includegraphics[width=0.44\textwidth]{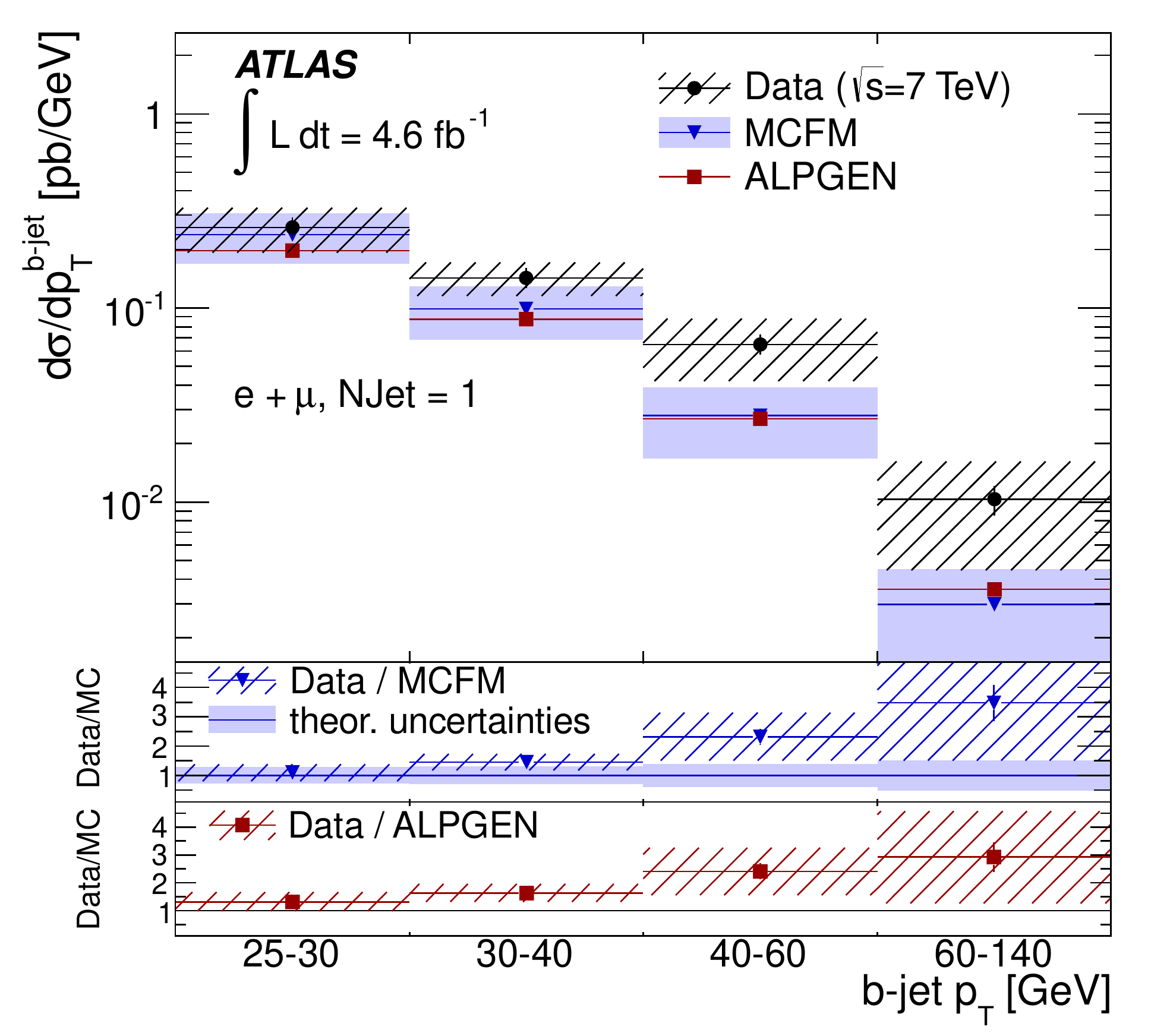}
\caption{
$\W+b(b)$ fiducial and differential cross sections at 7 TeV. Total cross sections in the 1, 2, and 1+2 jet exclusive bins with the statistical (inner error bar) and statistical plus systematic (outer error bar) uncertainty in the electron, muon, and combined electron plus muon channel (left). Differential $W+b$-jet cross-sections with total uncertainties as a function of the $b$-jet \pT in the 1-jet fiducial region,  compared to the \MCFM and \Alpgen predictions (right). Figures taken from~\cite{Aad:2013vka}.
\label{fig:VHF:Wb}}
\end{figure*}

Meanwhile, first measurements of $W+b$ productions with \LHC data were performed by \ATLAS~\cite{Aad:2011kp} with 7 TeV proton collision data. Events are required to have exactly one $b$-tagged jet  reducing significantly the top-quark background. Results are unfolded to a fiducial phase space at particle level, where $b$-jets are defined by the presence of a $b$-hadron associated to the jet, and are compared to QCD NLO predictions performed in the 5 flavor  scheme~\cite{Caola:2011pz} and to other leading order predictions. Results with a larger data set have been subsequently produced by \ATLAS~\cite{Aad:2013vka} allowing for an improved precision and differential measurements as a function of the $b$-jet \pT, shown in Fig.~\ref{fig:VHF:Wb}.

\begin{figure*}[htb] 
\centering
\includegraphics[width=0.47\textwidth]{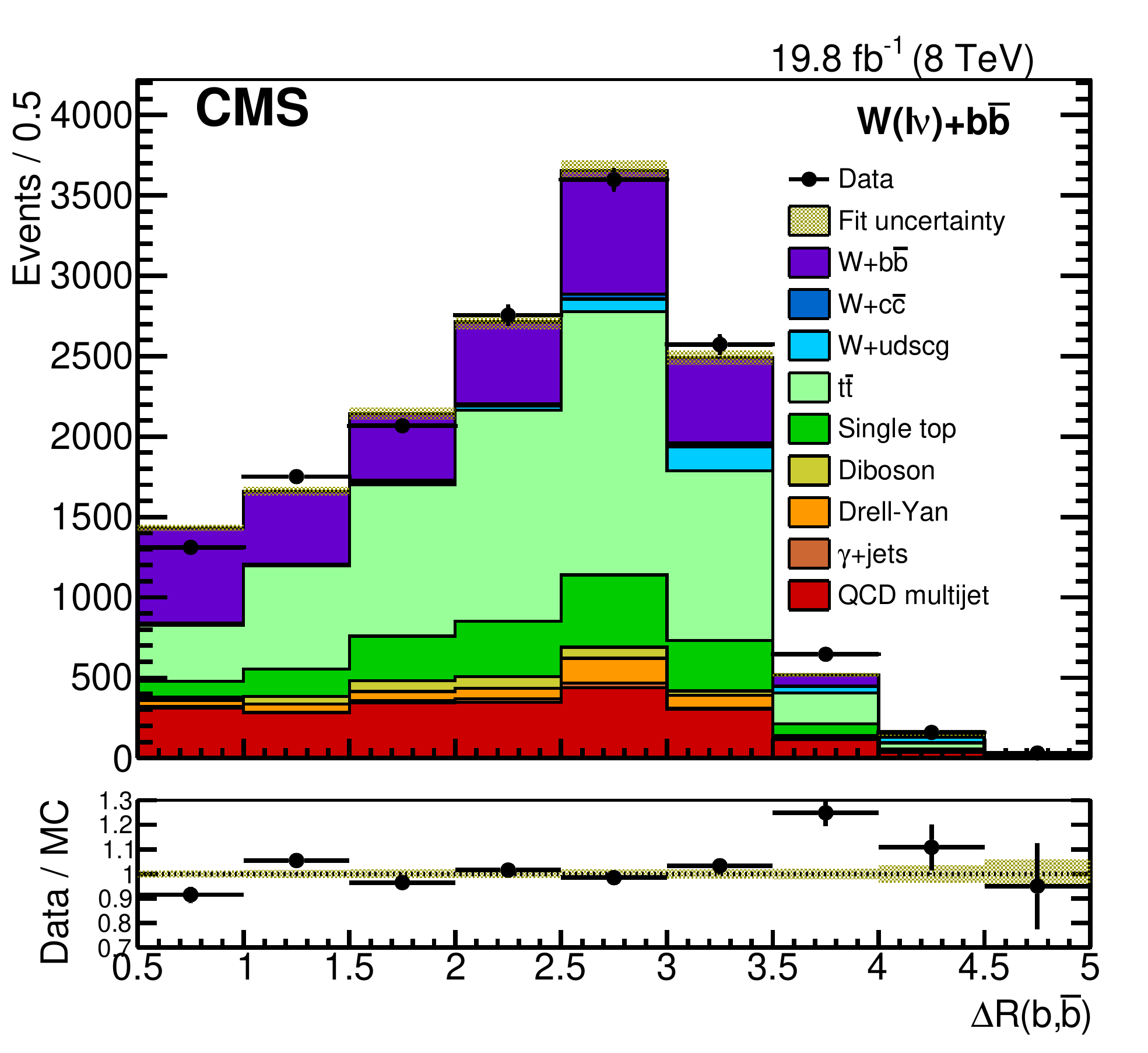}
\includegraphics[width=0.51\textwidth]{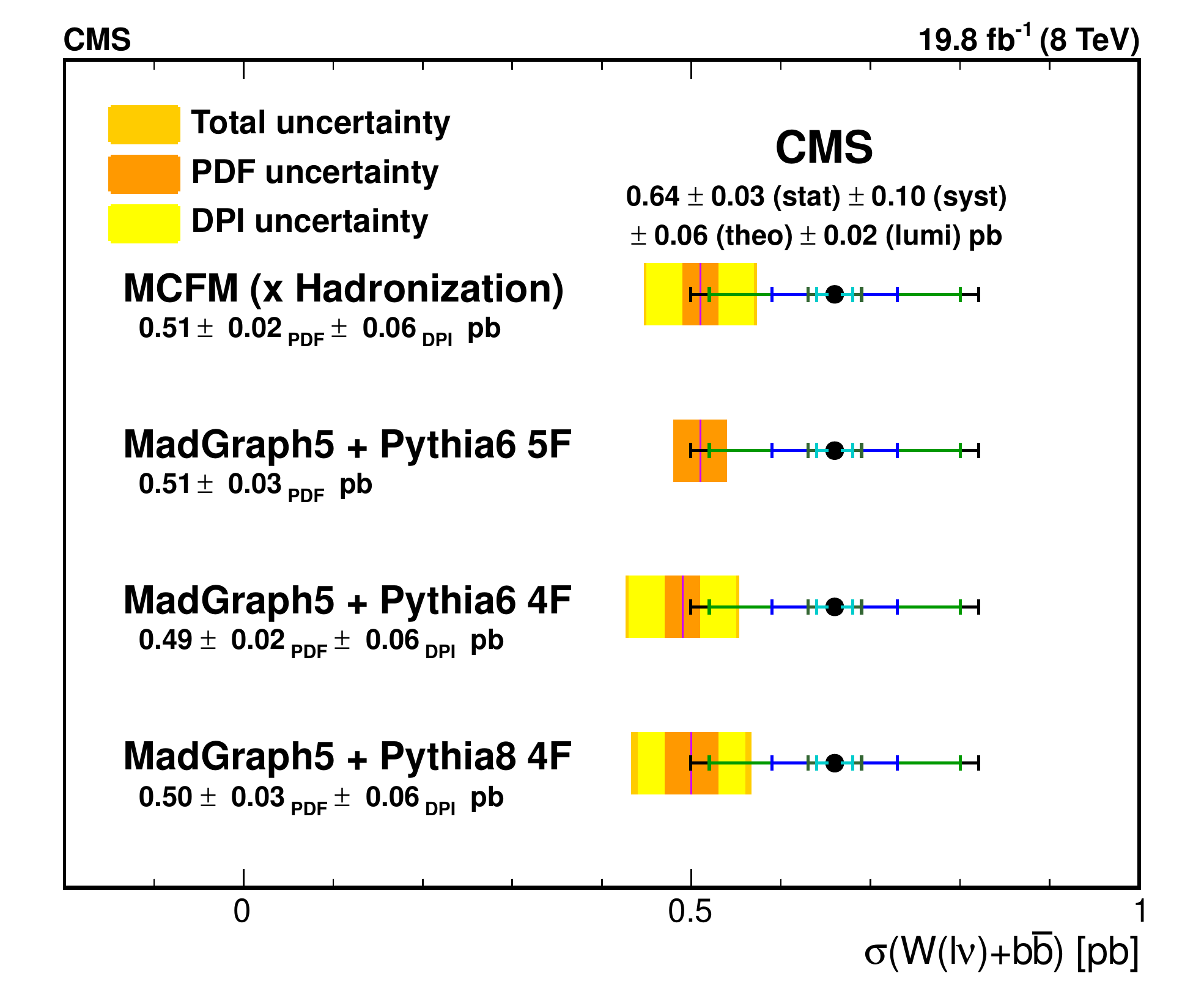}
\caption{ 
$W+bb$ production at 8 TeV. 
Post-fit $\Delta R(b,b)$ distribution (left).  Comparison between the measured cross section and various QCD predictions (right). Figures taken from~\cite{Chatrchyan:2013uza,Khachatryan:2016ipq}.
\label{fig:VHF:Wbb}}
\end{figure*}

Dedicated measurements of the production of a \W boson and two $b$-jets  have been performed by \CMS with both 7 TeV~\cite{Chatrchyan:2013uza} and 8 TeV~\cite{Khachatryan:2016ipq} proton collision data.
Events are required to have exactly two $b$-tagged jets, and top-quark background is reduced requiring no additional jets nor isolated electrons or muons.
Fig.~\ref{fig:VHF:Wbb} (left) shows the data and post-fit Monte~Carlo distributions for $\Delta R(b,b)$.
Results are given in terms of a fiducial $W+bb$ cross section and are in good agreement with several predictions from \MCFM, corrected for DPS and hadronization effects, and from \Madgraph+\Pythia, with different PDF flavor schemes, as shown in Fig.~\ref{fig:VHF:Wbb} (right).

It can be finally remarked that in the context of the measurements of the 
Higgs boson decays to bottom quarks in the associated $VH$ production mode, 
both the \ATLAS and \CMS collaborations determine normalization 
"scale factors" for the  \W/\Z $+ b(b)$ background sources, 
with respect to NLO QCD predictions,  and with a precision in the 10--20\% 
range~\cite{Aaboud:2018zhk,Sirunyan:2018kst}.
In this specific phase space of the Higgs to bottom-quark measurements, 
\CMS reported significantly large scale factors for $W+b(b)$ productions, 
up to a factor two with respect to the reference NLO QCD predictions, 
while \ATLAS normalizations are consistent with predictions.

\subsubsection{\texorpdfstring{$V+c$}{V+c}-quark productions}
\label{sec:VHF:exp:Vc}

First measurements of associated production of \Z bosons with charm-quark jets
were performed by \dzero~\cite{Abazov:2013hya} reporting an integrated fraction 
of $c$-jets of ~8\% with a 10\% relative uncertainty, and a ratio to $b$-jet production of about 4 with a 15\% relative uncertainty. The cross section ratios were also measured differentially as a function of jet and \Z boson transverse momenta showing significant deviations from existing perturbative QCD calculations and event generators predictions.

The first observation of $\Z+c$ production at \LHC was reported by  \LHCb 
in the forward region $ 2<y<4 $ with data from proton collisions at 7 TeV, 
making use of fully 
reconstructed $D^0$ and $D^\pm$ decays ~\cite{Aaij:2014hea}.

\begin{figure*}[htb] 
\centering
\includegraphics[width=0.54\textwidth,height=0.57\textwidth]{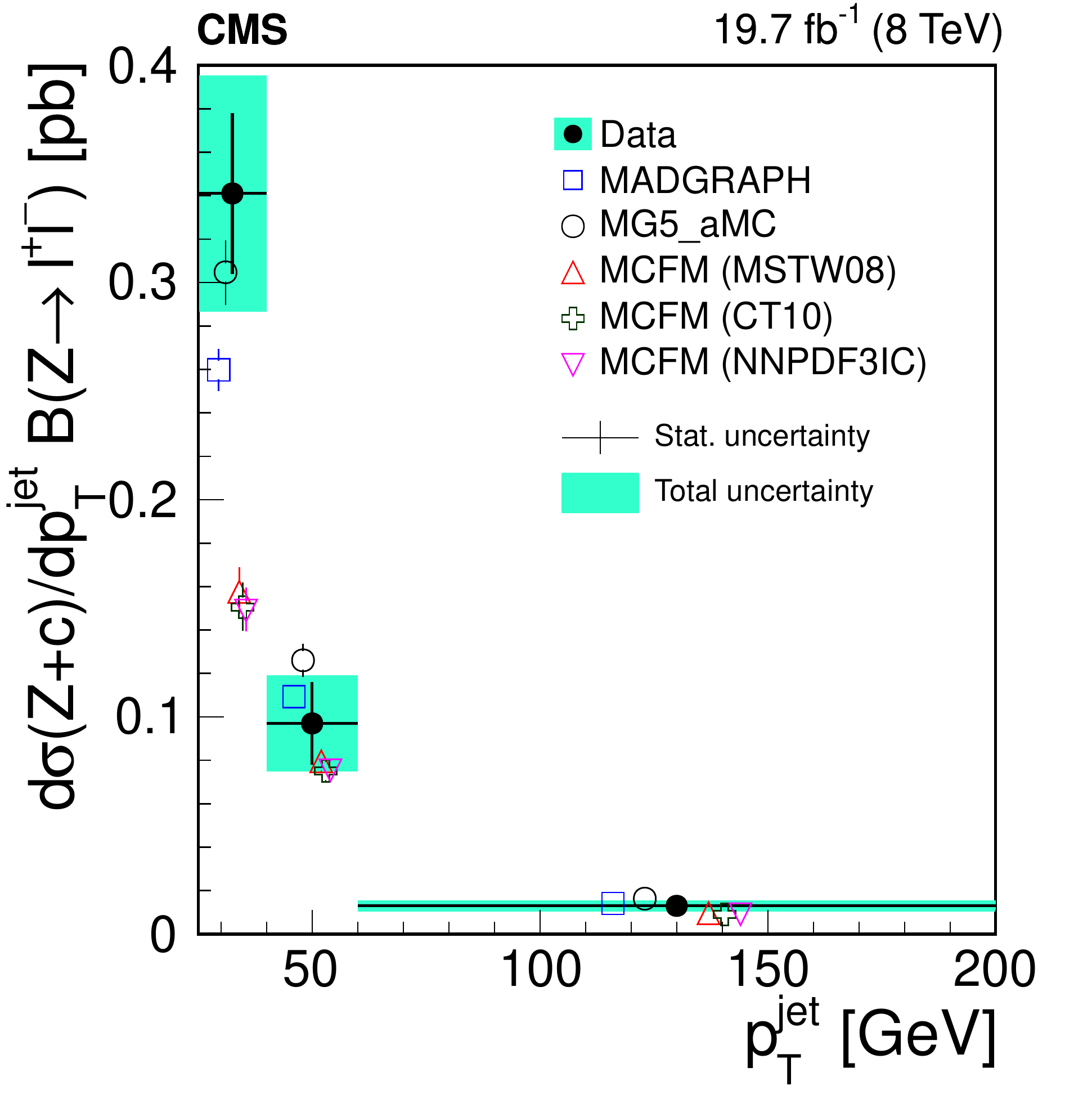}
\includegraphics[width=0.45\textwidth,height=0.60\textwidth]{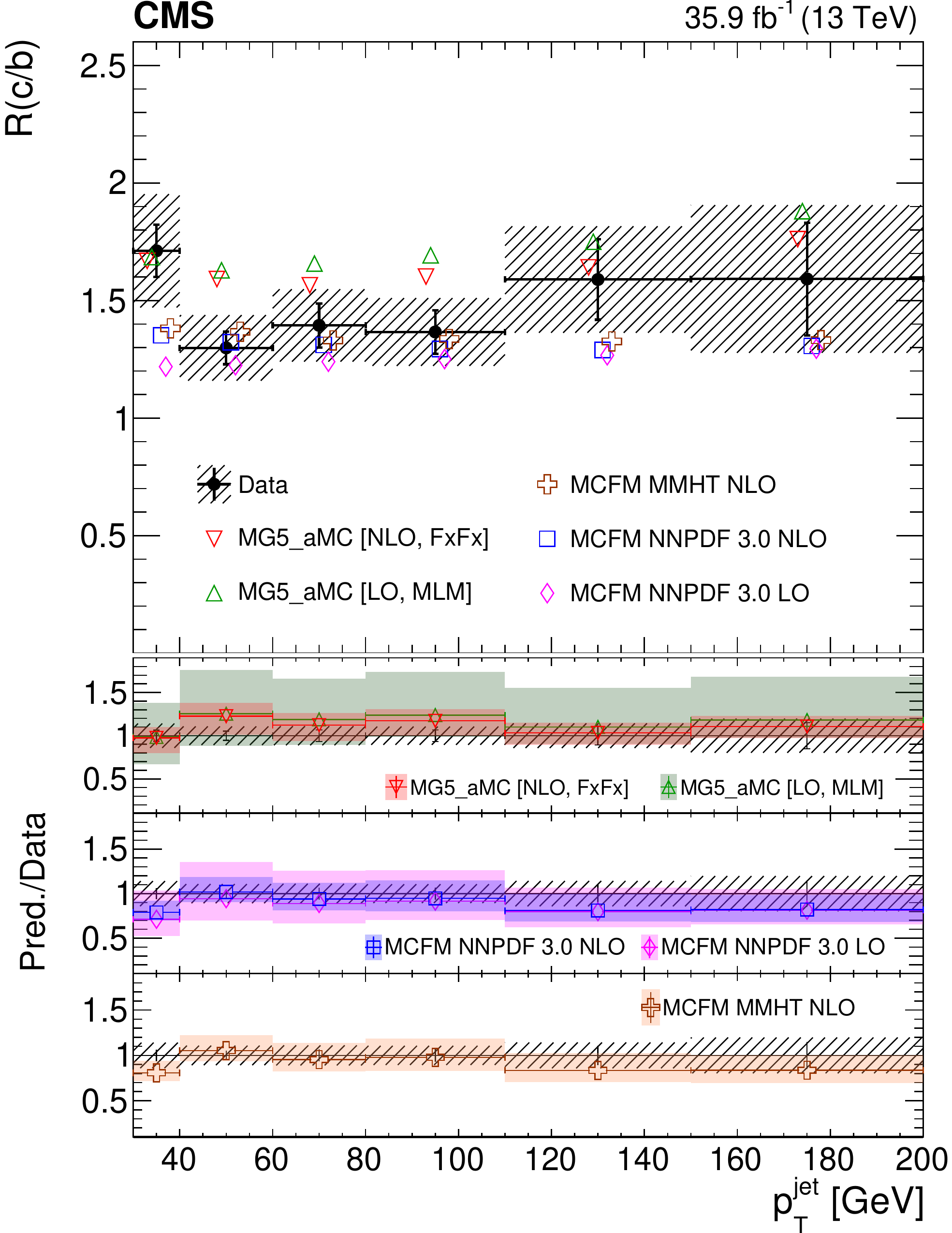}
\caption{ Differential $\Z+c$ cross section as a function of the transverse momentum of the $c$-jet at 8 TeV (left), figure taken from~\cite{Sirunyan:2017pob}.  ($\Z+c$)/($\Z+b$) cross section ratio as a function of the transverse momentum of the jet at 13 TeV (right), figure taken from~\cite{Sirunyan:2020lgh}.
\label{fig:VHF:Zc}}
\end{figure*}

The \CMS collaboration has performed a measurement of associated \Z + charm production in proton collisions at 8 TeV~\cite{Sirunyan:2017pob}.
The selection of event candidates relies on the identification of semileptonic decays of $c$ or $b$ hadrons with a muon in the final state and through the reconstruction of exclusive decay channels of D$^\pm$ and D$^{\ast\pm}$(2010) mesons. The total \Z + $c$ cross section is measured with a precision of 10\% while  the cross section ratio $\Z+c$/$\Z+b$ is determined to be $2.0\pm0.3$. Differential cross sections are measured as a function of the transverse momentum of the \Z boson and the heavy-flavor jet.
The measurements are in agreement with NLO QCD predictions, including parton shower development and non-perturbative effects. Results  in the highest transverse momentum regions are compatible with  predictions using PDF sets with no intrinsic charm component. 
Results with 13 TeV collision data have been released recently, 
making use of jet secondary vertex mass distributions to separate light, charm and bottom flavor components~\cite{Sirunyan:2020lgh,Sirunyan:2020cwc}.
Results are given in terms of $b$/light, $c$/light and $c/b$ production ratios,
both inclusively and differentially with respect to the \Z boson \pT and the jet \pT.  The experimental results are in reasonable agreement with current theoretical predictions that however carry a larger uncertainty.  
Figure~\ref{fig:VHF:Zc} shows $\Z + c$ cross section measurements at 8 TeV and $c/b$ cross section ratio measurements at 13 TeV, 
both as a function of the heavy-flavor jet \pT. 

Measurements of the production of a \W boson and charm quarks
are carried out determining the charge sign of the \W boson and the charm quark, and separating events with same-sign (SS) and opposite-sign (OS) charges.  
Contributions from $\W+$c  processes 
are inferred after performing a subtraction of SS events 
from OS ones, that effectively removes most background sources. 

Initial $\W + c$ measurements were performed by \cdf~\cite{Aaltonen:2007dm} 
and subsequently reached an overall 20\% precision~\cite{Aaltonen:2012wn} with results compatible with existing theoretical expectations. A 15\% constraint on the $ |V_{cs}|$  CKM quark mixing matrix element was also derived from these results.

\begin{figure*}[htb] 
\centering
\includegraphics[width=0.49\textwidth,height=0.45\textwidth]{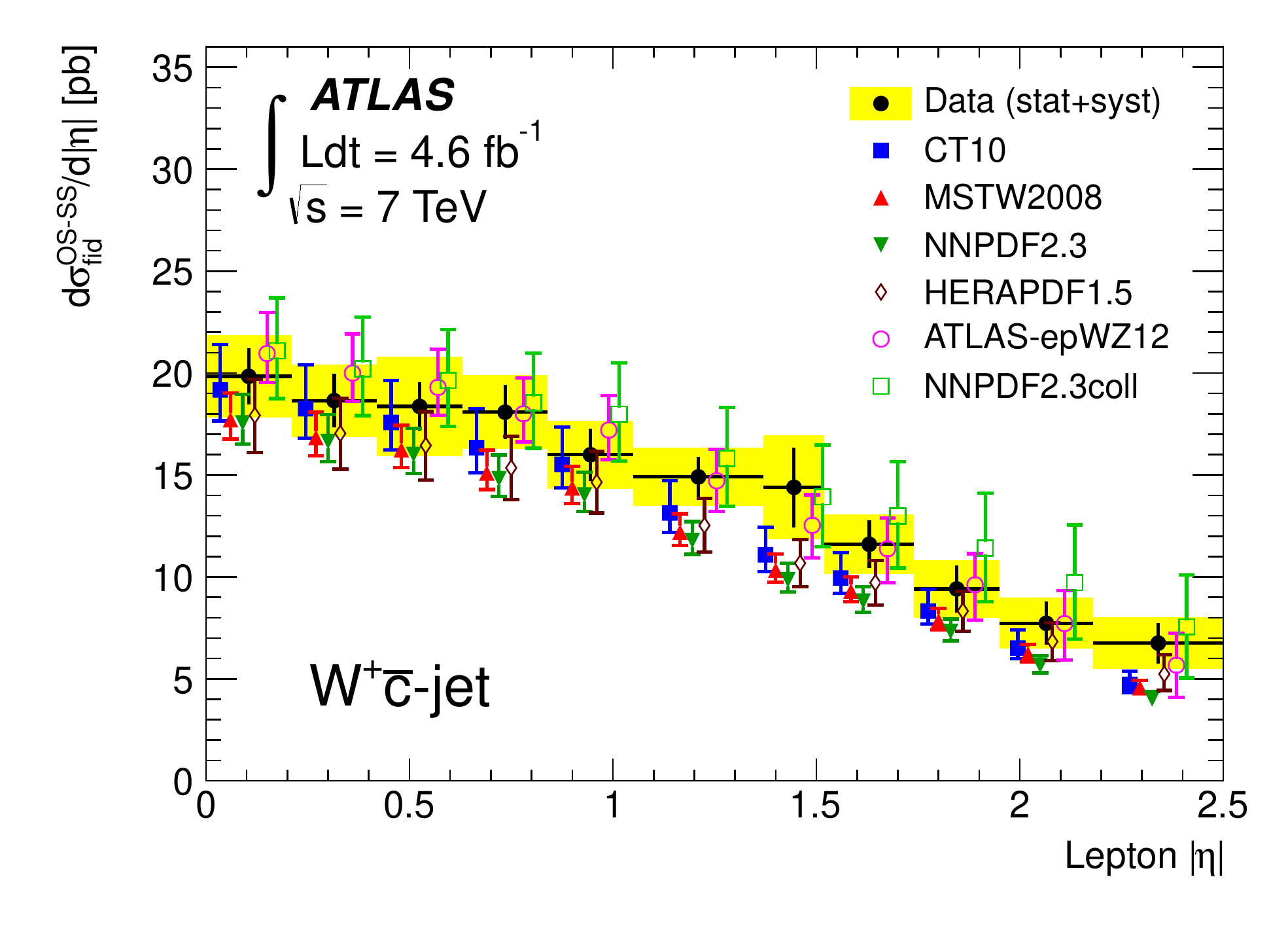}
\includegraphics[width=0.45\textwidth]{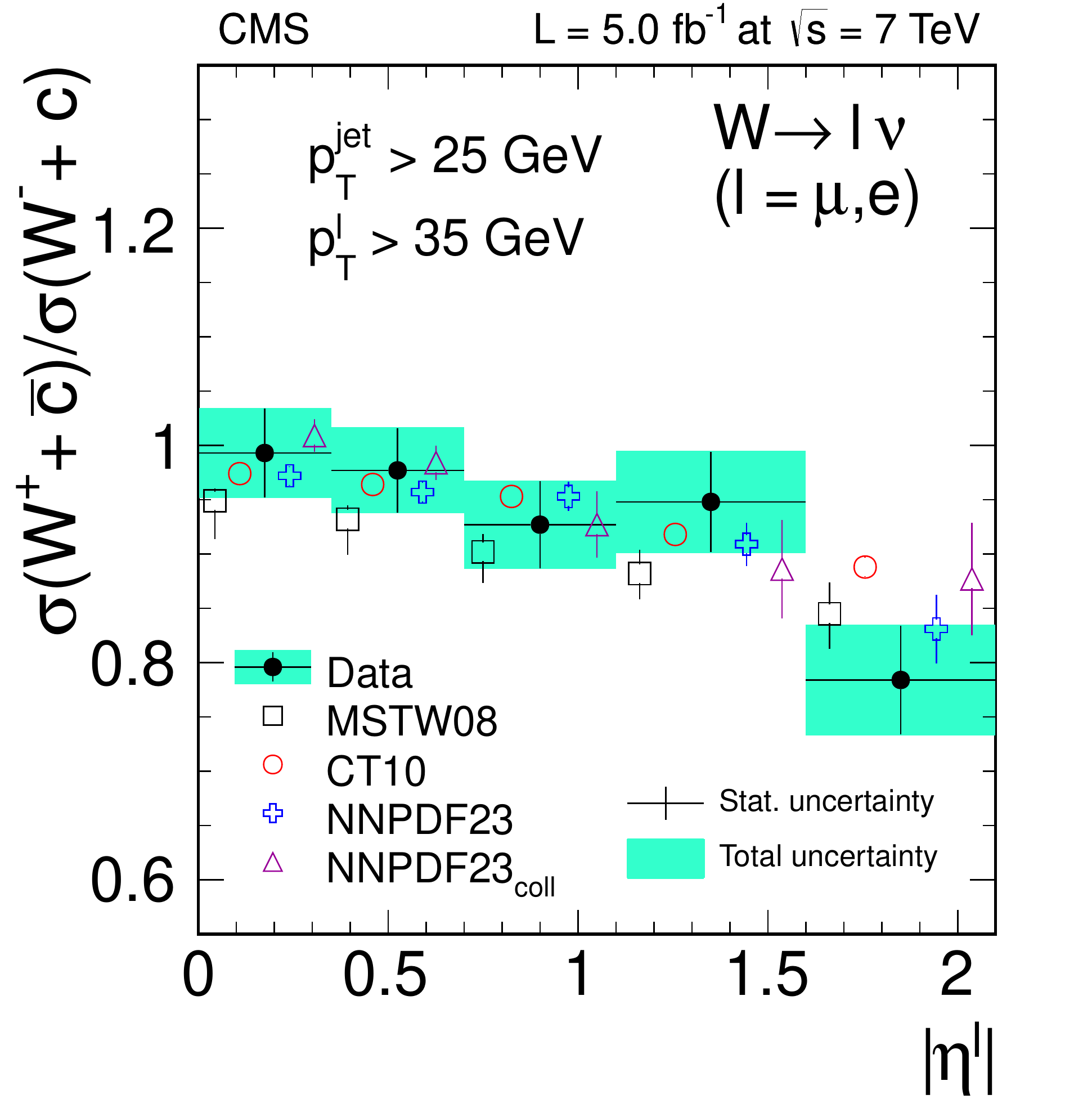}
\caption{ Measurements of $W + c$ production in proton collisions at 7 TeV.  The measured differential cross section (left), figure taken from~\cite{Aad:2014xca}, and the cross section ratio 
$\sigma(W^+ + c)/\sigma(W^- + c)$ as functions of the charged lepton $|\eta|$ (right) are compared to predictions obtained using various PDF sets, figure taken from~\cite{Chatrchyan:2013uja}. 
\label{fig:VHF:Wc}}
\end{figure*}

The first measurements of $\W + c$ productions at the \LHC were performed 
with 7 TeV proton collision data by  \CMS~\cite{Chatrchyan:2013uja},
where hadronic and inclusive semileptonic decays of charm hadrons are used
to select the presence of $c$-jets.
Cross sections and cross section ratios were measured inclusively to 
precisions of 3--7\%, and differentially with respect to the absolute value of the pseudorapidity of the charged lepton from the $W$-boson decay, shown in Fig~\ref{fig:VHF:Wc}. 
Results are directly sensitive to the strange quark content of the proton,  and are consistent with the predictions based on global fits of parton distribution functions (see Sec.~\ref{sec:QCD_interpretation}).

The \ATLAS collaboration  performed measurements of  $W + c$ production at 7 TeV where 
the charm quark is similarly tagged either by a semileptonic decay  or by the presence of a charmed meson~\cite{Aad:2014xca}. Results where also found in good agreement with theoretical predictions for the cross sections with different choices of the PDF set, with a preference for PDFs with an SU(3)-symmetric light-quark sea, as discussed in Sec.~\ref{sec:QCD_interpretation}.

\begin{figure}[htb] 
\includegraphics[width=0.49\textwidth]{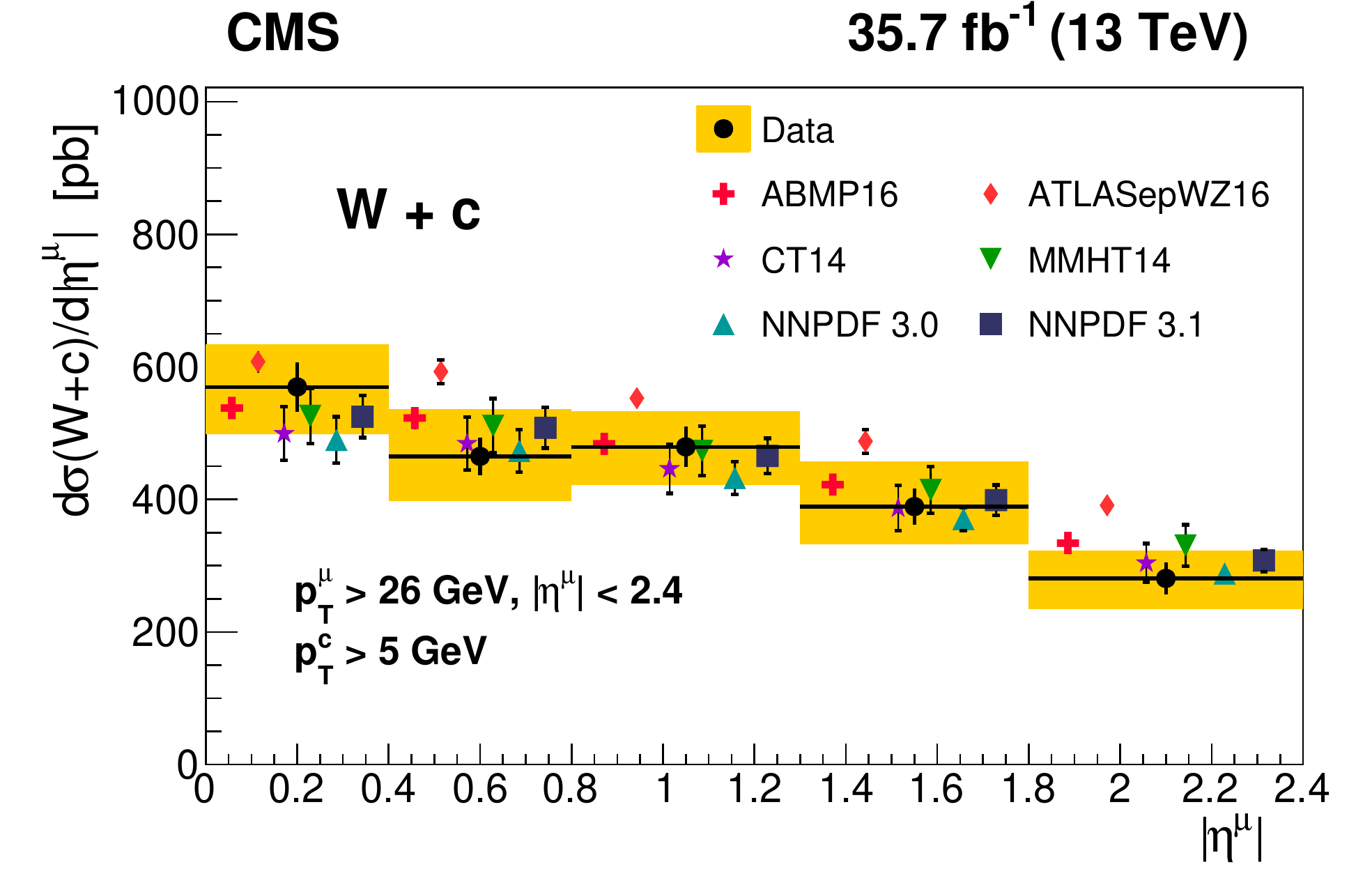}
\caption{
Differential cross sections of $W + c$
 production, measured as a function of the charged lepton 
 pseudorapidity at 13 TeV 
 The measurements are compared to the QCD predictions at NLO using different PDF sets. 
 Figure taken from~\cite{Sirunyan:2018hde}.
\label{fig:VHF:Wc2}}
\end{figure}

The \CMS collaboration has also produced $W + c$ measurements with 13 TeV collision data
using only charm quarks tagged via the full reconstruction of
$D$-mesons~\cite{Sirunyan:2018hde}.
Figure~\ref{fig:VHF:Wc2} shows differential 
$W + c$ cross sections measured at 13 TeV by \CMS. 

Measurements of $W$  boson productions in association with $b$- and $c$-quarks 
have also been carried out in the forward regions of proton collisions at 7  and 8 TeV by \LHCb ~\cite{Aaij:2015cha}. A dedicated secondary vertex tagger is used to identify and separate the presence of heavy-flavor jets. 
Results are generally in agreement with QCD predictions and 
do not  support a large contribution from intrinsic $b$-quark content in the  proton but the precision is not sufficient to rule out such a  contribution at O(10\%).


\section{Theoretical interpretations of data from vector boson production with associated jets}
\label{sec:QCD_interpretation}
 
The production of high statistics \Vjet Monte Carlo samples is key to much of the \Tevatron and the \LHC physics programs. The validation of MC event generators available with different levels of approximations in pQCD and different non-perturbative QCD model implementations require careful comparisons between predictions and with data. Several investigations into the tuning of related MC parameters and evaluations of uncertainties on their predictions have been carried out by both communities of experimentalists and theorists based on \Vjet processes. Examples of validation and tuning studies of \Vjet MC event generators are included in Refs.~\cite{ATL-PHYS-PUB-2016-001,ATL-PHYS-PUB-2016-003,Aad2014ZpT,Aad:2016ria,ATL-PHYS-PUB-2017-006,Cooper2012}.

\begin{figure}
\includegraphics[width=0.45\textwidth]{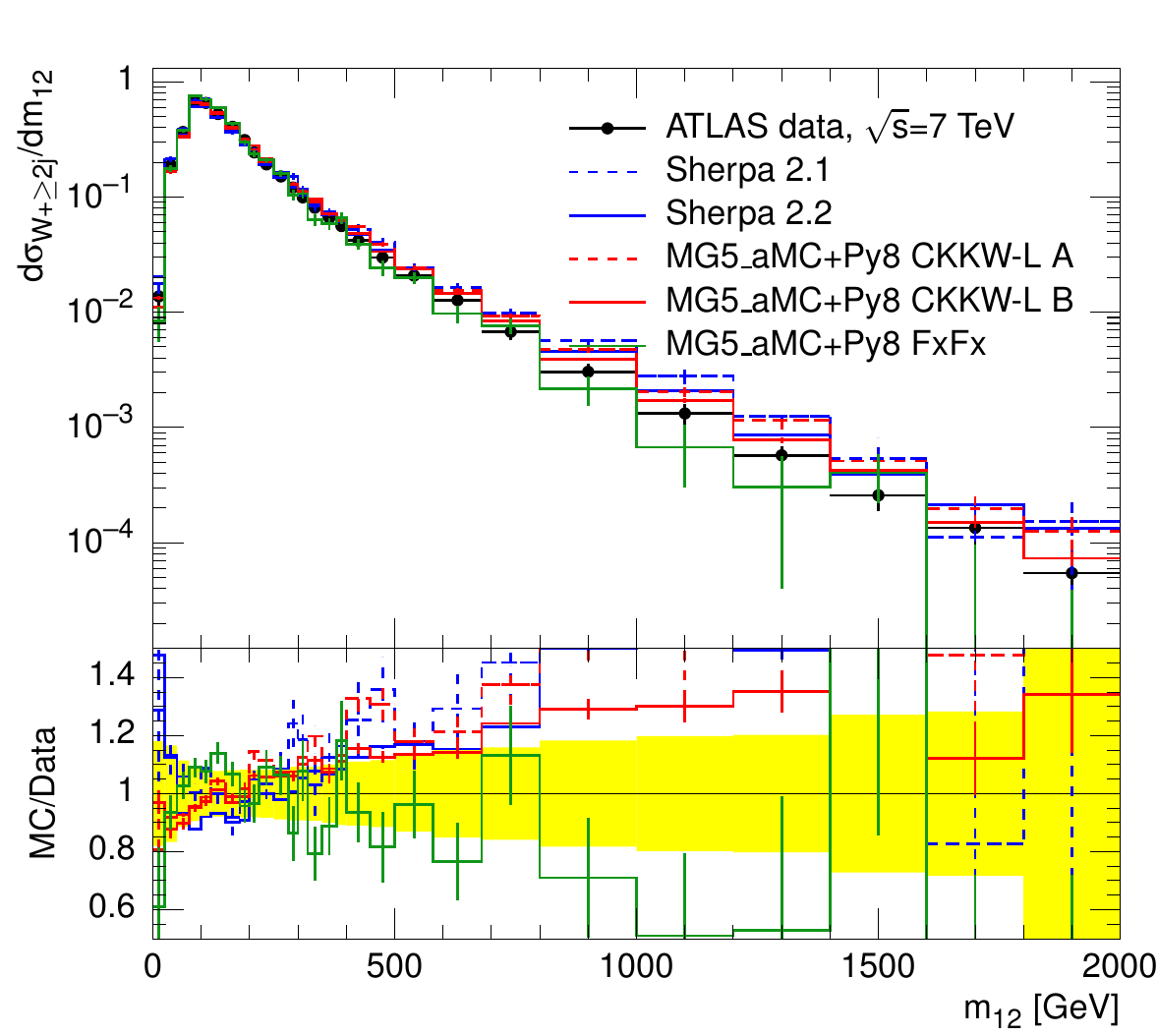}
\caption{\label{fig:ATLAS_MCtunes} Comparison between different MC generator models and data for events with a $W$ and at least two jets, in the distribution of the invariant mass of the two leading jets, at 7 TeV \pp collisions, figure taken from~\cite{ATL-PHYS-PUB-2016-003}. 
}
\end{figure}

An example is the comparison of the MC generators used by \LHC experiments in Run-2 analyses with \Vjet measurements performed at 7 TeV center-of mass energy in Run-1~\cite{ATL-PHYS-PUB-2016-003}. The same generators were then used to simulate events at the Run-2 center-of-mass energy of 13 TeV to investigate further the differences between the predictions and assess the theoretical uncertainties. Given the increase in cross section for the \WZjet production processes 
in Run-2 with respect to Run-1, it was important to carefully assess the accuracy of the MC generators in the new kinematic regime.
Predicted differential cross sections were compared to unfolded distributions in data using the \textsc{Rivet} package~\cite{BUCKLEY20132803,Bierlich:2019rhm}. 
Although an overall good description of the data is provided by all considered generators, as Fig.~\ref{fig:ATLAS_MCtunes} shows, some clear differences between prediction and data at 7 TeV were visible in some observables.
This type of studies has prompted a new tuning of model parameters~\cite{Aad2014ZpT} and improvements in the calculations. Comparisons were also made in phase space regions and for processes that became more relevant for Run-2 analyses due to the larger statistics, e.g.\ the production of vector bosons in association with heavy-flavor jets and for electroweak $V+2$ jets  production~\cite{ATL-PHYS-PUB-2017-006}. 
As a result of these studies the uncertainties associated to the normalization and shapes of the predictions of the MC generators for \Vjet processes have been routinely assessed on different MC event generators, including variations of matching and merging schemes, parton shower realisations as well as fragmentation and underlying event models, the strong coupling constant and PDFs.

\begin{figure}
\centering
\includegraphics[width=0.45\textwidth]{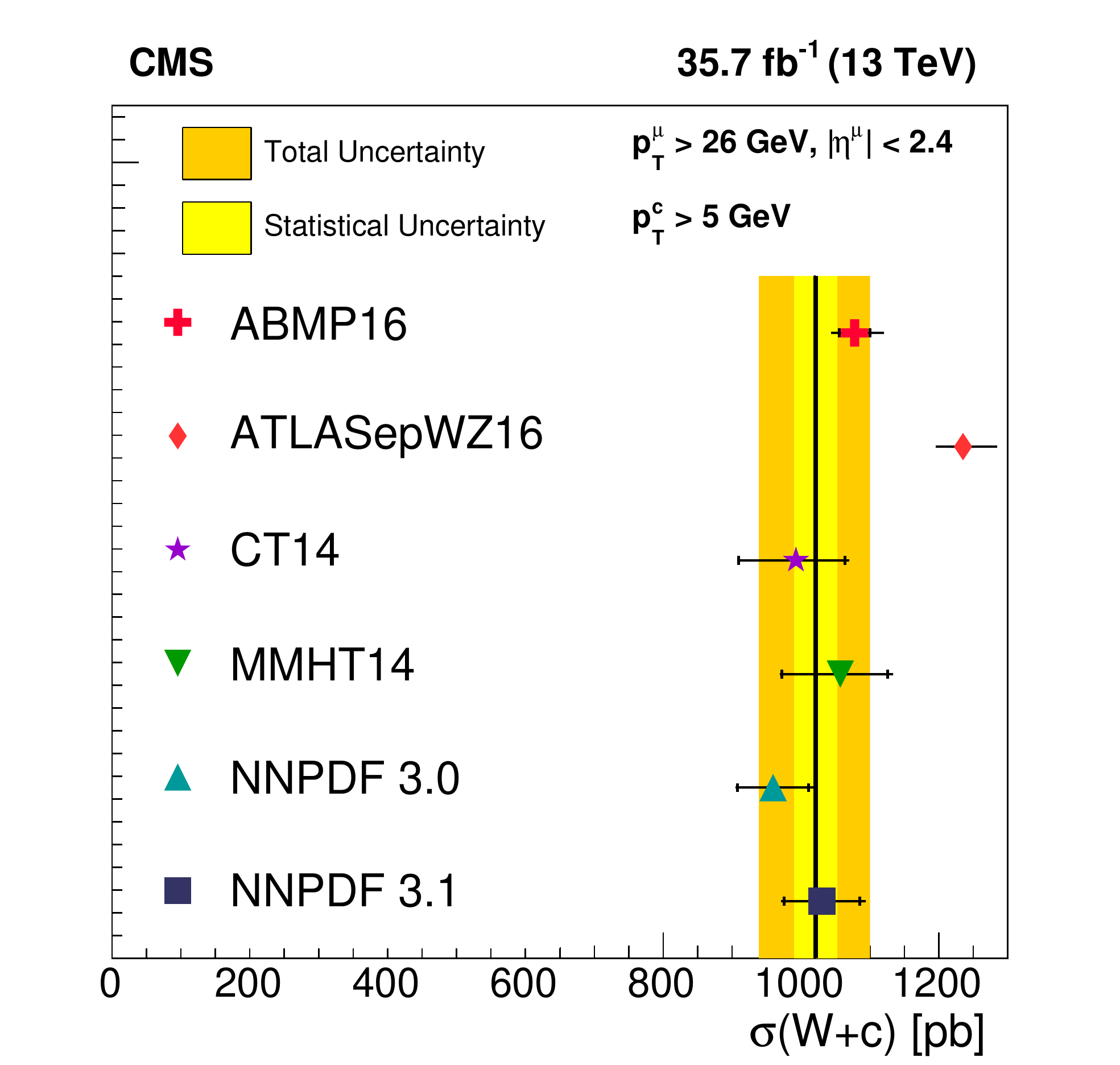}
\caption{\label{fig:CMS-Wc_PDF} The fiducial cross section for $W+c$ production at 13 TeV with the \CMS detector at the \LHC. The measurement is compared to predictions using several PDF sets. With the exception of ATLASepWZ16, which is obtained at NNLO in pQCD, all other PDF sets are obtained at NLO. Figure taken from~\cite{Sirunyan:2018hde}. 
}
\end{figure}

The understanding of the proton PDF, and specifically the flavor composition of the quark sea is important for the \LHC physics program as a whole. For instance the strange quark PDF has a direct impact on the measurement of the $W$ boson mass. In addition to measurements of charm  production  in  deep inelastic scattering experiments with neutrinos, the strange  quark  content  of  the  nucleon  can be  obtained  from the measurements of inclusive differential $W$ and $Z$ boson cross sections, $W ~+$ charm production as well as  \WZjets.
Inclusive differential $W$ and $Z$ boson cross sections at $\sqrt{s} = 7$ TeV
\cite{PhysRevD.85.072004,PhysRevLett.109.111806,PhysRevD.90.032004,Aaboud:2016btc}
allowed the strange content of the sea to be measured rather than assumed to be a fixed fraction of the light sea quarks. 
A QCD interpretation of inclusive $W$ and $Z$ boson production data by the ATLAS Collaboration together with data from deep-inelastic scattering at HERA presented in Ref~\cite{Aad:2012sb} shows the sensitivity to the light-quark sea composition of the proton at the LHC. The ratio of the strange-to-down sea quark distributions was determined to be consistent with one at momentum transfer squared $Q^2 = 1.9$ GeV${}^2$ and Bjorken $x = 0.023$, therefore supporting a symmetric composition of the light-quark sea at low $x$.
The \CMS $W + c$ measurement at $\sqrt{s}=7$ TeV~\cite{Chatrchyan:2013uja} identified processes where a $c$ quark is produced in association with a $W$ boson and was used for the determination of the strange-quark distribution in the proton. This analysis was followed recently by a measurement at 13 TeV center-of-mass energy~\cite{Sirunyan:2018hde}. The \CMS results point towards a strangeness suppression with respect to light sea-quark densities in agreement with measurements in neutrino scattering experiments. These results are hence in tension with \ATLAS studies based on the analysis of inclusive $W$ and $Z$ boson production~\cite{Aaboud:2016btc} and $W ~+$ charm production at $\sqrt{s} = 7$ TeV~\cite{Aad:2014xca}, which are found to be consistently and significantly better described by an unsuppressed strange sea at low-$x$ values. 
Figure~\ref{fig:CMS-Wc_PDF} shows the cross section measured by the \CMS experiment for the production of $W + c$, compared to various PDF fits, including the one by the \ATLAS collaboration (ATLASepWZ16) that makes use of inclusive $W$ and $Z$ cross section data as an input. 

The cross section ratio $R^{\pm}_c = \sigma(W^+ +\bar{c})/\sigma(W^−+c)$ can be sensitive to the $s$-$\bar{s}$ asymmetry in the PDFs that was suggested by neutrino data~\cite{Goncharov:2001qe}.
The results by \CMS (see Fig.~\ref{fig:VHF:Wc}) and \ATLAS (see Fig. 12 in Ref.~\cite{Aad:2014xca}) are compatible, within one standard deviation, with predictions obtained using PDF parameterisations with no asymmetry or a small asymmetry in the order of few percent. Those \LHC measurements of the $R^{\pm}_c$ ratio at 7 TeV were limited by statistical uncertainties. Given the far larger data set collected by \LHC experiments at higher center-of-mass energies, it will be very interesting to see such measurements repeated with greater precision.

\begin{figure}
\hspace*{-0.05\textwidth}
\includegraphics[width=0.52\textwidth]{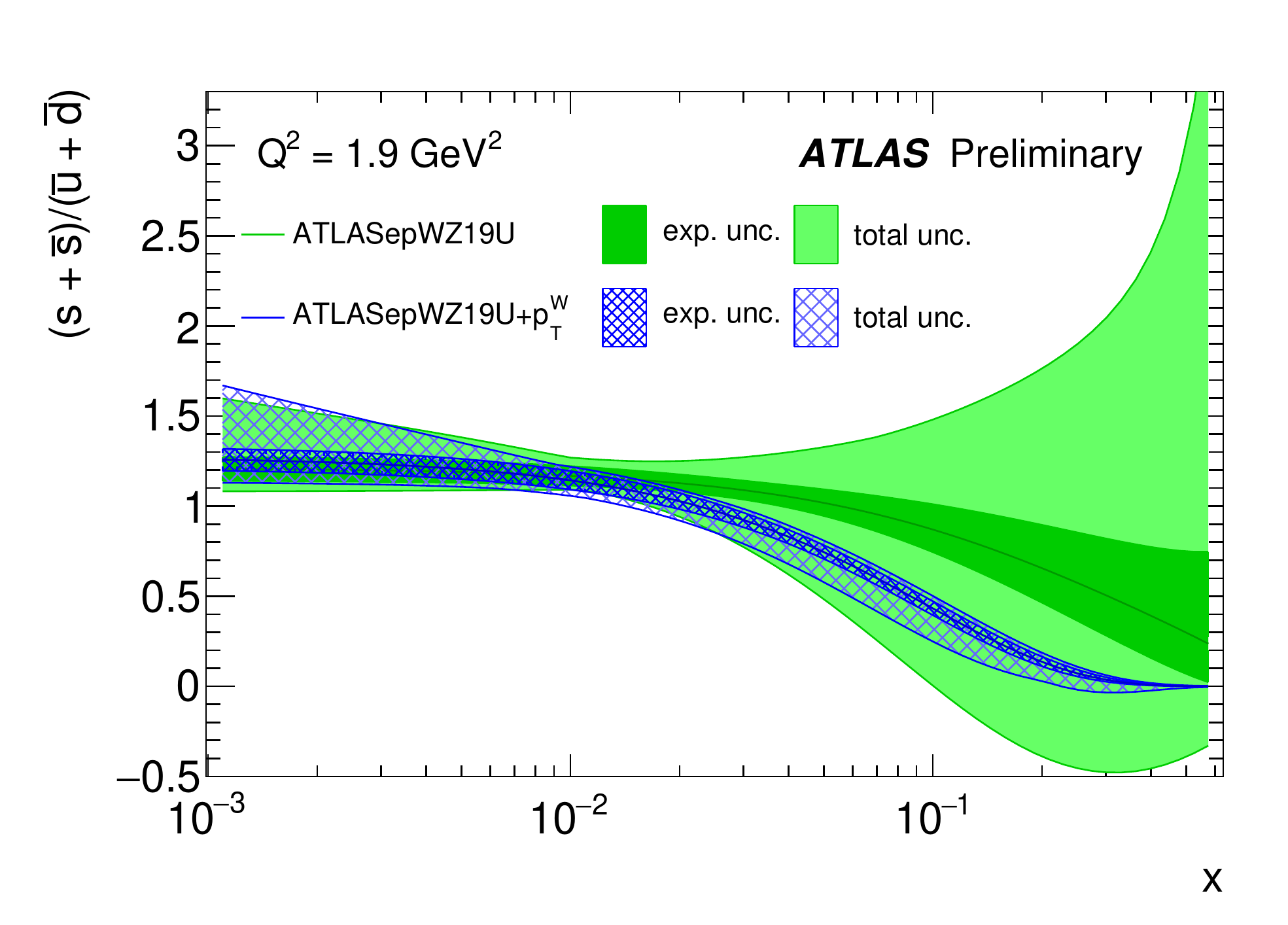}
\hspace*{-0.05\textwidth}
\includegraphics[width=0.52\textwidth]{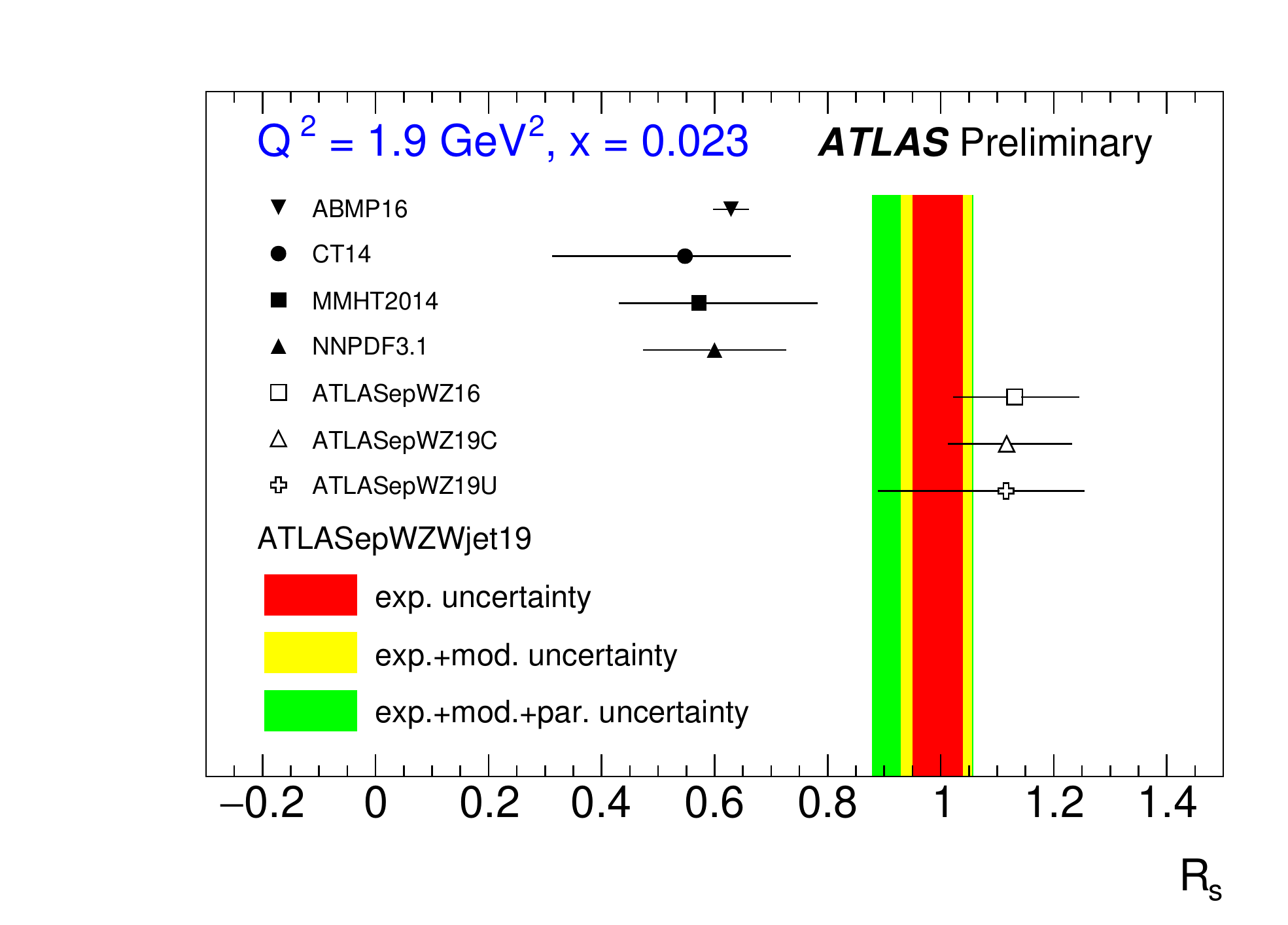}
\caption{\label{fig:ATLAS_sPDF} The $R_s$ distribution, evaluated at $Q^2$ = 1.9 GeV$^2$, as determined from a fit that includes \Wjet data as a function of W \pT, in comparison to a similar fit without \Wjet data as input (top). The $R_s$ ratio evaluated at $x=0.023$ and $Q^2=1.9$ GeV$^2$, for the \ATLAS PDF set that includes the \Wjet data as input, in comparison to other PDF sets (bottom). Figures taken from~\cite{ATL-PHYS-PUB-2019-016}. 
}
\end{figure}

As discussed in Sec.~\ref{sec:VLF:exp:ratios} the study of $W$ or $Z$ boson production with jets allows to access the high-$x$ region of the parton phase space that is of great importance for PDF fitting, as it is to-date poorly constrained by data and subject to non-perturbative effects with large uncertainties from phenomenological models.
The \Tevatron $W^{\pm}$ asymmetry data is not subject to such uncertainties, however the results from \CDF and \DO experiments are in tension~\cite{PhysRevLett.102.181801,D0:2014kma}. The production of \WZjets at the \LHC provides a new and independent data set that can be used as input to PDF fits to access partons at high $x$.
Given the \ATLAS and \CMS tension on the strange-quark PDF, it is therefore of particular interest to check the impact of the new \Wjet data on the strange-quark density.
Figure~\ref{fig:ATLAS_sPDF} shows the results of the  PDF analysis of the \ATLAS measurement of the W boson \pT spectrum~\cite{Aaboud:2017soa} in \Wjet events at a center-of-mass energy of 8 TeV, as  shown in Fig.~\ref{fig:ATLAS_Wjets_8TeV_chargeratio}, fitted together with \ATLAS inclusive $W$ and $Z$ production measurements at 7 TeV and HERA deep-inelastic scattering data. The PDF fit is performed at NNLO in pQCD, and was made possible by recent theoretical developments providing NNLO predictions for such processes~\cite{Boughezal:2015dva,Ridder:2015dxa}.
The fraction of the strange-quark density in the proton can be defined by the quantity $R_s=(s+\bar{s})/(\bar{u}+\bar{d})$, that is shown in Fig.~\ref{fig:ATLAS_sPDF}. The effect of the \Wjet data is most significant in the kinematic region $x > 0.02$, where the uncertainty is significantly lower and the fit results in the $R_s$ distribution that falls from about 1 at $x\approx 0.01$ to about 0.3 at $x \approx 0.1$.
At low $x$, i.e. $x < 0.023$, the fit with the \Wjet data is compatible with the unsuppressed strange-quark density found in previous \ATLAS analyses with different data sets.

Further measurements at the \LHC with even greater precision, 
including \WZjets, aided by the completion of the NNLO $W\!+c$ calculation, 
will help to understand this apparent tension 
between different experimental data sets in the determination 
of the strange-quark content of the proton.

\section{Conclusions and outlook}
\label{sec:conclusions}

This article has reviewed achievements in the understanding of the production of vector bosons in association with light- or heavy-flavor jets, with a focus on the \LHC results. 
These processes are of great importance for the success of physics programs at hadron colliders since they are major backgrounds to new physics searches and are ideal testing grounds for new calculations and models in the QCD and electroweak sectors of the Standard Model. 
This review has summarized theoretical techniques developed to describe experimental results in \ppbar collisions at the \Tevatron and in \pp collisions at the \LHC, and has highlighted a few of the several measurements that were carried out by the \CDF and \DO collaborations at the \Tevatron as well as the \ATLAS, \CMS and \LHCb collaborations at the \LHC at different center-or-mass energies. Detailed comparisons between the experimental results and cutting-edge predictions have been presented together with discussions of differences in phase spaces and production mechanisms between the \Tevatron and the \LHC. 

The modeling of \Vjet processes improved significantly at the \Tevatron, and measurements of such processes have prompted the development of high-order QCD calculations and new techniques for their modeling in MC generators. \Tevatron data sets were used to tune theoretical predictions and MC generators, and those tunes turned out to be also accurate at describing the first \Vjet data results at the \LHC Run-1, despite the large difference in center-of-mass energies. \Tevatron data still provides an important legacy for \Vjet analyses at the \LHC. The measurements carried out by the \LHC experimental collaborations, thanks to the greater statistics in \Vjet samples, have urged further developments in the theoretical description of such processes both in the QCD and the electroweak sectors. These processes have been at the center of both the so-called next-to-leading and next-to-next-to-leading order revolutions in perturbative QCD calculations and their implementations in MC generators. 

In several kinematic regions the experimental uncertainties are significantly smaller than the uncertainties in the predictions. Such a high experimental precision has allowed to test and constrain theoretical predictions and models, including parton density functions, in a broad kinematic region, including extreme regions of phase space, that are relevant for new physics searches.

Despite the great theoretical progress in the past decades and the many years in understanding of the \Vjet production mechanism there are still theory uncertainties that can be further reduced in future developments, for example higher order QCD and electroweak contributions in hard scattering matrix elements, parton-showers and their matching algorithms, or be better constrained by data like PDFs and underlying event modeling. Improvements in the understanding of these sources of uncertainties in \Vjet processes are critical for improvements in the precision of measurements and in the reach of searches at the \LHC and at future collider experiments.

\paragraph{Outlook}
\mbox{}\newline

Studies of \Vjet physics will necessarily continue in future \LHC runs and at possible future colliders, as the success of the physics programs of such experiments will critically rely on the good understanding of such processes. 
The higher and higher expected statistics, precision and extended phase space, e.g., to higher jet multiplicities or higher energy scales, will challenge theoretical predictions to perform calculations at higher orders in QCD, i.e., at ${\rm N^{3} LO}$,\footnote{
  A first result for lepton pair production via virtual photon exchange has been 
  presented in \cite{Duhr:2020seh,Duhr:2020sdp}.
} to systematically include higher-order electroweak corrections in MC generators together with mixed electroweak-QCD terms and to improve MC generators for more accurate estimation of the various sources of uncertainties in the modeling.

Studies of the QCD production of \Vjets will remain critical for the understanding of QCD dynamics as well as for a better understanding of electroweak corrections, as they will become more significant at higher energy scales. Experimental analyses are expected to become more sophisticated in studying statistical and systematic correlations between differential cross sections in the same and in different \Vjet processes, such that several observables can be used simultaneously as inputs to PDF global fits, MC tunes, and indirect searches for new physics.
With more accurate and precise theoretical predictions new measurements will become interesting, for example the extraction of the strong coupling $\alpha_s$ from jet rates in \Vjet events.\footnote{
  A precursor to this has been presented in \cite{Johnson:2017ttl}.
}

It is  expected that electroweak analyses will become predominant in the future to search for anomalies in the gauge structure of the Standard Model. 
A good understanding of vector boson fusion production of \Vjets is important for studies of Higgs production,
and for a thorough investigation of anomalies in the gauge couplings a comprehensive and simultaneous analysis of several electroweak processes will be beneficial. An example of an electroweak process that has not been investigated due to experimental challenges is the eletroweak production of two jets in association with a photon. This process will provide a new window in the studies of anomalous couplings. Developments in the separation of jets induced by quarks or gluons may have a significant impact in the discrimination of electroweak \Vjet processes from QCD-induced background in future analyses, and may impact the constraints on quark and gluon PDFs.  At higher center-of-mass energies the study of the emission of massive vector bosons collinearly with jets in \WZjets will become more important and the development and testing of electroweak showering models more relevant than it has been so far. 

As experimental results become more precise, preservation and sharing of analysis details are more important.
Future studies are expected to include multiple differential cross section measurements and such a large amount of experimental results will be a wealth for the understanding of the SM and beyond
SM physics. The analysis algorithms as well as the experimental results must be preserved for current and future generations.
The \LHC experimental data with their uncertainties are stored in the \HepData repository~\cite{Maguire:2017ypu}, which has become an essential tool for archiving detailed experimental results, for comparisons against MC predictions, and for tuning and constraining of theoretical models, e.g., underlying event and PDFs.
In recent years the \Rivet project (Robust Independent Validation of Experiment and Theory)~\cite{BUCKLEY20132803,Bierlich:2019rhm} has been asserted as \emph{the} repository for analysis algorithms (a combination of fiducial phase space definitions and physics objects). This project was originally intended as a toolkit for validation of MC event generators, however, thanks to its large (and ever growing) set of experimental analyses and its link to the \HepData repository for experimental data points from the collider experiments, it has become very useful also as a long term repository of analysis algorithms. \Rivet also provides useful algorithms to extract observable quantities from different MC generators in a model independent way, i.e., without prior knowledge of specific algorithm implementations or specific event record definitions. Such repositories and analysis tools will become more critical in multi-process fits, e.g., PDF, EFT, including for storing information about correlations between measurements.

\noindent
\section*{Acknowledgments}

The work of Alessandro Tricoli is supported by the U.S.A.\ Department of 
Energy under grant contact DE-SC0012704.
Marek Sch{\"o}nherr is funded by the Royal Society through 
a University Research Fellowship and acknowledges support 
from the European Unions Horizon 2020 research and innovation 
programme as part of the Marie Sklodowska-Curie Innovative 
Training Network MCnetITN3 (grant agreement no. 722104).


\bibliography{references}

\end{document}